\documentclass[a4paper, 11pt, oneside]{book}

\usepackage{amssymb}
\usepackage{amsmath}
\usepackage{amsthm}
\usepackage{hyperref}
\usepackage{graphicx}
\usepackage{amstext}
\usepackage{float}
\usepackage{protext}
\usepackage{url}
\usepackage{makeidx}
\usepackage{array}
\usepackage[english,ngerman]{babel}

\floatstyle{ruled}
\newfloat{protocolfloat}{tbp}{lop}[chapter]
\floatname{protocolfloat}{{Fig.}}
\newfloat{tabularfloat}{tbp}{lop}[chapter]
\floatname{tabularfloat}{Table}
\newfloat{picturefloat}{tbp}{lop}[chapter]
\floatname{picturefloat}{Plot}

\makeindex

\DeclareMathOperator*{\rank}{rank}

\DeclareMathOperator*{\im}{supp}
\DeclareMathOperator*{\spanv}{span}

\DeclareMathOperator*{\supp}{supp}

\theoremstyle{plain}
\newtheorem{theorem}{Theorem}[section]
\newtheorem{lemma}[theorem]{Lemma}
\newtheorem{corollary}[theorem]{Corollary}
\newtheorem{proposition}[theorem]{Proposition}
\theoremstyle{definition}
\newtheorem{definition}[theorem]{Definition}
\newtheorem{remark}[theorem]{Remark}

\newcommand*{\textfrac}[2]{{\textstyle \frac{#1}{#2}}}

\newcommand*{\eps}{\varepsilon}

\newcommand*{\epst}{\tilde{\eps}}

\newcommand*{\lambdat}{\tilde{\lambda}}

\newcommand*{\bz}{\mathbf{z}}

\newcommand*{\bzp}{\mathbf{z'}}
\newcommand*{\bzpp}{\mathbf{z''}}

\newcommand*{\cB}{\mathcal{B}}
\newcommand*{\cC}{\mathcal{C}}
\newcommand*{\cD}{\mathcal{D}}
\newcommand*{\cE}{\mathcal{E}}
\newcommand*{\cF}{\mathcal{F}}
\newcommand*{\cH}{\mathcal{H}}

\newcommand*{\cM}{\mathcal{M}}

\newcommand*{\cP}{\mathcal{P}}
\newcommand*{\cQ}{\mathcal{Q}}

\newcommand*{\cS}{\mathcal{S}}

\newcommand*{\cU}{\mathcal{U}}
\newcommand*{\cV}{\mathcal{V}}
\newcommand*{\cW}{\mathcal{W}}
\newcommand*{\cX}{\mathcal{X}}
\newcommand*{\cY}{\mathcal{Y}}
\newcommand*{\cZ}{\mathcal{Z}}

\newcommand*{\bbC}{\mathbb{C}}
\newcommand*{\bbN}{\mathbb{N}}
\newcommand*{\bbR}{\mathbb{R}}

\newcommand*{\Thetab}{\bar{\Theta}}

\newcommand*{\rhob}{\bar{\rho}}
\newcommand*{\rhot}{\tilde{\rho}}
\newcommand*{\sigmat}{\tilde{\sigma}}
\newcommand*{\Psih}{\hat{\Psi}}

\newcommand*{\psib}{\bar{\psi}}

\newcommand*{\sfrac}[2]{{\frac{#1}{#2}}}

\newcommand*{\slfrac}[2]{#1/#2}

\newcommand*{\hb}{\bar{h}}

\DeclareMathOperator*{\ExpE}{\mathbb{E}}

\newcommand*{\ket}[1]{| #1 \rangle}
\newcommand*{\bra}[1]{\langle #1 |}
\newcommand*{\spr}[2]{\langle #1 | #2 \rangle}
\newcommand*{\tr}{\mathrm{tr}}
\newcommand*{\proj}[1]{\ket{#1}\bra{#1}}

\newcommand*{\id}{\mathrm{id}}

\newcommand*{\freq}[1]{\lambda_{#1}}

\newcommand*{\Qb}{\bar{Q}}

\newcommand*{\bs}{\mathbf{s}}
\newcommand*{\bu}{\mathbf{u}}
\newcommand*{\bv}{\mathbf{v}}
\newcommand*{\bw}{\mathbf{w}}

\newcommand*{\bx}{\mathbf{x}}

\newcommand*{\by}{\mathbf{y}}

\newcommand*{\bc}{\mathbf{c}}
\newcommand*{\bC}{\mathbf{C}}
\newcommand*{\buh}{\mathbf{\hat{u}}}
\newcommand*{\bxh}{\mathbf{\hat{x}}}

\newcommand*{\Eb}{\bar{E}}
\newcommand*{\pb}{\bar{p}}
\newcommand*{\qb}{\bar{q}}
\newcommand*{\xb}{\bar{x}}
\newcommand*{\yb}{\bar{y}}
\newcommand*{\zb}{\bar{z}}
\newcommand*{\Pb}{\bar{P}}

\newcommand*{\Xb}{\bar{X}}
\newcommand*{\Yb}{\bar{Y}}
\newcommand*{\Zb}{\bar{Z}}
\newcommand*{\cXb}{\bar{\cX}}
\newcommand*{\cZb}{\bar{\cZ}}

\newcommand*{\Psit}{\tilde{\Psi}}

\newcommand*{\Hmin}{H_{\min}}
\newcommand*{\Hmax}{H_{\max}}

\newcommand*{\Upi}{\pi}

\newcommand*{\tabprotsep}{\vspace{3.5ex}}

\newcommand*{\rate}{\mathrm{rate}}

\newcommand*{\leak}{\mathrm{leak}}

\newcommand*{\distPE}{\mu}

\newcommand*{\distance}{$L_1$-distance}
\newcommand*{\Distance}{$L_1$-distance}

\begin{document}

\selectlanguage{english}

\begin{titlepage}
  \large\vspace*{-5ex}

  Diss.\ ETH No.\ 16242

  \begin{center}

    \vspace{8ex}

    {\Large\bf
      \rule[-1ex]{0pt}{3ex} Security of\\
      \rule[-1ex]{0pt}{3ex} Quantum Key Distribution\\
    }

    \vspace{7ex}

    A dissertation submitted to \\[3ex]%

    { SWISS FEDERAL INSTITUTE OF TECHNOLOGY \\
     ZURICH}\\[3ex]

    for the degree of\\
    Doctor of Natural Sciences

    \vspace{7ex}

    presented by\\[3ex]%

    {\bf Renato Renner\\
    Dipl.~Phys.~ETH}\\[3ex]%

    born December 11, 1974, in Lucerne\\
    citizen of Lucerne, LU, and Andermatt, UR, Switzerland

    \vspace{6ex}

    \mbox{} \\ accepted on the recommendation of\\[3ex]

    Prof.~Dr.~Ueli Maurer, examiner  \\
    Dr.~Charles H.\ Bennett, co-examiner\\

    \vspace{7ex}

    September 2005

    \vspace{4.8ex}


  \end{center}

\newpage

\thispagestyle{empty}

\mbox{}

\end{titlepage}
















\chapter*{Acknowledgments}

The research leading to this thesis was carried out under the
supervision of Ueli Maurer. I would like to thank him for his
encouragement and support. His lucid way of thinking and his
persistent quest for getting the ``right view'' had a major impact on
my research.  His door was always open to discuss both scientific and
non-scientific problems.

I also owe many thanks to Charles Bennett for investing his time in
studying my work and for being the co-examiner of this thesis.

During the last few years, I had a strong collaboration with Stefan
Wolf. It was always a pleasure working with him, both in Zurich and
during my visits to Montreal. He also introduced me to the research
group of Nicolas Gisin at the University of Geneva and, with that, to
the field of quantum cryptography.

I have undoubtedly benefited a lot from the stimulating discussions
with my collaborators.  In particular, I would like to thank Matthias
Christandl, Robert K\"onig, and Barbara Kraus for their valuable and
unique ideas. Moreover, I am grateful to all members of the
cryptography research group at ETH Zurich for their support (even
though they never got me to play tabletop football).  Particular
thanks go to my officemate Thomas Holenstein who gave me much valuable
advice on both research and TeX related questions.

I am also very grateful to Nicolas Gisin and the members of his
research group for many interesting and enjoyable discussions on
quantum cryptography. It was always a pleasure to visit them in
Geneva.


This thesis has improved substantially by the valuable comments of
those who read preliminary versions of it. In particular, I would like
to thank Robert K\"onig and Christopher Portmann for their helpful
suggestions and proofreading.  (The remaining errors and omissions are
entirely the author's responsibility.)

Finally and most importantly, I would like to thank my family for
their love and support. This thesis is dedicated to my wife Sophie and
my daughter Jill.

This research was partially supported by the Swiss National Science
Foundation, grant No. 2000-66716.01/1.

\chapter*{Abstract}

\emph{Quantum information theory} is an area of physics which studies
both fundamental and applied issues in quantum mechanics from an
information-theoretic viewpoint. The underlying techniques are,
however, often restricted to the analysis of systems which satisfy a
certain \emph{independence condition}. For example, it is assumed that
an experiment can be repeated independently many times or that a large
physical system consists of many virtually independent parts.
Unfortunately, such assumptions are not always justified.  This is
particularly the case for practical applications|e.g., in (quantum)
cryptography|where parts of a system might have an arbitrary and
unknown behavior.

We propose an approach which allows to study general physical systems
for which the above mentioned independence condition does not
necessarily hold.  It is based on an extension of various
information-theoretic notions. For example, we introduce new
uncertainty measures, called \emph{smooth min- and max-entropy}, which
are generalizations of the von Neumann entropy.  Furthermore, we
develop a quantum version of de Finetti's representation theorem, as
described below.

Consider a physical system consisting of $n$ parts. These might, for
instance, be the outcomes of $n$ runs of a physical experiment.
Moreover, assume that the joint state of this $n$-partite system can
be extended to an $(n+k)$-partite state which is symmetric under
permutations of its parts (for some $k \gg 1$).  The \emph{de Finetti
  representation theorem} then says that the original $n$-partite
state is, in a certain sense, close to a mixture of product states.
Independence thus follows (approximatively) from a symmetry condition.
This symmetry condition can easily be met in many natural situations.
For example, it holds for the joint state of $n$ parts which are
chosen at random from an arbitrary $(n+k)$-partite system.


As an application of these techniques, we prove the security of
\emph{quantum key distribution (QKD)}, i.e., secret key agreement by
communication over a quantum channel. In particular, we show that, in
order to analyze QKD protocols, it is generally sufficient to consider
so-called \emph{collective attacks}, where the adversary is restricted
to applying the same operation to each particle sent over the quantum
channel separately. The proof is generic and thus applies to known
protocols such as \emph{BB84} and \emph{B92} (where better bounds on
the secret-key rate and on the the maximum tolerated noise level of
the quantum channel are obtained) as well as to \emph{continuous
  variable} schemes (where no full security proof has been known).
Furthermore, the security holds with respect to a strong so-called
\emph{universally composable} definition.  This implies that the keys
generated by a QKD protocol can safely be used in any application,
e.g., for \emph{one-time pad} encryption|which, remarkably, is not the
case for most of the standard definitions.

\tableofcontents 



\newpage{}



\chapter{Introduction} \label{ch:intro}

\section{Motivation}

What is needed to establish a secret key between two spatially
separated parties?  Clearly, this question is of immediate interest
for practical cryptographic applications such as secure message
transmission.\footnote{For example, using one-time pad
  encryption~\cite{Vernam26}\index{one-time~pad~encryption}, the
  problem of secretly exchanging $\ell$ message bits reduces to the
  problem of distributing a secret key consisting of $\ell$ bits.}
More importantly, however, it is related to fundamental problems in
(classical and quantum) information theory.  Is information physical?
Is classical information distinct from quantum information?  In fact,
it turns out that the possibility of secret key agreement (over
insecure channels) strongly depends on the physical properties of
information and that there is indeed a fundamental difference between
classical and quantum information.

In this thesis, we address several basic question of quantum
information theory: What does secrecy mean in a quantum world?
(Chapter~\ref{ch:pre}) How can knowledge and uncertainty be
quantified?  (Chapter~\ref{ch:smooth}) What is the role of symmetry?
(Chapter~\ref{ch:sym}) Can any type of randomness be transformed into
uniform randomness?  (Chapter~\ref{ch:PA}) As we shall see, the
answers to these questions allow us to treat the problem of secret key
agreement in a very natural way
(Chapters~\ref{ch:QKD} and~\ref{ch:examples}).


\section{Quantum key distribution: general facts} \label{sec:QKD}

\subsubsection{Cryptographic setting}

We consider a setting where two distant parties, traditionally called
\emph{Alice}\index{Alice} and \emph{Bob}\index{Bob}, want to establish
a common \emph{secret key}\index{secret~key}, i.e., a string of random
bits which is unknown to an adversary, \emph{Eve}\index{Eve}.
Throughout this thesis, we focus on \emph{information-theoretic
  security}\index{information-theoretic~security}, which is actually
the strongest reasonable notion of security.\footnote{An example of a
  weaker level of security is \emph{computational
    security}\index{computational~security}, where one only requires
  that it is \emph{difficult} (i.e., time-consuming, but not
  impossible) for an adversary to compute information on the key.}  It
guarantees that an adversary does not get any information correlated
to the key, except with negligible probability.

For the following, we assume that Alice and Bob already have at hand
some means to exchange classical messages in an
\emph{authentic}\index{authentic~channel}
way.\footnote{\emph{Authentic} means that, upon receiving a message,
  Bob can verify whether the message was indeed sent by Alice, and
  vice-versa.} In fact, only relatively weak resources are needed to
turn a completely insecure communication channel into an authentic
channel.  For example, Alice and Bob might invoke an authentication
protocol (see, e.g., ~\cite{Stinso91,GemNao93}) for which they need a
short\footnote{The length of the key only grows logarithmically in the
  length of the message to be authenticated.}  initial key.  Actually,
as shown in~\cite{RenWol03c,RenWol04}, it is even sufficient for Alice
and Bob to start with only weakly correlated and partially secret
information (instead of a short secret key\index{secret~key}).

\subsubsection{Key agreement by quantum communication}

Under the sole assumption that Alice and Bob are connected by a
classical authentic communication channel, secret communication|and
thus also the generation of a secret key|is
impossible~\cite{Shanno49,Maurer93}\index{secret~key}. This changes
dramatically when quantum mechanics comes into the
game\index{quantum~mechanics}.  Bennett and Brassard \cite{BenBra84}
(see also \cite{Wiesner83}) were the first to propose a \emph{quantum
  key distribution
  (QKD)}\index{quantum~key~distribution}\index{QKD|see{quantum~key~distribution}}
scheme which uses communication over a (completely insecure) quantum
channel (in addition to the classical authentic
channel\index{authentic~channel}).  The scheme is commonly known as
the \emph{BB84 protocol}\index{BB84~protocol}.

Quantum key distribution is generally based on the impossibility to
observe a quantum mechanical system without changing its state. An
adversary trying to wiretap the quantum communication between Alice
and Bob would thus inevitably leave traces which can be detected.  A
quantum key distribution protocol thus achieves the following type of
security: As long as the adversary is passive, it generates an
(arbitrarily long) secret key.  On the other hand, if the adversary
tampers with the quantum channel, the protocol recognizes the attack
and aborts the computation of the key.\footnote{More precisely, it is
  guaranteed that the protocol does not abort as long as the adversary
  is passive (this is called \emph{robustness}\index{robustness}).
  Moreover, for any attack on the quantum channel, the probability
  that the protocol does not abort \emph{and} the adversary gets
  information on the generated key is negligible (see
  Section~\ref{sec:KAsec} for details).}  (Note that this is actually
the best one can hope for: As the quantum channel is completely
insecure, an adversary might always interrupt the quantum
communication between Alice and Bob, in which case it is impossible to
generate a secret key.)

\subsubsection{An example: the BB84 protocol}

\index{BB84~protocol|(}

To illustrate the main principle of quantum key distribution, let us
have a closer look at the \emph{BB84 protocol}.  It uses an encoding
of classical bits in \emph{qubits}\index{qubit}, i.e., two-level
quantum systems\footnote{For example, the classical bits might be
  encoded into the spin orientation of particles.}.  The encoding is
with respect to one of two different orthogonal bases, called the
\emph{rectilinear}\index{rectilinear~basis} and the \emph{diagonal
  basis}\index{diagonal~basis}.\footnote{See Section~\ref{sec:sixdesc}
  for a definition.}  These two bases are \emph{mutually
  unbiased}\index{mutually~unbiased~bases}, that is, a measurement in
one of the bases reveals no information on a bit encoded with respect
to the other basis.

In the first step of the protocol, Alice chooses $N$ random bits $X_1,
\ldots, X_N$, encodes each of these bits into qubits using at
random\footnote{In the original proposal of the BB84 protocol, Alice
  and Bob choose the two bases with equal probabilities.  However, as
  pointed out in~\cite{LoChAr05}, the efficiency of the protocol is
  increased if they select one of the two bases with probability
  almost one. In this case, the choices of Alice and Bob will coincide
  with high probability, which means that the number of bits to be
  discarded in the sifting\index{sifting} step is small.}  either the
rectilinear or the diagonal basis, and transmits them to Bob (using
the quantum channel).  Bob measures each of the qubits he receives
with respect to|a random choice of|either the rectilinear or the
diagonal basis to obtain classical bits $Y_i$. The pair of classical
bitstrings $X=(X_1, \ldots, X_N)$ and $Y=(Y_1, \ldots, Y_N)$ held by
Alice and
Bob after this step is called the \emph{raw key}\index{raw~key} pair.

The remaining part of the protocol is purely classical (in particular,
Alice and Bob only communicate classically).  First, Alice and Bob
apply a \emph{sifting}\index{sifting} step, where they announce their
choices of bases used for the encoding and the measurement,
respectively.  They discard all bits of their raw key for which the
encoding and measurement bases are not compatible.  Then Alice and Bob
proceed with a \emph{parameter estimation}\index{parameter~estimation}
step. They compare some (small) randomly chosen set of bits of their
raw key in order to get a guess for the \emph{error
  rate}\index{error~rate}, i.e., the fraction of positions $i$ in
which $X_i$ and $Y_i$ disagree.  If the error rate is too large|which
might indicate the presence of an adversary|Alice and Bob abort the
protocol.

Let $X'$ and $Y'$ be the remaining parts of the raw
keys\index{raw~key} (i.e., the bits of $X$ and $Y$ that have neither
been discarded in the sifting\index{sifting} step nor used for
parameter estimation).  These strings are now used for the actual
computation of the final key. In an \emph{information
  reconciliation}\index{information~reconciliation} step, Alice sends
certain error correcting information on $X'$ to Bob.\footnote{The
  information reconciliation step might also be interactive.}  This,
together with $Y'$, allows him to compute a guess for $X'$. (Note
that, because of the parameter estimation step, it is guaranteed that
$X'$ and $Y'$ only differ in a limited number of positions.) In the
final step of the protocol, called \emph{privacy
  amplification}\index{privacy~amplification}, Alice and Bob use
\emph{two-universal hashing}\index{two-universal~hashing}\footnote{See
  Section~\ref{sec:twohash} for a definition of two-universality.} to
turn the (generally only partially secret) string $X'$ into a shorter
but secure key.

The security of the BB84 protocol is based on the fact that an
adversary, ignorant of the actual encoding bases used by Alice, cannot
gain information about the encoded bits without disturbing the qubits
sent over the quantum channel. If the disturbance is too large, Alice
and Bob will observe a high error rate and abort the protocol in the
parameter estimation step. On the other hand, if the disturbance is
below a certain threshold, then the strings $X'$ and $Y'$ held by
Alice and Bob are sufficiently correlated and secret in order to
distill a secret key.

In order to prove security, one thus needs to quantify the amount of
information that an adversary has on the raw key\index{raw~key}, given
the disturbance measured by Alice and Bob. It is a main goal of this
thesis to develop the information-theoretic techniques which are
needed for this analysis. (See also Section~\ref{sec:proofsketch} for
a sketch of the security proof.)

\index{BB84~protocol|)}

\subsubsection{Alternative protocols}

Since the invention of quantum cryptography, a considerable effort has
been taken to get a better understanding of its theoretical
foundations as well as to make it more
practical\index{practical~implementation}. In the course of this
research, a large variety of alternative QKD protocols has been
proposed.  Some of them are very efficient with respect to the
\emph{secret-key rate}\index{secret~key~rate@secret-key~rate}, i.e.,
the number of key bits generated per channel
use~\cite{Bruss98,BecGis99}.  Others are designed to cope with high
channel noise or noise in the detector, which makes them more suitable
for practical implementations~\cite{SARG04}.

The structure of these protocols is mostly very similar to the BB84
protocol described above.  For example, the \emph{six-state
  protocol}\index{six-state~protocol} proposed
in~\cite{Bruss98,BecGis99} uses \emph{three} different bases for the
encoding (i.e., \emph{six} different states), but otherwise is
identical to the BB84 protocol.  On the other hand, the \emph{B92
  protocol}\index{B92~protocol}~\cite{Bennet92} is based on an
encoding with respect to only \emph{two} non-orthogonal states.

\subsubsection{QKD over noisy channels}

\index{noisy~channel|(}

Any realistic quantum channel is subject to intrinsic noise. Alice and
Bob will thus observe errors even if the adversary is passive.
However, as these errors are not distinguishable from errors caused by
an attack, the distribution of a secret key can only be successful if
the noise level of the channel is sufficiently low.


As an example, consider the BB84 protocol described above. In the
parameter estimation step, Alice and Bob compute a guess for the error
rate\index{error~rate} and abort the protocol if it exceeds a certain
threshold. Hence, the scheme only generates a key if the noise level
of the channel is below this threshold.

The amount of noise tolerated by a QKD scheme is an important measure
for its practicability\index{practical~implementation}. In fact, in an
implementation, the level of noise inevitably depends on the distance
between Alice and Bob (i.e., the length of the optical fiber, for an
implementation based on photons). To characterize the efficiency of
QKD schemes, one thus often considers the relation between the channel
noise and the secret-key rate (see plots in
Chapter~\ref{ch:examples}).  Typically, the secret-key rate decreases
with increasing noise level and becomes zero as soon as the noise
reaches a certain bound, called the \emph{maximum tolerated channel
  noise}\index{maximum~tolerated~channel~noise}.

\index{noisy~channel|)}

\subsubsection{Quantum key distribution and distillation}

Assume that Alice and Bob have access to some correlated quantum
systems (e.g., predistributed pairs of entangled particles). A
\emph{quantum key distillation}\index{quantum~key~distillation}
protocol allows them to transform this correlation into a common
secret key, while using only classical authentic communication.

As explained below, a quantum key \emph{distribution}
(QKD)\index{quantum~key~distribution} protocol can generally be
transformed into a key \emph{distillation} protocol in such a way that
security of the latter implies security of the first.  This is very
convenient for security proofs, as key distillation only involves
quantum states (instead of quantum channels) which are easier to
analyze (see \cite{Ekert91,BeBrMe92}).






The connection between key distillation and key distribution protocols
is based on the following observation: Let $X$ be a classical value
chosen according to a distribution $P_X$ and let $\ket{\phi^x}$ be a
quantum encoding of $X$.  This situation could now equivalently be
obtained by the following two-step process: (i) prepare a bipartite
quantum state $\ket{\Psi} := \sum_{x} \sqrt{P_X(x)} \ket{x} \otimes
\ket{\phi^x}$, where $\{\ket{x}\}_{x}$ is some orthonormal basis of
the first subsystem; (ii) measure the first part of $\ket{\Psi}$ with
respect to the basis $\{\ket{x}\}_{x}$.  In fact, it is easy to verify
that the outcome $X$ is distributed according to $P_X$ and that the
remaining quantum system contains the correct encoding of
$X$.


To illustrate how this observation applies to QKD, consider a protocol
where Alice uses the quantum channel to transmit an encoding
$\ket{\phi^x}$ of some randomly chosen value $X$ to Bob (as, e.g., in
the first step of the BB84 protocol described above).  According to
the above discussion, this can equivalently be achieved as
follows:\footnote{More generally, any arbitrary protocol step can be
  replaced by a coherent quantum operation followed by some
  measurement.}  First, Alice locally prepares the bipartite state
$\ket{\Psi}$ defined above, keeps the first half of it, and sends the
second half over the quantum channel to Bob.  Second, Alice measures
the quantum system she kept to get the classical value $X$. (Such a
protocol is sometimes called an
\emph{entanglement-based}\index{entanglement-based~protocol} scheme.)

Note that, after the use of the quantum channel|but before the
measurement|Alice and Bob share some (generally entangled) quantum
state. The remaining part of the key distribution protocol is thus
actually a quantum key distillation protocol.  Hence, if this key
distillation protocol\index{quantum~key~distillation} is secure (for
any predistributed entanglement) then the original quantum key
distribution\index{quantum~key~distribution} protocol is secure (for
any arbitrary attack of Eve).



\section{Contributions} \label{sec:contr}

This thesis makes two different types of contributions. First, we
introduce various concepts and prove results which are of general
interest in quantum information theory and cryptography.\footnote{For
  example, our result on privacy amplification against quantum
  adversaries is not only useful to prove the security of QKD. It has
  also found interesting applications within other fields of
  cryptography, as for instance in the context of multi-party
  computation (see, e.g., \cite{DFSS05} for a result on bit
  commitment).} These contributions are summarized in
Section~\ref{sec:contrgen} below.  Second, we apply our techniques to
QKD in order to derive a general security criterion.  Some aspects and
implications of this result are discussed in
Section~\ref{sec:contrsec}.

\subsection{New notions in quantum information theory} \label{sec:contrgen}

\subsubsection{Smooth min- and max-entropies as generalizations of von Neumann entropy}

The von Neumann entropy\index{von~Neumann~entropy}, as a measure for
the uncertainty on the state of a quantum system, plays an important
role in quantum information theory. This is mainly due to the fact
that it characterizes fundamental information-theoretic tasks such as
\emph{randomness extraction}\index{randomness~extraction} or
\emph{data compression}\index{data~compression}.  For example, the von
Neumann entropy\index{von~Neumann~entropy} of a source emitting
quantum states can be interpreted as the minimum space needed to
encode these states such that they can later be reconstructed with
arbitrarily small error. However, any such interpretation of the von
Neumann entropy only holds asymptotically in situations where a
certain underlying experiment is repeated many times independently.
For the above example, this means that the encoding is over many
(sufficiently independent) outputs of the source.

In the context of cryptography, where an adversary might corrupt parts
of a system in an arbitrary way, this independence can often not be
guaranteed. The von Neumann entropy is thus usually not an appropriate
measure|e.g., to quantify the uncertainty of an adversary|unless we
put some severe restrictions on her capabilities (e.g., that her
attack consists of many independent repetitions of the same action).

In this thesis, we introduce two entropy measures, called \emph{smooth
  min-}\index{smooth~min-entropy} and
\emph{max-entropy}\index{smooth~max-entropy}, which can be seen as
generalizations of the von Neumann entropy. While smooth min-entropy
quantifies the amount of uniform randomness that can be
\emph{extracted} from a quantum system, the smooth max-entropy
corresponds to the length of an optimal \emph{encoding} of the
system's state. Unlike the von Neumann entropy, however, this
characterization applies to arbitrary situations|including those for
which there is no underlying independently repeated experiment.

In the special case of many \emph{independent
  repetitions}\index{independent~repetitions}\index{product~state}
(that is, if the system's state is described by a density operator
which has product form), smooth min- and max-entropy both reduce to
the von Neumann entropy, as expected.  Moreover, smooth min- and
max-entropy inherit most of the properties known from the von Neumann
entropy, as for example the strong subadditivity.  (We refer to
Section~\ref{sec:outline} for a summary of these results.) On the
other hand, because the von Neumann entropy is a special case of
smooth min- and max-entropy, its properties follow directly from the
corresponding properties of the smooth min- or max-entropy.
Interestingly, some of the proofs are surprisingly easy in this
general case. For example, the strong subadditivity of the smooth
min-entropy follows by a very short argument (cf.\ 
Lemma~\ref{lem:Hinfcondclasse} and
Lemma~\ref{lem:Hinfcondclassesmooth}). Note that this immediately
gives a simple proof for the strong
subadditivity\index{strong~subadditivity} of the von Neumann entropy.

\subsubsection{De Finetti representation
  theorem for finite symmetric quantum states}



\index{de~Finetti~representation|(}

An $n$-partite density operator $\rho_{n}$ is said to be
\emph{$N$-exchangeable}\index{exchangeability}\index{finite~exchangeability},
for $N \geq n$, if it is the partial state (i.e., $\rho_n =
\tr_k(\rho_N)$) of an $N$-partite density operator $\rho_{N}$ which is
invariant under permutations of the subsystems.  Moreover, $\rho_{n}$
is \emph{infinitely-exchangeable}\index{infinite~exchangeability} if
it is $N$-exchangeable for all $N \geq n$. The \emph{quantum de
  Finetti representation theorem}~\cite{HudMoo76} (which is the
quantum version of a theorem in probability theory named after its
inventor Bruno de Finetti\footnote{See~\cite{deFine93} for a
  collection of de Finetti's original papers.}) makes a fundamental
statement on such symmetric operators.\footnote{See \cite{CaFuSh02}
  for a nice proof of the quantum de Finetti theorem based on its
  classical analogue.}  Namely, it says that any
infinitely-exchangeable operator $\rho_{n}$ can be written as a convex
combination (i.e., a \emph{mixture}) of product
operators,\index{product~state}
\[
\rho_{n} = \int_{\sigma} \sigma^{\otimes n} \nu(\sigma) \ .
\]

We generalize the quantum de Finetti representation theorem for
\emph{infinitely} exchangeable operators to the \emph{finite}
case.\footnote{The result presented in this thesis is different from
  the one proposed in a previous paper~\cite{KoeRen05} (see
  Section~\ref{sec:outline} for more details).}  More precisely, we
show that the above formula still holds approximatively if $\rho_{n}$
is only $N$-exchangeable for, some finite $N$ which is sufficiently
larger than $n$.  (We refer to Section~\ref{sec:outline} below for a
more detailed description of this statement.)

The de Finetti representation theorem turns out to be a useful tool in
quantum information theory. In fact, symmetric (and exchangeable)
states play an important role in many applications. For example, the
operator describing the joint state of $n$ particles selected at
random from a set of $N$ particles is $N$-exchangeable.  Hence,
according to our finite version of the de Finetti representation
theorem, the analysis of such states can be reduced to the analysis of
product states|which is often much easier than the general case.
Following this idea, we will use the finite de Finetti representation
theorem to argue that, for proving the security of a QKD scheme
against arbitrary attacks, it suffices to consider attacks that have a
certain product structure (so-called \emph{collective attacks}, cf.\ 
Section~\ref{sec:collcoh}).

\index{de~Finetti~representation|)}

\subsubsection{Universal security of keys in a quantum world}

\index{universal~security|(} \index{$\eps$-security|(}

In quantum cryptography, the security of a secret key $S$ is typically
defined with respect to the classical information $W$ that an
adversary might obtain when measuring her quantum system $\cH_E$. More
precisely, $S$ is said to be secure if, for any measurement of the
adversary's system $\cH_E$, the resulting outcome $W$ gives virtually
no information on $S$. Although this definition looks quite strong, we
shall see that it is not sufficient for many applications, e.g., if
the key $S$ is used for one-time pad
encryption\index{one-time~pad~encryption} (see
Section~\ref{sec:univ}).

We propose a security definition which overcomes this problem.
Roughly speaking, we say that a key $S$ is \emph{$\eps$-secure} if,
except with probability $\eps$, $S$ is equal to a \emph{perfect
  key}\index{perfect~key} which is uniformly distributed and
completely independent of the adversary's quantum system.  In
particular, our security definition is \emph{universal} in the sense
that an $\eps$-secure key can safely be used in any application,
except with probability $\eps$.\footnote{Hence, our security
  definition fits into general frameworks concerned with the universal
  security of quantum protocols, as proposed by Ben-Or and
  Mayers~\cite{BenMay04} and Unruh~\cite{Unruh04} (see
  Section~\ref{sec:univ} for more details).}

\index{universal~security|)} \index{$\eps$-security|)}

\subsubsection{Security of privacy amplification against quantum adversaries}

\index{privacy~amplification|(}

Let $X$ be a classical random variable on which an adversary has some
partial information. \emph{Privacy amplification} is the art of
transforming this partially secure $X$ into a fully secure key $S$,
and has been studied extensively for the case where the adversary's
information is purely classical. It has been
shown~\cite{BeBrRo88,ImLeLu89,BBCM95} that it is always possible to
generate an $\ell$-bit key $S$ which is secure against any adversary
whose uncertainty on $X$|measured in terms of the \emph{collision
  entropy}\index{collision~entropy}\footnote{The \emph{collision
    entropy}, also called \emph{R\'enyi entropy of order
    two}\index{R\'enyi~entropy~of~order~two}, of a probability
  distribution $P_X$ is the negative binary logarithm of its collision
  probability $\sum_x P_X(x)^2$.}|is sufficiently larger than~$\ell$.

We generalize this classical privacy amplification theorem to include
\emph{quantum} adversaries who might hold information on $X$ encoded
in the state of a quantum system.  We show that, similar to the
classical result, $X$ can be transformed into a key of length $\ell$
which is secure\footnote{We prove security according to the strong
  definition proposed in Section~\ref{sec:univ}|i.e., the security is
  \emph{universal}.}  if the uncertainty of the adversary on $X$|this
time measured in terms of the \emph{smooth
  min-entropy}\index{smooth~min-entropy}|is at least roughly $\ell$.
Because the smooth min-entropy is generally larger than the collision
entropy, this also implies the above classical result.

Our privacy amplification theorem is optimal with respect to the
maximum length $\ell$ of the extractable secret key|i.e., smooth
min-entropy completely characterizes the number of secret key bits
that can be generated from a partially secret string (up to some small
constant). This also improves our previous
results~\cite{KoMaRe05,RenKoe05} which are only optimal in certain
special cases.\footnote{The result proven in~\cite{RenKoe05} is
  optimal if the density operator describing the initial string
  together with the adversary's quantum information has product form.}

\index{privacy~amplification|)}

\subsection{Properties and implications of the security result} \label{sec:contrsec}

We provide a simple and general\footnote{The security criterion is
  \emph{general} in the sense that it applies to virtually all known
  protocols. Note that this stands in contrast to previous security
  proofs, which are mostly designed for specific protocols.}
criterion for the security of QKD against \emph{any} attack allowed by
the laws of quantum
physics.\index{security~proof}\index{security~criterion} The following
is a summary of the most important properties and consequences of this
result.  (For a more detailed description of the security criterion
and a proof sketch, we refer to Section~\ref{sec:secsum} below.)


\subsubsection{Coherent attacks are not stronger than collective
  attacks} \label{sec:collcoh}

An adversary might in principle apply an arbitrary operation on the
quantum states exchanged between Alice and Bob. In the case of the
most general, so-called \emph{coherent
  attacks}\index{coherent~attack}, this operation could involve all
subsystems (particles) simultaneously, which makes it (seemingly)
difficult to analyze.  One thus often considers a restricted class of
attacks, called \emph{collective
  attacks}\index{collective~attack}~\cite{BihMor97a,BihMor97b}, where
the adversary is assumed to apply the same transformation to each of
the subsystems that is sent over the channel.\footnote{An even more
  restricted type of attacks are the so-called
  \emph{individual~attacks}\index{individual attack} where,
  additionally, the adversary is supposed to apply some fixed
  measurement operation to each of the subsystems sent through the
  channel. In particular, this measurement cannot depend on the
  classical information that Alice and Bob exchange for error
  correction and privacy amplification. As shown in~\cite{BeMoSm96},
  such individual attacks are generally weaker than collective
  attacks. Hence, security against individual attacks does not imply
  full security.}
A natural and long-standing open question in this context is whether
security against collective attacks implies full security (see, e.g.,
\cite{BBBGM02}).  Our result immediately answers this question in the
positive, that is, coherent attacks cannot be more powerful than
collective attacks.\footnote{This statement holds for virtually any
  QKD protocol; the only requirement is that the protocol is symmetric
  under permutations of the channel
  uses\index{permutation-invariant~protocol} (see
  Section~\ref{sec:secsum} for more details).}

\subsubsection{Security of practical implementations}

Because of technical limitations, practical implementations of QKD are
subject to many imperfections. In addition to noisy channels, these
might include faulty sources\footnote{For example, it is difficult to
  design sources that emit perfect single-photon pulses.}  or detector
losses.\index{practical~implementation} Because of its generality, our
security criterion can be used for the analysis of such practical
settings.\footnote{As there is no restriction on the structure of the
  underlying Hilbert space, the security criterion applies to any
  modeling of the physical system which is used for the quantum
  communication between Alice and Bob.}

\subsubsection{Keys generated by QKD can safely be used in applications}

The security result holds with respect to a so-called
\emph{universal}\index{universal~security} security definition. This
guarantees that the key generated by a QKD protocol can safely be used
in applications such as for one-time pad encryption.  (As mentioned
above, this is not necessarily the case for many of the standard
security definitions.)

\subsubsection{Improved bounds on the efficiency of concrete protocols}

Our security result applies to protocols which could not be analyzed
with previously known techniques (e.g., a reduction to entanglement
purification\index{entanglement~purification} schemes, as proposed
in~\cite{ShoPre00}). In particular, it allows to compute the key rates
for new variants of known protocols.\footnote{E.g., we will analyze
  protocols that use an alternative method for the processing of the
  raw key.}  For example, we propose an improved version of the
six-state protocol and show that it is more efficient than previous
variants.  Moreover, we derive new bounds on the maximum tolerated
channel noise\index{maximum~tolerated~channel~noise} of the BB84 or
the six-state protocol with one-way post-processing.

\subsubsection{Explicit bounds on the security of finite keys}

The security criterion gives explicit (non-asymptotic) bounds on the
secrecy and the length of keys generated from any (finite) number of
invocations of the quantum channel.  Moreover, it applies to schemes
which use arbitrary (not necessarily optimal) subprotocols for
information reconciliation\index{information~reconciliation}. This is
in contrast to most known security results which|with a few
exceptions\footnote{See, e.g., \cite{InLuMa01} for a nice and very
  careful explicit analysis of the BB84 protocol.}|only hold
asymptotically for large key sizes and for asymptotically optimal
information reconciliation.

\section{Related work}



The techniques developed in this thesis are partly motivated by ideas
known from classical information theory and, in particular,
cryptography (e.g., classical de Finetti-style theorems, privacy
amplification against classical adversaries, or universally composable
security). For a discussion of these notions and their relation to our
results we refer to Section~\ref{sec:contr}. In the following, we
rather focus on work related to the security of QKD.

Since Bennett and Brassard proposed the first QKD protocol in 1984
\cite{BenBra84}, it took more than a decade until
Mayers~\cite{Mayers96} proved that the scheme is secure against
arbitrary attacks.\footnote{See also~\cite{Mayers01} for an improved
  version of Mayers' proof.}\index{security~proof} This result was
followed by various alternative proofs (see, e.g., \cite{ChReEk04}
or~\cite{LoChAr05} for an overview).

One of the most popular proof techniques was proposed by Shor and
Preskill~\cite{ShoPre00}, based on ideas of Lo and
Chau~\cite{LoCha99}.  It uses a connection between key distribution
and entanglement
purification~\cite{BBPSSW96}\index{entanglement~purification} pointed
out by Ekert~\cite{Ekert91} (see also~\cite{BeBrMe92}). The proof
technique of Shor and Preskill was later refined and applied to other
protocols (see, e.g., \cite{GotLo03,TaKoIm03}).


In~\cite{ChReEk04}, we have presented a general method for proving the
security of QKD which does not rely on entanglement purification.
Instead, it is based on a result on the security of privacy
amplification in the context of quantum
adversaries~\cite{KoMaRe05,RenKoe05}\index{privacy~amplification}.
Later, this method has been extended and applied to prove the security
of new variants of the BB84 and the six-state
protocol~\cite{ReGiKr05,KrGiRe05}.\footnote{In~\cite{ReGiKr05,KrGiRe05}
  we use an alternative technique (different from the quantum de
  Finetti theorem) to show that collective attacks are equivalent to
  coherent attacks for certain QKD protocols.}  The security proof
given in this thesis is based on ideas developed in these papers.

Our new approach for proving the security of QKD has already found
various applications. For example, it is used for the analysis of
protocols based on continuous systems\index{continuous~variable~QKD}
as well as to improve the analysis of known (practical)
protocols\index{practical~implementation} exploiting the fact that an
adversary cannot control the noise in the physical devices owned by
Alice and Bob (see, e.g., \cite{Grossha05,NavAci05,Lo05}).

\section{Outline of the thesis} \label{sec:outline}

The following is a brief summary of the main results obtained in each
chapter.

\subsubsection{Chapter~\ref{ch:pre}: Preliminaries}

The first part of this chapter (Section~\ref{sec:mathdecr}) is
concerned with the representation of physical (cryptographic) systems
as mathematical objects. We briefly review the density operator
formalism which is used to describe quantum mechanical systems.
Moreover, we present some variant of this formalism which is useful
when dealing with physical systems that consist of both classical and
quantum parts.

The second part of Chapter~\ref{ch:pre} (Section~\ref{sec:univ}) is
devoted to the security definition for secret keys\index{secret~key}.
We first argue that many of the widely used definitions are
problematic|in the sense that they do not imply the security of
applications such as one-time pad
encryption\index{one-time~pad~encryption}. Then, as a solution to this
problem, we introduce a so-called \emph{universal}
security\index{universal~security} definition for secret keys and
discuss its properties.

\subsubsection{Chapter~\ref{ch:smooth}: Smooth min- and max-entropy}

This chapter introduces and studies \emph{smooth
  min-entropy}\index{smooth~min-entropy} $\Hmin^\eps$ and \emph{smooth
  max-entropy}\index{smooth~max-entropy} $\Hmax^\eps$, which both are
entropy measures for density operators.  We first discuss some basic
properties (Sections~\ref{sec:convRenyi} and~\ref{sec:smoothRenyi})
which are actually very similar to those of the von Neumann entropy
(Theorem~\ref{thm:Hmincalc}).  For example, the smooth min-entropy is
\emph{strongly subadditive}\index{strong~subadditivity}, that
is,\footnote{We use a slightly simplified notation in this summary.
  For example, we write $\Hmin^\eps(A|B)$ to denote the smooth
  min-entropy of a state $\rho_{A B}$ given the second subsystem
  (instead of $\Hmin^\eps(\rho_{A B}|B)$ which is used in the
  technical part).}
\begin{equation} \label{eq:sumHsubadd}
  \Hmin^\eps(A|B C) \leq \Hmin^\eps(A|B) \ ,
\end{equation}
and it obeys an inequality which can be interpreted as a \emph{chain
  rule}\index{chain~rule},
\begin{equation} \label{eq:sumHchain}
  \Hmin^\eps(A B|C) \leq \Hmin^\eps(A|B C) + \Hmax(B) \ .
\end{equation}
Moreover, if the states in the subsystems $\cH_A$ and $\cH_C$ are
independent conditioned on a classical value $Y$ then
\begin{equation} \label{eq:sumHdataproc}
  \Hmin^\eps(A Y|C) \geq \Hmin^\eps(Y|C) + \Hmin(A|Y) \ .
\end{equation}

The second part of Chapter~\ref{ch:smooth}
(Section~\ref{sec:smoothprod}) treats the special case where the
density operators have product form\index{product~state}. In this
case, smooth min- and max-entropy both reduce to the von Neumann
entropy.\index{von~Neumann~entropy} Formally, the smooth min-entropy
$\Hmin^\eps(A^n | B^n)$ of a product state $\rho_{A^n B^n} = \sigma_{A
  B}^{\otimes n}$ satisfies
\begin{equation} \label{eq:sumprod}
  \lim_{n \to \infty} \frac{1}{n} \Hmin^\eps(A^n|B^n) 
=
  H(A|B) \ ,
\end{equation}
where $H(A|B) = H(\sigma_{A B}) - H(\sigma_B)$ is the (conditional)
von Neumann entropy\index{conditional~von~Neumann~entropy} evaluated
for the operator $\sigma_{A B}$ (cf.\ Theorem~\ref{thm:Hmincondrep}
and Corollary~\ref{cor:Hmincondrepclass}).


\subsubsection{Chapter~\ref{ch:sym}: Symmetric states} 

This chapter is concerned with \emph{symmetric}\index{symmetric~state}
states, that is, states on $n$-fold product system $\cH^{\otimes n}$
that are invariant under permutations of the subsystems. We first show
that any permutation-invariant\index{permutation-invariant~state}
density operator has a symmetric
purification\index{symmetric~purification}, which allows us to
restrict our attention to the analysis of \emph{pure} symmetric states
(Section~\ref{sec:sympur}).

The main result of this section is a finite version of the
\emph{quantum de Finetti representation
  theorem}\index{de~Finetti~representation}
(Section~\ref{sec:symrepr}). It says that symmetric states can be
approximated by a convex combination of states which have ``almost''
product form (cf.\ Theorem~\ref{thm:symmix})\index{product~state}.
Formally, if $\rho_{n + k}$ is a permutation-invariant operator on $N
= n+k$ subsystems $\cH$, then the partial state $\rho_n$ on
$\cH^{\otimes n}$ (obtained by tracing over $k$ subsystems) is
approximated by a mixture of operators $\rho_n^{\sigma}$, i.e.,
\begin{equation} \label{eq:sumconv}
  \rho_n 
\approx
  \int_{\sigma} \rho_n^{\sigma} \nu(\sigma) \ ,
\end{equation}
where the integral ranges over all density operators $\sigma$ on
\emph{one single} subsystem $\cH$ and $\nu$ is some probability
measure on these operators. Roughly speaking, the states
$\rho_n^{\sigma}$ are superpositions of states which, on at least
$n-r$ subsystems, for some small $r$, have product form
$\sigma^{\otimes n-r}$. Moreover, the distance\footnote{The distance
  is measured with respect to the \distance{}, as defined in
  Section~\ref{sec:tracedistintro}.}  between the left and the right
hand side of the approximation~\eqref{eq:sumconv} decreases
exponentially fast in $r$ and $k$.\footnote{Note that this version of
  the finite quantum de Finetti representation theorem|although the
  same in spirit|is distinct from the the one proposed
  in~\cite{KoeRen05}: In~\cite{KoeRen05}, the decomposition is with
  respect to \emph{perfect} $n$-fold product states $\sigma^{\otimes
    n}$|instead of states $\rho_n^{\sigma}$ which are products on only
  $n-r$ subsystems|but the approximation is not exponential.}
 
The properties of the states $\rho_n^{\sigma}$ occurring in the convex
combination~\eqref{eq:sumconv} are similar to those of perfect product
states $\sigma^{\otimes n}$. The main result of
Section~\ref{sec:smsym} can be seen as a generalization
of~\eqref{eq:sumprod}. It states that, for a state $\rho_{A^n
  B^n}^{\sigma_{A B}}$ which has almost product form $\sigma_{A
  B}^{\otimes n}$ (in the sense defined above, where $\sigma_{A B}$ is
a bipartite operator on $\cH_A \otimes \cH_B$) the smooth min-entropy
is given by
\begin{equation} \label{eq:sumHsym}
  \lim_{n \to \infty} \frac{1}{n} \Hmin^\eps(A^n|B^n) = H(A|B) 
\end{equation}
(see Theorem~\ref{thm:Renyisym}\footnote{Note that
  Theorem~\ref{thm:Renyisym} only implies one direction ($\geq$). The
  other direction ($\leq$) follows from a similar argument for the
  smooth max-entropy, which is an upper bound on the smooth
  min-entropy.}).

Analogously, in Section~\ref{sec:symstat}, we show that states
$\rho_n^\sigma$ which have almost product form $\sigma^{\otimes n}$
lead to similar statistics as perfect product states $\sigma^{\otimes
  n}$ if they are measured with respect to a product measurement.
Formally, let $P_Z$ be the distribution of the outcomes when measuring
$\sigma$ with respect to a POVM $\cM$.  Moreover, let $\freq{\bz}$ be
the statistics (i.e., the frequency
distribution)\index{frequency~distribution} of the outcomes $\bz =
(z_1, \ldots, z_n)$ of the product measurement $\cM^{\otimes n}$
applied to $\rho_n^\sigma$.  Then
\begin{equation} \label{eq:symstatsum}
  \lim_{n \to \infty} \freq{\bz} 
=
  P_Z
\end{equation}
(cf. Theorem~\ref{thm:symstat}).

\subsubsection{Chapter~\ref{ch:PA}: Privacy amplification}

This chapter is on privacy amplification\index{privacy~amplification}
in the context of quantum adversaries. The main result is an explicit
expression for the secrecy of a key $S$ which is computed from an only
partially secure string $X$ by \emph{two-universal
  hashing}\index{two-universal~hashing}\footnote{That is, $S$ is the
  output $f(X)$ of a function $f$ which is randomly chosen from a
  so-called \emph{two-universal} family of hash functions (see
  Section~\ref{sec:twohash} for a definition).}  (Theorem~\ref{thm:pa}
and Corollary~\ref{cor:pasmooth}). The result implies that the key $S$
is secure under the sole condition that its length $\ell$ is bounded
by
\begin{equation} \label{eq:sumsecPA}
  \ell \lessapprox \Hmin^\eps(X|E)
\end{equation}
where $\Hmin^\eps(X|E)$ denotes the smooth min-entropy of $X$ given
the adversary's initial information.\index{smooth~min-entropy}

\subsubsection{Chapter~\ref{ch:QKD}: Security of QKD}

This chapter is devoted to the statement and proof of our main result
on the security of QKD.  In particular, it contains an expression for
the key rate for a general class of protocols in terms of simple
entropic quantities (Theorem~\ref{thm:main} and
Corollary~\ref{cor:main}).\index{security~proof}\index{secret~key~rate@secret-key~rate}
(We refer to Section~\ref{sec:secsum} for an overview on this result
and its proof.)

\subsubsection{Chapter~\ref{ch:examples}: Examples}

As an illustration, we apply the general result of
Chapter~\ref{ch:QKD} to specific types of QKD protocols.  The focus is
on schemes which are based on two-level systems.  In particular, we
analyze different versions of the six-state\index{six-state~protocol}
QKD protocol and compute explicit values for their rates (see
Plots~\ref{pl:SixSnoNoise}--\ref{pl:sixSoptrate}).

\section{Outline of the security analysis of QKD} \label{sec:secsum}

\index{security~proof|(}

The following is a summary of our main result on the security of
quantum key distillation which|according to the discussion in
Section~\ref{sec:QKD}|also implies the security of quantum key
distribution.\index{quantum~key~distillation}\index{quantum~key~distribution}
Moreover, we give a sketch of the security proof, which is based on
the technical results summarized in Section~\ref{sec:outline} above.
(For a complete description of the security result and the full proof,
we refer to Chapter~\ref{ch:QKD}.)


\subsection{Protocol}

We start with a brief characterization of the general type of quantum
key distillation protocols to which our security proof applies. For
this, we assume that Alice and Bob start with $N$ bipartite quantum
systems $\cH_A \otimes \cH_B$ (describing, e.g., pairs of entangled
particles). The protocol then runs through the following steps in
order to transform this initial entanglement between Alice and Bob
into a common secret key.\index{secret~key}

\begin{itemize}
\item \emph{Parameter estimation:}\index{parameter~estimation} Alice
  and Bob sacrifice some small number, say $m$, subsystems $\cH_A
  \otimes \cH_B$ in order to estimate their average correlation. For
  this, they both apply measurements with respect to different bases
  and publicly announce the outcomes (using the authentic classical
  communication channel)\index{authentic~channel}.  Depending on the
  resulting statistics, they either decide to proceed with the
  computation of the key or to abort the protocol.
\item \emph{Measurement:}\index{measurement} Alice and Bob both apply
  measurements to their parts of the remaining subsystems $\cH_A
  \otimes \cH_B$ to obtain a pair of \emph{raw keys}\index{raw~key}.
  (Note that these raw keys are generally only weakly correlated and
  partially secure.)
\item \emph{Block-wise processing:}\index{block-wise~processing} Alice
  and Bob might\footnote{In many protocols, this step is omitted,
    i.e., Alice and Bob directly proceed with information
    reconciliation.}  further process their raw key pair in order to
  improve its correlation or secrecy.  We assume that this processing
  acts on $n$ blocks of size $b$ individually. For example, Alice and
  Bob might invoke a so-called \emph{advantage
    distillation}\index{advantage~distillation} protocol (see
  Section~\ref{sec:AD}) whose purpose is to single out blocks of the
  raw key\index{raw~key} that are highly correlated. We denote by
  $X^n$ and $Y^n$ the strings held by Alice and Bob after this step.
\item \emph{Information
    reconciliation:}\index{information~reconciliation} The purpose of
  this step is to transform the (possibly only weakly correlated) pair
  of strings $X^n$ and $Y^n$ into a pair of identical strings.
  Typically, Alice sends certain error correcting information on $X^n$
  to Bob which allows him to compute a guess $\hat{X}^n$ of~$X^n$.
\item \emph{Privacy amplification:}\index{privacy~amplification} Alice
  and Bob use two-universal hashing\index{two-universal~hashing} to
  transform their strings $X^n$ and $\hat{X}^n$ into secret keys of
  length $\ell$.
\end{itemize}

Additionally, we assume that the action of the protocol is invariant
under permutations of the $N$ input
systems.\index{permutation-invariant~protocol} This does not restrict
the generality of our results, because any protocol can easily be
turned into a permutation-invariant one: Before starting with the
parameter estimation, Alice and Bob simply have to (publicly) agree on
a random permutation which they use to reorder their subsystems (see
Section~\ref{sec:proofsketch} below for more details).

\subsection{Security criterion}

\index{security~criterion|(}


The security of a key distillation scheme depends on the actual choice
of various protocol parameters which we define in the following:


\begin{itemize}
\item $\Gamma$ is the set of states on \emph{single} subsystems which
  are not filtered by the \emph{parameter
    estimation}\index{parameter~estimation} subprotocol: More
  precisely, $\Gamma$ contains all density operators $\sigma_{A B}$
  such that, when starting with the product state $\rho_{A^N B^N} :=
  \sigma_{A B}^{\otimes N}$, the protocol does \emph{not} abort.
\item $\cE_{X Y \Eb\leftarrow A^b B^b E^b}$ is the CPM\footnote{See
    Section~\ref{sec:qdefs} for a definition of completely positive
    maps (CPM).} on $b$ subsystems which describes the
  \emph{measurement}\index{measurement} together with the
  \emph{block-wise processing}\index{block-wise~processing} on blocks
  of size $b$.
\item $n$ is the \emph{number of blocks} of size $b$ that are used for
  the actual computation of the key (i.e., the number of blocks of
  subsystems that are left after the parameter estimation step).
\item $\ell$ denotes the length of the final key generated in the
  \emph{privacy amplification}\index{privacy~amplification} step.
\end{itemize}

In addition, the security of the scheme depends on the efficiency of
the \emph{information
  reconciliation}\index{information~reconciliation} subprotocol, i.e.,
the amount of information that is leaked to Eve.\index{leakage}
However, for this summary, we assume that Alice and Bob use an
\emph{optimal}\footnote{In Section~\ref{sec:IR}, we show that optimal
  information reconciliation protocols exist.} information
reconciliation protocol. In this case, the leakage is roughly equal to
the entropy of $X^n$ given $Y^n$.\footnote{We refer to
  Chapter~\ref{ch:QKD} for the general result which deals with
  arbitrary|not necessarily optimal|information reconciliation
  schemes.}


We are now ready to formulate a general security criterion for quantum
key distillation (cf.\ Theorem~\ref{thm:main}): The scheme described
above is secure (for any initial state) if\footnote{The approximation
  $\lessapprox$ in~\eqref{eq:summain} indicates that the criterion
  holds asymptotically for increasing $n$. We refer to
  Chapter~\ref{ch:QKD} for a non-asymptotic result.}
\begin{equation} \label{eq:summain}
  \frac{\ell}{n}
\lessapprox
  \min_{\sigma_{A B} \in \Gamma} 
    H(X|E) - H(X|Y) \ ,
\end{equation}
where the minimum ranges over all states $\sigma_{A B}$ contained in
the set $\Gamma$ defined above and where $H(X|E)$ and $H(X|Y)$ are the
(conditional) von Neumann entropies\index{von~Neumann~entropy} of
\[
  \sigma_{X Y E} 
:= 
  \cE_{X Y \Eb \leftarrow A^b B^b E^b}(\sigma_{A B E}^{\otimes b})
\]
where $\sigma_{A B E}$ is a purification of $\sigma_{A B}$. Note that,
because the operators $\sigma_{A B}$ are on \emph{single} subsystems,
formula~\eqref{eq:summain} is usually fairly easy to evaluate for
concrete protocols
(cf.\ Chapter~\ref{ch:examples}).

Typically, the number $m$ of subsystem that are sacrificed for
parameter estimation is small compared to the total number $N$ of
initial subsystems. Hence, the number $n$ of blocks of size $b$ that
can be used for the actual computation of the key is roughly given by
$n \approx \frac{N}{b}$.\footnote{This is also true for QKD protocols
  with a sifting\index{sifting} step (where Alice and Bob discard the
  subsystems for which they have used incompatible encoding and
  decoding bases).  In fact, as mentioned in Section~\ref{sec:QKD}, if
  Alice and Bob choose one of the bases with probability close to one,
  the fraction of positions lost in the sifting step is small.}  The
criterion~\eqref{eq:summain} can thus be turned into an expression for
the \emph{key rate}\index{secret~key~rate@secret-key~rate} of the
protocol (i.e., the number of key bits generated per channel use):
\[
  \mathrm{rate}
=
  \frac{1}{b} \min_{\sigma_{A B} \in \Gamma} 
    H(X|E) - H(X|Y) \ .
\]

\index{security~criterion|)}



\subsection{Security proof} \label{sec:proofsketch}



We need to show that, for \emph{any} initial state shared by Alice and
Bob, the probability that the protocol generates an insecure key is
negligible.\footnote{Note that the protocol might abort if the initial
  state held by Alice and Bob is not sufficiently correlated.} Roughly
speaking, the proof consists of two parts. In the first (Steps~1--2)
we argue that we can restrict our analysis to a much smaller set of
initial states, namely those that have (almost) product
form\index{product~state}.  In the second part (Steps 3--5) we show
that for each such state either of the following holds: (i)~there is
not sufficient correlation between Alice and Bob in which case the
protocol aborts during the parameter estimation or (ii)~a measurement
applied to the state generates an outcome with sufficient entropy such
that the key computed from it is secure.

\subsubsection{Step~1: Restriction to permutation-invariant initial states}

As we assumed that the protocol is invariant under permutations of the
input systems, we can equivalently think of a protocol which starts
with the following symmetrization
step:\index{permutation-invariant~protocol} Alice chooses a
permutation $\pi$ at random and announces it to Bob, using the
(insecure) classical communication channel. Then Alice and Bob both
permute the order of their $N$ subsystems according to $\pi$.
Obviously, the state $\rho_{A^N B^N}$ of Alice and Bob's system after
this symmetrization step (averaged over all choices of $\pi$) is
invariant under permutations.\index{permutation-invariant~state}

Because the state $\rho_{A^N B^N}$ is invariant under permutations, it
has a purification $\rho_{A^N B^N E^N}$ (with an auxiliary system
$\cH_E^{\otimes N}$) which is symmetric as well (cf.\ 
Lemma~\ref{lem:sympurification}).\index{symmetric~purification} As the
pure state $\rho_{A^N B^N E^N}$ cannot be correlated with anything
else (cf.\ Section~\ref{sec:prodpur}) we can assume without loss of
generality that the knowledge of a potential adversary is fully
described by the auxiliary system.


\subsubsection{Step~2: Restriction to (almost) product states}



Because $\rho_{A^N B^N E^N}$\index{permutation-invariant~state} is
invariant under permutations, it is, according to our finite version
of the de Finetti representation\index{de~Finetti~representation}
theorem approximated by a mixture of states which have ``almost''
product form~$\sigma_{A B E}^{\otimes N}$\index{product~state}|in the
sense described by formula~\eqref{eq:sumconv}.


\subsubsection{Step~3: Smooth min-entropy of Alice and Bob's raw keys}

Assume for the moment that the joint initial state $\rho_{A^{N} B^{N}
  E^{N}}$ held by Alice, Bob, and Eve has perfect product form
$\sigma_{A B E}^{\otimes N}$.
As Alice and Bob's measurement operation (including the block-wise
processing) $\cE_{X Y \leftarrow A B}$ acts on $n$ blocks of size $b$
individually, the density operator $\rho_{X^n Y^n E^n}$ which
describes the situation before the information reconciliation step is
given by
\[
  \rho_{X^n Y^n E^n} 
= 
  (\cE_{X Y \leftarrow A B} \otimes \id_{E})^{\otimes n}
    (\rho_{A^{b n} B^{b n} E^{b n}}) \ ,
\]
where $X^n$ and $Y^n$ is Alice and Bob's raw key\index{raw~key},
respectively.  Consequently, $\rho_{X^n Y^n E^n}$ is the product of
operators of the form
\begin{equation} \label{eq:sigmaXYE}
  \sigma_{X Y E} 
= 
  (\cE_{X Y \leftarrow A B} \otimes \id_{E})(\sigma_{A B E}^{\otimes b}) \ .
\end{equation}
By~\eqref{eq:sumprod}, the smooth min-entropy of $\rho_{X^n E^n}$ is
approximated in terms of the von Neumann entropy of $\sigma_{X E}$,
i.e.,\index{smooth~min-entropy}
\begin{equation} \label{eq:Hminappr}
  \Hmin^\eps(X^n|E^n) 
\gtrapprox
  n H(X|E) \ . 
\end{equation}

Using~\eqref{eq:sumHsym}, this argument can easily be generalized to
states $\rho_{A^{N} B^{N} E^{N}}$ which have almost product form.

\subsubsection{Step~4: Smooth min-entropy after information reconciliation}

In the information reconciliation
step\index{information~reconciliation}, Alice sends error correcting
information $C$ about $X^n$ to Bob, using the authentic classical
communication channel. Eve might wiretap this communication which
generally decreases the smooth min-entropy of $X^n$ from her point of
view.

As mentioned above, we assume for this summary that the information
reconciliation subprotocol is optimal with respect to the amount of
information leaked\index{leakage} to Eve. It follows from classical
coding theory that the number of bits that Alice has to send to Bob in
order to allow him to compute her value $X^n$ is given by the Shannon
entropy of $X^n$ conditioned on Bob's knowledge $Y^n$. Formally, if
$\rho_{X^n Y^n}$ has product form $\sigma_{X Y}^{\otimes n}$ then the
communication $C$ satisfies
\begin{equation} \label{eq:prleak}
  \Hmax(C) - \Hmin(C|X^n) \approx n H(X|Y) \ , 
\end{equation}
where $H(X|Y)$ is the Shannon entropy of $X$ given $Y$, evaluated for
the probability distribution defined by $\sigma_{X Y}$.  (Note that
the entropy difference on the left hand side can be interpreted as a
measure for the information that $C$ gives on $X^n$.)

Let us now compute a lower bound on the smooth min-entropy of $X^n$
given Eve's knowledge after the information reconciliation step.  By
the chain rule~\eqref{eq:sumHchain}, we have
\[
  \Hmin^\eps(X^n|C E^n)
\geq
  \Hmin^\eps(X^n C | E^n) - \Hmax(C) \ .
\]
Moreover, because $C$ is computed from $X^n$, we can apply
inequality~\eqref{eq:sumHdataproc}, i.e.,
\[
  \Hmin^\eps(X^n C| E^n)
\geq
  \Hmin^\eps(X^n | E^n)  + \Hmin(C|X^n) \ .
\]
Combining this with~\eqref{eq:prleak} gives
\[
\begin{split}
  \Hmin^\eps(X^n|C E^n)
& \geq 
  \Hmin^\eps(X^n | E^n) - \bigl(\Hmax(C) - \Hmin(C|X^n) \bigr)  \\ 
& \approx
  \Hmin^\eps(X^n | E^n) -  n H(X|Y) \ .
\end{split}
\]
Finally, using the approximation~\eqref{eq:Hminappr} for
$\Hmin^\eps(X^n | E^n)$, we conclude
\begin{equation} \label{eq:Hminasym}
  \Hmin^\eps(X^n|C E^n)
\gtrapprox
  n H(X|E) - n H(X|Y) \ .
\end{equation}

\subsubsection{Step~5: Security of the key generated by privacy amplification}

To argue that the key generated in the final privacy
amplification\index{privacy~amplification} step is secure, we apply
criterion~\eqref{eq:sumsecPA}.  Because the adversary has access to
both the quantum system and the classical communication $C$, this
security criterion reads
\begin{equation} \label{eq:ellbound}
  \ell \lessapprox \Hmin^\eps(X^n|E^n C)
\end{equation}
where $\ell$ is the length of the key.

As shown in Step~2, the state $\rho_{A^N B^N E^N}$ has almost product
form $\sigma_{A B E}^{\otimes N}$. Hence, according
to~\eqref{eq:symstatsum}, the statistics obtained by Alice and Bob in
the parameter estimation\index{parameter~estimation} step corresponds
to the statistics that they would obtain if they started with a
perfect product state $\sigma_{A B}^{\otimes N}$.  We conclude that,
by the definition of the set $\Gamma$, the protocol aborts whenever
$\sigma_{A B} \notin \Gamma$.

To bound the smooth min-entropy of the string held by Alice before
privacy amplification, it thus suffices to
evaluate~\eqref{eq:Hminasym} for all states $\sigma_{A B}$ contained
in $\Gamma$. Formally,
\[
  \frac{1}{n} \Hmin^\eps(X^n|C E^n)
\gtrapprox
  \min_{\sigma_{A B E}} H(X|E) - H(X|Y) \ .
\]
where the minimum is over all (pure) states $\sigma_{A B E}$ such that
$\sigma_{A B} \in \Gamma$ and where $H(X|E)$ and $H(X|Y)$ are the
entropies of the state $\sigma_{X Y E}$ given by~\eqref{eq:sigmaXYE}.
Combining this with criterion~\eqref{eq:ellbound} concludes the proof.

\index{security~proof|)}



\chapter{Preliminaries} \label{ch:pre}

\section{Representation of physical systems} \label{sec:mathdecr}

\subsection{Density operators, measurements, and operations} \label{sec:qdefs}

\newcommand*{\NN}{\mathcal{P}}

Quantum mechanics\index{quantum~mechanics}, like any other physical
theory, allows us to make certain predictions about the behavior of
physical systems.  These are, however, not deterministic|a system's
initial state merely determines a probability distribution over all
possible outcomes of an observation.\footnote{With his famous
  statement ``Gott w\"urfelt nicht,'' Einstein expressed his doubts
  about the completeness of such a non-deterministic theory.}

Mathematically, the state\index{quantum~state} of a quantum mechanical
system with $d$ degrees of freedom is represented by a normalized
nonnegative\footnote{An operator $\rho$ on $\cH$ is
  \emph{nonnegative}\index{nonnegative~operator} if it is hermitian
  and has nonnegative eigenvalues.}  operator $\rho$, called
\emph{density operator}\index{density~operator}, on a $d$-dimensional
Hilbert space $\cH$.  The normalization is with respect to the trace
norm\index{trace~norm}, i.e., $\|\rho\|_1 = \tr(\rho) = 1$.  In the
following, we denote by $\NN(\cH)$\label{lb:nonneg} the set of
nonnegative operators on $\cH$, i.e., $\rho$ is a density operator on
$\cH$ if and only if $\rho \in \NN(\cH)$ and $\tr(\rho) = 1$.

Any observation of a quantum system corresponds to a
\emph{measurement}\index{measurement} and is represented
mathematically as a \emph{positive operator valued measure
  (POVM)}\index{positive~operator~valued~measure}\index{POVM|see{positive~operator~valued~measure}},
i.e., a family $\cM = \{M_w\}_{w \in \cW}$ of nonnegative operators
such that $\sum_{w \in \cW} M_w = \id_{\cH}$.  The theory of quantum
mechanics postulates that the probability distribution $P_W$ of the
outcomes when measuring a system in state $\rho$ with respect to $\cM$
is given by $P_W(w) := \tr(M_w \rho)$.

Consider a physical system whose state $\rhob^z$ depends on the value
$z$ of a classical random variable $Z$ with distribution $P_Z$. For an
observer which is ignorant of the value of $Z$, the state $\rho$ of
the system is given by the convex combination\footnote{Because a
  measurement is a linear mapping from the set of density operators to
  the set of probability distributions, this is consistent with the
  above description. In particular, the distribution of the outcomes
  resulting from a measurement of $\rho$ is the convex combination of
  the distributions obtained from measurements of $\rhob^z$.}
\begin{equation} \label{eq:summdec}
  \rho = \sum_{z \in \cZ} P_Z(z) \rhob^z \ .
\end{equation}

The decomposition~\eqref{eq:summdec} of a density operator $\rho$ is
generally not unique. Consider for example the \emph{fully mixed
  state}\index{fully~mixed~state} defined by $\rho :=
\frac{1}{\dim(\cH)} \id_{\cH}$. In the case of a two-level system,
$\rho$ might represent a photon which is polarized horizontally or
vertically with equal probabilities; but the same operator $\rho$
might also represent a photon which is polarized according to one of
the two diagonal directions with equal probabilities. In fact, the two
settings cannot be distinguished by any measurement.

A physical process is most generally described by a linear mapping
$\cE$, called a \emph{quantum operation}\index{quantum~operation},
which takes the system's initial state $\rho$ to its final state
$\rho'$.\footnote{A measurement can be seen as a special case of a
  quantum operation where the outcome is classical (see
  Section~\ref{sec:quantclass}).}  Mathematically, a quantum operation
$\cE$ is a \emph{completely positive map
  (CPM)}\index{completely~positive~map}\index{CPM|see{completely~positive~map}}\footnote{\emph{Complete
    positivity} means that any extension $\cE \otimes \id$ of the map
  $\cE$, where $\id$ is the identity map on the set of hermitian
  operators on some auxiliary Hilbert space $\cH''$, maps nonnegative
  operators to nonnegative operators.  Formally, $(\cE \otimes
  \id)(\rho) \in \NN(\cH' \otimes \cH'')$ for any $\rho \in \NN(\cH
  \otimes \cH'')$.} from the set of hermitian operators on a Hilbert
space $\cH$ to the set of hermitian operators on another Hilbert space
$\cH'$.  Additionally, in order to ensure that the image $\cE(\rho)$
of a density operator $\rho$ is again a density operator, $\cE$ must
be \emph{trace-preserving}\index{trace-preserving~map}, i.e.,
$\tr(\cE(\rho)) = \tr(\rho)$, for any $\rho \in \NN(\cH)$. It can be
shown that any CPM $\cE$ can be written as
\begin{equation} \label{eq:qop}
  \cE(\rho) = \sum_{w \in \cW} E_w \rho E_w^\dagger
\end{equation} 
where $\{E_w\}_{w \in \cW}$ is a family of linear operators from $\cH$
to $\cH'$. On the other hand, any mapping of the form~\eqref{eq:qop}
is a CPM.\footnote{This is in fact a direct consequence of
  Lemma~\ref{lem:nnprod}.} Moreover, it is trace-preserving if and
only if $\sum_{w \in \cW} E_w^\dagger E_w = \id_\cH$.

As we have seen, the state of a quantum system might depend on some
classical event $\Omega$ (e.g., that $Z$ takes a certain value $z$).
In this context, it is often convenient to represent both the
probability $\Pr[\Omega]$ of $\Omega$ and the state $\rhob^\Omega$ of
the system conditioned on $\Omega$ as one single mathematical object,
namely the nonnegative operator $\rho^{\Omega} := \Pr[\Omega] \cdot
\rhob^{\Omega}$.\footnote{The probability of the event $\Omega$ is
  then equal to the trace of $\rhob^\Omega$, i.e., $\Pr[\Omega] =
  \tr(\rho^{\Omega})$, and the system's state conditioned on $\Omega$
  is $\rhob^\Omega = \frac{1}{\Pr[\Omega]} \rho^{\Omega}$.}  For this
reason, we formulate most statements on quantum states in terms of
general (not necessarily normalized) nonnegative operators.
Similarly, we often consider general (not necessarily
trace-preserving) CPMs\index{completely~positive~map} $\cE$. The
quantity $\tr(\cE(\rho))$ can then be interpreted as the probability
that the process represented by $\cE$ occurs when starting with a
system in state $\rho$.



\subsection{Product systems and purifications} \label{sec:prodpur}

To analyze complex physical systems, it is often convenient to
consider a partitioning into a number of subsystems. This is
particularly useful if one is interested in the study of operations
that act on the parts of the system individually.\footnote{This is
  typically the case in the context of cryptography, where various
  parties control separated subsystems.}  Mathematically, the
partition of a quantum system into subsystems induces a product
structure on the underlying Hilbert space. For example, the state of a
bipartite system is represented as a density operator $\rho_{A B}$ on
a product space $\cH_A \otimes \cH_B$. The state of one part of a
product system is then obtained by taking the corresponding partial
trace of the overall state, e.g., $\rho_A = \tr_B(\rho_{A B})$ for the
first part of a bipartite system.

A density operator $\rho$ on $\cH$ is said to be
\emph{pure}\index{pure~state} if it has rank\footnote{The
  \emph{rank}\index{rank~of~an~operator} of a hermitian operator $S$,
  denoted $\rank(S)$, is the dimension of the
  \emph{support}\index{support~of~an~operator} $\im(S)$, i.e., the
  space spanned by the eigenvectors of $S$ with nonzero eigenvalues.}
one, that is, $\rho = \proj{\theta}$, for some $\ket{\theta} \in \cH$.
If it is normalized, $\rho$ is a projector\footnote{A hermitian
  operator $P$ is said to be a \emph{projector}\index{projector} if $P
  P = P$.} onto $\ket{\theta}$.  A pure density operator can only be
decomposed trivially, i.e., for any decomposition of the
form~\eqref{eq:qop}, $\rhob^z = \rho$ holds for all $z \in \cZ$.
According to the above interpretation, one could say that a pure state
contains no classical randomness, that is, it cannot be correlated
with any other system.

The fact that a pure state cannot be correlated with the environment
plays a crucial role in cryptography. It implies, for example, that
the randomness obtained from the measurement of a pure state is
independent of any other system and thus guaranteed to be secret.
More generally, let $\rho_A$ be an arbitrary operator on $\cH_A$ and
let $\rho_{A E}$ be a \emph{purification}\index{purification} of
$\rho_A$, i.e., $\rho_{A E}$ is a pure state on a product system
$\cH_A \otimes \cH_E$ such that $\tr_E(\rho_{A E}) = \rho_A$. Then,
because $\rho_{A E}$ is uncorrelated with any other system, the
partial system $\cH_E$ comprises everything that might possibly be
correlated with the system $\cH_A$ (including the knowledge of a
potential adversary).

\subsection{Quantum and classical systems} \label{sec:quantclass}

Consider a classical random variable $Z$ with distribution $P_Z$ on
some set $\cZ$. In a quantum world, it is useful to view $Z$ as a
special case of a quantum system\index{classical~system}.  For this,
one might think of the classical values $z \in \cZ$ as being
represented by orthogonal\footnote{The orthogonality of the states
  $\ket{z}$ guarantees that they can be distinguished perfectly, as
  this is the case for classical values.}  states $\ket{z}$ on some
Hilbert space $\cH_\cZ$. The state $\rho_Z$ of the quantum system is
then defined by
\begin{equation} \label{eq:fullclass}
  \rho_Z = \sum_{z \in \cZ} P_Z(z) \proj{z} \ .
\end{equation}
We say that $\rho_Z$ is the \emph{operator
  representation}\index{operator~representation} of the classical
distribution $P_Z$ \emph{(with respect to the basis $\{\ket{z}\}_{z
    \in \cZ}$)}.\footnote{This definition can easily be generalized to
  multi-partite nonnegative (not necessarily normalized) functions
  (e.g., $P_{X Y} \in \NN(\cX \times \cY)$, where $\NN(\cX \times
  \cY)$ denotes the set of nonnegative
  functions\index{nonnegative~function} on $\cX \times \cY$) in which
  case one gets nonnegative operators on product systems (e.g.,
  $\rho_{X Y} \in \NN(\cH_X \otimes \cH_Y)$).}

On the other hand, any operator $\rho_Z$ can be written in the
form~\eqref{eq:fullclass} where $P_Z(z)$ are the \emph{eigenvalues} of
$\rho_Z$ and $\ket{z}$ are the corresponding \emph{eigenvectors}. The
right hand side of~\eqref{eq:fullclass} is called the \emph{spectral
  decomposition}\index{spectral~decomposition} of $\rho_Z$. Moreover,
we say that $P_Z$ is the \emph{probability distribution defined by}
$\rho_Z$.

This notion can be extended to hybrid settings where the state
$\rhob_A^z$ of a quantum system $\cH_A$ depends on the value $z$ of a
classical random variable $Z$ (see, e.g., \cite{DevWin05}).
The joint state of the system is then given by
\begin{equation} \label{eq:classform}
  \rho_{A Z} = \sum_{z \in \cZ} \rho_A^z \otimes \proj{z} \ ,
\end{equation}
where $\rho_A^z := P_Z(z) \rhob_A^z$.

We can also go in the other direction: If a density operator has the
form~\eqref{eq:classform}, for some basis $\{\ket{z}\}_{z \in \cZ}$,
then the first subsystem can be interpreted as the representation of a
classical random variable $Z$.  This motivates the following
definition: An operator $\rho_{A Z} \in \NN(\cH_A \otimes \cH_Z)$ is
said to be \emph{classical with respect to $\{\ket{z}\}_{z \in
    \cZ}$}\index{classical~system} if there exists a family
$\{\rho_A^z\}_{z \in \cZ}$ of operators on $\cH_A$, called
\emph{(non-normalized) conditional
  operators}\index{conditional~state}, such that $\rho_{A Z}$ can be
written in the form~\eqref{eq:classform}.  Moreover, we say that
$\rho_{A Z}$ is \emph{classical on $\cH_Z$} if there exists a basis
$\{\ket{z}\}_{z \in \cZ}$ of $\cH_Z$ such that $\rho_{A Z}$ is
classical with respect to $\{\ket{z}\}_{z \in \cZ}$.\footnote{The
  operators $\rho_A^z$, for $z \in \cZ$, are uniquely defined by
  $\rho_{A Z}$ and the basis $\{\ket{z}\}_{z \in \cZ}$. Moreover,
  because $\rho_{A Z}$ is nonnegative, the operators $\rho_{A}^z$, for
  $z \in \cZ$, are also nonnegative.}


A similar definition can be used to characterize quantum operations
(i.e., CPMs)\index{quantum~operation}\index{completely~positive~map}
whose outcomes are partly classical: A CPM $\cE$ from $\cH$ to $\cH_A
\otimes \cH_Z$ is said to be \emph{classical with respect to
  $\{\ket{z}\}_{z \in \cZ}$}\index{classical~operation} (or simply
\emph{classical on $\cH_Z$}) if it can be written as
\[
  \cE(\sigma) 
= 
  \sum_{z \in \cZ} \cE^z(\sigma) \otimes \proj{z} \ ,
\]
where, for any $z \in \cZ$, $\cE^z$ is a CPM from $\cH$ to $\cH_A$.
Note that a measurement on $\cH$ with outcomes in $\cZ$ can be seen as
a CPM from $\cH$ to $\cH_Z$ which is classical on $\cH_Z$.


\subsection{Distance between states} \label{sec:tracedistintro}

Intuitively, we say that two states of a physical system are
\emph{similar} if any observation of them leads to identical results,
except with small probability. For two operators $\rho, \rho' \in
\NN(\cH)$ representing the state of a quantum system, this notion of
similarity is captured by the \emph{\distance{}}\index{\distance{}},
i.e., the trace norm\footnote{The \emph{trace norm}\index{trace~norm}
  $\| S \|_1$ of a hermitian operator $S$ on $\cH$ is defined by $\| S
  \|_1 := \tr(|S|)$.}  $\|\rho - \rho'\|_1$ of the difference between
$\rho$ and $\rho'$.\footnote{The \distance{} between two operators is
  closely related to the \emph{trace distance}\index{trace~distance},
  which is usually defined with an additional factor~$\frac{1}{2}$.}
The \distance{} for operators can be seen as the quantum version of
the \distance{} for probability distributions \footnote{The
  \distance{} between classical probability distributions is also
  known as \emph{variational distance}\index{variational~distance} or
  \emph{statistical distance}\index{statistical~distance} (which are
  often defined with an additional factor~$\frac{1}{2}$).} (or, more
generally, nonnegative functions), which is defined by $\|P - P'\|_1
:= \sum_{z} |P(z) - P'(z)|$, for $P, P' \in \NN(\cZ)$.  In particular,
if $\rho$ and $\rho'$ are operator
representations\index{operator~representation} of probability
distributions $P$ and $P'$, respectively, then the \distance{} between
$\rho$ and $\rho'$ is equal to the \distance{} between $P$ and $P'$.

Under the action of a quantum operation, the \distance{} between two
density operators $\rho$ and $\rho'$ cannot increase (cf.
Lemma~\ref{lem:distdecr}).  Because any measurement can be seen as a
quantum operation, this immediately implies that the distance $\|P -
P'\|_1$ between the distributions $P$ and $P'$ obtained from
(identical) measurements of two density operators $\rho$ and $\rho'$,
respectively, is bounded by $\|\rho - \rho'\|_1$.

The following proposition provides a very simple interpretation of the
\distance{}: If two probability distributions $P$ and $P'$ have
\distance{} at most $2 \eps$, then the two settings described by $P$
and $P'$, respectively, cannot differ with probability more
than~$\eps$.

\begin{proposition} \label{pro:vardistevent}
  Let $P, P' \in \NN(\cX)$ be probability distributions. Then there
  exists a joint distribution $P_{X X'}$ such that $P$ and $P'$ are
  the marginals of $P_{X X'}$ (i.e., $P = P_X$, $P' = P_{X'} $) and,
  for $(x,x')$ chosen according to $P_{X X'}$,
  \[
  \Pr_{(x,x')}[x \neq x'] \leq \frac{1}{2} \|P - P'\|_1  \ .
  \]
\end{proposition}

In particular, if the \distance{} between two states is bounded by $2
\eps$, then they cannot be distinguished with probability more than
$\eps$.

\section{Universal security of secret keys} \label{sec:univ}

\index{universal~security|(} 
\index{secret~key|(}
 
Cryptographic primitives (e.g., a secret key or an authentic
communication channel) are often used as components within a more
complex system. It is thus natural to require that the security of a
cryptographic scheme is not compromised when it is employed as part of
another system.  This requirement is captured by the notion of
\emph{universal security}.  Roughly
speaking, we say that a cryptographic primitive is \emph{universally
  secure} if it is secure in \emph{any} arbitrary context. For
example, the universal security of a secret key $S$ implies that any
bit of $S$ remains secret even if some other part of $S$ is given to
an adversary.

In the past few years, universal security has attracted a lot of
interest and led to important new definitions and proofs (see, e.g.,
the so-called \emph{universal
  composability}\index{universal~composability} framework of
Canetti~\cite{Canetti01} or Pfitzmann and Waidner~\cite{PfiWai00a}).
Recently, Ben-Or and Mayers~\cite{BenMay04} and Unruh~\cite{Unruh04}
have generalized Canetti's notion of universal composability to the
quantum world.

Universal security definitions are usually based on the idea of
characterizing the security of a \emph{real} cryptographic scheme by
its distance to an \emph{ideal} system which (by definition) is
perfectly secure.  For instance, a secret key $S$ is said to be secure
if it is close to a \emph{perfect key}\index{perfect~key} $U$, i.e., a
uniformly distributed string which is independent of the adversary's
information.  As we shall see, such a definition immediately implies
that any cryptosystem which is proven secure when using a perfect key
$U$\index{perfect~key} remains secure when $U$ is replaced by the
(real) key~$S$.

\subsection{Standard security definitions are not universal}

Unfortunately, many security definitions that are commonly used in
quantum cryptography are not universal. For instance, the security of
the key $S$ generated by a QKD scheme is typically defined in terms of
the mutual information $I(S;W)$ between $S$ and the classical outcome
$W$ of a measurement of the adversary's system (see, e.g.,
\cite{LoCha99,ShoPre00,NieChu00,GotLo03,LoChAr05} and also the
discussion in~\cite{BHLMO05} and~\cite{RenKoe05}).  Formally, $S$ is
said to be secure if, for some small $\eps$,
\begin{equation} \label{eq:defI}
  \max_{W} I(S;W) \leq \eps \ ,
\end{equation}
where the maximum ranges over all measurements on the adversary's
system with output $W$.  Such a definition|although it looks
reasonable|does, however, not guarantee that the key $S$ can safely be
used in applications. Roughly speaking, the reason for this flaw is
that criterion~\eqref{eq:defI} does not account for the fact that an
adversary might wait with the measurement of her system until she
learns parts of the key. (We also refer to~\cite{RenKoe05}
and~\cite{BHLMO05} for a more detailed discussion and an analysis of
existing security definitions with respect to this
concern.\footnote{Note that the conclusions in~\cite{BHLMO05} are
  somewhat different to ours: It is shown that existing privacy
  conditions of the form~\eqref{eq:defI} \emph{do} imply universal
  security, which seems to contradict the counterexample sketched
  below.  However, the result of~\cite{BHLMO05} only holds if the
  parameter $\eps$ in~\eqref{eq:defI} is exponentially small in the
  key size, which is not the case for most of the existing protocols.
  (In fact, the security parameter $\eps$ can only be made
  exponentially small at the expense of decreasing the key rate
  substantially.)})

Let us illustrate this potential problem with a concrete example:
Assume that we would like to use an $n$-bit key $S = (S_1, \ldots,
S_n)$ as a one-time pad\index{one-time~pad~encryption} to encrypt an
$n$-bit message $M = (M_1, \ldots, M_n)$.\footnote{That is, the
  ciphertext $C = (C_1, \ldots, C_n)$ is the bit-wise XOR of $S$ and
  $M$, i.e., $C_i = S_i \oplus M_i$.}  Furthermore, assume that an
adversary is interested in the $n$th bit $M_n$ of the message, but
already knows the first $n-1$ bits $M_1, \ldots, M_{n-1}$. Upon
observing the ciphertext, the adversary can easily
determine\footnote{Note that $S_i = M_i \oplus C_i$.} the first $n-1$
bits of $S$. Hence, in order to guarantee the secrecy of the $n$th
message bit $M_n$, we need to ensure that the adversary still has no
information on the $n$th key bit $S_n$, even though she already knows
all previous key bits $S_1, \ldots, S_{n-1}$. This requirement,
however, is not implied by the above definition. Indeed, for any
arbitrary $\eps>0$ and $n$ depending on $\eps$, it is relatively easy
to construct examples which satisfy~\eqref{eq:defI} whereas an
adversary|once she knows the first $n-1$ bits of the key|can determine
the $n$th bit $S_n$ with certainty. For an explicit construction and
analysis of such examples, we refer to~\cite{Bariska05}.\footnote{This
  phenomenon has also been studied in other contexts (see, e.g.,
  \cite{DHLST04,HLSW04}) where it is called as \emph{locking of
    classical correlation}\index{locking}.}


\subsection{A universal security definition} \label{sec:univdef}

Consider a key $S$ distributed according to $P_S$ and let $\rho_E^s$
be the state of the adversary's system given that $S$ takes the value
$s$, for any element $s$ of the \emph{key space}\index{key~space}
$\cS$.  According to the discussion in Section~\ref{sec:quantclass},
the joint state of the classical key $S$ and the adversary's quantum
system can be represented by the density operator
\[
  \rho_{S E} := \sum_{s \in \cS} P_S(s) \proj{s} \otimes \rho_E^s \ ,
\]
where $\{\ket{s}\}_{s \in \cS}$ is an orthonormal basis of some
Hilbert space $\cH_S$. We say that $S$ is \emph{$\eps$-secure with
  respect to $\cH_E$}\index{$\eps$-security} if
\begin{equation} \label{eq:secdef}
  \sfrac{1}{2} \bigl\| \rho_{S E} - \rho_U \otimes \rho_E \bigr\|_1 \leq \eps \ ,
\end{equation}
where $\rho_U = \sum_{s \in \cS} \frac{1}{|\cS|} \proj{s}$ is the
fully mixed state on $\cH_S$.

The universal security of a key $S$ satisfying this definition follows
from a simple argument: Criterion~\eqref{eq:secdef} guarantees that
the \emph{real} situation described by $\rho_{S E}$ is
$\eps$-close|with respect to the \distance{}|to an \emph{ideal}
situation where $S$ is replaced by a perfect key\index{perfect~key}
$U$ which is uniformly distributed and independent of the state of the
system $\cH_E$. Moreover, since the \distance{} cannot increase when
applying a quantum operation (cf.~Lemma~\ref{lem:distdecr}), this also
holds for any further evolution of the world (where, e.g., the key is
used as part of a larger cryptographic system). In fact, it follows
from the discussion in Section~\ref{sec:tracedistintro} that an
$\eps$-secure key can be considered \emph{identical} to an
\emph{ideal} (perfect) key\index{perfect~key}|except with
probability~$\eps$.\footnote{For this statement to hold, it is crucial
  that the criterion~\eqref{eq:secdef} is formulated in terms of the
  \distance{} (instead of other distance measures such as the
  fidelity).}  In particular, an $\eps$-secure\index{$\eps$-security}
key is secure within any reasonable framework providing universal
composability (e.g., \cite{BenMay04}
or~\cite{Unruh04}).\footnote{These frameworks are usually based on the
  so-called \emph{simulatability
    paradigm}\index{simulatability~paradigm}. That is, a real
  cryptosystem is said to be \emph{as secure as} an ideal cryptosystem
  if any attack to the real scheme can be \emph{simulated} by an
  attack to the ideal scheme (see also~\cite{MaReHo04}).  It is easy
  to see that our security criterion is compatible with this paradigm:
  Consider a (real) key agreement protocol and assume that, for any
  possible attack of the adversary, the final key
  satisfies~\eqref{eq:secdef}.  The adversary's quantum state after
  the attack is then almost independent of the key, that is, the
  adversary could simulate virtually all her information without even
  interacting with the cryptosystem. The real key agreement protocol
  is thus \emph{as secure as} an ideal key agreement scheme which, by
  definition, does not leak any information at all.}

The security of a key according to~\eqref{eq:secdef} also implies
security with respect to most of the standard security definitions in
quantum cryptography. For example, if $S$ is $\eps$-secure with
respect to $\cH_E$ then the mutual information between $S$ and the
outcome of any measurement applied to the adversary's system is small
(whereas the converse is often not true, as discussed above).  In
particular, if the adversary is purely classical, \eqref{eq:secdef}
reduces to a classical security definition which has been proposed in
the context of information-theoretically secure key agreement (see,
e,g., \cite{DziMau04}).

\index{universal~security|)}
\index{secret~key|)}

\chapter{(Smooth) Min- and Max-Entropy} \label{ch:smooth}

Entropy measures are indispensable tools in classical and quantum
information theory. They quantify \emph{randomness}, that is, the
uncertainty that an observer has on the state of a (quantum) physical
system.  In this chapter, we introduce two entropic quantities, called
\emph{smooth~min-entropy}\index{smooth~min-entropy} and
\emph{smooth~max-entropy}\index{smooth~max-entropy}. As we shall see,
these are useful to characterize randomness with respect to
fundamental information-theoretic tasks such as the extraction of
uniform randomness or data compression.\footnote{\emph{Randomness
    extraction}\index{randomness~extraction} is actually privacy
  amplification and is the topic of Chapter~\ref{ch:PA}. \emph{Data
    compression}\index{data~compression} is closely related to
  information reconciliation which is treated in
  Section~\ref{sec:IR}.}  Moreover, smooth min- and max-entropies have
natural properties which are similar to those known from the von
Neumann entropy and its classical special case, the Shannon
entropy\footnote{The \emph{Shannon entropy}\index{Shannon~entropy} of
  a probability distribution $P$ is defined by $H(P) := - \sum_{x}
  P(x) \log P(x)$, where $\log$\label{lb:binlog} denotes the binary
  logarithm.  Similarly, the \emph{von Neumann
    entropy}\index{von~Neumann~entropy} of a density operator $\rho$
  is $H(\rho) := -\tr(\rho \log \rho )$.}
(Sections~\ref{sec:convRenyi} and~\ref{sec:smoothRenyi}). In fact, for
product states\index{product~state}, smooth min- and max-entropy are
asymptotically equal to the von Neumann entropy
(Section~\ref{sec:smoothprod}).

Smooth min- and max-entropies are actually families of entropy
measures parameterized by some nonnegative real number $\eps$, called
\emph{smoothness}\index{smoothness}. In applications, the smoothness
is related to the error probability of certain information-theoretic
tasks and is thus typically chosen to be small. We first consider the
``non-smooth'' special case where $\eps = 0$
(Section~\ref{sec:convRenyi}). This is the basis for the general
definition where the smoothness $\eps$ is arbitrary
(Section~\ref{sec:smoothRenyi}).

We will introduce a \emph{conditional} version of smooth min- and
max-entropy. It is defined for bipartite operators $\rho_{A B}$ on
$\cH_A \otimes \cH_B$ and measures the uncertainty on the state of the
subsystem $\cH_A$ given access to the subsystem $\cH_B$. Unlike the
\emph{conditional von Neumann
  entropy}\index{conditional~von~Neumann~entropy} $H(A|B) := H(\rho_{A
  B}) - H(\rho_B)$, however, it cannot be written as a difference
between two ``unconditional'' entropy measures.

To illustrate our definition of (conditional) min- and max-entropy,
let us, as an analogy, consider an alternative formulation of the
conditional von Neumann entropy $H(A|B)$. Let
\begin{equation} \label{eq:vNtwo}
  H(\rho_{A B}|\sigma_{B})
:=
  -\tr\bigl(\rho_{A B} 
    (\log \rho_{A B} - \log \id_A \otimes \sigma_B)\bigr) \ ,
\end{equation}
for some state $\sigma_B$ on $\cH_B$. This quantity can be rewritten
as
\[
  H(\rho_{A B}|\sigma_{B})
=
  H(\rho_{A B}) - H(\rho_B) - D(\rho_{B}\|\sigma_B) \ ,
\]
where $D(\rho_B\|\sigma_B)$ is the relative entropy\footnote{The
  \emph{relative entropy}\index{relative~entropy} $D(\rho\|\sigma)$ is
  defined by $D(\rho\|\sigma) := \tr(\rho \log \rho) - \tr(\rho \log
  \sigma)$.}  of $\rho_B$ to $\sigma_B$.  Because
$D(\rho_B\|\sigma_B)$ cannot be negative, this expression takes its
maximum for $\sigma_B = \rho_B$, in which case it is equal to
$H(A|B)$.  We thus have
\begin{equation} \label{eq:vNsup}
  H(A|B)
=
  \sup_{\sigma_B} H(\rho_{A B}|\sigma_B) \ ,
\end{equation}
where the supremum ranges over all density operators $\sigma_B$ on
$\cH_B$. 

The definitions of (smooth) min- and max-entropies are inspired by
this approach. We first introduce a quantity which corresponds
to~\eqref{eq:vNtwo} (cf.\ Definitions~\ref{def:minmaxentr}
and~\ref{def:smoothminmaxentr}) and then define our entropy measures
by a formula of the form~\eqref{eq:vNsup}
(Definitions~\ref{def:Hminmaxentrgen}
and~\ref{def:smoothminmaxentrgen}).

\section{Min- and max-entropy} \label{sec:convRenyi}

\index{min-entropy|(}
\index{max-entropy|(}

This section introduce a ``non-smooth'' version of min- and
max-entropy. It is the basis for the considerations in
Section~\ref{sec:smoothRenyi}, where these entropy measures are
generalized.  The focus is on min-entropy, which is used extensively
in the remaining part of the thesis.  However, most of the properties
derived in the following also hold for max-entropy.

\subsection{Definition of min- and max-entropy}

\begin{definition} \label{def:minmaxentr}
  Let $\rho_{A B} \in \NN(\cH_A \otimes \cH_B)$ and $\sigma_B \in
  \NN(\cH_B)$. The \emph{min-entropy of $\rho_{A B}$ relative to
    $\sigma_B$} is
  \[
    \Hmin(\rho_{A B}|\sigma_B)
  :=
    - \log \lambda
  \]
  where $\lambda$ is the minimum real number such that $\lambda \cdot
  \id_A \otimes \sigma_B - \rho_{A B}$ is nonnegative.  The
  \emph{max-entropy of $\rho_{A B}$ relative to $\sigma_B$} is
  \[
    \Hmax(\rho_{A B}|\sigma_B)
  :=
    \log \tr\bigl((\id_{A} \otimes \sigma_B) \rho_{A B}^0\bigr)
  \]
  where $\rho_{A B}^0$ denotes the projector onto the support of
  $\rho_{A B}$.
\end{definition}


\begin{definition} \label{def:Hminmaxentrgen}
  Let $\rho_{A B} \in \NN(\cH_A \otimes \cH_B)$. The
  \emph{min-entropy} and the \emph{max-entropy of $\rho_{A B}$ given
    $\cH_B$} are
  \begin{align*}
    \Hmin(\rho_{A B}|B) & := \sup_{\sigma_B} \Hmin(\rho_{A B}|\sigma_B) \\
    \Hmax(\rho_{A B}|B) & := \sup_{\sigma_B} \Hmax(\rho_{A B}|\sigma_B) \ ,
  \end{align*}
  respectively, where the supremum ranges over all $\sigma_B \in
  \NN(\cH_B)$ with $\tr(\sigma_B) = 1$.
\end{definition}

\begin{remark} \label{rem:Hminlambda}
  It follows from Lemma~\ref{lem:maxcond} that the min-entropy of
  $\rho_{A B}$ relative to $\sigma_B$, for $\sigma_B$ invertible, can
  be written as
\[
  \Hmin(\rho_{A B}|\sigma_B)
=
  - \log \lambda_{\max}\bigl( 
      (\id_A \otimes \sigma_B^{-1/2}) \rho_{A B} 
     (\id_A \otimes \sigma_B^{-1/2})
  \bigr) \ ,
\]
where $\lambda_{\max}(\cdot)$ denotes the maximum eigenvalue of the
argument.
\end{remark}

If $\cH_B$ is the trivial space $\bbC$, we simply write
$\Hmin(\rho_{A})$ and $\Hmax(\rho_{A})$ to denote the min- and the
max-entropy of $\rho_A$, respectively. In particular,
\begin{align*}
  \Hmin(\rho_A) & = - \log \lambda_{\max}(\rho_A) \\
  \Hmax(\rho_A) & = \log \rank(\rho_A) \ .
\end{align*}


\subsubsection{The classical analogue}

\index{classical~system|(}
 
The above definitions can be specialized canonically to classical
probability distributions.\footnote{Similarly, the Shannon entropy can
  be seen as the classical special case of the von Neumann entropy.}
More precisely, for $P_{X Y} \in \NN(\cX \times \cY)$ and $Q_Y \in
\NN(\cY)$, we have
  \begin{align*}
    \Hmin(P_{X Y} | Q_Y) & := \Hmin(\rho_{X Y} | \sigma_Y) \\
    \Hmax(P_{X Y} | Q_Y) & := \Hmax(\rho_{X Y} | \sigma_Y)  
  \end{align*}
  where $\rho_{X Y}$ and $\sigma_Y$ are the operator
  representations\index{operator~representation} of $P_{X Y}$ and
  $Q_Y$, respectively (cf.  Section~\ref{sec:quantclass}).

\begin{remark} \label{rem:classminmaxentr}
  Let $P_{X Y} \in \NN(\cX \times \cY)$ and $Q_Y \in \NN(\cY)$.
  Then\footnote{The \emph{support}\index{support~of~a~function} of a
    nonnegative function $f \in \NN(\cX)$, denoted $\supp(f)$, is the
    set of values $x \in \cX$ such that $f(x) > 0$.}
\begin{align*}
    \Hmin(P_{X Y}|Q_Y)
  & =
    - \log \max_{y \in \supp(Q_Y)} 
      \max_{x \in \cX} \frac{P_{X Y}(x, y)}{Q_Y(y)} 
  \\
    \Hmax(P_{X Y}|Q_Y) 
  & = 
    \log \sum_{y \in \cY} Q_Y(y) 
      \cdot \bigl| \supp(P^y_{X}) \bigr| \ ,
\end{align*}
where $P^y_X$ denotes the function $P^y_X: \, x \mapsto P_{X Y}(x,y)$.
In particular,
\[
  \Hmax(P_{X Y}|Y)
=
  \log \max_{y \in \cY} \bigl|\supp(P^y_{X}) \bigr| \ .
\]
\end{remark}

\index{classical~system|)}

\subsection{Basic properties of min- and max-entropy}
  
\subsubsection{Min-entropy cannot be larger than max-entropy}

The following lemma gives a relation between min- and max-entropy. It
implies that, for a density operator $\rho_{A B}$, the min-entropy
cannot be larger than the max-entropy.

\begin{lemma} \label{lem:Hminmax}
  Let $\rho_{A B} \in \NN(\cH_A \otimes \cH_B)$ and $\sigma_B \in
  \NN(\cH_B)$. Then
  \[
    \Hmin(\rho_{A B}|\sigma_B) + \log \tr(\rho_{A B})
  \leq
    \Hmax(\rho_{A B}|\sigma_B) \ .
  \]
\end{lemma}

\begin{proof}
  Let $\rho_{A B}^0$ be the projector onto the support of $\rho_{A B}$
  and let $\lambda \geq 0$ such that $\Hmin(\rho_{A B}|\sigma_B) = -
  \log \lambda$, i.e., $\lambda \cdot \id_A \otimes \sigma_B - \rho_{A
    B}$ is nonnegative.  Using the fact that the trace of the product
  of two nonnegative operators is nonnegative
  (Lemma~\ref{lem:trprod}), we have
  \[
    \tr\bigl(\lambda \cdot (\id_A \otimes \sigma_B) \rho_{A B}^0\bigr)
  -
    \tr(\rho_{A B}) \\
  =
    \tr\bigl((\lambda \cdot \id_A \otimes \sigma_B  - \rho_{A B}) 
      \rho_{A B}^0\bigr)
  \geq
    0 \ .
  \]
  Hence,
  \[
    \log \tr\bigl((\id_A \otimes \sigma_B) \rho_{A B}^0\bigr) 
  \geq
    \log \tr(\rho_{A B}) - \log \lambda \ .
  \]
  The assertion then follows by the definition of the max-entropy and
  the choice of~$\lambda$.
\end{proof}

\subsubsection{Additivity of min- and max-entropy}

\index{additivity|(}

The von Neumann entropy of a state which consists of two independent
parts is equal to the sum of the entropies of each part, i.e.,
$H(\rho_{A} \otimes \rho_{A'}) = H(\rho_{A}) + H(\rho_{A'})$.  This
also holds for min- and max-entropy.
 
\begin{lemma} \label{lem:HRindadd}
  Let $\rho_{A B} \in \NN(\cH_A \otimes \cH_B)$, $\sigma_B \in
  \NN(\cH_B)$ and, similarly, $\rho_{A' B'} \in \NN(\cH_{A'} \otimes
  \cH_{B'})$, $\sigma_{B'} \in \NN(\cH_{B'})$. Then
  \begin{align*}
    \Hmin(\rho_{A B} \otimes \rho_{A' B'} | \sigma_{B} \otimes \sigma_{B'})
  & =
    \Hmin(\rho_{A B} | \sigma_B) + \Hmin(\rho_{A' B'} | \sigma_{B'}) 
\\
    \Hmax(\rho_{A B} \otimes \rho_{A' B'} | \sigma_{B} \otimes \sigma_{B'})
  & =
    \Hmax(\rho_{A B} | \sigma_B) + \Hmax(\rho_{A' B'} | \sigma_{B'}) 
    \ .
  \end{align*}
\end{lemma}

\begin{proof}
  The statement follows immediately from
  Definition~\ref{def:minmaxentr}.
\end{proof}

\index{additivity|)}

\subsubsection{Strong subadditivity}

\index{strong~subadditivity|(}

The von Neumann entropy is subadditive, i.e., $H(A|B C) \leq H(A |
B)$, which means that the entropy cannot increase when conditioning on
an additional subsystem.  This property can be generalized to min- and
max-entropy.

\begin{lemma} \label{lem:Hinfcondclasse}
  Let $\rho_{A B C} \in \NN(\cH_A \otimes \cH_B \otimes \cH_C)$ and
  $\sigma_{B C} \in \NN(\cH_B \otimes \cH_C)$. Then
  \begin{align*}
    \Hmin(\rho_{A B C}|\sigma_{B C})
  & \leq
    \Hmin(\rho_{A B}|\sigma_B) 
  \\
    \Hmax(\rho_{A B C}|\sigma_{B C}) 
  & \leq 
    \Hmax(\rho_{A B}|\sigma_B) \ .
  \end{align*}
\end{lemma}

Note that, for min-entropy, the statement follows directly from the
more general fact that the entropy cannot decrease under certain
quantum operations (cf.\ Lemma~\ref{lem:Hinfclassop}).

\begin{proof}
  Let $\lambda \geq 0$ such that $-\log \lambda = \Hmin(\rho_{A B
    C}|\sigma_{B C})$, i.e., $\lambda \cdot \id_A \otimes \sigma_{B C}
  - \rho_{A B C}$ is nonnegative. Because the operator obtained by
  taking the partial trace of a nonnegative operator is nonnegative,
  $\lambda \cdot \id_A \otimes \sigma_{B} - \rho_{A B}$ is also
  nonnegative.  This immediately implies $- \log \lambda \leq
  \Hmin(\rho_{A B}|\sigma_B)$ and thus concludes the proof of the
  statement for min-entropy.
  
  To show that the assertion also holds for max-entropy, let $\rho_{A
    B}^0$ and $\rho_{A B C}^0$ be the projectors on the support of
  $\rho_{A B}$ and $\rho_{A B C}$, respectively. Because the support
  of $\rho_{A B C}$ is contained in the tensor product of the support
  of $\rho_{A B}$ and $\cH_C$ (cf.\ Lemma~\ref{lem:tensprodimage}),
  the operator $\rho_{A B}^0 \otimes \id_C - \rho_{A B C}^0$ is
  nonnegative. Moreover, because the trace of the product of two
  nonnegative operators is nonnegative (cf.\ Lemma~\ref{lem:trprod}),
  we find
  \begin{multline*}
    \tr\bigl((\id_A \otimes \sigma_{B}) \rho_{A B}^0\bigr)
  -
    \tr\bigl((\id_{A} \otimes \sigma_{B C} ) \rho_{A B C}^0\bigr) \\
  =
    \tr\bigl((\id_{A} \otimes \sigma_{B C})
      (\rho_{A B}^0 \otimes \id_C - \rho_{A B C}^0) \bigr)
  \geq
    0 \ .
  \end{multline*}
  The assertion then follows by the definition of the max-entropy.
\end{proof}

Note that the strong subadditivity of the max-entropy together with
Lemma~\ref{lem:Hminmax} implies that $\Hmin(\rho_{A B}|\sigma_B) \leq
\Hmax(\rho_A)$, for density operators $\rho_{A B}$ and $\sigma_B$.

\index{strong~subadditivity|)}

\subsubsection{Conditioning on classical information}

\index{classical~system|(}

The min- and max-entropies of states which are partially classical can
be expressed in terms of the min- and max-entropies of the
corresponding conditional operators (see
Section~\ref{sec:quantclass}).

\begin{lemma} \label{lem:Hminclasscondr}
  Let $\rho_{A B Z} \in \NN(\cH_A \otimes \cH_B \otimes \cH_Z)$ and
  $\sigma_{B Z} \in \NN(\cH_B \otimes \cH_Z)$ be classical with
  respect to an orthonormal basis $\{\ket{z}\}_{z \in \cZ}$ of
  $\cH_Z$, and let $\rho_{A B}^z$ and $\sigma_B^z$ be the
  corresponding (non-normalized) conditional operators. Then
  \begin{align*}
    \Hmin(\rho_{A B Z}|\sigma_{B Z})
  & =
    \inf_{z \in \cZ} \Hmin(\rho_{A B}^z|\sigma_{B}^z) 
  \\
    \Hmax(\rho_{A B Z}|\sigma_{B Z}) 
   & = 
    \log \sum_{z \in \cZ} 2^{\Hmax(\rho_{A B}^z|\sigma_B^z)} \ .
  \end{align*}
\end{lemma}

\begin{proof}
  Because the vectors $\ket{z}$ are mutually orthogonal, the equivalence
  \begin{multline} \label{eq:classcondeq}
      \lambda \cdot \id_A \otimes \sigma_{B Z} - \rho_{A B Z} 
    \in \NN(\cH_A \otimes \cH_B \otimes \cH_Z) \\
  \iff
    \forall z \in \cZ : \, 
      \lambda \cdot \id_A \otimes \sigma_B^z - \rho_{A B}^z 
    \in \NN(\cH_A \otimes \cH_B)
  \end{multline}
  holds for any $\lambda \geq 0$.  The assertion for the min-entropy
  then follows from the fact that the negative logarithm of the
  minimum $\lambda$ satisfying the left hand side and the right hand
  side of~\eqref{eq:classcondeq} are equal to the quantities
  $\Hmin(\rho_{A B Z}|\sigma_{B Z})$ and $\inf_{z \in \cZ}
  \Hmin(\rho_{A B}^z|\sigma_{B}^z)$, respectively.
  
  To prove the statement for the max-entropy, let $\rho_{A B Z}^0$ and
  $(\rho_{A B}^z)^0$, for $z \in \cZ$, be projectors onto the support
  of $\rho_{A B Z}$ and $\rho_{A B}^z$, respectively. Because the
  vectors $\ket{z}$ are mutually orthogonal, we have
  \[
    \rho_{A B Z}^0 = \sum_{z \in \cZ} (\rho_{A B}^z)^0 \otimes \proj{z} \ ,
  \]
  and thus
  \[
    \tr\bigl( (\id_A \otimes \sigma_{B Z}) \rho_{A B Z}^0 \bigr)
  =
    \sum_{z \in \cZ} 
      \tr\bigl( (\id_A \otimes \sigma_{B}^z) (\rho_{A B}^z)^0 \bigr) \ .
  \]
  The assertion then follows by the definition of the max-entropy.
\end{proof}

\index{classical~system|)}

\subsubsection{Classical subsystems have nonnegative min-entropy}

\index{classical~system|(}

Similarly to the conditional von Neumann entropy, the min- and
max-entropies of entangled systems can generally be negative. This is,
however, not the case for the entropy of a classical subsystem.
Lemma~\ref{lem:classnn} below implies that
\[
  \Hmin(\rho_{X C}|\rho_C)
\geq
  0 \ ,
\]
for any density operator $\rho_{X C}$ which is classical on the first
subsystem\footnote{To see this, let $\cH_B$ be the trivial space
  $\bbC$ and set $\sigma_C = \rho_C$.}.  By Lemma~\ref{lem:Hminmax},
the same holds for max-entropy.

\begin{lemma} \label{lem:classnn}
  Let $\rho_{X B C} \in \NN(\cH_X \otimes \cH_B \otimes \cH_C)$ be
  classical on $\cH_X$ and let $\sigma_C \in \NN(\cH_C)$.  Then
  \[
    \Hmin(\rho_{X B C}|\sigma_C) \geq \Hmin(\rho_{B C}|\sigma_C) \ .
  \]
\end{lemma}

\begin{proof}
  Let $\lambda \geq 0$ such that $-\log \lambda = \Hmin(\rho_{B
    C}|\sigma_C)$. Because $\rho_{X B C}$ is classical on $\cH_X$,
  there exists an orthonormal basis $\{\ket{x}\}_{x \in \cX}$ and a
  family $\{\rho_{B C}^x\}_{x \in \cX}$ of operators on $\cH_B \otimes
  \cH_C$ such that $\rho_{X B C} = \sum_{x \in \cX} \proj{x} \otimes
  \rho_{B C}^x$. By the definition of $\lambda$, the operator
  \[
    \lambda \cdot \id_B \otimes \sigma_C - \sum_{x \in \cX} \rho_{B C}^x
  =
    \lambda \cdot \id_B \otimes \sigma_C - \rho_{B C}
  \]
  is nonnegative. Hence, for any $x \in \cX$, the operator $\lambda
  \cdot \id_B \otimes \sigma_C - \rho_{B C}^x$ must also be
  nonnegative. This implies that the operator
  \[
    \lambda \cdot \id_{X B} \otimes \sigma_C - \rho_{X B C}
  =
    \sum_{x \in \cX}
      \lambda \cdot \proj{x} \otimes \id_B \otimes \sigma_C
      - \proj{x} \otimes \rho_{B C}^x 
  \]
  is nonnegative as well. We thus have $-\log \lambda \leq
  \Hmin(\rho_{X B C}|\sigma_C)$, from which the assertion follows.
\end{proof}

\index{classical~system|)}

\index{max-entropy|)}

\subsection{Chain rules for min-entropy}

\index{chain~rule|(}

The chain rule for the von Neumann entropy reads $H(A B | C) = H(A | B
C) + H(B|C)$. In particular, since $H(B|C)$ cannot be larger than
$H(B)$, we have $H(A B | C) \leq H(A |B C) + H(B)$.  The following
lemma implies that a similar statement holds for min-entropy, namely,
\[
  \Hmin(\rho_{A B C}|C) \leq \Hmin(\rho_{A B C} | B C) + \Hmax(\rho_B) \ .
\]

\begin{lemma} \label{lem:Halphachain}
  Let $\rho_{A B C} \in \NN(\cH_A \otimes \cH_B \otimes \cH_C)$,
  $\sigma_C \in \NN(\cH_C)$, and let $\sigma_B \in \NN(\cH_B)$ be the
  fully mixed state on the support of $\rho_B$. Then
  \[
    \Hmin(\rho_{A B C} | \sigma_C) 
  =
    \Hmin(\rho_{A B C} | \sigma_B \otimes \sigma_C) + \Hmax(\rho_B) \ .
  \]
\end{lemma}

\begin{proof}
  Let $\cH_{B'} := \im(\rho_B)$ be the support of $\rho_B$ and let
  $\lambda \geq 0$. The operator $\sigma_B$ can then be written as
  $\sigma_B = \frac{1}{\rank(\rho_B)} \id_{B'}$, where $\id_{B'}$ is
  the identity on $\cH_{B'}$.  Hence, because the support of $\rho_{A
    B C}$ is contained in $\cH_A \otimes \cH_{B'} \otimes \cH_C$ (cf.
  Lemma~\ref{lem:tensprodimage}), the operator $\lambda \cdot \id_{A}
  \otimes \sigma_B \otimes \sigma_C - \rho_{A B C}$ is nonnegative if
  and only if the operator $\lambda \cdot \frac{1}{\rank(\rho_B)}
  \cdot \id_{A} \otimes \id_{B} \otimes \sigma_C - \rho_{A B C}$ is
  nonnegative.  The assertion thus follows from the definition of the
  min-entropy and the fact that $\Hmax(\rho_B) = \log \rank(\rho_B)$.
\end{proof}

\index{chain~rule|)}

\subsubsection{Data processing}

\index{data~processing|(}

Let $A$, $Y$, and $C$ be random variables such that $A \leftrightarrow
Y \leftrightarrow C$ is a \emph{Markov chain}\index{Markov~chain},
i.e., the conditional probability distributions $P_{A C|Y=y}$ have
product form $P_{A|Y=y} \times P_{C|Y=y}$.  The uncertainty on $A$
given $Y$ is then equal to the uncertainty on $A$ given $Y$ and $C$,
that is, in terms of Shannon entropy, $H(A|Y) = H(A|Y C)$. Hence, by
the chain rule, we get the equality $H(A Y| C) = H(Y|C) + H(A|Y)$.

The same equality also holds for quantum states $\rho_{A Y C}$ on
$\cH_A \otimes \cH_Y \otimes \cH_C$ which are classical on $\cH_Y$ and
where, analogously to the Markov condition, the conditional density
operators $\rhob_{A C}^y$ have product form, i.e., $\rhob_{A C}^y =
\rhob_A^y \otimes \rhob_C^y$. The following lemma generalizes this
statement to min-entropy.

\begin{lemma} \label{lem:Hclasschain}
  Let $\rho_{A Y C} \in \NN(\cH_A \otimes \cH_Y \otimes \cH_C)$ be
  classical with respect to an orthonormal basis $\{\ket{y}\}_{y \in
    \cY}$ of $\cH_Y$ such that the corresponding conditional operators
  $\rho_{A C}^y$, for any $y \in \cY$, have product form and let
  $\sigma_C \in \NN(\cH_C)$.  Then
  \[
    \Hmin(\rho_{A Y C} | \sigma_C) 
  \geq
    \Hmin(\rho_{Y C} | \sigma_C)
    + \Hmin(\rho_{A Y} | \rho_Y) \ .
  \]
\end{lemma}

\begin{proof}
  For any $y \in \cY$, let $p_y := \tr(\rho_{A C}^y)$ and let
  $\rhob_{A C}^y := \frac{1}{p_y} \rho_{A C}^y$ be the normalization
  of $\rho_{A C}^y$. The operator $\rho_{A Y C}$ can then be written
  as
  \[
    \rho_{A Y C} 
  = 
    \sum_{y \in \cY} 
      p_y \cdot \rhob_A^y \otimes \proj{y} \otimes \rhob_C^y \ .
  \]
  Let $\lambda, \lambda' \geq 0$ such that $-\log \lambda =
  \Hmin(\rho_{Y C} | \sigma_C)$, $-\log \lambda' = \Hmin(\rho_{A Y} |
  \rho_Y)$.  Because the vectors $\ket{y}$ are mutually orthogonal, it
  follows immediately from the definition of the min-entropy that the
  operators $\lambda \cdot \sigma_C - p_y \cdot \rhob_{C}^y$ and
  $\lambda' \cdot \id_A - \rhob_{A}^y$ are nonnegative, for any $y \in
  \cY$.  Consequently, the operator
  \begin{multline*}
    \lambda \cdot \lambda' \cdot \id_A
      \otimes \id_Y \otimes \sigma_C
    - \rho_{A Y C} \\
  = 
    \sum_{y \in \cY} 
      \lambda \cdot \lambda' 
      \cdot \id_A \otimes \proj{y} \otimes \sigma_C 
       - p_y \cdot \rhob_A^y \otimes \proj{y} \otimes \rhob_C^y 
  \end{multline*}
  is nonnegative as well.  This implies
  \[
    \Hmin(\rho_{A Y C}|\sigma_{C})
  \geq 
    -\log (\lambda \cdot \lambda') 
  =
    - \log\lambda - \log \lambda' 
  \]
  from which the assertion follows by the definition of $\lambda$ and
  $\lambda'$. 
\end{proof}

\index{data~processing|)}

\subsection{Quantum operations can only increase min-entropy}

The min-entropy can only increase when applying quantum operations.
Because the partial trace is a quantum operation, this general
statement also implies the first assertion of
Lemma~\ref{lem:Hinfcondclasse} (strong
subadditivity\index{strong~subadditivity}).

\begin{lemma} \label{lem:Hinfclassop}
  Let $\rho_{A B} \in \NN(\cH_A \otimes \cH_B)$, $\sigma_{B} \in
  \NN(\cH_B)$, $\sigmat_{B'} \in \NN(\cH_{B'})$ and let $\cE$ be a CPM
  from $\cH_A \otimes \cH_B$ to $\cH_{A'} \otimes \cH_{B'}$ such that
  $\id_{A'} \otimes \sigmat_{B'} - \cE(\id_A \otimes \sigma_B)$ is
  nonnegative.  Then, for $\rhot_{A' B'} := \cE(\rho_{A B})$,
  \[
    \Hmin(\rhot_{A' B'}|\sigmat_{B'})
  \geq
    \Hmin(\rho_{A B}|\sigma_B) \ .
  \]  
\end{lemma}

\begin{proof}
  Let $\lambda \geq 0$ such that $-\log \lambda = \Hmin(\rho_{A
    B}|\sigma_B)$, that is, the operator $\lambda \cdot \id_A \otimes
  \sigma_B - \rho_{A B}$ is nonnegative.  Because $\cE$ is a quantum
  operation, the operator $\lambda \cdot \cE(\id_A \otimes \sigma_B) -
  \cE(\rho_{A B})$ is also nonnegative. Combining this with the
  assumption that $\id_{A'} \otimes \sigmat_{B'} - \cE(\id_A \otimes
  \sigma_B)$ is nonnegative, we conclude that the operator
  \begin{multline*}
    \lambda \cdot \id_{A'} \otimes \sigmat_{B'}- \rhot_{A' B'} \\
  =
    \lambda \bigl(
      \id_{A'} \otimes \sigmat_{B'} - \cE(\id_A \otimes \sigma_B)
    \bigr)
    + \lambda \cdot \cE(\id_A \otimes \sigma_B) - \rhot_{A' B'} 
  \end{multline*}
  is also nonnegative. The assertion then follows by the definition of
  the min-entropy.
\end{proof}

\subsection{Min-entropy of superpositions}

Let $\{\ket{x}\}_{x \in \cX}$ be an orthonormal basis on $\cH_X$, let
$\{\ket{\psi^x}\}_{x \in \cX}$ be a family of vectors on $\cH_A
\otimes \cH_B \otimes \cH_E$, and define
\begin{align} \label{eq:rhoABEdef}
  \rho_{A B E} & := \proj{\psi} \qquad 
  \text{where $\ket{\psi} := \sum_{x \in \cX} \ket{\psi^x}$}
\\ \label{eq:rhobABEXdef}
  \rhot_{A B E X} & := \sum_{x \in \cX} \proj{\psi^x} \otimes \proj{x} \ .
\end{align}
Note that, if the states $\ket{\psi^x}$ are orthogonal then $\rhot_{A
  B E X}$ can be seen as the state resulting from an orthogonal
measurement of $\rho_{A B E}$ with respect to the projectors along
$\ket{\psi^x}$.  While $\rho_{A B E}$ is a \emph{superposition}
(linear combination) of vectors~$\ket{\psi^x}$, $\rhot_{A B E}$ is a
\emph{mixture} of vectors~$\ket{\psi^x}$. The following lemma gives a
lower bound on the min-entropy of $\rho_{A B E}$ in terms of the
min-entropy of $\rhot_{A B E}$.

\begin{lemma} \label{lem:Hinfcondlowbound}
  Let $\rho_{A B E}$ and $\rhot_{A B E X}$ be defined
  by~\eqref{eq:rhoABEdef} and~\eqref{eq:rhobABEXdef}, respectively,
  and let $\sigma_B \in \NN(\cH_B)$. Then
  \[
    \Hmin(\rho_{A B}|\sigma_B)
  \geq
    \Hmin(\rhot_{A B}|\sigma_B) - \Hmax(\rhot_X) \ .
  \]
\end{lemma}

\begin{proof}
  Assume without loss of generality that, for all $x \in \cX$,
  $\ket{\psi^x}$ is not the zero vector. This implies $\Hmax(\rhot_X)
  = \log |\cX|$.  Moreover, let $\lambda \geq 0$ such that $-\log
  \lambda = \Hmin(\rhot_{A B}|\sigma_B)$. It then suffices to show
  that the operator
  \begin{equation} \label{eq:Hinfcondlowboundt}
    \lambda \cdot |\cX| \cdot \id_A \otimes \sigma_B - \rho_{A B} 
  \end{equation}
  is nonnegative.
  
  Let $\ket{\theta} \in \cH_A \otimes \cH_B$.  By linearity, we have
  \begin{equation} \label{eq:sumxxp}
    \bra{\theta}  
    \rho_{A B} 
    \ket{\theta}
  =
    \bra{\theta} \tr_E(\proj{\psi}) \ket{\theta} \\
  =
    \sum_{x, x' \in \cX} 
    \bra{\theta} \tr_E(\ket{\psi^x}\bra{\psi^{x'}}) \ket{\theta} \ .
  \end{equation}  
  Let $\{\ket{z}\}_{z \in \cZ}$ be an orthonormal basis of $\cH_E$ and
  define $\ket{\theta, z} := \ket{\theta} \otimes \ket{z}$. Then, by
  the Cauchy-Schwartz inequality, for any $x, x' \in \cX$,
  \[
  \begin{split}
    |\bra{\theta} \tr_E(\ket{\psi^x}\bra{\psi^{x'}}) \ket{\theta}|
   & =
    \bigl|\sum_{z \in \cZ} 
      \spr{\theta,z}{\psi^x} \spr{\psi^{x'}}{\theta,z} \bigr| \\
   & \leq
     \sqrt{
       \sum_{z \in \cZ} \spr{\theta,z}{\psi^x} \spr{\psi^x}{\theta,z}
     }
     \sqrt{
       \sum_{z \in \cZ} \spr{\theta,z}{\psi^{x'}} \spr{\psi^{x'}}{\theta,z}
     } \\
   & =
     \sqrt{\bra{\theta} \tr_E(\proj{\psi^x}) \ket{\theta} 
       \bra{\theta} \tr_E(\proj{\psi^{x'}}) \ket{\theta}} \ .
  \end{split}
  \]
  Combining this with~\eqref{eq:sumxxp} and using Jensen's inequality,
  we find
  \[
  \begin{split}
    \bra{\theta}  
    \rho_{A B} 
    \ket{\theta}
  & \leq
    \sum_{x, x' \in \cX} 
      \sqrt{\bra{\theta} \tr_E(\proj{\psi^x}) \ket{\theta}
        \bra{\theta} \tr_E(\proj{\psi^{x'}}) \ket{\theta}} \\
  & \leq
    |\cX| \sqrt{\sum_{x, x' \in \cX} 
      \bra{\theta} \tr_E(\proj{\psi^x}) \ket{\theta} 
        \bra{\theta} \tr_E(\proj{\psi^{x'}})  \ket{\theta}} \\
  & =
    |\cX| \sum_{x \in \cX} 
      \bra{\theta} \tr_E(\proj{\psi^x}) \ket{\theta} \\
  & =
    |\cX| \cdot
    \bra{\theta} 
    \rhot_{A B} 
    \ket{\theta}  \ .
  \end{split}
  \]  
  By the choice of $\lambda$, the operator $\lambda \cdot \id_A
  \otimes \sigma_B - \rhot_{A B}$ is nonnegative. Hence $\bra{\theta}
  \rhot_{A B} \ket{\theta} \leq \lambda \bra{\theta} \id_A \otimes
  \sigma_B \ket{\theta}$ and thus, by the above inequality,
  $\bra{\theta} \rho_{A B} \ket{\theta} \leq \lambda \cdot |\cX| \cdot
  \bra{\theta} \id_A \otimes \sigma_B \ket{\theta}$.  Because this is
  true for any vector $\ket{\theta}$, we conclude that the operator
  defined by~\eqref{eq:Hinfcondlowboundt} is nonnegative.
\end{proof}


\begin{lemma} \label{lem:Hinfcondlowboundp} 
  Let $\rho_{A B E}$, $\rhot_{A B E X}$ be defined
  by~\eqref{eq:rhoABEdef} and \eqref{eq:rhobABEXdef}, respectively,
  and let $\sigma_{B X} \in \NN(\cH_B \otimes \cH_\cX)$. Then
  \[
    \Hmin(\rho_{A B}|\sigma_B)
  \geq
    \Hmin(\rhot_{A B X}|\sigma_{B X}) - \Hmax(\rhot_X) \ .
  \] 
\end{lemma}

\begin{proof}
  The assertion follows from Lemma~\ref{lem:Hinfcondlowbound} together
  with Lemma~\ref{lem:Hinfcondclasse}.
\end{proof}

\index{min-entropy|)}

\section{Smooth min- and max-entropy} \label{sec:smoothRenyi}

\index{smooth~min-entropy|(} \index{smooth~max-entropy|(}

\index{smoothness|(}

The min-entropy and the max-entropy, as defined in the previous
section, are discontinuous in the sense that a slight modification of
the system's state might have a large impact on its entropy.  To
illustrate this, consider for example a classical random variable $X$
on the set $\{0, \ldots, n-1\}$ which takes the values $0$ and $1$
with probability almost one half, i.e., $P_X(0) = P_X(1) =
\frac{1-\eps}{2}$, for some small $\eps > 0$, whereas the other values
have equal probabilities, i.e., $P_X(x) = \frac{\eps}{n-2}$, for all
$x > 1$.  Then, by the definition of the max-entropy, $\Hmax(P_X) =
\log n$. On the other hand, if we slightly change the probability
distribution $P_X$ to some probability distribution $\Pb_X$ such that
$\Pb_X(x) = 0$, for all $x > 1$, then $\Hmax(\Pb_X) = 1$. In
particular, for $n$ large, $\Hmax(P_X) \gg \Hmax(\Pb_X)$, while $\|P_X
- \Pb_X\|_1 \leq \eps$.

We will see later (cf.\ Section~\ref{sec:IR}) that the max-entropy
$\Hmax(P_X)$ can be interpreted as the minimum number of bits needed
to encode $X$ in such a way that its value can be recovered from the
encoding without errors. The above example is consistent with this
interpretation.  Indeed, while we need at least $\log n$ bits to store
a value $X$ distributed according to $P_X$, one single bit is
sufficient to store a value distributed according to $\Pb_X$.
However, for most applications, we allow some small error probability.
For example, we might want to encode $X$ in such a way that its value
can be recovered with probability $1-\eps$.  Obviously, in this case,
one single bit is sufficient to store $X$ even if it is distributed
according to $P_X$.

The example illustrates that, given some probability distribution
$P_X$, one might be interested in the maximum (or minimum) entropy of
any distribution $\Pb_X$ which is close to $P_X$. This idea is
captured by the notion of smooth min- and max-entropy.

\index{smoothness|)}

\subsection{Definition of smooth min- and max-entropy}

The definition of smooth min- and max-entropy is based on the
``non-smooth'' version (Definition~\ref{def:minmaxentr}).

\begin{definition} \label{def:smoothminmaxentr}
  Let $\rho_{A B} \in \NN(\cH_A \otimes \cH_B)$, $\sigma_B \in
  \NN(\cH_B)$, and $\eps \geq 0$. The \emph{$\eps$-smooth min-entropy}
  and the \emph{$\eps$-smooth max-entropy of $\rho_{A B}$ relative to
    $\sigma_B$} are
  \begin{align*}
    \Hmin^\eps(\rho_{A B}|\sigma_B)
  & :=
    \sup_{\rhob_{A B}} \Hmin(\rhob_{A B}|\sigma_B)
  \\
    \Hmax^\eps(\rho_{A B}|\sigma_B)
  & :=
    \inf_{\rhob_{A B}} \Hmax(\rhob_{A B}|\sigma_B) \ ,
  \end{align*}
  where the supremum and infimum ranges over the set $\cB^\eps(\rho_{A
    B})$ of all operators $\rhob_{A B} \in \NN(\cH_A \otimes \cH_B)$
  such that $\| \rhob_{A B} - \rho_{A B} \|_1 \leq \tr(\rho_{A B})
  \cdot \eps$ and $\tr(\rhob_{A B}) \leq \tr(\rho_{A B})$.
\end{definition}


\begin{definition} \label{def:smoothminmaxentrgen}
  Let $\rho_{A B} \in \NN(\cH_A \otimes \cH_B)$ and let $\eps \geq 0$.
  The \emph{$\eps$-smooth min-entropy} and the \emph{$\eps$-smooth
    max-entropy of $\rho_{A B}$ given $\cH_B$} are
  \begin{align*}
    \Hmin^\eps(\rho_{A B}|B) 
  & := 
    \sup_{\sigma_B} \Hmin^\eps(\rho_{A B}|\sigma_B) 
\\
    \Hmax^\eps(\rho_{A B}|B) 
  & := 
    \sup_{\sigma_B} \Hmax^\eps(\rho_{A B}|\sigma_B) \ ,
  \end{align*}
  where the supremum ranges over all $\sigma_B \in \NN(\cH_B)$ with
  $\tr(\sigma_B) = 1$.
\end{definition} 

Note that, similar to the description in Section~\ref{sec:convRenyi},
these definitions can be specialized to classical probability
distributions.  \index{classical~system}

\index{smooth~max-entropy|)}

\subsubsection{Evaluating the suprema and infima}

\begin{remark}
  If the Hilbert space $\cH_A \otimes \cH_B$ has finite dimension,
  then the set of operators $\rhob_{A B} \in \cB^\eps(\cH_A \otimes
  \cH_B)$ as well as the set of operators $\sigma_B \in \NN(\cH_B)$
  with $\tr(\sigma_B) = 1$ is compact.  Hence, the infima and suprema
  in the above definitions can be replaced by minima and maxima,
  respectively.
\end{remark}

\begin{remark} \label{rem:Hinfex}
  The supremum in the definition of the smooth min-entropy
  $\Hmin^\eps(\rho_{A B}|\sigma_B)$
  (Definition~\ref{def:smoothminmaxentr}) can be restricted to the set
  of operators $\rhob_{A B} \in \cB^\eps(\rho_{A B})$ with
  $\im(\rhob_{A B}) \subseteq \im(\rho_A) \otimes \im(\sigma_B)$.
  
  Additionally, to compute $\Hmin^\eps(\rho_{A B Z}|\sigma_{B Z})$
  where $\rho_{A B Z}$ and $\sigma_{B Z}$ are classical with respect
  to an orthonormal basis $\{\ket{z}\}_{z \in \cZ}$ on a subsystem
  $\cH_Z$, it is sufficient to take the supremum over operators
  $\rhob_{A B Z} \in \cB^\eps(\rho_{A B Z})$ which are classical with
  respect to $\{\ket{z}\}_{z \in \cZ}$.
  
  Similarly, to compute $\Hmin^\eps(\rho_{X A B}|\sigma_B)$ where
  $\rho_{X A B}$ is classical on a subsystem $\cH_X$, the supremum can
  be restricted to states $\rhob_{X A B} \in \cB^\eps(\rho_{X A B})$
  which are classical on $\cH_X$.
\end{remark}

\begin{proof}

  For the first statement, we show that any operator $\rhob_{A B} \in
  \cB^\eps(\rho_{A B})$ can be transformed to an operator
  $\cE(\rhob_{A B}) \in \cB^\eps(\rho_{A B})$ which has at least the
  same amount of min-entropy as $\rhob_{A B}$ and, additionally, has
  support on $\im(\rho_A) \otimes \im(\sigma_B)$.
  
  Let $\cE$ be the operation on $\cH_A \otimes \cH_B$ defined by
  \[
    \cE(\rhob_{A B}) 
  := 
    (\rho_A^0 \otimes \id_B) 
    \rhob_{A B} (\rho_A^0 \otimes \id_B) \ .
  \]
  Because the operator $\id_A \otimes \sigma_B - \cE(\id_A \otimes
  \sigma_B)$ is nonnegative, Lemma~\ref{lem:Hinfclassop} implies that
  the min-entropy can only increase under the action of $\cE$.
  Moreover, $\im(\rho_{A B}) \subseteq \im(\rho_A) \otimes \cH_B$
  (cf.\ Lemma~\ref{lem:tensprodimage}) and thus $\cE(\rho_{A B}) =
  \rho_{A B}$.  Because $\cE$ is a projection, the \distance{} cannot
  increase under the action of $\cE$ (cf.\ Lemma~\ref{lem:distdecr}),
  i.e., 
  \[
    \bigl\| \cE(\rhob_{A B}) - \rho_{A B} \bigr\|_1 
  =
    \bigl\| \cE(\rhob_{A B} - \rho_{A B}) \bigr\|_1 
  \leq
    \bigl\| \rhob_{A B} - \rho_{A B} \bigr\|_1  
  \leq
    \tr(\rho_{A B}) \cdot \eps \ .
  \]
  We thus have $\rhob_{A B} \in \cB^\eps(\rho_{A B})$. The assertion
  then follows because we can assume that $\im(\rhob_{A B})$ is
  contained in $\cH_A \otimes \im(\sigma_B)$ (otherwise, the
  min-entropy is arbitrarily negative and the statement is trivial)
  and thus $\im(\cE(\rhob_{A B})) \subseteq \im(\rho_A) \otimes
  \im(\sigma_B)$.
  
  The statements for $\rho_{A B Z}$ and $\rho_{X A B}$ are proven
  similarly.
\end{proof}


\begin{remark} \label{rem:classsigmaexsmooth}
  Let $\rho_{A B Z} \in \NN(\cH_A \otimes \cH_B \otimes \cH_Z)$ be
  classical with respect to an orthonormal basis $\{\ket{z}\}_{z \in
    \cZ}$ of $\cH_Z$. Then the supremum in the definition of the
  min-entropy $\Hmin^\eps(\rho_{A B Z} | B Z)$ can be restricted to
  operators $\sigma_{B Z} \in \NN(\cH_B \otimes \cH_Z)$ which are
  classical with respect to $\{\ket{z}\}_{z \in \cZ}$.
\end{remark}

\begin{proof}  
  We show that for any $\rhob'_{A B Z} \in \cB^\eps(\rho_{A B Z})$ and
  $\sigma'_{B Z} \in \NN(\cH_B \otimes \cH_Z)$ with $\tr(\sigma'_{B
    Z}) = 1$ there exists $\rhob_{A B Z} \in \cB^\eps(\rho_{A B Z})$
  and $\sigma_{B Z} \in \NN(\cH_B \otimes \cH_Z)$ with $\tr(\sigma_{B
    Z}) = 1$ such that $\sigma_{B Z}$ is classical with respect to
  $\{\ket{z}\}_{z \in \cZ}$ and $\Hmin(\rhob_{A B Z}| \sigma_{B Z})
  \geq \Hmin(\rhob'_{A B Z}| \sigma'_{B Z})$.
  
  Let thus $\rhob'_{A B Z} \in \cB^\eps(\rho_{A B Z})$ and $\sigma'_{B
    Z} \in \NN(\cH_B \otimes \cH_Z)$ be fixed.  Define $\rhob_{A B Z}
  := (\id_{A B} \otimes \cE_Z)(\rhob'_{A B Z})$ and $\sigma_{B Z} :=
  (\id_B \otimes \cE_Z)(\sigma'_{B Z})$ where $\cE_Z$ is the
  projective measurement operation on $\cH_Z$, i.e., 
  \[
    \cE_Z(\rho) := \sum_{z \in \cZ} \proj{z}  \rho \proj{z} \ .
  \]
  
  Note that $\sigma_{B Z}$ is classical with respect to
  $\{\ket{z}\}_{z \in \cZ}$ and, because $\cE_Z$ is trace-preserving,
  $\tr(\sigma_{B Z}) =\tr(\sigma'_{B Z}) = 1$. Similarly,
  $\tr(\rhob_{A B Z}) = \tr(\rhob'_{A B Z})$. Moreover, because
  $(\id_{A B} \otimes \cE_Z)(\rho_{A B Z}) = \rho_{A B Z}$ and because
  the distance{} can only decrease when applying $\id_{A B} \otimes
  \cE_Z$ (cf.\ Lemma~\ref{lem:distdecr}), we have 
  \[
    \| \rhob_{A B Z} - \rho_{A B Z} \|_1 
  \leq 
    \| \rhob'_{A B Z} - \rho_{A B Z} \|_1
  \] 
  which implies $\rhob_{A B Z} \in \cB^\eps(\rho_{A B Z})$.  Finally,
  using Lemma~\ref{lem:Hinfclassop}, we find $\Hmin(\rhob_{A B Z}|
  \sigma_{B Z}) \geq \Hmin(\rhob'_{A B Z}| \sigma'_{B Z})$.
\end{proof}

\subsection{Basic properties of smooth min-entropy}

\subsubsection{Superadditivity}

\index{superadditivity|(}

The following is a generalization of (one direction of)
Lemma~\ref{lem:HRindadd} to smooth min-entropy.

\begin{lemma} \label{lem:Hminindaddsmooth}
  Let $\rho_{A B} \in \NN(\cH_A \otimes \cH_B)$, $\sigma_B \in
  \NN(\cH_B)$ and, similarly, $\rho_{A' B'} \in \NN(\cH_{A'} \otimes
  \cH_{B'})$, $\sigma_{B'} \in \NN(\cH_{B'})$, and let $\eps, \eps'
  \geq 0$. Then
  \[
    \Hmin^{\eps+\eps'}(
      \rho_{A B} \otimes \rho_{A' B'} | \sigma_{B} \otimes \sigma_{B'}
    )
  \geq
    \Hmin^\eps(\rho_{A B} | \sigma_B) 
    + \Hmin^{\eps'}(\rho_{A' B'} | \sigma_{B'}) \ .
  \]
\end{lemma}

\begin{proof}
  For any $\nu > 0$, there exist $\rhob_{A B} \in \cB^\eps(\rho_{A
    B})$ and $\rhob_{A' B'} \in \cB^{\eps'}(\rho_{A' B'})$ such that
  \begin{align*}
    \Hmin(\rhob_{A B}|\sigma_B) 
  & >
    \Hmin^\eps(\rho_{A B}|\sigma_B) - \nu 
  \\
    \Hmin(\rhob_{A' B'}|\sigma_{B'}) 
  & >
    \Hmin^{\eps'}(\rho_{A' B'}|\sigma_{B'}) - \nu \ .
  \end{align*}
  Hence, by Lemma~\ref{lem:HRindadd},
  \[
    \Hmin(\rhob_{A B} \otimes \rhob_{A' B'} | \sigma_B \otimes \sigma_{B'})
  >
    \Hmin^\eps(\rho_{A B}|\sigma_B)
    + \Hmin^{\eps'}(\rho_{A' B'}|\sigma_{B'}) - 2 \nu \ .
  \]
  Because this holds for any $\nu > 0$, it remains to verify that
  $\rhob_{A B} \otimes \rhob_{A' B'} \in \cB^{\eps + \eps'}(\rho_{A B}
  \otimes \rho_{A' B'})$. This is however a direct consequence of the
  triangle inequality, i.e.,
  \begin{multline*}
    \bigl\| 
      \rhob_{A B} \otimes \rhob_{A' B'} - \rho_{A B} \otimes \rho_{A' B'} 
    \bigr\|_1 \\
  \leq
    \tr(\rhob_{A' B'}) \cdot \bigl\| \rhob_{A B} - \rho_{A B} \bigr\|_1 
    + \tr(\rho_{A B}) \cdot \bigl\| \rhob_{A' B'} - \rho_{A' B'} \bigr\|_1 \\
  \leq    
    \tr(\rho_{A B} \otimes \rho_{A' B'}) (\eps + \eps') \ .
  \end{multline*}
\end{proof}

\index{superadditivity|)}

\subsubsection{Strong subadditivity}

\index{strong~subadditivity|(}

The following statement is a generalization of
Lemma~\ref{lem:Hinfcondclasse} to smooth min-entropy.

\begin{lemma} \label{lem:Hinfcondclassesmooth}
  Let $\rho_{A B C} \in \NN(\cH_A \otimes \cH_B \otimes \cH_C)$,
  $\sigma_{B C} \in \NN(\cH_B \otimes \cH_C)$, and let $\eps \geq 0$.
  Then
  \[
    \Hmin^\eps(\rho_{A B C}|\sigma_{B C})
  \leq
    \Hmin^\eps(\rho_{A B}|\sigma_B) \ .
  \]
\end{lemma}

\begin{proof}
  For any $\nu >0$, there exists $\rhob_{A B C} \in \cB^\eps(\rho_{A B
    C})$ such that 
  \[
    \Hmin(\rhob_{A B C}|\sigma_{B C}) 
  \geq
    \Hmin^\eps(\rho_{A B C}|\sigma_{B C}) - \nu \ .
  \]
  Hence, by Lemma~\ref{lem:Hinfcondclasse}, applied to the operator
  $\rhob_{A B C}$,
  \[
    \Hmin(\rhob_{A B}|\sigma_B)
  \geq 
    \Hmin^\eps(\rho_{A B C}|\sigma_{B C}) - \nu \ .
  \]
  Because this holds for any $\nu > 0$, it remains to show that
  $\rhob_{A B} \in \cB^\eps(\rho_{A B})$. This is however a direct
  consequence of the fact that the \distance{} cannot increase when
  taking the partial trace (cf.\ Lemma~\ref{lem:distdecr}), i.e.,
  \[
    \|\rhob_{A B}- \rho_{A B}\|_1 
  \leq 
    \|\rhob_{A B C}- \rho_{A B C}\|_1 \leq \tr(\rho_{A B C}) \cdot \eps \ .
   \]
\end{proof}

\index{strong~subadditivity|)}

\subsubsection{Conditioning on classical information}

\index{classical~system|(}

The following lemma generalizes (one direction of)
Lemma~\ref{lem:Hminclasscondr} to smooth min-entropy.

\begin{lemma} \label{lem:Hminsmoothclasscondr}
  Let $\rho_{A B Z} \in \NN(\cH_A \otimes \cH_B \otimes \cH_Z)$ and
  $\sigma_{B Z} \in \NN(\cH_B \otimes \cH_Z)$ be classical with
  respect to an orthonormal basis $\{\ket{z}\}_{z \in \cZ}$ of
  $\cH_Z$, let $\rho_{A B}^z$ and $\sigma_B^z$ be the corresponding
  (non-normalized) conditional operators, and let $\eps \geq 0$.  Then
  \[
    \Hmin^\eps(\rho_{A B Z}|\sigma_{B Z})
  \geq
    \inf_{z \in \cZ} \Hmin^\eps(\rho_{A B}^z|\sigma_{B}^z) \ .
  \]
\end{lemma}

\begin{proof}
  For any $\nu > 0$ and $z \in \cZ$, there exists $\rhob_{A B}^z \in
  \cB^\eps(\rho_{A B}^z)$ such that
  \[
    \Hmin(\rhob_{A B}^z|\sigma_B^z) 
  \geq 
    \Hmin^\eps(\rho_{A B}^z|\sigma_B^z) 
    - \nu \ .
  \]
  Let
  \[
    \rhob_{A B Z}
  :=
    \sum_{z \in \cZ} \rhob_{A B}^z \otimes \proj{z} \ .
  \]
  Using Lemma~\ref{lem:Hminclasscondr}, we find
  \begin{equation} \label{eq:Hinfminone}
    \Hmin(\rhob_{A B Z}|\sigma_{B Z})
  =
    \inf_{z \in \cZ} \Hmin(\rhob_{A B}^z|\sigma_{B}^z) 
  \geq
    \inf_{z \in \cZ} \Hmin^\eps(\rho_{A B}^z|\sigma_B^z) 
    - \nu \ .
  \end{equation}
  Because this holds for any value of $\nu > 0$, it suffices to verify
  that $\rhob_{A B Z} \in \cB^\eps(\rho_{A B Z})$. This is however a
  direct consequence of
  \[
    \bigl\| \rhob_{A B Z} - \rho_{A B Z} \bigr\|_1
  =
    \sum_{z \in \cZ} \bigl\| \rhob_{A B}^z - \rho_{A B}^z \bigr\|_1
  \leq
    \sum_{z \in \cZ} \tr(\rho_{A B}^z) \cdot \eps 
  =
    \tr(\rho_{A B Z}) \cdot \eps \ ,
  \]
  where the first equality follows from Lemma~\ref{lem:distclass}.
\end{proof}

\index{classical~system|)}

\subsection{Chain rules for smooth min-entropy}

\index{chain~rule|(}

The following lemma generalizes (one direction of)
Lemma~\ref{lem:Halphachain} to smooth min-entropy.

\begin{lemma} \label{lem:Halphachainsmooth}
  Let $\rho_{A B C} \in \NN(\cH_A \otimes \cH_B \otimes \cH_C)$,
  $\sigma_C \in \NN(\cH_C)$, let $\sigma_B \in \NN(\cH_B)$ be the
  fully mixed state on the support of $\rho_B$, and let $\eps \geq 0$.
  Then
  \[
    \Hmin^\eps(\rho_{A B C} | \sigma_C) 
  \leq
    \Hmin^\eps(\rho_{A B C} | \sigma_B \otimes \sigma_C) + \Hmax(\rho_B) \ .
  \]
\end{lemma}

\begin{proof}
  According to Remark~\ref{rem:Hinfex}, for any $\nu > 0$, there exists
  $\rhob_{A B C} \in \cB^\eps(\rho_{A B C})$ such that
  \begin{equation} \label{eq:Hchainone}
    \Hmin(\rhob_{A B C}|\sigma_C) 
  \geq 
    \Hmin^\eps(\rho_{A B C}|\sigma_C) - \nu
  \end{equation}
  and $\im(\rhob_{A B C}) \subseteq \im(\rho_{A B}) \otimes \cH_C =
  \im(\rho_{A B} \otimes \id_C)$. Hence, from
  Lemma~\ref{lem:tensimage}, $\im(\rhob_B) \subseteq \im(\rho_B)$.
  Consequently, the operator $\rhob_B$ is arbitrarily close to an
  operator whose support is equal to the support of $\rho_B$. By
  continuity, we can thus assume without loss of generality that
  $\im(\rhob_B) = \im(\rho_B)$, that is,
  \begin{equation} \label{eq:Hchaintwo}
    \Hmax(\rhob_B)
  =
    \Hmax(\rho_B) \ .
  \end{equation}
  Moreover, since $\rhob_{A B C} \in \cB^\eps(\rho_{A B C})$, we have
  \begin{equation} \label{eq:Hchainthree}
    \Hmin^\eps(\rho_{A B C}|\sigma_B \otimes \sigma_C) 
  \geq
    \Hmin(\rhob_{A B C}|\sigma_B \otimes \sigma_C) \ .
  \end{equation}
  
  Finally, because $\sigma_B$ is the fully mixed state on $\im(\rho_B)
  = \im(\rhob_{B})$, Lemma~\ref{lem:Halphachain}, applied to the state
  $\rhob_{A B C}$, gives
  \[
    \Hmin(\rhob_{A B C} | \sigma_C)
  =
    \Hmin(\rhob_{A B C} | \sigma_B \otimes \sigma_C) + \Hmax(\rhob_B) \ .
  \]
  Combining this with~\eqref{eq:Hchainone}, \eqref{eq:Hchaintwo},
  and~\eqref{eq:Hchainthree} concludes the proof.
\end{proof}

\index{chain~rule|)}

\subsubsection{Data processing}

\index{data~processing|(}

The following lemma is a generalization of Lemma~\ref{lem:Hclasschain}
to smooth min-entropy.

\begin{lemma} \label{lem:Hclasschainsmooth}
  Let $\rho_{A Y C} \in \NN(\cH_A \otimes \cH_Y \otimes \cH_C)$ be
  classical with respect to an orthonormal basis $\{\ket{y}\}_{y \in
    \cY}$ of $\cH_Y$ such that the corresponding conditional operators
  $\rho_{A C}^y$, for any $y \in \cY$, have product form, let
  $\sigma_C \in \NN(\cH_C)$, and let $\eps \geq 0$. Then
  \[
    \Hmin^\eps(\rho_{A Y C} | \sigma_C) 
  \geq
    \Hmin^\eps(\rho_{Y C} | \sigma_C)
    + \Hmin(\rho_{A Y} | \rho_Y) \ .
  \]
\end{lemma}

\begin{proof}
  For any $y \in \cY$, let $p_y := \tr(\rho_{A C}^y)$ and define
  $\rhot_{A}^y := \frac{1}{p_y} \rho_{A}^y$.  Because $\rho_{A C}^y$
  has product form, we have
  \[
    \rho_{A Y C} 
  = 
    \sum_{y \in \cY} \rhot_A^y \otimes \proj{y} \otimes \rho_C^y
    \ .
  \]
  According to Remark~\ref{rem:Hinfex}, for any $\nu > 0$, there exists
  a nonnegative operator $\rhob_{Y C} \in \cB^\eps(\rho_{Y C})$ such
  that
  \begin{equation} \label{eq:Hclasschainone}
    \Hmin(\rhob_{Y C}|\sigma_C) 
  \geq 
    \Hmin^\eps(\rho_{Y C}|\sigma_C) - \nu
  \end{equation}
  where $\rhob_{Y C}$ is classical with respect to $\{\ket{y}\}_{y \in
    \cY}$, that is, $\rhob_{Y C} = \sum_{y \in \cY} \proj{y} \otimes
  \rhob_{C}^y$, for some family $\{\rhob_{C}^y\}_{y \in \cY}$ of
  conditional operators on $\cH_C$.  Let $\rhob_{A Y C} \in \NN(\cH_A
  \otimes \cH_Y \otimes \cH_C)$ be defined by
  \[
    \rhob_{A Y C}
  :=
    \sum_{y \in \cY}
      \rhot_{A}^y \otimes \proj{y} \otimes \rhob_C^y \ .
  \]
  Because the operators $\rhot_A^y$ are normalized, we have
  \[
  \begin{split}
    \bigl\| \rhob_{A Y C} - \rho_{A Y C} \bigr\|_1
  & =
    \sum_{y}  \bigl\| 
      \rhot_A^y \otimes \rhob_C^y - \rhot_A^y \otimes \rho_C^y 
    \bigr\|_1 \\
  & =
    \sum_{y} \bigl\| \rhob_C^y - \rho_C^y \bigr\|_1 \\
  & =
    \bigl\| \rhob_{Y C} - \rho_{Y C} \bigr\|_1 \ ,
  \end{split}
  \]
  where the first and the last equality follow from
  Lemma~\ref{lem:distclass}. Because $\rhob_{Y C} \in \cB^\eps(\rho_{Y
    C})$, this implies $\rhob_{A Y C} \in \cB^\eps(\rho_{A Y C})$ and
  thus
  \begin{equation} \label{eq:Hclasschaintwo}
    \Hmin^\eps(\rho_{A Y C} | \sigma_C)
  \geq
    \Hmin(\rhob_{A Y C} | \sigma_C) \ .
  \end{equation}
  Moreover, using Lemma~\ref{lem:Hminclasscondr} and the fact that,
  for any $y \in \cY$, the operators $\rhot_A^y$ and $\rho_A^y$ only
  differ by a factor $p_y$, we have
  \begin{equation} \label{eq:Hclasschainthree}
  \begin{split}
    \Hmin(\rhob_{A Y} | \rhob_Y)
  & =
    \inf_{y \in \cY} \Hmin(\rhot_{A}^y|\tr(\rhot_A^y)) \\
  & =
    \inf_{y \in \cY} \Hmin(\rho_{A}^y|\tr(\rho_A^y)) \\
  & = 
    \Hmin(\rho_{A Y} | \rho_Y) \ .
  \end{split}
  \end{equation}
  Finally, applying Lemma~\ref{lem:Hclasschain} to the state $\rhob_{A
    Y C}$ gives
  \[
    \Hmin(\rhob_{A Y C} | \sigma_C) 
  \geq
    \Hmin(\rhob_{Y C} | \sigma_C)
    + \Hmin(\rhob_{A Y} | \rhob_Y) \ .
  \]
  Combining this with~\eqref{eq:Hclasschainone},
  \eqref{eq:Hclasschaintwo}, and~\eqref{eq:Hclasschainthree} concludes
  the proof.
\end{proof}

\index{data~processing|)}

\subsection{Smooth min-entropy of superpositions}

The following statement generalizes Lemma~\ref{lem:Hinfcondlowboundp}.

\begin{lemma} \label{lem:smoothHinfcondlowbound}
  Let $\rho_{A B E}$, $\rhot_{A B E X}$ be defined
  by~\eqref{eq:rhoABEdef} and \eqref{eq:rhobABEXdef}, respectively,
  for mutually orthogonal vectors $\ket{\psi^x}$, let $\sigma_{B X}
  \in \NN(\cH_B \otimes \cH_\cX)$, and let $\eps \geq 0$. Then
  \[
    \Hmin^\eps(\rho_{A B}|\sigma_B)
  \geq
    \Hmin^{\epst}(\rhot_{A B X}|\sigma_{B X}) - \Hmax(\rhot_X) \ ,
  \]
  where $\epst = \frac{\eps^2}{6 |\cX|}$.
\end{lemma}

\begin{proof}
  By Remark~\ref{rem:Hinfex}, for any $\nu > 0$, there exists an
  operator $\bar{\rhot}_{A B X} \in \cB^{\epst}(\rhot_{A B X})$ which
  is classical with respect to the basis $\{\ket{x}\}_{x \in \cX}$
  such that
  \begin{equation} \label{eq:rhoABXentrcond}
    \Hmin(\bar{\rhot}_{A B X}|\sigma_{B X})
  \geq
    \Hmin^{\epst}(\rhot_{A B X}|\sigma_{B X}) - \nu \ .
  \end{equation}
  
  Let $\{\bar{\rhot}_{A B}^x\}_{x \in \cX}$ be the family of
  conditional operators defined by $\bar{\rhot}_{A B X}$ and
  $\{\ket{x}\}_{x \in \cX}$, i.e., $\bar{\rhot}_{A B X} = \sum_{x \in
    \cX} \bar{\rhot}_{A B}^x \otimes \proj{x}$.  According to
  Lemma~\ref{lem:puredistbound}, for any $x \in \cX$, there exists a
  purification $\proj{\psib^x}$ of $\bar{\rhot}_{A B}^x$ such that
  \[
    \| \ket{\psi^x} - \ket{\psib^x} \| 
  \leq 
    \sqrt{\bigl\|\rhot_{A B}^x - \bar{\rhot}_{A B}^x\bigr\|_1} \ .
  \]
  Let $\ket{\psib} := \sum_{x \in \cX} \ket{\psib^x}$ and define
  $\rhob_{A B E} := \proj{\psib}$. By the triangle inequality, we
  find
  \[
    \bigl\|\ket{\psi} - \ket{\psib}\bigr\|
  \leq
    \sum_{x \in \cX} \bigl\| \ket{\psi^x} - \ket{\psib^x} \bigr\|
  \leq
    \sum_{x \in \cX} 
      \sqrt{\bigl\|\rhot_{A B}^x - \bar{\rhot}_{A B}^x \bigr\|_1} \ .
  \]
  Hence, with Jensen's inequality,
  \[
  \begin{split}
    \bigl\|\ket{\psi} - \ket{\psib}\bigr\|
  & \leq
    \sqrt{|\cX| 
      \sum_{x \in \cX} \bigl\|\rhot_{A B}^x - \bar{\rhot}_{A B}^x \bigr\|_1
    } \\
  & =
    \sqrt{|\cX| \cdot \bigl\| \rhot_{A B X} - \bar{\rhot}_{A B X} \bigr\|_1}
   \ ,
  \end{split}
  \]
  where the equality follows from Lemma~\ref{lem:distclass}.  Because
  the vectors $\ket{\psi^x}$ are orthogonal, we have $\tr(\rhot_{A B
    X}) = \tr(\rho_{A B})$. Consequently, since $\bar{\rhot}_{A B X}
  \in \cB^{\epst}(\rhot_{A B X})$, we obtain
  \begin{equation} \label{eq:psidiffbound}
    \bigl\|\ket{\psi} - \ket{\psib}\bigr\|
  \leq 
    \sqrt{|\cX| \cdot \epst \cdot \tr(\rhot_{A B X})}
  =
    \sqrt{|\cX| \cdot \epst \cdot \tr(\rho_{A B})} \ .
  \end{equation}
    
  Assume without loss of generality that $|\cX| \cdot \epst \leq
  \frac{1}{6}$ (otherwise, the assertion is trivial). Then, because
  $\sqrt{\tr(\rho_{A B})} = \|\ket{\psi}\|$, we have
  \[
  \begin{split}
    \| \ket{\psi} \| + \| \ket{\psib} \|
  & \leq
    2 \| \ket{\psi} \|
    + \|\ket{\psi} - \ket{\psib} \| \\
  & \leq
    2 \sqrt{\tr(\rho_{A B})} + \sqrt{\sfrac{1}{6} \tr(\rho_{A B})} 
  <
    \sqrt{6 \, \tr(\rho_{A B})} \ .
  \end{split}
  \]
  and thus, by Lemma~\ref{lem:distpurebound},
  \[
    \bigl\| 
      \rho_{A B} - \rhob_{A B}
    \bigr\|_1
  \leq
    \sqrt{6 \, \tr(\rho_{A B})} \cdot \bigl\|
      \ket{\psi} - \ket{\psib}
    \bigr\|
  \leq
    \tr(\rho_{A B}) \cdot \eps \ ,
  \]
  where the last inequality follows from~\eqref{eq:psidiffbound}. This
  implies
  \begin{equation} \label{eq:Hinftypcomp}
    \Hmin^\eps(\rho_{A B}|\sigma_{B}) 
  \geq 
    \Hmin(\rhob_{A B}|\sigma_{B}) \ .
  \end{equation}
  
  Note that $\bar{\rhot}_{A B X}$ can be seen as the operator obtained
  by taking the partial trace of
  \[
    \bar{\rhot}_{A B E X} 
  := 
    \sum_{x \in \cX} \proj{\psib^x} \otimes \proj{x} \ .
  \]  
  We can thus apply Lemma~\ref{lem:Hinfcondlowboundp} to the operators
  $\rhob_{A B E}$ and $\bar{\rhot}_{A B E X}$, which gives
  \[
    \Hmin(\rhob_{A B}|\sigma_B)
  \geq 
    \Hmin(\bar{\rhot}_{A B X} | \sigma_{B X}) - \Hmax(\bar{\rhot}_X) \ .
  \]
  Finally, because the support of $\bar{\rhot}_X$ is contained in the
  support of $\rhot_X$, we have $\Hmax(\bar{\rhot}_X) \leq
  \Hmax(\rhot_X)$ and thus
  \[
    \Hmin(\rhob_{A B}|\sigma_B) 
  \geq 
    \Hmin(\bar{\rhot}_{A B X}| \sigma_{B X}) - \Hmax(\rhot_X) \ .
  \]
  Combining this with~\eqref{eq:Hinftypcomp}
  and~\eqref{eq:rhoABXentrcond} concludes the proof.
\end{proof}

\subsection{Smooth min-entropy calculus}

The properties proven so far are formulated in terms of the smooth
min-entropy $H(\rho_{A B}|\sigma_B)$ relative to an operator
$\sigma_B$ (Definition~\ref{def:smoothminmaxentr}).  The following
theorem translates these statements to conditional smooth min-entropy
$H(\rho_{A B}|B)$ (Definition~\ref{def:smoothminmaxentrgen}).

\begin{theorem} \label{thm:Hmincalc}
  Let $\eps, \eps' \geq 0$. Then the following inequalities hold:
  \begin{itemize}
  \item (Super-)additivity: \index{additivity}\index{superadditivity}
    \begin{equation} \label{eq:findadd}
      \Hmin^{\eps + \eps'}(\rho_{A B} \otimes \rho_{A' B'}|B B') 
    \geq
      \Hmin^\eps(\rho_{A B}|B) + \Hmin^{\eps'}(\rho_{A' B'}|B') \ ,
    \end{equation}
    for $\rho_{A B} \in \NN(\cH_A \otimes \cH_B)$ and $\rho_{A' B'}
    \in \NN(\cH_{A'} \otimes \cH_{B'})$.
  \item Strong subadditivity: \index{strong~subadditivity}
    \begin{equation} \label{eq:fsubadd}
      \Hmin^\eps(\rho_{A B C}|B C) 
    \leq
      \Hmin^\eps(\rho_{A B}|B) \ ,
    \end{equation}
    for $\rho_{A B C} \in \NN(\cH_A \otimes \cH_B \otimes \cH_C)$.
  \item Conditioning on classical information:
    \begin{equation} \label{eq:fclasscond}
      \Hmin^\eps(\rho_{A B Z}|B Z)
    \geq
      \inf_{z \in \cZ} \Hmin^\eps(\rhob_{A B}^z|B) \ ,
    \end{equation} 
    for $\rho_{A B Z} \in \NN(\cH_A \otimes \cH_B \otimes \cH_Z)$
    normalized and classical on $\cH_Z$, and for normalized
    conditional operators $\rhob_{A B}^z$.
  \item Chain rule: \index{chain~rule}
    \begin{equation} \label{eq:fchain}
      \Hmin^\eps(\rho_{A B C} | C) 
    \leq 
      \Hmin^\eps(\rho_{A B C} | B C) + \Hmax(\rho_B) \ ,
    \end{equation}
    for $\rho_{A B C} \in \NN(\cH_A \otimes \cH_B \otimes \cH_C)$.
  \item Data processing: \index{data~processing}
    \begin{equation} \label{eq:fdataproc}
      \Hmin^\eps(\rho_{A Y C} | C) 
    \geq 
      \Hmin^\eps(\rho_{Y C} | C) + \Hmin(\rho_{A Y} | \rho_Y) \ ,
    \end{equation}
    for $\rho_{A Y C} \in \NN(\cH_A \otimes \cH_Y
    \otimes \cH_C)$ classical on $\cH_Y$ such that the conditional
    operators $\rho_{A C}^y$ have product form.  
  \end{itemize}
\end{theorem}

\begin{proof}
  The statements follow immediately from
  Lemmata~\ref{lem:Hminindaddsmooth}, \ref{lem:Hinfcondclassesmooth},
  \ref{lem:Hminsmoothclasscondr}, \ref{lem:Halphachainsmooth},
  and~\ref{lem:Hclasschainsmooth}. 
\end{proof}

\section{Smooth min- and max-entropy of products} \label{sec:smoothprod}

\index{product~state|(}

In this section, we show that the smooth min- and max-entropies of
product states are asymptotically equal to the von Neumann entropy.
In a first step, we consider a purely classical situation, i.e., we
prove that the smooth min- and max-entropies of a sequence of
independent and identically distributed random variables can be
expressed in terms of Shannon entropy\index{Shannon~entropy} (which is
the classical analogue of the von Neumann
entropy\index{von~Neumann~entropy}).  Then, in a second step, we
generalize this statement to quantum states
(Section~\ref{sec:prodquant}).

\index{smooth~min-entropy|)}

\subsection{The classical case} \label{sec:prodclass}

The proof of the main result of this section
(Theorem~\ref{thm:classprodentr}) is based on a Chernoff style bound
(Theorem~\ref{thm:probchernoff}) which is actually a variant of the
asymptotic equipartition property
(AEP)\index{asymptotic~equipartition}\index{AEP|see{asymptotic
    equipartition}} known from information theory (see, e.g.,
\cite{CovTho91}).  It states that, with high probability, the negative
logarithm of the probability of an $n$-tuple of values chosen
according to a product distribution $P^n$ is close to the Shannon
entropy\index{Shannon~entropy} of~$P^n$\index{Shannon~entropy}.

\subsubsection{Typical sequences and their probabilities}

\index{typical~sequence}

\begin{lemma} \label{lem:Etbound}
  Let $P_{X Y} \in \NN(\cX \times \cY)$ be a probability distribution.
  Then, for any $t \in \bbR$ with $|t| \leq \frac{1}{\log(|\cX|+3)}$,
  \[
    \log \ExpE_{x,y}\bigl[P_{X|Y}(x,y)^{-t}\bigr]  
  \leq 
    t H(X|Y) + \sfrac{1}{2} t^2 \log(|\cX|+3)^2 \ ,
  \]
  where the expectation is taken over pairs $(x,y)$ chosen according
  to $P_{X Y}$.
\end{lemma}

\begin{proof}
  For any $t \in \bbR$, let $r_t$ be the function on the open interval
  $(0, \infty)$ defined by
  \begin{equation} \label{eq:rtdef}
    r_t(z) := z^t - t \ln z -1 \ .
  \end{equation}
  We will use several properties of this function proven in
  Appendix~\ref{sec:rt}.

  For any $x \in \cX$ and $y \in \cY$, let $p_{x,y}:=P_{X|Y}(x,y)$.
  If $p_{x,y} > 0$ then
  \[
    p_{x,y}^{-t}
  =
    r_t\bigl(\sfrac{1}{p_{x,y}}\bigr) + t \ln \sfrac{1}{p_{x,y}} + 1
  \leq
    r_t\bigl(\sfrac{1}{p_{x,y}}+3\bigr) + t \ln \sfrac{1}{p_{x,y}}
    + 1 \ ,
  \]
  where the inequality holds because $r_t$ is monotonically increasing
  on the interval $[1, \infty)$ (Lemma~\ref{lem:rtincr}) and
  $\frac{1}{p_{x,y}} = \frac{P_Y(y)}{P_{X Y}(x,y)} \geq 1$.  Because
  $\frac{1}{p_{x,y}} + 3 \in [4, \infty)$ and because $r_t$ is concave
  on this interval (Lemma~\ref{lem:rtconc} which can be applied
  because $t \in [-\frac{1}{2}, \frac{1}{2}]$), Jensen's inequality
  leads to
  \[  
  \begin{split}
    \ExpE_{x,y}\bigl[p_{x,y}^{-t}\bigr]
  & \leq
    \ExpE_{x,y}\Bigl[r_t\bigl(\sfrac{1}{p_{x,y}} + 3\bigr) \Bigr] 
    + t \ExpE_{x,y} \Bigl[\ln\sfrac{1}{p_{x,y}}\Bigr] + 1 \\
  & \leq
    r_t\Bigl(\ExpE_{x,y}\bigl[\sfrac{1}{p_{x,y}}+3\bigr]\Bigr) 
    + t (\ln 2) \ExpE_{x,y} \Bigl[\log \sfrac{1}{p_{x,y}}\Bigr] + 1 
    \ ,
  \end{split}
  \]
  where $\ExpE_{x,y}[\cdot]$ denotes the expectation with respect to
  $(x,y)$ chosen according to the distribution $P_{X Y}$.  Because
  $\ExpE_{x,y}[\frac{1}{p_{x,y}}] = \sum_{x,y} P_{X Y}(x,y)
  \frac{P_{Y}(y)}{P_{X Y}(x,y)} = |\cX|$ and $\ExpE_{x,y}[\log
  \frac{1}{p_{x,y}}] = H(X|Y)$, we obtain
  \[
    \ExpE_{x,y}\bigl[p_{x,y}^{-t}\bigr]
  \leq
    r_t(|\cX|+3) + t (\ln 2) H(X|Y) + 1 \ .
  \]
  Furthermore, because $\log a \leq \frac{1}{\ln 2}(a-1)$,
  \[
    \log \ExpE_{x,y}\bigl[p_{x,y}^{-t}\bigr]
  \leq
    \sfrac{1}{\ln 2} r_t\bigl(|\cX|+3\bigr) + t H(X|Y) \ .
  \]
  Finally, together with Lemma~\ref{lem:rtbound}, since $|t| \leq
  \frac{1}{\log(|\cX|+3)}$, we conclude
  \[
    \log \ExpE_{x,y}\bigl[p_{x,y}^{-t}\bigr]
  \leq
    \bigl(\sfrac{1}{\ln 2} - 1 \bigr) t^2 \log(|\cX|+3)^2 + t H(X|Y) \ .
  \]
  The assertion follows because $\frac{1}{\ln2} - 1 \leq
  \frac{1}{2}$.
\end{proof}

\newcommand*{\entrm}{\gamma}

\begin{lemma} \label{lem:exptZbound}
  Let $P_{X Y} \in \NN(\cX \times \cY)$ be a probability distribution
  and let $\entrm$ be the function on $\cX \times \cY$ defined by
  \[
    \entrm(x,y) := - \log P_{X|Y}(x,y) - H(X|Y) \ .
  \]
  Then, for any $t \in \bbR$ with $|t| \leq \frac{1}{\log(|\cX|+3)}$,
  \[
    \ExpE_{x,y}\bigl[2^{t \entrm(x,y)}\bigr] 
  \leq 
    2^{\frac{1}{2} t^2 \log(|\cX|+3)^2} \ .
  \]
\end{lemma}

\begin{proof}
  The assertion follows directly from Lemma~\ref{lem:Etbound}, that is,
  \[
  \begin{split}
    \ExpE_{x,y}\bigl[2^{t \entrm(x,y)}\bigr] 
  & =
    2^{-t H(X|Y)} \ExpE_{x,y}\bigl[P_{X|Y}(x,y)^{-t}\bigr] \\
  & \leq
    2^{-t H(X|Y)} 
    \cdot  2^{t H(X|Y) + \frac{1}{2} t^2 \log(|\cX|+3)^2} \ .  \qedhere
  \end{split}
  \]
\end{proof}

\begin{theorem} \label{thm:probchernoff}
  Let $P_{X Y} \in \NN(\cX \times \cY)$ be a probability distribution
  and let $n \in \bbN$.  Then, for any $\delta \in [0, \log |\cX|]$
  and $(\bx, \by)$ chosen according to $P_{X^n Y^n} := (P_{X Y})^n$,
  \[
    \Pr_{\bx,\by} \bigl[
        -\log P_{X^n|Y^n}(\bx,\by)
        \geq  n \bigl(H(X|Y) + \delta\bigr) \bigr] 
  \leq 
    2^{- \frac{n \delta^2}{2 \log(|\cX|+3)^2} }  \ ,
  \] 
  and, similarly,
  \[
    \Pr_{\bx,\by} \bigl[
        - \log P_{X^n|Y^n}(\bx, \by) 
        \leq  n \bigl(H(X|Y) - \delta\bigr) \bigr] 
  \leq 
    2^{- \frac{n \delta^2}{2 \log(|\cX|+3)^2} }  \ .
  \]   
\end{theorem}

\begin{proof}
  Let $\bx=(x_1, \ldots, x_n)$, $\by=(y_1, \ldots, y_n)$, and let
  $\entrm$ be the function defined in Lemma~\ref{lem:exptZbound} for
  the probability distribution $P_{X Y}$. Then
  \begin{equation} \label{eq:zxysum}
    \sum_{i=1}^n \entrm(x_i, y_i)
  =
    - \log P_{X^n|Y^n}(\bx, \by)
    - n H(X|Y) \ .
  \end{equation}
  Using Markov's inequality, for any $t > 0$,
  \begin{equation} \label{eq:HsMarkov}
  \begin{split}
    \Pr_{\bx,\by}\bigl[\sum_{i=1}^n \entrm(x_i, y_i) \geq n \delta\bigr]
  & =
    \Pr_{\bx,\by}
      \bigl[2^{t \sum_{i=1}^n \entrm(x_i, y_i)} \geq 2^{t n \delta}\bigr] \\
  & \leq
    \frac{\ExpE_{\bx,\by}
      \bigl[2^{t \sum_{i=1}^n \entrm(x_i, y_i)}\bigr]}{2^{t n \delta}} \ .
  \end{split}
  \end{equation}
  Moreover, because the pairs $(x_i, y_i)$ are chosen independently,
  \[
  \begin{split}
    \ExpE_{\bx,\by}\bigl[ 2^{t \sum_{i=1}^n \entrm(x_i, y_i)} \bigr]
  & =
    \ExpE_{\bx,\by}\bigr[ \prod_{i=1}^n 2^{t \entrm(x_i, y_i)} \bigr] \\
  & =
    \prod_{i=1}^n  \ExpE_{x_i, y_i}\bigr[2^{t \entrm(x_i, y_i)} \bigr]  \\
  & \leq  
    \bigl( 2^{\frac{1}{2} t^2 \log(|\cX|+3)^2} \bigr)^n \ ,
  \end{split}
  \]
  where the inequality follows from Lemma~\ref{lem:exptZbound}, for
  any $|t| \leq \frac{1}{\log(|\cX|+3)}$.  Combining this
  with~\eqref{eq:HsMarkov} gives
  \[
    \Pr_{\bx,\by}\bigl[\sum_{i=1}^n \entrm(x_i, y_i) \geq n \delta\bigr]
  \leq
    2^{\frac{1}{2} n t^2 \log(|\cX|+3)^2 - t n \delta} \ .
  \]
  With $t := \frac{\delta}{\log(|\cX|+3)^2}$ (note that $t \leq
  \frac{1}{\log(|\cX| +3)}$ because $\delta \leq \log |\cX|$), we
  conclude
  \[
    \Pr_{\bx,\by}\bigl[\sum_{i=1}^n \entrm(x_i, y_i) \geq n \delta\bigr]
  \leq
    2^{- \frac{n \delta^2}{2 \log(|\cX|+3)^2} } \ .
  \]
  The first inequality of the lemma then follows
  from~\eqref{eq:zxysum}.
 
  Similarly, if $t < 0$,
  \[
  \begin{split}
    \Pr_{\bx,\by}\bigl[\sum_{i=1}^n \entrm(x_i, y_i) \leq -n \delta\bigr]
  & =
    \Pr_{\bx,\by}
      \bigl[2^{t \sum_{i=1}^n \entrm(x_i, y_i)} \geq 2^{- t n \delta}\bigr] \\
  & \leq
    \frac{\ExpE_{\bx,\by}
      \bigl[2^{t \sum_{i=1}^n \entrm(x_i, y_i)}\bigr]}{2^{- t n \delta}}
   \ ,
  \end{split}
  \]
  and thus
  \[
    \Pr_{\bx,\by}\bigl[\sum_{i=1}^n \entrm(x_i, y_i) \leq -n \delta\bigr]
  \leq
    2^{\frac{1}{2} n t^2 \log(|\cX|+3)^2 + t n \delta} \ .
  \]
  The second inequality follows with $t :=
  -\frac{\delta}{\log(|\cX|+3)^2}$.
\end{proof}

\index{smooth~min-entropy|(} \index{smooth~max-entropy|(}

\subsubsection{Asymptotic equality of smooth entropy and Shannon entropy}

\index{Shannon~entropy}

\begin{theorem} \label{thm:classprodentr} 
  Let $P_{X Y} \in \NN(\cX \times \cY)$ be a probability distribution
  and let $n \in \bbN$.  Then, for any $\eps \geq 0$ and $P_{X^n Y^n}
  := (P_{X Y})^n$,
  \begin{align*}
    \frac{1}{n} \Hmax^\eps(P_{X^n Y^n}|P_{Y^n})  
  & \leq
    H(X|Y) + \delta
  \\
    \frac{1}{n} \Hmin^\eps(P_{X^n Y^n}|P_{Y^n})  
  & \geq
    H(X|Y) - \delta \ ,
  \end{align*}
  where 
  $\delta := \log(|\cX|+3) \sqrt{\frac{2 \log(\slfrac{1}{\eps})}{n}}
  $.
\end{theorem}

\begin{proof}
  We first prove the bound on the (classical) smooth max-entropy
  $\Hmax^\eps(P_{X^n Y^n}|P_{Y^n})$. For any $\by \in \cY^n$ with
  $P_{Y^n}(\by) > 0$, let $\cXb_{\by}$ be the set of all $n$-tuples
  $\bx \in \cX^n$ such that
  \[
    -\log P_{X^n|Y^n}(\bx,\by)
  \leq 
    n \bigl( H(X|Y) + \delta \bigr) \ .
  \]
  Furthermore, let $P_{\Xb^n \Yb^n}$ be the nonnegative function on
  $\cX^n \times \cY^n$ defined by
  \begin{equation} \label{eq:PXbYbdef}
    P_{\Xb^n \Yb^n}(\bx, \by)
  =
    \begin{cases}
      P_{X^n Y^n}(\bx, \by) & \text{if $\bx \in \cXb_{\by}$} \\
      0 & \text{otherwise.}
    \end{cases}
  \end{equation}
  We can assume without loss of generality that $\delta \leq \log
  |\cX|$ (otherwise, the statement is trivial). Hence, by the first
  inequality of Theorem~\ref{thm:probchernoff}, $\Pr_{\bx,\by}[\bx
  \notin \cXb_{\by}] \leq \eps$. This implies $\|P_{X^n Y^n} -
  P_{\Xb^n \Yb^n} \|_1 \leq \eps$ and thus
  \begin{equation} \label{eq:Hzeropred}
    \Hmax^\eps(P_{X^n Y^n}|P_{Y^n})
  \leq
    \Hmax(P_{\Xb^n \Yb^n}|P_{Y^n}) \ .
  \end{equation}
  
  For any fixed $\by:=(y_1, \ldots, y_n) \in \cY^n$ with $P_{Y^n}(\by)
  > 0$,
  \[
    1 
  \geq 
    \sum_{\bx \in \cXb_{\by}} 
      \prod_{i=1}^n P_{X|Y}(x_i,y_i) 
  \geq 
    |\cXb_{\by}| 2^{- n (H(X|Y) + \delta)} \ ,
  \]
  where the second inequality follows from the definition of the set
  $\cXb_{\by}$. Consequently, we have $|\cXb_{\by}| \leq 2^{n (H(X|Y)
    + \delta)}$.  Moreover, by the definition of $P_{\Xb^n \Yb^n}$,
  the support of the function $\bx \mapsto P_{\Xb^n \Yb^n}(\bx,\by)$
  is contained in $\cXb_{\by}$.  Hence, using
  Remark~\ref{rem:classminmaxentr},
  \[
    \Hmax(P_{\Xb^n \Yb^n}|P_{Y^n})
  \leq
    \log \bigl( \sum_{\by \in \cY^n} P_{Y^n}(\by) \cdot |\cXb_{\by}| \bigr)
  \leq
    n \bigl(H(X|Y) + \delta \bigr) \ .
  \]
  Combining this with~\eqref{eq:Hzeropred} proves the first inequality of
  the lemma.
  
  To prove the bound on the min-entropy $\Hmin^\eps(P_{X^n
    Y^n}|P_{Y^n})$, let $\cXb_{\by}$, for any $\by \in \cY^n$ with
  $P_{Y^n}(\by) > 0$, be the set of $n$-tuples $\bx \in \cX^n$ such
  that
  \[
    - \log P_{X^n|Y^n}(\bx,\by)
  \geq 
    n \bigl( H(X|Y) - \delta \bigr) \ ,
  \]
  and let again $P_{\Xb^n \Yb^n}$ be defined by~\eqref{eq:PXbYbdef}.
  By the second inequality of Theorem~\ref{thm:probchernoff},
  $\Pr_{\bx,\by}[\bx \notin \cXb_{\by}] \leq \eps$, which, similarly
  to the previous argument, implies
  \begin{equation} \label{eq:Hinfpred}
    \Hmin^\eps(P_{X^n Y^n}|P_{Y^n})
  \geq
    \Hmin(P_{\Xb^n \Yb^n}|P_{Y^n}) \ .
  \end{equation}
  Moreover, using Remark~\ref{rem:classminmaxentr}
  \[
  \begin{split}
    \Hmin(P_{\Xb^n \Yb^n}|P_{Y^n})
  & =
    - \log \max_{\by \in \supp(P_{Y^n})} \max_{\bx \in \cX^n} 
        \frac{P_{\Xb^n \Yb^n}(\bx, \by)}{P_{Y^n}(\by)} \\
  & =
    - \log \max_{\by \in \supp(P_{Y^n})} \max_{\bx \in \cXb_{\by}} 
        \frac{P_{X^n Y^n}(\bx, \by)}{P_{Y^n}(\by)} \\
  & \geq
    n \bigl( H(X|Y) - \delta \bigr) \ ,
  \end{split}
  \]
  where the inequality follows from the definition of the set
  $\cXb_{\by}$. Combining this with~\eqref{eq:Hinfpred} proves the
  second inequality of the lemma.
\end{proof}

Because the min-entropy $\Hmin(P_{X^n Y^n}|P_{Y^n})$ cannot be larger
than the max-entropy $\Hmax(P_{X^n Y^n}|P_{Y^n})$ (cf.\ 
Lemma~\ref{lem:Hminmax}), Theorem~\ref{thm:classprodentr} implies that
\begin{equation} \label{eq:productapprox}
  \sfrac{1}{n} \Hmin^\eps(P_{X^n Y^n}|P_{Y^n}) 
\approx 
  \sfrac{1}{n} \Hmax^\eps(P_{X^n Y^n}|P_{Y^n}) 
\approx 
\sfrac{1}{n} H(X^n|Y^n) \ ,
\end{equation}
where asymptotically, for increasing $n$, the approximation becomes an
equality.

\begin{remark}
  It is easy to see that Theorem~\ref{thm:classprodentr} can be
  generalized to probability distributions $P_{X^n Y^n}$ which are the
  product of not necessarily identical distributions $P_{X_i Y_i}$.
  That is, for any distribution of the form $P_{X^n Y^n} =
  \prod_{i=1}^n P_{X_i Y_i}$, the
  approximation~\eqref{eq:productapprox} still holds.
\end{remark}

\index{smooth~max-entropy|)}

\subsection{The quantum case} \label{sec:prodquant}

The following theorem and its corollary can be seen as a quantum
version of Theorem~\ref{thm:classprodentr} for smooth min-entropy
(where the Shannon entropy is replaced by the von Neumann
entropy\index{von~Neumann~entropy}).  The proof essentially follows
the same line as the classical argument described above.\footnote{An
  alternative method to prove the statement $\frac{1}{n}
  \Hmin^\eps(\rho_{A B}^{\otimes n}|\rho_{B}^{\otimes n}) \gtrapprox
  H(\rho_{A B}) - H(\rho_B)$ is to use a chain rule of the form
  $\Hmin^\eps(\rho_{A B}^{\otimes n}|\rho_{B}^{\otimes n}) \gtrapprox
  \Hmin^\eps(\rho_{A B}^{\otimes n}) - \Hmax^\eps(\rho_B^{\otimes
    n})$. The entropies on the right hand side of this inequality can
  be rewritten as the entropies of the classical probability
  distributions defined by the eigenvalues of $\rho_{A B}^{\otimes n}$
  and $\rho_B^{\otimes n}$, respectively.  The desired bound then
  follows from the classical Theorem~\ref{thm:classprodentr}.
  However, the results obtained with such an alternative method are
  less tight and less general than Theorem~\ref{thm:Hmincondrep}.} A
similar argument shows that the statement also holds for smooth
max-entropy.

\begin{theorem} \label{thm:Hmincondrep}
  Let $\rho_{A B} \in \NN(\cH_A \otimes \cH_B)$, $\sigma_B \in
  \NN(\cH_B)$ be density operators, and let $n \in \bbN$. Then, for
  any $\eps \geq 0$,
  \[
    \frac{1}{n} \Hmin^\eps(\rho_{A B}^{\otimes n}|\sigma_B^{\otimes n}) 
  \geq 
    H(\rho_{A B}) - H(\rho_B) - D(\rho_{B}\|\sigma_B) - \delta \ ,
  \]
  where $\delta := 2 \log\bigl(\rank(\rho_A)+\tr(\rho_{A B}^2 (\id_A
  \otimes \sigma_B^{-1}))+2\bigr)
  \sqrt{\frac{\log(\slfrac{1}{\eps})}{n} +1}$.
\end{theorem}

\begin{proof}
  Define $H(\rho_{A B} | \sigma_{B}) := H(\rho_{A B}) - H(\rho_B) -
  D(\rho_{B}\|\sigma_B)$.  We show that there exists a density
  operator $\rhob_{A^n B^n} \in \cB^\eps(\rho_{A B}^{\otimes n})$ such
  that
  \begin{equation} \label{eq:maxr}
    \Hmin(\rhob_{A^n B^n}|\sigma_{B}^{\otimes n})
  \geq
    n H(\rho_{A B}|\sigma_B) -  n \delta \ .
  \end{equation}
  According to the definition of min-entropy, this is equivalent to
  saying that the operator $\lambda \cdot (\id_A \otimes
  \sigma_B)^{\otimes n} - \rhob_{A^n B^n}$ is nonnegative, for
  $\lambda \geq 0$ such that $-\log \lambda = n H(\rho_{A B} |
  \sigma_{B}) - n \delta$.
  
  Let 
  \[
    (\id_A \otimes \sigma_B)^{\otimes n}
  =
    \sum_{\bz \in \cZ^n} q_{\bz} \proj{\bz}
  \]
  be a spectral decomposition of $(\id_A \otimes \sigma_B)^{\otimes
    n}$. We can assume without loss of generality that there exists an
  order relation on the values $\cZ^n$ such that $q_{\bz} \geq
  q_{\bzp}$, for any $\bz \geq \bzp$. For any $\bz \in \cZ$, let
  $B_{\bz}$ be the projector defined by
  \[
    B_{\bz} := \sum_{\bzp: \, \bzp \geq \bz} \proj{\bzp} \ .
  \]
  Moreover, let $\beta_{\bz}$, for $\bz \in \cZ^n$, be nonnegative
  coefficients such that, for any $\bzp \in \cZ^n$,
  \[
    \sum_{\bz: \, \bz \leq \bzp} \beta_{\bz} = q_{\bzp} \ .
  \]
  Note that the spectral decomposition above can then be rewritten as
  \begin{equation} \label{eq:betaproj}
    (\id_A \otimes \sigma_B)^{\otimes n}
  =
    \sum_{\bz \in \cZ^n} \beta_{\bz} B_{\bz} \ .
  \end{equation}
  
  Let 
  \[
    \rho_{A B}^{\otimes n} = \sum_{\bx \in \cX^n} p_{\bx} \proj{\bx}
  \]
  be a spectral decomposition of $\rho_{A B}^{\otimes n}$. In the
  following, we denote by $\infty$ an element which is larger than any
  element of $\cZ^n$. Moreover, let $p_{\bx, \bz}$, for $\bx \in
  \cX^n$ and $\bz \in \cZ^n \cup \{\infty\}$, be nonnegative
  coefficients such that, for any $\bzp \in \cZ^n$,
  \begin{align*}
    \sum_{\bz : \, \bz \leq \bzp} p_{\bx,\bz} 
  & =
    \min(p_{\bx}, \lambda q_{\bzp})
\\
    \sum_{\bz \in \cZ^n \cup \{\infty\}} p_{\bx,\bz} 
  & = 
    p_{\bx} \ .
  \end{align*}
    
  We show that inequality~\eqref{eq:maxr} holds for the operator
  \[
    \rhob_{A^n B^n}
  :=
    \sum_{\bx \in \cX^n} 
      \sum_{\bz \in \cZ^n} p_{\bx,\bz} B_{\bz} \proj{\bx} B_{\bz} \ .
  \]  
  
  Note first that, by the definition of $p_{\bx, \bz}$
  and~$\beta_{\bz}$, we have $p_{\bx,\bz} \leq \lambda \beta_{\bz}$,
  for any $\bx \in \cX^n$ and $\bz \in \cZ^n$, that is, the operator
  \[
    \sum_{\bz \in \cZ^n} \lambda \beta_{\bz} B_{\bz} B_{\bz}
  - \rhob_{A^n B^n}
  = 
    \sum_{\bz \in \cZ^n} \sum_{\bx \in \cX^n}
      (\lambda \beta_{\bz} - p_{\bx, \bz}) 
      B_{\bz} \proj{\bx}  B_{\bz} 
  \]
  is nonnegative. Using~\eqref{eq:betaproj} and the fact that the
  operators $B_{\bz}$ are projectors, we conclude that the operator
  \[
    \lambda \cdot (\id_A \otimes \sigma_B)^{\otimes n}
  - \rhob_{A^n B^n}
  =
    \sum_{\bz \in \cZ^n} \lambda \beta_{\bz} B_{\bz}
  - \rhob_{A^n B^n}
  \]
  is nonnegative, which implies~\eqref{eq:maxr}. It thus remains to be
  proven that $\rhob_{A^n B^n} \in \cB^\eps(\rho_{A B}^{\otimes n})$.
  
  Using the above definitions and the convention that $B_\infty$ is
  the zero matrix, we have
  \[
  \begin{split}
    \bigl\| \rho_{A B}^{\otimes n} - \rhob_{A^n B^n} \bigr\|_1
  & =
    \bigl\|
      \sum_{\bx \in \cX^n} \sum_{\bz \in \cZ^n \cup \{\infty\}} 
        p_{\bx,\bz} \bigl(\proj{\bx} - B_{\bz} \proj{\bx} B_{\bz} \bigr) 
    \bigr\|_1 \\
  & \leq
    \sum_{\bx \in \cX^n} \sum_{\bz \in \cZ^n \cup \{\infty\}} p_{\bx,\bz}
      \bigl\|\proj{\bx} - B_{\bz} \proj{\bx} B_{\bz} \bigr\|_1 \ .
  \end{split}
  \]
  We can use Lemma~\ref{lem:disttracebound} to bound the trace
  distance on the right hand side of this inequality, that is,
  \[
    \bigl\|\proj{\bx} - B_{\bz} \proj{\bx} B_{\bz} \bigr\|_1
  \leq
    2 \sqrt{1-\tr(B_{\bz} \proj{\bx} B_{\bz})} \ .
  \]
  Because $\rho_{A B}$ is a density operator, the nonnegative
  coefficients $p_{\bx, \bz}$ sum up to one. We can thus apply
  Jensen's inequality which gives
  \begin{equation} \label{eq:trrootbound}
  \begin{split}
    \bigl\| \rho_{A B}^{\otimes n} - \rhob_{A^n B^n} \bigr\|_1
  & \leq
    2 \sum_{\bx \in \cX^n} \sum_{\bz \in \cZ^n \cup \{\infty\}} p_{\bx,\bz}
      \sqrt{1-\tr(B_{\bz} \proj{\bx} B_{\bz})} \\
  & \leq
    2 \sqrt{
      \sum_{\bx \in \cX^n} \sum_{\bz \in \cZ^n \cup \{\infty\}} 
        p_{\bx,\bz} (1-\tr(B_{\bz} \proj{\bx} B_{\bz}))
    } \\
  & =
    2 \sqrt{1-\tr(\rhob_{A^n B^n})} \ .
  \end{split}
  \end{equation}
  The trace in the square root can be rewritten as
  \[
  \begin{split}
    \tr(\rhob_{A^n B^n})
  & =
    \sum_{\bzp \in \cZ^n} \bra{\bzp} \Bigl(
      \sum_{\bx \in \cX^n} 
        \sum_{\bz \in \cZ^n} p_{\bx,\bz} B_{\bz} \proj{\bx} B_{\bz} 
    \Bigr) \ket{\bzp} \\
  & =
    \sum_{\bzp \in \cZ^n} \sum_{\bz: \, \bz \leq \bzp}
      \sum_{\bx \in \cX^n}
        p_{\bx,\bz} |\spr{\bzp}{\bx}|^2 \ .
  \end{split}
  \]
  Because the terms in the sum are all nonnegative, the sum can only
  become smaller if we restrict the set of values $\bx$ over which the
  sum is taken. Consequently,
  \[
    \tr(\rhob_{A^n B^n})  
  \geq
    \sum_{\bzp \in \cZ^n}
      \sum_{\bx: \, p_{\bx} \leq \lambda q_{\bzp}} 
        |\spr{\bzp}{\bx}|^2 \sum_{\bz: \, \bz \leq \bzp}
        p_{\bx,\bz}  \ .
  \]
  By the definition of $p_{\bx, \bz}$, we have $\sum_{\bz: \, \bz \leq
    \bzp} p_{\bx, \bz} = p_{\bx}$, for any $(\bx, \bz')$ such that
  $p_{\bx} \leq \lambda q_{\bzp}$, and hence
  \[
    \tr(\rhob_{A^n B^n})
  \geq
    \sum_{(\bx,\bzp): \, p_{\bx} \leq \lambda q_{\bzp}}
        p_{\bx} |\spr{\bzp}{\bx}|^2 \ .
  \]  
  Because $\sum_{\bz, \bx} p_{\bx} |\spr{\bz}{\bx}|^2 = 1$, 
  this inequality can be rewritten as
  \[
    1 -  \tr(\rhob_{A^n B^n})
  \leq
    \sum_{(\bx,\bz): \, p_{\bx} > \lambda q_{\bz}} p_{\bx} |\spr{\bz}{\bx}|^2 
  \]
  Recall that we need to prove that $\rhob_{A^n B^n} \in
  \cB^\eps(\rho_{A B}^{\otimes n})$. Hence,
  combining~\eqref{eq:trrootbound} with the above bound on
  $\tr(\rhob_{A^n B^n})$, it remains to be shown that
  \begin{equation} \label{eq:tracelowbound}
    \sum_{(\bx,\bz): \, p_{\bx} > \lambda q_{\bz}} p_{\bx} |\spr{\bz}{\bx}|^2 
  \leq
    \Bigl(\frac{\eps}{2}\Bigr)^2 \ .
  \end{equation}

  Let
  \[
    \id_A \otimes \sigma_B
  =
    \sum_{\zb \in \cZb} \qb_{\zb} \proj{\zb}
  \]
  and
  \[
    \rho_{A B}
  =
    \sum_{\xb \in \cXb} \pb_{\xb} \proj{\xb}
  \]
  be spectral decompositions of $\id_A \otimes \sigma_B$ and $\rho_{A
    B}$, respectively. Moreover, let $P_{\Xb \Zb}$ be the probability
  distribution defined by
  \[
    P_{\Xb \Zb}(\xb, \zb)
  :=
    \pb_{\xb} |\spr{\zb}{\xb}|^2 \ .
  \]
   
  Note that $\ket{\bx}$ and $p_{\bx}$, as used above, can be defined
  as $\ket{\bx} := \bigotimes_{i=1}^n \ket{x_i}$ and $p_{\bx} =
  p_{(x_1, \ldots, x_n)} := \prod_{i=1}^n \pb_{x_i}$. Similarly, we
  can set $\ket{\bz} := \bigotimes_{i=1}^n \ket{z_i}$ and $q_{\bz} =
  q_{(z_1, \ldots, z_n)} := \prod_{i=1}^n \qb_{z_i}$.  Then, the left
  hand side of~\eqref{eq:tracelowbound} can be rewritten as
  \begin{equation} \label{eq:probexp}
  \begin{split}  
    \sum_{(\bx,\bz): \, p_{\bx} > \lambda q_{\bz}} p_{\bx} |\spr{\bz}{\bx}|^2 
  & =
     \Pr_{\bx, \bz}[ p_{\bx} > \lambda q_{\bz} ]\\
  & = \Pr_{\bx, \bz}[ 
      - \log p_{\bx} + \log q_{\bz} < -\log \lambda 
    ] \\
  & = \Pr_{\bx, \bz}\Bigl[ 
      \sum_{i=1}^n - \log \pb_{x_i} + \log \qb_{z_i} < -\log\lambda
    \Bigr]
  \end{split}
  \end{equation}
  for $(\bx, \bz)$ chosen according to the probability distribution
  $(P_{\Xb \Zb})^n$. 
  
  By the definition of $H(\rho_{A B} | \sigma_{B})$, we have
  \[
  \begin{split}
    H(\rho_{A B} | \sigma_{B})
  & =
    - \tr(\rho_{A B} \log \rho_{A B} ) 
    + \tr(\rho_{A B} \log \id_A \otimes \sigma_B ) \\
  & =
    \sum_{\xb, \zb} 
      \pb_{\xb} |\spr{\zb}{\xb}|^2 
      \bigl( \log \frac{1}{\pb_{\xb}} - \log \frac{1}{\qb_{\zb}} \bigr) \\
  & =
    \ExpE_{\xb, \zb}[- \log \pb_{\xb} + \log \qb_{\zb}] \ ,
  \end{split}
  \]  
  for $(\xb, \yb)$ chosen according to $P_{\Xb \Yb}$. According to
  Birkhoff's theorem (cf.\ Theorem~\ref{thm:Birkhoff}) there exist
  nonnegative coefficients $\mu_{\pi}$ parameterized by the bijections
  $\pi$ from $\cX$ to $\cZ$ such that $\sum_{\pi} \mu_{\pi} = 1$ and
  $|\spr{\zb}{\xb}|^2 = \sum_{\pi} \mu_\pi \delta_{\zb, \pi(\xb)}$.
  The identity above can thus be rewritten as
  \begin{equation} \label{eq:entrAB}
    H(\rho_{A B} | \sigma_{B})
  =
    \sum_{\xb, \zb} \pb_{\xb} 
      |\spr{\zb}{\xb}|^2 \log \frac{\qb_{\zb}}{\pb_{\xb}} 
  =
    \sum_{\pi} \mu_\pi
      \sum_{\xb} \pb_{\xb} \log \frac{\qb_{\pi(\xb)}}{\pb_{\xb}} \ .
  \end{equation}  
  
  For $(\xb, \zb)$ chosen according to $P_{\Xb \Yb}$,
  \[
    \ExpE_{\xb, \zb}\bigl[2^{-t (\log \pb_{\xb} - \log \qb_{\zb})}\bigr] 
  =
   \sum_{\xb, \zb}
      \pb_{\xb} |\spr{\zb}{\xb}|^2 
      \bigl( \frac{\pb_{\xb}}{\qb_{\zb}} \bigr)^{-t   } 
  =
    \sum_{\pi} \mu_\pi 
      \sum_{\xb} \pb_{\xb} 
        \Bigl( \frac{\pb_{\xb}}{\qb_{\pi(\xb)}} \Bigr)^{-t} \ .
  \]  
  For any $t \in \bbR$, let $r_t$ be the function defined
  by~\eqref{eq:rtdef}. The last term in the sum above can then be
  bounded by
  \[
  \begin{split}
    \Bigl( \frac{\pb_{\xb}}{\qb_{\pi(\xb)}} \Bigr)^{-t}
  & =
    r_t\Bigl(\frac{\qb_{\pi(\xb)}}{\pb_{\xb}}\Bigr) 
    + t \ln\frac{\qb_{\pi(\xb)}}{\pb_{\xb}} + 1 \\
  & \leq
    r_{|t|}\Bigl(
      \frac{\qb_{\pi(\xb)}}{\pb_{\xb}} + \frac{\pb_{\xb}}{\qb_{\pi(\xb)}} + 2
    \Bigr)  
    + t \ln \frac{\qb_{\pi(\xb)}}{\pb_{\xb}} + 1 
  \end{split}
  \]
  where the inequality follows from the fact that, for all $z > 0$,
  $r_t(z) \leq r_{|t|}(z+\frac{1}{z})$ (Lemma~\ref{lem:rtzz}) and the
  fact that $r_t$ is monotonically increasing (Lemma~\ref{lem:rtincr})
  on the interval $[1, \infty)$.  Because
  $\frac{\qb_{\pi(\xb)}}{\pb_{\xb}} + \frac{\pb_{\xb}}{\qb_{\pi(\xb)}}
  + 2 \in [4, \infty)$ and because $r_t$ is concave on this interval
  (Lemma~\ref{lem:rtconc}) we can apply Jensen's inequality, which
  gives
  \begin{multline} \label{eq:expRSbound}
    \ExpE_{\xb, \zb}[2^{-t (\log \pb_{\xb} - \log \qb_{\zb}) }] 
  \leq
      r_{|t|}\Bigl(\sum_{\pi} \mu_\pi \sum_{\xb} \pb_{\xb} 
        \bigl(\frac{\qb_{\pi(\xb)}}{\pb_{\xb}}
        + \frac{\pb_{\xb}}{\qb_{\pi(\xb)}} + 2 \bigr)\Bigr) \\
    + t (\ln 2) \sum_{\pi} \mu_\pi 
        \sum_{\xb} \pb_{\xb}
          \log \frac{\qb_{\pi(\xb)}}{\pb_{\xb}}
    + 1 \ . 
  \end{multline}
  Note that $\sum_{\zb} \qb_{\zb} = \tr(\id_A \otimes \sigma_B) =
  \dim(\cH_A)$. As we can assume without loss of generality that
  $\cH_A$ is restricted to the support of $\rho_A$, we have
  \[
    \sum_{\pi} \mu_{\pi} \sum_{\xb} \qb_{\pi(\xb)} = \rank(\rho_{A}) \ .
  \]
  Moreover, 
  \[
    \sum_{\pi} \mu_{\pi} \sum_{\xb} \frac{\pb_{\xb}^2}{\qb_{\pi(\xb)}}
  =
    \sum_{\xb, \zb} |\spr{\xb}{\zb}|^2 \pb_{\xb}^2 \qb_{\zb}^{-1} 
  = 
    \tr\bigl(\rho_{A B}^2 (\id_A \otimes \sigma_B^{-1})\bigr) \ .
  \]
  Hence, together with~\eqref{eq:entrAB}, the
  bound~\eqref{eq:expRSbound} can be rewritten as
  \[
    \ExpE_{\xb, \zb}[2^{- t (\log \pb_{\xb} - \log \qb_{\zb}) }]
  \leq
    r_{|t|}(\gamma  + 2) 
    + t (\ln 2) H(\rho_{A B}|\sigma_B) 
    + 1 \ ,
  \]
  where $\gamma := \rank(\rho_A) + \tr\bigl(\rho_{A B}^2 (\id_A
  \otimes \sigma_B^{-1})\bigr)$. Furthermore, using the fact that
  $\log a \leq \frac{1}{\ln 2}(a-1)$ we find
  \[
  \begin{split}
    \ExpE_{\xb, \zb}[2^{-t (\log \pb_{\xb} - \log \qb_{\zb})}] 
  & \leq
   2^{\log\bigl(r_{|t|}(\gamma+2) 
     + t (\ln 2) H(\rho_{A B}|\sigma_B) + 1\bigr)}   \\
  & \leq
    2^{\frac{1}{\ln 2} r_{|t|}(\gamma+2) 
    + t H(\rho_{A B}|\sigma_B)} \ .
  \end{split}
  \]
  With Lemma~\ref{lem:rtbound}, we conclude
  \begin{equation} \label{eq:expcalc}
    \ExpE_{\xb, \zb}[2^{t (
      -\log \pb_{\xb} + \log \qb_{\zb} - H(\rho_{A B}|\sigma_B))}]
  \leq 
    2^{(\frac{1}{\ln 2}-1) t^2 \log(\gamma+2)^2} 
  \leq
    2^{\frac{1}{2}  t^2 \log(\gamma+2)^2} \ .
  \end{equation}
  
  Let now $w(\bx, \bz) := \sum_{i=1}^n (- \log \pb_{x_i} + \log
  \qb_{z_i} - H(\rho_{A B}|\sigma_B))$. Because the expectation of the
  product of independent values is equal to the product of the
  expectation of these values, we have, for $(\bx, \bz)$ chosen
  according to $(P_{\Xb \Zb})^n$,
  \[
    \ExpE_{\bx, \bz}[2^{t w(\bx, \bz)}]
  =
    \ExpE_{\xb, \zb}
      [2^{t (-\log \pb_{\xb} + \log \qb_{\zb} - H(\rho_{A B}|\sigma_B))}]^n \ .
  \]
  Hence, by Markov's inequality, for any $t \leq 0$,
  \[
  \begin{split}
    \Pr_{\bx, \bz}[w(\bx, \bz) \leq - n \delta]
  & =
    \Pr_{\bx, \bz}[2^{t w(\bx, \bz)} \geq 2^{- t n \delta}] \\
  & \leq 
    \frac{\ExpE_{\bx, \bz}[2^{t w(\bx, \bz)}]}{2^{- t n \delta}} \\
  & =
    \frac{\ExpE_{\xb, \zb}
      [2^{t (-\log \pb_{\xb} + \log \qb_{\zb} - H(\rho_{A B}|\sigma_B))}]^n}
      {2^{- t n \delta}}
  \end{split}
  \]
  and thus, using~\eqref{eq:expcalc},
  \[
    \Pr_{\bx, \bz}[w(\bx, \bz) \leq - n \delta]
  \leq
    2^{\frac{1}{2}  t^2 n \log(\gamma+2)^2 
      + t n \delta} \ . 
  \]
  Consequently, with $t:= - \frac{\delta}{\log(\gamma+2)^2}$,
  \begin{multline*}
    \Pr_{\bx, \bz}\bigl[
      \sum_{i=1}^n - \log \pb_{x_i} + \log \qb_{z_i} < n H(\rho_{A B}|\sigma_B) - n \delta
    \bigr] \\
  \leq
    \Pr_{\bx, \bz}\bigl[w(\bx, \bz) \leq - n \delta\bigr] 
  \leq 
    2^{-\frac{n \delta^2}
      {2 \log(\gamma+2)^2}} 
  \leq
    \Bigl(\frac{\eps}{2}\Bigr)^2 \ .    
  \end{multline*}
  Combining this with~\eqref{eq:probexp}
  implies~\eqref{eq:tracelowbound} and thus concludes the proof.
\end{proof}

The following corollary specializes Theorem~\ref{thm:Hmincondrep} to
the case where the first part of the state $\rho_{A B} = \rho_{X B}$
is classical and where $\sigma_B = \rho_B$.

\begin{corollary} \label{cor:Hmincondrepclass}
  Let $\rho_{X B} \in \NN(\cH_X \otimes \cH_B)$ be a density operator
  which is classical on $\cH_X$.  Then, for any $\eps \geq 0$,
  \[
    \sfrac{1}{n} \Hmin^\eps(\rho_{X B}^{\otimes n}|\rho_B^{\otimes n}) 
  \geq 
    H(\rho_{X B}) - H(\rho_B) -  \delta \ ,
  \]
  where $\delta := \bigl( 2 \Hmax(\rho_X)+ 3 \bigr)
  \sqrt{\frac{\log(\slfrac{1}{\eps})}{n} + 1}$.
\end{corollary}

\begin{proof}
  Assume without loss of generality that $\rho_B$ is invertible (the
  general statement then follows by continuity). Because the operator
  \[
    \id_X \otimes \rho_B  - \rho_{X B} 
  =
    \sum_{x \in \cX} \id_X \otimes \rho_B^x - \proj{x} \otimes \rho_B^x
  \]
  is nonnegative, we can apply Lemma~\ref{lem:prodpos} which gives
  \[
    \lambda_{\max}\bigl( 
      \rho_{X B}^{1/2} (\id_X \otimes \rho_B^{-1}) \rho_{X B}^{1/2}
    \bigr)
  \leq 
    1 \ .
  \]
  Hence, since $\rho_{X B}$ is normalized,
  \[
    \tr\bigl(\rho_{X B}^2 (\id_X \otimes \rho_B^{-1})\bigr)
  =
    \tr\bigl(\rho_{X B}
      \rho_{X B}^{1/2} (\id_X \otimes \rho_B^{-1}) \rho_{X B}^{1/2}
    \bigr)
  \leq
    1 \ . 
  \]
  Using the fact that, for any $a \geq 2$, $\log(a+3) \leq \log
  a+\frac{3}{2}$, we thus have
  \[
  \begin{split}
    \log\Bigl(
      \rank(\rho_X)+\tr\bigl(\rho_{X B}^2 (\id_X \otimes \sigma_B^{-1})\bigr)+2
    \Bigr)
  & \leq
    \log\bigl(\rank(\rho_X) + 3\bigr) \\
  & \leq
    \log \rank(\rho_X) + \sfrac{3}{2} \\
  & =
    \Hmax(\rho_X) + \sfrac{3}{2} \ .
  \end{split}
  \]
  The assertion then follows directly from
  Theorem~\ref{thm:Hmincondrep} with $\rho_{A B} := \rho_{X B}$ and
  $\sigma_B := \rho_B$.
\end{proof}

\index{smooth~min-entropy|)} 
\index{product~state|)}

\chapter{Symmetric States} \label{ch:sym}

\newcommand*{\Sym}[2]{\mathrm{Sym}(#1^{\otimes {#2}})}
\newcommand*{\SymR}[4]{\mathrm{Sym}(#1^{\otimes {#2}},
  {\ket{#3}^{\otimes #4}})}

\index{symmetric~state|(}
\index{permutation-invariant~state|(}

The state of an $n$-partite quantum system is said to be
\emph{symmetric} or \emph{per\-mu\-tation-invariant} if it is
unchanged under reordering of the subsystems. Such states have nice
properties which are actually very similar to those of product states.


The chapter is organized as follows: We first review some basic
properties of symmetric subspaces of product spaces
(Section~\ref{sec:symdef}) and show that any permutation-invariant
density operator has a purification in such a space
(Section~\ref{sec:sympur}). Next, we state our main result on the
structure of symmetric states, which generalizes the so-called
\emph{de Finetti representation
  theorem}\index{de~Finetti~representation}
(Section~\ref{sec:symrepr}).  Based on this result, we derive
expressions for the smooth min-entropy\index{smooth~min-entropy}
(Section~\ref{sec:smsym}) and the measurement statistics
(Section~\ref{sec:symstat}) of symmetric states.

\section{Definition and basic properties} \label{sec:symdef}

\index{symmetric~subspace|(}

\subsection{Symmetric subspace of $\cH^{\otimes n}$}

Let $\cH$ be a Hilbert space and let $\cS_n$ be the set of
permutations on $\{1, \ldots, n\}$\index{permutation}. For any $\pi
\in \cS_n$, we denote by the same letter $\Upi$ the unitary operation
on $\cH^{\otimes n}$ which permutes the $n$ subsystems, that is,
\[
  \Upi (\ket{\theta_1} \otimes \cdots \otimes \ket{\theta_n})
:=
  \ket{\theta_{\pi^{-1}(1)}} \otimes \cdots \otimes \ket{\theta_{\pi^{-1}(n)}} \ ,
\]
for any $\ket{\theta_1}, \ldots, \ket{\theta_n} \in \cH$.

\begin{definition} 
  Let $\cH$ be a Hilbert space and let $n \geq 0$. The \emph{symmetric
    subspace $\Sym{\cH}{n}$ of $\cH^{\otimes
      n}$} is the subspace of $\cH^{\otimes
    n}$ spanned by all vectors which are invariant under permutations
  of the subsystems, that is,
  \[
    \Sym{\cH}{n}
  :=
    \bigl\{ 
      \ket{\Psi} \in \cH^{\otimes n} : \, \Upi \ket{\Psi}  = \ket{\Psi} 
    \bigr\}\ .
  \]
\end{definition}

\begin{remark} \label{rem:symsub}
  For any $n', n'' \geq 0$,
\[
    \Sym{\cH}{n'+n''} \subseteq \Sym{\cH}{n'} \otimes \Sym{\cH}{n''} \ .
\]
\end{remark}

Lemma~\ref{lem:thetarep} below provides an alternative
characterization of the symmetric subspace
$\Sym{\cH}{n}$.

\begin{lemma} \label{lem:thetarep}
  Let $\cH$ be a Hilbert space and let $n \geq 0$. Then
  \[
    \Sym{\cH}{n}
  =
    \spanv 
      \bigl\{ \ket{\theta}^{\otimes n} : \, \ket{\theta} \in \cH \bigr\} \ .
  \]
\end{lemma}

\begin{proof}
  For a proof of this statement, we refer to the standard literature
  on symmetric functions or representation
  theory\index{representation~theory} (see, e.g., \cite{WalGoo00}).
\end{proof}


\newcommand*{\ntuplefreq}[2]{\Lambda^{#1}_{#2}}
\newcommand*{\freqset}[2]{\cQ^{#1}_{#2}}

\subsubsection{A basis of the symmetric subspace}

Let $\bx = (x_1, \ldots, x_n)$ be an $n$-tuple of elements from $\cX$.
The \emph{frequency distribution}\index{frequency~distribution}
$\freq{\bx}$ of $\bx$ is the probability distribution on $\cX$ defined
by the relative number of occurrences of each symbol, that is,
\[
  \freq{\bx}(x) := \sfrac{1}{n} \bigl| \{ i : \, x_i = x \} \bigr| \ ,
\]
for any $x \in \cX$.  In the following, we denote by
$\freqset{\cX}{n}$ the set of frequency distributions of $n$-tuples on
$\cX$, also called \emph{types with denominator $n$ on
  $\cX$}\index{type}.  Moreover, for any type $Q \in
\freqset{\cX}{n}$, we denote by $\ntuplefreq{Q}{n}$ the corresponding
\emph{type class}\index{type~class}, i.e., the set of all $n$-tuples
$\bx = (x_1, \ldots, x_n)$ with frequency distribution $\freq{\bx} =
Q$.

Let $\{\ket{x}\}_{x \in \cX}$ be an orthonormal basis of $\cH$. For
any $Q \in \freqset{\cX}{n}$, we define the vector $\ket{\Theta^Q}$ on
$\Sym{\cH}{n}$ by
\begin{equation} \label{eq:symsdef}
    \ket{\Theta^{Q}}
  :=
    \frac{1}{\sqrt{|\Lambda^{Q}_n|}} \sum_{(x_1, \ldots, x_n) \in \ntuplefreq{Q}{n}} 
        \ket{x_1} \otimes \cdots \otimes \ket{x_n} \ ,
\end{equation}
where, according to Lemma~\eqref{lem:Lambdasize}, $| \ntuplefreq{Q}{n} |=
\frac{n!}{\prod_x (n Q(x))!}$.

The vectors $\ket{\Theta^{Q}}$, for $Q \in \freqset{\cX}{n}$, are
mutually orthogonal and normalized.  We will see below (cf.\ 
Lemma~\ref{lem:symbasisrest}) that the family $\{\ket{\Theta^{Q}}\}_{Q
  \in \freqset{\cX}{n}}$ is a basis of $\Sym{\cH}{n}$. In particular,
if $\cH$ has dimension $d$, then $\dim( \Sym{\cH}{n}) =
|\freqset{\cX}{n}| = \binom{n+d-1}{n}$ (cf.  Lemma~\ref{lem:cQsize}).

\newcommand*{\Vrep}[4]{\cV(#1^{\otimes #2}, \ket{#3}^{\otimes #4})}

\subsection{Symmetric subspace along product states}

Let $\cH$ be a Hilbert space, let $\ket{\theta} \in \cH$ be fixed, and
let $0 \leq m \leq n$.  We denote by $\Vrep{\cH}{n}{\theta}{m}$ the
set of vectors $\ket{\Psi} \in \cH^{\otimes n}$ which, after some
reordering of the subsystems, are of the form $\ket{\theta}^{\otimes
  m} \otimes \ket{\Psit}$, that is,
\begin{equation} \label{eq:Vrepdef}
  \Vrep{\cH}{n}{\theta}{m}
:=
  \bigl\{ 
    \Upi (\ket{\theta}^{\otimes m} \otimes \ket{\Psit}) : \,
    \pi \in \cS_n, \,  \ket{\Psit} \in \cH^{\otimes n-m}
  \bigr\} \ .
\end{equation}

We will be interested in the subspace of $\Sym{\cH}{n}$ which only
consists of linear combinations of vectors from
$\Vrep{\cH}{n}{\theta}{m}$.

\begin{definition}
  Let $\cH$ be a Hilbert space, let $\ket{\theta} \in \cH$, and let $0
  \leq m \leq n$. The \emph{symmetric subspace
    $\SymR{\cH}{n}{\theta}{m}$ of $\cH^{\otimes n}$ along
    $\ket{\theta}^{\otimes m}$} is 
  \[
    \SymR{\cH}{n}{\theta}{m}
  :=
    \Sym{\cH}{n} \cap \spanv \Vrep{\cH}{n}{\theta}{m} \ ,
  \]
  where $\Vrep{\cH}{n}{\theta}{m}$ denotes the subset of $\cH^{\otimes n}$
  defined by~\eqref{eq:Vrepdef}.
\end{definition}

Note that $\SymR{\cH}{n}{\theta}{m} \subseteq \Sym{\cH}{n}$, where
equality holds if $m = 0$. In Section~\ref{sec:smsym}
and~\ref{sec:symstat}, we shall see that, if $r:=n-m$ is small
compared to $n$, then the states in $\SymR{\cH}{n}{\theta}{m}$ have
similar properties as product states $\ket{\theta}^{\otimes n}$.

\begin{lemma} \label{lem:symbasisrest}
  Let $\cH$ be a Hilbert space with orthonormal basis $\{\ket{x}\}_{x
    \in \cX}$, let $\ket{\theta} := \ket{\xb}$ for some $\xb \in \cX$,
  and let $0 \leq m \leq n$.  Then the family
  \[
    \cB := \{\ket{\Theta^Q}\}_{Q \in \freqset{\cX}{n}: \, Q(\xb) \geq \frac{m}{n}}
  \]
  of vectors $\ket{\Theta^Q}$ defined by~\eqref{eq:symsdef}\ is an
  orthonormal basis of $\SymR{\cH}{n}{\theta}{m}$.
\end{lemma}

Note that, for $m=0$, Lemma~\ref{lem:symbasisrest} implies that the
family $\{\ket{\Theta^Q}\}_{Q \in \freqset{\cX}{n}}$ is an orthonormal
basis of $\Sym{\cH}{n}$.

\begin{proof}
  For any $Q \in \freqset{\cX}{n}$, the vector $\ket{\Theta^Q}$ is
  invariant under permutations of the subsystems, that is,
  $\ket{\Theta^Q} \in \Sym{\cH}{n}$. Moreover, if $Q(\xb) \geq
  \frac{m}{n}$ then the sum on the right hand side
  of~\eqref{eq:symsdef} only runs over $n$-tuples which contain at
  least $m$ symbols $\xb$, that is, each term of the sum is contained
  in the set $\Vrep{\cH}{n}{\theta}{m}$ defined by~\eqref{eq:Vrepdef}
  and hence $\ket{\Theta^Q} \in \spanv \Vrep{\cH}{n}{\theta}{m}$. This
  proves that all vectors $\ket{\Theta^Q} \in \cB$ are contained in
  $\SymR{\cH}{n}{\theta}{m}$. Moreover, the vectors $\ket{\Theta^Q}$
  are mutually orthogonal and normalized.
 
  It remains to be shown that $\SymR{\cH}{n}{\theta}{m}$ is spanned by
  the vectors $\ket{\Theta^Q} \in \cB$.  Let thus $\ket{\Psi} \in
  \SymR{\cH}{n}{\theta}{m}$ be fixed. Since $\{\ket{x}\}_{x \in \cX}$
  is a basis of $\cH$, there exist coefficients $\alpha_{\bx}$, for
  $\bx = (x_1, \ldots, x_n) \in \cX^n$, such that
  \[
    \ket{\Psi}
  =
    \sum_{\bx \in \cX^n} \alpha_{\bx} \ket{x_1} \otimes \cdots \otimes \ket{x_n}
      \ .
  \]
  Because $\ket{\Psi}$ is invariant under permutations of the
  subsystems, the coefficients $\alpha_{\bx}$ can only depend on the
  frequency distribution $\freq{\bx}$.  This implies that there exist
  coefficients $\beta_{Q}$ such that
  \[
    \ket{\Psi}
  =
    \sum_{Q \in \freqset{\cX}{n}} \beta_{Q} \ket{\Theta^Q} \ .
  \]
  To conclude the proof, we need to verify that this sum can be
  restricted to frequency distributions $Q$ such that $Q(\xb) \geq
  \frac{m}{n}$. Observe that, for any $Q \in \freqset{\cX}{n}$ with
  $Q(\xb) < \frac{m}{n}$, the vector $\ket{\Theta^Q}$ is orthogonal to
  any vector in $\Vrep{\cH}{n}{\theta}{m}$ and thus also to any vector
  in $\SymR{\cH}{n}{\theta}{m}$. The corresponding coefficient
  $\beta_Q$ must thus be zero.
\end{proof}


Any vector $\ket{\Psi} \in \SymR{\cH}{n}{\theta}{m}$ can be written as
a linear combination of at most\footnote{$h$ denotes the \emph{binary
    Shannon entropy
    function}\index{binary~Shannon~entropy}\index{Shannon~entropy}
  defined by $h(p) := - p \log(p) - {(1-p)} {\log(1-p)}$.}
$2^{n h(\slfrac{m}{n})}$ vectors from the set
$\Vrep{\cH}{n}{\theta}{m}$ defined by~\eqref{eq:Vrepdef}.

\begin{lemma} \label{lem:symspacebin}
  Let $\ket{\Psi} \in \SymR{\cH}{n}{\theta}{m}$.  Then there exists an
  orthonormal family $\{\ket{\Psi^{s}}\}_{s \in \cS}$ of vectors from
  $\Vrep{\cH}{n}{\theta}{m}$ with cardinality $|\cS| \leq 2^{n
    h(\slfrac{m}{n})}$ such that $\ket{\Psi} \in \spanv
  \{\ket{\Psi^{s}}\}_{s \in \cS}$.
\end{lemma}

\begin{proof} 
  Let $\{\ket{x}\}_{x \in \cX}$ be an orthonormal basis of $\cH$ such
  that $\ket{\xb} = \ket{\theta}$. For any $n$-tuple $\bx = (x_1,
  \ldots, x_n) \in \cX^n$, we denote by $\ket{\bx}$ the vector
  $\ket{x_1} \otimes \cdots \otimes \ket{x_n}$.  Because $\ket{\Psi}
  \in \spanv \Vrep{\cH}{n}{\theta}{m}$, there exist coefficients
  $\beta_{\bx}$, for $\bx \in \cX^n$, such that
  \begin{equation} \label{eq:Psisum}
    \ket{\Psi}
  =
    \sum_{\bx: \, \freq{\bx}(\xb) \geq \frac{m}{n}}
      \beta_{\bx} \ket{\bx}  \ .
  \end{equation}
  
  Let $\cS$ be the set of all subsets $s \subseteq \{1, \ldots, n\}$
  of cardinality $|s| = m$.  Moreover, for any $\bx = (x_1, \ldots,
  x_n) \in \cX^n$ with $\freq{\bx}(\xb) \geq \frac{m}{n}$, let $s(\bx)
  \in \cS$ be a set of $m$ indices from $\{1, \ldots, n\}$ such that
  $i \in s(\bx) \implies x_i = \xb$.  Finally, for any $s \in \cS$,
  let
  \begin{equation} \label{eq:Psisdef}
    \ket{\Psi^s} := \sum_{\bx: \, s(\bx) = s} \beta_{\bx} \ket{\bx} \ .
  \end{equation}
  The sum in~\eqref{eq:Psisum} can then be rewritten as $\ket{\Psi} =
  \sum_{s \in \cS} \ket{\Psi^s}$, that is, $\ket{\Psi} \in \spanv
  \{\ket{\Psi^{s}}\}_{s \in \cS}$.  Moreover, Lemma~\ref{lem:binsize}
  implies $|\cS| \leq 2^{n h(\slfrac{m}{n})}$.
  
  It remains to be shown that $\{\ket{\Psi^{s}}\}_{s \in \cS}$ is an
  orthonormal family of vectors from $\Vrep{\cH}{n}{\theta}{m}$. Let
  thus $s \in \cS$ be fixed and let $\pi$ be a permutation such that
  $\pi(s) = \{1, \ldots, m\}$. Hence, for any $\bx$ with $s(\bx) = s$,
  the vector $\Upi \ket{\bx}$ has the form $\ket{\theta}^{\otimes m}
  \otimes \ket{\Psit}$, for some $\ket{\Psit} \in \cH^{\otimes n-m}$.
  By the definition~\eqref{eq:Psisdef}, the same holds for $\Upi
  \ket{\Psi^s}$, i.e., $\ket{\Psi^s} \in \Vrep{\cH}{n}{\theta}{m}$.
  Furthermore, because for distinct $s, s' \in \cS$, the sum
  in~\eqref{eq:Psisdef} runs over disjoint sets of $n$-tuples $\bx$,
  and because the vectors $\ket{\bx}$ are mutually orthogonal, the
  states $\ket{\Psi^{\bs}}$ are also mutually orthogonal. The
  assertion thus follows by normalizing the vectors $\ket{\Psi^s}$.
\end{proof}

\index{symmetric~subspace|)}

\section{Symmetric purification} \label{sec:sympur}

\index{symmetric~purification|(}

An operator $\rho_n$ on $\cH^{\otimes n}$ is called
\emph{permutation-invariant}\index{permutation-invariant~state} if
$\Upi \rho_n \Upi^\dagger = \rho_n$, for any permutation $\pi \in
\cS_n$. For example, the pure state $\rho_n=\proj{\Psi}$, for some
vector $\ket{\Psi}$ of the symmetric subspace of $\cH^{\otimes n}$, is
permutation-invariant.  More generally, any mixture of symmetric pure
states is permutation-in\-var\-iant.

The converse, however, is not always true. Consider for example the
fully mixed state $\rho_2$ on $\cH^{\otimes 2}$ where $\dim(\cH) = 2$.
Because this operator can be written as $\rho_2 = \sigma^{\otimes 2}$,
it is invariant under permutations. However, $\rho_2$ has rank $4$,
whereas the symmetric subspace of $\cH^{\otimes 2}$ only has dimension
$3$.  Consequently, $\rho_2$ is not a mixture of symmetric pure
states.

Lemma~\ref{lem:sympurification} below establishes another connection
between permutation-invariant operators and symmetric pure states. We
show that any permuta\-tion-invariant operator $\rho_n$ on
$\cH^{\otimes n}$ has a purification on the symmetric
subspace\index{symmetric~subspace} of $(\cH \otimes \cH)^{\otimes n}$.

To prove this result, we need a technical lemma which states that a
fully entangled state on two subsystems is unchanged when the same
unitary operation is applied to both subsystems.

\begin{lemma} \label{lem:invariantstate}
  Let $\{\ket{x}\}_{x \in \cX}$ be an orthonormal family of vectors on
  a Hilbert space $\cH$ and define 
  \[
    \ket{\Psi} 
  := 
    \sum_{x \in \cX} \ket{x} \otimes \overline{\ket{x}} \ ,
  \]
  where, for any $x \in \cX$, $\overline{\ket{x}}$ denotes the complex
  conjugate of $\ket{x}$ (with respect to some basis of $\cH$). Let
  $U$ be a unitary operation on the subspace spanned by
  $\{\ket{x}\}_{x \in \cX}$ and let $\overline{U}$ be its complex
  conjugate. Then
  \[
    (U \otimes \overline{U}) \ket{\Psi} = \ket{\Psi} \ .
  \]
\end{lemma}

\begin{proof}  
  A simple calculation shows that, for any $x, x' \in \cX$,
  \begin{align*}
    \bigl(\bra{x} \otimes \overline{\bra{x'}}\bigr) \ket{\Psi}
  & =
    \delta_{x,x'}
  \\
    \bigl(\bra{x} \otimes \overline{\bra{x'}}\bigr) 
      \bigl(U \otimes \overline{U}\bigr)\ket{\Psi}
  & =
    \delta_{x,x'} \ .
  \end{align*}
  The assertion follows because, obviously, $\{\ket{x} \otimes
  \overline{\ket{x'}}\}_{x, x' \in \cX}$ is a basis of the subspace of
  $\cH \otimes \cH$ that contains $\ket{\Psi}$.
\end{proof}

\begin{lemma} \label{lem:sympurification}
  Let $\rho_n \in \NN(\cH^{\otimes n})$ be permutation-invariant. Then
  there exists a purification of $\rho_n$ on $\Sym{(\cH \otimes
    \cH)}{n}$.
\end{lemma}

\newcommand*{\avect}{\phi}

\begin{proof}
  Let $\{\ket{x}\}_{x \in \cX}$ be an (orthonormal) eigenbasis of
  $\rho_n$ and let $\Lambda$ be the set of eigenvalues of $\rho_n$.
  For any $\lambda \in \Lambda$, let $\cH_\lambda$ be the
  corresponding eigenspace of $\rho_n$, i.e., $\rho_n \ket{\avect} =
  \lambda \ket{\avect}$, for any $\ket{\avect} \in \cH_\lambda$.
  
  Because $\rho_n$ is invariant under permutations, we have $
  \Upi^\dagger \rho_n \Upi \ket{\avect} = \lambda \ket{\avect}$, for
  any $\ket{\avect} \in \cH_\lambda$ and $\pi \in \cS_n$.  Applying
  the unitary operation $\Upi$ to both sides of this equality gives
  $\rho_n \Upi \ket{\avect} = \lambda \Upi \ket{\avect}$, that is,
  $\Upi \ket{\avect} \in \cH_\lambda$.  This proves that the
  eigenspaces $\cH_\lambda$ of $\rho_n$ are invariant under
  permutations.
  
  For any $\ket{\phi} \in \cH^{\otimes n}$, we denote by
  $\overline{\ket{\phi}}$ the complex conjugate of $\ket{\phi}$ with
  respect to some product basis on $\cH^{\otimes n}$.  Moreover, for
  any eigenvalue $\lambda \in \Lambda$, let
  \[
    \ket{\Psi^\lambda} 
  := 
    \sum_{x \in \cX_\lambda}
      \ket{x} \otimes \overline{\ket{x}} \ ,
  \]
  where $\cX_{\lambda} := \{x \in \cX : \, \ket{x} \in
  \cH_{\lambda}\}$, i.e., $\{\ket{x}\}_{x \in \cX_{\lambda}}$ is an
  orthonormal basis of the eigenspace $\cH_{\lambda}$.  Finally, we
  define the vector $\ket{\Psi} \in \cH^{\otimes n} \otimes
  \cH^{\otimes n}$ by
  \[
    \ket{\Psi} 
  :=
    \sum_{\lambda \in \Lambda} \sqrt{\lambda} \ket{\Psi^\lambda} \ .
  \]
  It is easy to verify that the operator obtained by taking the
  partial trace of $\proj{\Psi}$ satisfies
  \[
    \tr_{\cH^{\otimes n}}(\proj{\Psi})
  = 
    \sum_{\lambda \in \Lambda} \sum_{x \in \cX_\lambda} 
      \lambda \proj{x} 
  = 
    \rho_n \ ,
  \]
  i.e., $\proj{\Psi}$ is a purification of $\rho_n$. It thus remains
  to be shown that $\ket{\Psi}$ is symmetric.
  
  Let $\pi \in \cS_n$ be a fixed permutation. Note that its complex
  conjugate $\overline{\Upi}$ is equal to $\Upi$.  (Recall that we
  defined the complex conjugate with respect to a product basis of
  $\cH^{\otimes n}$.) Moreover, because $\Upi$ is unitary on
  $\cH^{\otimes n}$ and, additionally, for any $\lambda \in \Lambda$,
  the subspace $\cH_\lambda$ is invariant under $\Upi$, the
  restriction of $\Upi$ to $\cH_\lambda$ is unitary as well. Hence, by
  Lemma~\ref{lem:invariantstate},
  \[
    (\Upi \otimes \Upi) \ket{\Psi^\lambda} 
  =
    (\Upi \otimes \overline{\Upi}) \ket{\Psi^\lambda} 
  = 
    \ket{\Psi^\lambda}
  \]
  and thus, by linearity, 
  \[
    (\Upi \otimes \Upi) \ket{\Psi}
  = 
    \sum_{\lambda \in \Lambda} 
      \sqrt{\lambda} (\Upi \otimes \Upi) \ket{\Psi^\lambda} 
  =
    \sum_{\lambda \in \Lambda} \sqrt{\lambda} \ket{\Psi^\lambda} 
  = 
    \ket{\Psi} \ .
  \]
  Because this holds for any permutation $\pi$ on $\cH^{\otimes n}$,
  we conclude $\ket{\Psi} \in \Sym{(\cH \otimes \cH)}{n}$.
\end{proof}

\index{symmetric~purification|)}

\section{De Finetti representation} \label{sec:symrepr}

\index{de~Finetti~representation|(}

While any product state $\rho_n = \sigma^{\otimes n}$ on $\cH^{\otimes
  n}$ is permutation-invariant, the converse is not true in general.
Nevertheless, as we shall see, the properties of permutation-invariant
states $\rho_n$ are usually very similar to those of product states.

The quantum de Finetti representation theorem makes this connection
explicit. In its basic version, it states that any density operator
$\rho_n$ on $\cH^{\otimes n}$ which is \emph{infinitely}
exchangeable\index{infinite~exchangeability}\index{exchangeability},
i.e., $\rho_n$ is the partial state of a permutation-invariant
operator $\rho_{n+k}$ on $n+k$ subsystems, for \emph{all} $k \geq 0$,
can be written as a mixture of product states $\sigma^{\otimes n}$.

In this section, we generalize the quantum de Finetti representation
to the \emph{finite}\index{finite~exchangeability} case, where
$\rho_n$ is only $(n+k)$-exchangeable, i.e., $\rho_n$ is the partial
state of a permutation-invariant operator $\rho_{n+k}$ on $n+k$
subsystems, for some \emph{fixed} $k \geq 0$.
Theorem~\ref{thm:symmix} below states that any pure density operator
$\rho_n$ on $\cH^{\otimes n}$ which is $(n+k)$-exchangeable is close
to a mixture of states $\rhob_n^{\ket{\theta}}$ which have almost
product form $\ket{\theta}^{\otimes n}$, for $\ket{\theta} \in \cH$.
More precisely, for any $\ket{\theta}$, $\rhob_n^{\ket{\theta}}$ is a
pure state of the symmetric subspace of $\cH^{\otimes n}$ along
$\ket{\theta}^{\otimes n-r}$, for some small $r \geq 0$.  Because of
Lemma~\ref{lem:sympurification}, this statement also holds for mixed
states $\rho_n$.

The proof of Theorem~\ref{thm:symmix} is based on the following lemma
which states that the uniform mixture of product states
$(\proj{\theta})^{\otimes n}$, for all normalized vectors
$\ket{\theta} \in \cS_1(\cH) := \{\ket{\theta} \in \cH:
\|\ket{\theta}\| = 1\}$, is equal to the fully mixed state on the
symmetric subspace of $\cH^{\otimes n}$.

\begin{lemma} \label{lem:SymPOVM}
  Let $\cH$ be a $d$-dimensional Hilbert space and let $n \geq 0$.
  Then
  \[
    \int_{\cS_1(\cH)} (\proj{\theta})^{\otimes n} \omega(\ket{\theta})
  =
    {\textstyle \binom{n+d-1}{n}^{-1}} \cdot \id_{\Sym{\cH}{n}} \ ,
  \]
  where $\omega$ denotes the uniform probability measure on the unit
  sphere $\cS_1(\cH)$.
\end{lemma}

Lemma~\ref{lem:SymPOVM} can be proven using techniques from
representation theory\index{representation~theory}, in particular,
Schur's Lemma\index{Schur's~Lemma} (see, e.g., \cite{WalGoo00}). In
the following, however, we propose an alternative proof.

\newcommand*{\constc}{c}
\newcommand*{\vecta}{u}
\newcommand*{\vectb}{v}
\newcommand*{\vectc}{w}

\begin{proof}
  Let 
  \[
  T := \int_{\cS_1(\cH)} (\proj{\theta})^{\otimes n} \omega(\ket{\theta}) \ .
  \]
  We first show that $T = \constc \cdot \id_{\Sym{\cH}{n}}$ for some
  constant $\constc$.
  
  Because the space $\Sym{\cH}{n}$ is spanned by vectors of the form
  $\ket{\theta}^{\otimes n}$ (cf.\ Lemma~\ref{lem:thetarep}), it is
  sufficient to show that, for any $\ket{\vecta}, \ket{\vectb} \in
  \cS_1(\cH)$,
  \begin{equation} \label{eq:SymPOVMp}
    \bra{\vecta}^{\otimes n} T \ket{\vectb}^{\otimes n} 
  = 
    \bra{\vecta}^{\otimes n} \constc \cdot   \id_{\Sym{\cH}{n}}
      \ket{\vectb}^{\otimes n} \ .
  \end{equation}
  
  Let thus $\ket{\vecta}, \ket{\vectb} \in \cS_1(\cH)$ be fixed and
  define $\alpha := \spr{\vecta}{\vectb}$ and
  $\ket{\vectc}:=\ket{\vectb}-\alpha \ket{\vecta}$, i.e.,
  $\spr{\vecta}{\vectc} = 0$. Then
  \begin{equation} \label{eq:intTall}
  \begin{split}
    \bra{\vecta}^{\otimes n} T \ket{\vectb}^{\otimes n}
  & =
    \int_{\cS_1(\cH)} \spr{\vecta}{\theta}^n \spr{\theta}{\vectb}^n
  \omega(\ket{\theta}) \\
  & =
    \int_{\cS_1(\cH)} 
      \spr{\vecta}{\theta}^n 
        \bigl(\alpha \spr{\theta}{\vecta} + \spr{\theta}{\vectc}\bigr)^n 
      \omega(\ket{\theta}) \ .
  \end{split}
  \end{equation}
  Note that, for any $m \in \{0, \ldots, n\}$,
  \[
    \int_{\cS_1(\cH)} 
      \spr{\vecta}{\theta}^n \spr{\theta}{\vecta}^{n-m} \spr{\theta}{\vectc}^{m} 
      \omega(\ket{\theta})
  =
    \int_{\cS_1(\cH)} 
      |\spr{\vecta}{\theta}|^{2 (n-m)} 
      \spr{\vecta}{\theta}^{m} \spr{\theta}{\vectc}^{m} \omega(\ket{\theta}) \ .
  \]
  Because, for any fixed value of $\spr{\vecta}{\theta}$, the integral
  runs over all phases of $\spr{\theta}{\vectc}$ (recall that
  $\ket{\vecta}$ and $\ket{\vectc}$ are orthogonal) and because the
  probability measure $\omega$ is invariant under unitary operations,
  this expression equals zero for any $m > 0$. The integral on the
  right hand side of~\eqref{eq:intTall} can thus be rewritten as
  \begin{equation} \label{eq:intTsingle}
  \begin{split}
    \bra{\vecta}^{\otimes n} T \ket{\vectb}^{\otimes n}
  & =
    \int_{\cS_1(\cH)} 
      \alpha^n |\spr{\vecta}{\theta}|^{2 n}  \omega(\ket{\theta}) \\
  & =
    \spr{\vecta}{\vectb}^n 
      \int_{\cS_1(\cH)} |\spr{\vecta}{\theta}|^{2 n} \omega(\ket{\theta}) \ .
  \end{split}
  \end{equation}
  Using again the fact that the probability measure $\omega$ is
  invariant under unitary operations, we conclude that the integral on
  the right hand side cannot depend on the vector $\ket{\vecta}$,
  i.e., it is equal to a constant $\constc$.  This
  implies~\eqref{eq:SymPOVMp} and thus proves that $T = \constc \cdot
  \id_{\Sym{\cH}{n}}$.
  
  To determine the value of $\constc$,\footnote{Alternatively, the
    constant~$\constc$ can be computed by an explicit evaluation of
    the integral on the right hand side of~\eqref{eq:intTsingle}.
    Remarkably, this can be used to prove Lemma~\ref{lem:thetarep}:
    Observe first that, by the arguments given in the proof,
    $\constc^{-1}$ must be equal to the dimension of the space spanned
    by the vectors of the form $\ket{\theta}^{\otimes n}$.  On the
    other hand, the explicit computation of $\constc$ shows that
    $\constc^{-1}$ equals $\binom{n+d-1}{n}$, which is the dimension
    of $\Sym{\cH}{n}$.  Because the space spanned by the vectors
    $\ket{\theta}^{\otimes n}$ is a subspace of $\Sym{\cH}{n}$, it
    follows that these spaces are equal.}  observe that
  \begin{equation} \label{eq:cdet}
    \tr(T)
  =
    \int_{\cS_1(\cH)} \tr\bigl((\proj{\theta})^{\otimes n}\bigr) 
    \omega(\ket{\theta})
  =
    \int_{\cS_1(\cH)} \omega(\ket{\theta})
  =
    1 \ ,
  \end{equation}
  where the last equality holds because $\omega$ is a probability
  measure on $\cS_1(\cH)$.  On the other hand, we have $\tr(T) =
  \constc \cdot \dim(\Sym{\cH}{n})$. Hence, $\constc^{-1} =
  \dim(\Sym{\cH}{n}) = \binom{n+d-1}{n}$, which concludes the proof.
\end{proof}

We are now ready to state and prove a de Finetti style representation
theorem. Note that Theorem~\ref{thm:symmix} is restricted to
\emph{pure} symmetric states. The statement for general
permutation-invariant states then follows because any such state has a
symmetric purification (see Lemma~\ref{lem:sympurification}).

\begin{theorem} \label{thm:symmix}
  Let $\rho_{n+k}$ be a pure density operator on $\Sym{\cH}{n+k}$ and
  let $0 \leq r \leq n$.  Then there exists a measure $\nu$ on
  $\cS_1(\cH)$ and, for each $\ket{\theta} \in \cS_1(\cH)$, a pure
  density operator $\rhob^{\ket{\theta}}_n$ on
  $\SymR{\cH}{n}{\theta}{n-r}$ such that
  \[
    \Bigl\|
      \tr_k(\rho_{n+k}) 
      -  \int_{\cS_1(\cH)} \rhob_n^{\ket{\theta}} \, \nu(\ket{\theta}) 
    \Bigr\|_1
  \leq 
    2 e^{-\frac{k (r+1)}{2(n+k)} + \frac{1}{2} \dim(\cH) \ln k} \ .
  \]
\end{theorem}

\begin{proof}
  Because the density operator $\rho_{n+k}$ is pure, we have
  $\rho_{n+k} = \proj{\Psi}$ for some $\ket{\Psi} \in \Sym{\cH}{n+k}$.
  For any $\ket{\theta} \in \cS_1(\cH)$, let
  \[
    \ket{\Psi^{\ket{\theta}}}
  :=
   \sqrt{{\textstyle \binom{k+d-1}{k}}} 
     \cdot \bigl(\id_{\cH}^{\otimes n} \otimes \bra{\theta}^{\otimes k}\bigr) 
     \cdot \ket{\Psi} \ ,
  \]
  where $d := \dim(\cH)$. Because $\Sym{\cH}{n+k}$ is a subspace of
  $\Sym{\cH}{n} \otimes \Sym{\cH}{k}$ (see Remark~\ref{rem:symsub}),
  $\ket{\Psi^{\ket{\theta}}}$ is contained in $\Sym{\cH}{n}$. Let
  $\rho^{\ket{\theta}}_n:=\proj{\Psi^{\ket{\theta}}}$, let
  $P^{\ket{\theta}}$ be the projector onto the subspace
  $\SymR{\cH}{n}{\theta}{n-r}$, and define
  \[
    \rhob^{\ket{\theta}}_n 
  := 
    \textfrac{1}{p(\ket{\theta})} 
      P^{\ket{\theta}} \rho^{\ket{\theta}}_n P^{\ket{\theta}} \ ,
  \]
  where $p(\ket{\theta}) := \tr(P^{\ket{\theta}} \rho^{\ket{\theta}}_n
  P^{\ket{\theta}})$, i.e., $\rhob^{\ket{\theta}}_n$ is normalized
  and, because $\rho^{\ket{\theta}}_n$ has rank one, it is also pure.
  Finally, let $\nu$ be the measure defined by $\nu := p \cdot
  \omega$, where $\omega$ is the uniform probability measure on
  $\cS_1(\cH)$.  It then suffices to show that
  \begin{equation} \label{eq:symmixdist}
    \delta := \Bigl\| \tr_k(\rho_{n+k}) 
    - \int_{\cS_1(\cH)} P^{\ket{\theta}} \rho^{\ket{\theta}}_n
    P^{\ket{\theta}} 
      \, \omega(\ket{\theta}) \Bigr\|_1 
  \leq 
    2 e^{-\frac{k (r+1)}{2(n+k)} + \frac{1}{2} d \ln k} \ .
  \end{equation}
  
  By the definition of $\rho^{\ket{\theta}}_n$, we have
  \begin{equation} \label{eq:rhothetaproj}
    \rho^{\ket{\theta}}_n
  =
    \proj{\Psi^{\ket{\theta}}} 
  = 
    {\textstyle \binom{k+d-1}{k}} \cdot \tr_k\bigl(
      \id_{\cH}^{\otimes n} \otimes (\proj{\theta})^{\otimes k} 
        \cdot \proj{\Psi} \bigr) \ ,
  \end{equation}
  and thus, by Lemma~\ref{lem:SymPOVM}, 
  \begin{align*}
    \int_{\cS_1(\cH)} \rho^{\ket{\theta}}_n \, \omega(\ket{\theta}) 
  & =
    {\textstyle \binom{k+d-1}{k}} \cdot \int_{\cS_1(\cH)} \tr_k\bigl( 
      \id_{\cH}^{\otimes n} \otimes (\proj{\theta})^{\otimes k}
      \cdot \proj{\Psi}
    \bigr) \,  \omega(\ket{\theta})  \\ 
  &= 
    \tr_k\bigl( 
      \id_{\cH}^{\otimes n} \otimes \id_{\Sym{\cH}{k}}
      \cdot \proj{\Psi}
    \bigr)  \ .
  \end{align*}
  Since $\Sym{\cH}{n+k}$ is a subspace of $\cH^{\otimes n} \otimes
  \Sym{\cH}{k}$, the vector $\ket{\Psi}$ is contained in $\cH^{\otimes
    n} \otimes \Sym{\cH}{k}$.  The operation $\id_{\cH}^{\otimes n}
  \otimes \id_{\Sym{\cH}{k}}$ in the above expression thus leaves
  $\proj{\Psi}$ unchanged. Because $\tr_k( \proj{\Psi}) =
  \tr_k(\rho_{n+k})$, we conclude
  \begin{equation} \label{eq:rhointp}
    \int_{\cS_1(\cH)} \rho^{\ket{\theta}}_n 
      \, \omega(\ket{\theta})
  =
    \tr_k(\rho_{n+k})  \ .
  \end{equation}
  
  Using this representation of $\tr_k(\rho_{n+k})$ and the triangle
  inequality, the distance $\delta$ defined by~\eqref{eq:symmixdist}
  can be bounded by
  \[
    \delta
  \leq
    \int_{\cS_1(\cH)} 
      \bigl\| \rho^{\ket{\theta}}_n - P^{\ket{\theta}} \rho^{\ket{\theta}}_n
      P^{\ket{\theta}} \bigr\|_1 \, 
      \, \omega(\ket{\theta}) \ .
  \]
  Because the operators $P^{\ket{\theta}}$ are projectors, we can
  apply Lemma~\ref{lem:disttracebound} to bound the distance between
  $\rho^{\ket{\theta}}_n$ and $P^{\ket{\theta}} \rho^{\ket{\theta}}_n
  P^{\ket{\theta}}$, which gives
  \[
    \delta
  \leq
    2 \int_{\cS_1(\cH)} 
      \sqrt{\tr(\rho_n^{\ket{\theta}})} 
      \sqrt{\tr(\rho_n^{\ket{\theta}}) - \tr(P^{\ket{\theta}} \rho_n^{\ket{\theta}})} 
    \, \omega(\ket{\theta}) \ .
  \]
  To bound the integral on the right hand side, we use the
  Cauchy-Schwartz inequality for the scalar product defined by
  $\spr{f}{g} := \int_{\cS_1(\cH)} f(\ket{\theta}) g(\ket{\theta}) \,
  \omega(\ket{\theta})$, i.e.,
  \[
    \delta
  \leq
    2 \sqrt{
      \int_{\cS_1(\cH)} 
        \tr(\rho^{\ket{\theta}}_n) \, \omega( \ket{\theta})
      }
      \sqrt{
      \int_{\cS_1(\cH)} 
        \bigl(
          \tr(\rho^{\ket{\theta}}_n) 
          - \tr(P^{\ket{\theta}} \rho^{\ket{\theta}}_n)
        \bigr) \, \omega( \ket{\theta}) \ .
    }
  \]
  Because of~\eqref{eq:rhointp}, the first integral on the right hand
  side equals $\tr(\rho_{n+k}) = 1$, that is,
  \begin{equation} \label{eq:rhorhobdistbound}
    \delta
  \leq
    2 \sqrt{
      \int_{\cS_1(\cH)} 
        \bigl(\tr(\rho^{\ket{\theta}}_n) 
        - \tr(P^{\ket{\theta}}\rho^{\ket{\theta}}_n)\bigr) 
      \, \omega(\ket{\theta}) 
    } \ .
  \end{equation}
  Let $\bar{P}^{\ket{\theta}}$ be the projector orthogonal to
  $P^{\ket{\theta}}$, i.e., $\bar{P}^{\ket{\theta}} :=
  \id_{\Sym{\cH}{n}} - P^{\ket{\theta}}$.
  With~\eqref{eq:rhothetaproj}, the term in the integral can be
  rewritten as
  \begin{equation} \label{eq:trrhorhobdiff}
  \begin{split}
    \tr(\rho^{\ket{\theta}}_n) - \tr(P^{\ket{\theta}} \rho^{\ket{\theta}}_n) 
  & =
    \tr(\bar{P}^{\ket{\theta}} \rho_n^{\ket{\theta}}) \\
  & =
    {\textstyle \binom{k+d-1}{k}} \cdot \tr\bigl(
      \bar{P}^{\ket{\theta}} \otimes (\proj{\theta})^{\otimes k}
        \cdot \proj{\Psi} 
    \bigr)  \ .
  \end{split}
  \end{equation}
  
  Let $\ket{\theta} \in \cH$ be fixed and let $\{\ket{x}\}_{x \in
    \cX}$ be an orthonormal basis of $\cH$ with $\ket{\xb} =
  \ket{\theta}$, for some $\xb \in \cX$. Moreover, for all frequency
  distributions $Q \in \freqset{\cX}{n}$ and $\Qb \in
  \freqset{\cX}{n+k}$, let $\ket{\Theta_n^Q}$ and
  $\ket{\Thetab_{n+k}^{\Qb}}$ be the vectors in $\Sym{\cH}{n}$ and
  $\Sym{\cH}{n+k}$, respectively, defined by~\eqref{eq:symsdef}.
  
  According to Lemma~\ref{lem:symbasisrest}, the family of vectors
  $\ket{\Theta_n^Q}$, for all $Q \in \freqset{\cX}{n}$, is an
  orthonormal basis of $\Sym{\cH}{n}$.  Moreover, the subfamily where
  $Q(\xb) \geq \frac{n-r}{n}$ is a basis of
  $\SymR{\cH}{n}{\theta}{n-r}$.  Consequently, the projector
  $\bar{P}^{\ket{\theta}}$ on the space orthogonal to
  $\SymR{\cH}{n}{\theta}{n-r}$ can be written as
  \[
    \bar{P}^{\ket{\theta}} 
  = 
    \sum_{Q: \, Q(\xb) <\frac{n-r}{n}} \proj{\Theta_n^{Q}} \ .
  \]
  Identity~\eqref{eq:trrhorhobdiff} then reads
  \begin{equation} \label{eq:trdiffp}
    \tr(\rho^{\ket{\theta}}_n) - \tr(P^{\ket{\theta}} \rho^{\ket{\theta}}_n) 
  =
    {\textstyle \binom{k+d-1}{k}} \sum_{Q: \, Q(\xb) < \frac{n-r}{n}} 
      \Bigl|
        \bigl(\bra{\Theta_n^Q} \otimes \bra{\theta}^{\otimes k}\bigr)
        \cdot \ket{\Psi}
      \Bigr |^2 \ .
  \end{equation}
  
  Because the family of vectors $\ket{\Thetab_{n+k}^{\Qb}}$, for $\Qb
  \in \freqset{\cX}{n+k}$, is a basis of the symmetric subspace
  $\Sym{\cH}{n+k}$ (see again Lemma~\ref{lem:symbasisrest}) there
  exist coefficients $\alpha_{\Qb}$ such that
  \begin{equation} \label{eq:Psilincomb}
    \ket{\Psi} 
  = 
    \sum_{\Qb} \alpha_{\Qb} \ket{\Thetab_{n+k}^{\Qb}} \ ,
  \end{equation}
  where the sum runs over all $\Qb \in \freqset{\cX}{n+k}$.
  
  It is easy to verify that, for any $Q \in \freqset{\cX}{n}$ and $\Qb
  \in \freqset{\cX}{n+k}$, the scalar product $(\bra{\Theta^Q_n}
  \otimes\bra{\theta}^{\otimes k}) \cdot \ket{\Thetab_{n+k}^{\Qb}}$
  equals zero unless
  \begin{equation} \label{eq:QQb}
    (n+k) \Qb(x) 
  = 
    \begin{cases} 
       n Q(x) + k & \text{if $x = \xb$} \\
       n Q(x) & \text{otherwise}
    \end{cases}
  \end{equation}
  holds for all $x \in \cX$, in which case
  \begin{equation} \label{eq:Thetaspr}
     \bigl(\bra{\Theta_n^Q} \otimes \bra{\theta}^{\otimes k}\bigr)
     \cdot \ket{\Thetab_{n+k}^{\Qb}}
  =
     \sqrt{\frac{\frac{n!}{\prod_x (n Q(x))! }}
                 {\frac{(n+k)!}{\prod_x ((n+k) \Qb(x))!}}
          } 
  =
    \sqrt{\frac{n! (n Q(\xb)+k)!}{(n+k)!(n Q(\xb))!}} \ .
  \end{equation}
  Let $Q \in \freqset{\cX}{n}$ with $Q(\xb) < \frac{n-r}{n}$ and let
  $\Qb \in \freqset{\cX}{n+k}$ such that~\eqref{eq:QQb} holds. Then,
  from~\eqref{eq:Psilincomb} and~\eqref{eq:Thetaspr},
  \[
    \Bigl|
      \bigl(\bra{\Theta_n^Q} \otimes \bra{\theta}^{\otimes k}\bigr)
      \cdot \ket{\Psi}
    \Bigr|^2
  =
    |\alpha_{\Qb}|^2 \frac{n! (n Q(\xb) + k)!}{(n+k)!(n Q(\xb))!}
  \leq
    |\alpha_{\Qb}|^2 D_{n,k,r} 
  \]
  where $D_{n,k,r} := \frac{n! (n+k-r-1)!}{(n+k)!(n-r-1)!}$. Note that
  $Q(\xb) < \frac{n-r}{n}$ implies $\Qb(\xb) < \frac{n+k -r}{n+k}$.
  Consequently, from~\eqref{eq:trdiffp},
  \[
    \tr(\rho^{\ket{\theta}}_n) - \tr(P^{\ket{\theta}} \rho^{\ket{\theta}}_n) 
  \leq
    {\textstyle \binom{k+d-1}{k}} \cdot D_{n,k,r}
      \sum_{\Qb: \, \Qb(\xb)< \frac{n+k-r}{n+k}}  
        |\alpha_{\Qb}|^2 
  \leq
    {\textstyle \binom{k+d-1}{k}} \cdot D_{n,k,r}  \ ,
   \]
   where the last inequality follows from the fact that $\sum_{\Qb}
   |\alpha_{\Qb}|^2 = \|\ket{\Psi}\|^2 = \tr(\rho_{n+k}) = 1$. The
   term $D_{n,k,r}$ can be bounded by
  \[
  \begin{split}
    D_{n,k,r}
  & =
    \frac{(n-r) (n-r+1) \cdots (n+k-r-1)}{(n+1) (n+2) \cdots (n+k)} \\
  & \leq
    \left(\frac{n+k-r-1}{n+k}\right)^k \\
  & =
    \left(1- \frac{r+1}{n+k} \right)^k \ .
  \end{split}
  \]
  Defining $\beta:=\frac{r+1}{n+k}$ and using the fact that, for any
  $\beta \in [0,1]$, $(1-\beta)^{1/\beta} \leq e^{-1}$, we find
  \[
    D_{n,k,r}
  \leq
    (1-\beta)^k
  =
    \left((1-\beta)^{1/\beta}\right)^{\beta k} 
  \leq
    e^{-\beta k} \ .
  \]
  Finally, because for any $k \geq 2$ (note that, for $k < 2$, the
  assertion is trivial) $\binom{k+d-1}{k} \leq k^d$, we have
  \[
    \tr(\rho_n^{\ket{\theta}}) - \tr(P^{\ket{\theta}} \rho_n^{\ket{\theta}}) 
  \leq
    k^d e^{-k \frac{r+1}{n+k}} \ .
  \]
  Inserting this into~\eqref{eq:rhorhobdistbound}, the
  bound~\eqref{eq:symmixdist} follows because $\omega(\ket{\theta})$
  is a probability measure on $\cS_1(\cH)$.
\end{proof}

If the symmetric state $\rho_{n+k}$ on $\Sym{\cH}{n+k}$ has some
additional structure then the set of states that contribute to the
mixture in the expression of Theorem~\ref{thm:symmix} can be
restricted.  Remark~\ref{rem:addcond} below treats the case where the
subspaces $\cH = \cH_A \otimes \cH_B$ are bipartite systems and where
the partial state on $\cH_A^{\otimes n+k}$ has product form.




\begin{remark} \label{rem:addcond}
  Let $\cH := \cH_A \otimes \cH_B$ be a bipartite Hilbert space, let
  $\rho_{A^{n+k} B^{n+k}}$ be a pure density operator on
  $\Sym{\cH}{n}$ such that $\rho_{A^{n+k}} = \sigma_A^{\otimes n+k}$,
  let $0 \leq r \leq n$, and let $\nu$ be the measure defined by
  Theorem~\ref{thm:symmix}. Then, for any $\delta \geq 0$, the set
  \[
    \overline{\Gamma^\delta} 
  :=
    \bigl\{\ket{\theta} \in \cS_1(\cH) : \, 
      \| \tr_B(\proj{\theta}) - \sigma_{A} \|_1 > \delta \bigr\} 
  \]
  has at most weight $ \nu(\overline{\Gamma^\delta}) \leq e^{-
    \frac{1}{4} k \delta^2 +  \dim(\cH) \ln k}$.
\end{remark}

\begin{proof}
  Let $\ket{\Psi} \in \Sym{\cH}{n+k}$ and $\rho^{\ket{\theta}}_n \in
  \NN(\Sym{\cH}{n})$ as defined in the proof of
  Theorem~\ref{thm:symmix}. It then suffices to show that
  \begin{equation} \label{eq:symmixrest}
    \int_{\overline{\Gamma^\delta}} 
      \tr(\rho^{\ket{\theta}}_n) \omega(\ket{\theta}) 
  \leq
    e^{- \frac{1}{4} k \delta^2 + d\ln k} \ ,
  \end{equation}
  where $\omega$ is the uniform probability measure on the unit sphere
  $\cS_1(\cH)$ and $d := \dim(\cH)$.
  
  Let $\ket{\theta} \in \overline{\Gamma^\delta}$ be fixed, i.e., $\|
  \tr_B(\proj{\theta}) - \sigma_{A} \|_1 > \delta$.  Then,
  by~\eqref{eq:rhothetaproj},
  \[
  \begin{split}
    \tr(\rho^{\ket{\theta}}_n)
  & =
    {\textstyle \binom{k+d-1}{k}} \cdot \tr\bigl(
      \id_{\cH}^{\otimes n} \otimes (\proj{\theta})^{\otimes k} 
        \cdot \proj{\Psi} 
    \bigr) \\
  & =
    {\textstyle \binom{k+d-1}{k}} \cdot \tr\bigl(
      (\proj{\theta})^{\otimes k} \cdot \rho_{A^k B^k} 
    \bigr) \ ,
  \end{split}
  \]
  where $\rho_{A^k B^k} := \tr_n(\rho_{A^{n+k} B^{n+k}}) =
  \tr_n(\proj{\Psi})$. Since the fidelity cannot decrease when taking
  the partial trace (cf.\ Lemma~\ref{lem:fidincr}) we get
  \[
  \begin{split}
    \tr\bigl(
      (\proj{\theta})^{\otimes k} \rho_{A^k B^k} 
    \bigr)    
  & =
    F\bigl( \rho_{A^k B^k}, (\proj{\theta})^{\otimes k} \bigr)^2 \\
  & \leq
    F\bigl(\rho_{A^k}, \tr_B(\proj{\theta})^{\otimes k}\bigr)^2 \\
  & =
    F\bigl(\sigma_A^{\otimes k}, \tr_B(\proj{\theta})^{\otimes k}\bigr)^2 \\
  & =
    F\bigl(\sigma_{A}, \tr_B(\proj{\theta})\bigr)^{2 k} \ .
  \end{split}
  \]
  Because, by Lemma~\ref{lem:distfidbound}, 
  \[
    F\bigl( \sigma_{A}, \tr_B(\proj{\theta})\bigr)^2
  \leq 
    1 - \sfrac{1}{4} \bigl\| \sigma_{A} - \tr_B(\proj{\theta})  \bigr\|_1^2
  <
    1 - \sfrac{\delta^2}{4} \ ,
  \]
  we conclude
  \[
    \tr(\rho^{\ket{\theta}}_n) 
  \leq 
    {\textstyle \binom{k+d-1}{k}} \cdot \bigl( 1- \sfrac{\delta^2}{4} \bigr)^{k}
  \leq
    k^d e^{k \ln(1-\frac{\delta^2}{4} )} 
  \leq
    e^{- \frac{1}{4} k \delta^2 + d \ln k } \ ,
  \]  
  where we have used $\ln(1-a) \leq -a$, for $a \in [0,1]$.
  Inequality~\eqref{eq:symmixrest} then follows because $\omega$ is a
  probability measure.
\end{proof}

\index{de~Finetti~representation|)}

\section{Smooth min-entropy of symmetric states} \label{sec:smsym}

\index{smooth~min-entropy|(}

\newcommand*{\denss}{\sigma}

Let $\ket{\theta} \in \cH$, let $\cE$ be a quantum operation from
$\cH$ to $\cH_X \otimes \cH_B$, and define $\rho_{X^n B^n} :=
\cE^{\otimes n}(\proj{\Psi})$, for $\ket{\Psi} :=
\ket{\theta}^{\otimes n}$.  Obviously, $\rho_{X^n B^n}$ has product
form, i.e., $\rho_{X^n B^n} = \sigma_{X B}^{\otimes n}$, where
$\sigma_{X B} = \cE(\proj{\theta})$.  Hence, as demonstrated in
Section~\ref{sec:smoothprod} (Corollary~\ref{cor:Hmincondrepclass}),
the smooth min-entropy of such a product state can be expressed in
terms of the von Neumann entropy, that is,
\begin{equation} \label{eq:minappr}
  \sfrac{1}{n} \Hmin(\rho_{X^n B^n}|B^n) 
\gtrapprox 
  H(\sigma_{X B}) - H(\sigma_B) \ .
\end{equation}

Theorem~\ref{thm:Renyisym} below states that this still holds if the
product state $\ket{\Psi} := \ket{\theta}^{\otimes n}$ is replaced by
a state in the symmetric subspace of $\cH^{\otimes n}$ along
$\ket{\theta}^{\otimes n-r}$, for some $r \ll n$.

\begin{theorem} \label{thm:Renyisym}
  Let $0 \leq r \leq \frac{1}{2} n$, let $\ket{\theta} \in \cH$ and
  $\ket{\Psi} \in \SymR{\cH}{n}{\theta}{n-r}$ be normalized, and let
  $\cE$ be a trace-preserving CPM from $\cH$ to $\cH_X \otimes \cH_B$
  which is classical on $\cH_X$. Define $\rho_{X^n B^n} :=
  \cE^{\otimes n}(\proj{\Psi})$ and $\denss_{X B} :=
  \cE(\proj{\theta})$. Then, for any $\eps \geq 0$,
  \[
    \frac{1}{n} \Hmin^\eps(\rho_{X^n B^n}|B^n) 
  \geq 
    H(\denss_{X B})  - H(\denss_{B}) - \delta \ ,
  \]
  where $\delta := \bigl(\frac{5}{2} \Hmax(\rho_X) + 4 \bigr)
  \sqrt{\frac{2 \log(\slfrac{4}{\eps})}{n} + h(\slfrac{r}{n})}$.
\end{theorem}

\begin{proof}
  According to Lemma~\ref{lem:symspacebin}, there exists a family
  $\{\ket{\Psi^{s}}\}_{s \in \cS}$ of orthonormal vectors from
  $\Vrep{\cH}{n}{\theta}{n-r}$ of size $|\cS| \leq 2^{n
    h(\slfrac{r}{n})}$ such that
  \begin{equation} \label{eq:PsisumRenyi}
    \ket{\Psi} 
  =
    \sum_{s \in \cS} \gamma_{s} \ket{\Psi^s} \ ,
  \end{equation}
  where $\gamma_s$ are coefficients with $\sum_{s \in \cS}
  |\gamma_s|^2 = 1$.
  
  Let $\{E_w\}_{w \in \cW}$ be the family of operators from $\cH$ to
  $\cH_X \otimes \cH_B$ defined by the CPM $\cE$, i.e., $\cE(\sigma) =
  \sum_{w \in \cW} E_w \sigma E_w^\dagger$, for any operator $\sigma$
  on $\cH$. Moreover, let $\cH_W$ be a Hilbert space with orthonormal
  basis $\{\ket{w}\}_{w \in \cW}$ and let $U$ be the operator from
  $\cH$ to $\cH_X \otimes \cH_B \otimes \cH_W$ defined by
  \[
    U := \sum_{w \in \cW} E_w \otimes \ket{w} \ .
  \]
  Because $\cE$ is trace-preserving, i.e., $\sum_{w} E_w^\dagger E_w =
  \id_\cH$, we have $U^\dagger U = \id_{\cH}$, that is, $U$ is
  unitary.  Furthermore, for any operator $\sigma$ on $\cH$,
  \begin{equation} \label{eq:trUE}
    \tr_W(U \sigma U^\dagger) = \cE(\sigma) \ .
  \end{equation}
  
  Let $\ket{\Phi} := U^{\otimes n} \ket{\Psi}$ and, similarly, for any
  $s \in \cS$, let $\ket{\Phi^{s}} := U^{\otimes n} \ket{\Psi^{s}}$.
  Then, using~\eqref{eq:PsisumRenyi},
  \[
    \ket{\Phi} 
  =
    \sum_{s \in \cS} \gamma_{s} \ket{\Phi^{s}} \ .
  \]
  Because $U$ is unitary and the vectors $\ket{\Psi^{s}}$ are
  orthonormal, the vectors $\ket{\Phi^{s}}$ are orthonormal as well.
  Moreover, using~\eqref{eq:trUE},
  \[
    \rho_{X^n B^n} 
  = 
    \cE^{\otimes n}(\proj{\Psi})
  =
    \tr_{W^n}(U^{\otimes n} \proj{\Psi} (U^\dagger)^{\otimes n})
  = 
    \tr_{W^n}(\proj{\Phi}) \ .
  \]
 
  Let $\rhot_{X^n B^n}^{s} := \tr_{W^n}(\proj{\Phi^{s}})$ and define
  the operator $\rhot_{X^n B^n S}$ on $\cH_X^{\otimes n} \otimes
  \cH_B^{\otimes n} \otimes \cH_S$ by
  \[
    \rhot_{X^n B^n S} 
  := 
    \sum_{s \in \cS} |\gamma_{s}|^2 \rhot_{X^n B^n}^{s} \otimes \proj{s} \ ,
  \]
  where $\cH_S$ is a Hilbert space with orthonormal basis
  $\{\ket{s}\}_{s \in \cS}$.  Lemma~\ref{lem:smoothHinfcondlowbound}
  then allows us to express the smooth min-entropy of $\rho_{X^n B^n}$
  in terms of the smooth min-entropy of $\rhot_{X^n B^n S}$. Moreover,
  by Lemma~\ref{lem:Hminsmoothclasscondr}, the smooth min-entropy of
  $\rhot_{X^n B^n S}$ is lower bounded by the min-entropy of the
  operators $\rhot_{X^n B^n}^s$, that is,
  \[
  \begin{split}
    \Hmin^{\eps}(\rho_{X^n B^n}|\rhot_{B^n}) 
  & \geq 
    \Hmin^{\epst}(\rhot_{X^n B^n S}|\rhot_{B^n S})  - \Hmax(\rhot_{S}) \\
  & \geq
    \min_{s \in \cS} 
      \Hmin^{\epst}(\rhot_{X^n B^n}^{s}|\rhot_{B^n}^{s}) 
      - \Hmax(\rhot_{S})\ ,
  \end{split}
  \]
  where $\epst = \frac{\eps^2}{6 |\cS|}$. Using the fact that $|\cS|
  \leq 2^{n h(\slfrac{r}{n})}$, we find
  \begin{equation} \label{eq:Hinfmins}
    \Hmin^{\eps}(\rho_{X^n B^n}|\rhot_{B^n}) 
  \geq
    \min_{s \in \cS} 
      \Hmin^{\epst}(\rhot_{X^n B^n}^{s}|\rhot_{B^n}^{s}) 
    - n h(\slfrac{r}{n}) 
  \end{equation}
  and
  \begin{equation} \label{eq:epspbound}
    \log(\slfrac{1}{\epst})  \leq \log(\slfrac{2}{\eps}) + \log 6 + n h(\slfrac{r}{n}) \ .
  \end{equation}
  
  Let us now compute the min-entropies of the operators $\rhot_{X^n
    B^n}^{s}$, for $s \in \cS$. Since $\ket{\Psi^s} \in
  \Vrep{\cH}{n}{\theta}{n-r}$, the vector $\ket{\Psi^{s}}$, after some
  appropriate reordering of the subsystems, has the form
  $\ket{\Psi^{s}} = \ket{\theta}^{\otimes n-r} \otimes
  \ket{\hat{\Psi}^{s}}$, for some $\ket{\hat{\Psi}^{s}} \in
  \cH^{\otimes r}$.  Hence, the same holds for the vector
  $\ket{\Phi^{s}}$, i.e.,
  \[
    \ket{\Phi^{s}}
  =
    U^{\otimes n} \ket{\Psi^{s}}
  =
    \bigl(U \ket{\theta}\bigr)^{\otimes n-r} 
    \otimes U^{\otimes r}\ket{\hat{\Psi}^{s}} \ .
  \]
  Consequently, from~\eqref{eq:trUE} and the definition of $\denss_{X
    B}$,
  \[
  \begin{split}
    \rhot_{X^n B^n}^{s} 
  & =
    \bigl(\tr_{W}(U \proj{\theta} U^\dagger)\bigr)^{\otimes n-r}
    \otimes \tr_{W^r}\bigl(
      U^{\otimes r}\proj{\hat{\Psi}^{s}} (U^\dagger)^{\otimes r}
    \bigr) \\
  & =
    \denss_{X B}^{\otimes n-r}
    \otimes \hat{\rho}^s_{X^{r} B^{r}} \ ,
  \end{split}
  \]
  where $\hat{\rho}^s_{X^{r} B^{r}} := \cE^{\otimes
    r}(\proj{\hat{\Psi}^{s}})$. Because $\cE$ is classical on $\cH_X$,
  $\hat{\rho}_{X^{r} B^{r}}$ is also classical on $\cH_X^{\otimes r}$.
  Using the superadditivity of the smooth min-entropy
  (Lemma~\ref{lem:Hminindaddsmooth}) and the fact that the min-entropy
  of a classical subsystem cannot be negative
  (Lemma~\ref{lem:classnn}) we find
  \begin{equation} \label{eq:Hinfsplit}
  \begin{split}
    \Hmin^{\epst}(\rhot_{X^n B^n}^{s}|\rhot_{B^n}^{s})
  & \geq
    \Hmin^{\epst}\bigl(\denss_{X B}^{\otimes n-r}
        \big|\denss_{B}^{\otimes n-r}\bigr)
  +
    \Hmin(\hat{\rho}^s_{X^{r} B^{r}}|\hat{\rho}_{B^{r}}) \\
  & \geq
    \Hmin^{\epst}\bigl(\denss_{X B}^{\otimes n-r}
        \big|\denss_{B}^{\otimes n-r}\bigr) \ .
  \end{split}
  \end{equation} 
  Furthermore, because $\denss_{X B}$ is classical on $\cH_X$, we can
  use Corollary~\ref{cor:Hmincondrepclass} to bound the smooth
  min-entropy of the product state in terms of the von Neumann
  entropy,
  \begin{multline*} 
     \Hmin^{\epst}\bigl(\denss_{X B}^{\otimes n-r}
        \big|\denss_{B}^{\otimes n-r}\bigr)
  \geq
    (n-r) \bigl( H(\denss_{X B}) - H(\denss_B) - \delta' \bigr) \\
  \geq
    n \bigl( H(\denss_{X B}) - H(\denss_B) \bigr) - r \Hmax(\rho_X) 
    - (n-r) \delta'
  \end{multline*}
  with $\delta' := \bigl(2 \Hmax(\rho_X) + 3 \bigr)
  \sqrt{\frac{\log(\slfrac{1}{\epst})+1}{n-r}}$.  Together
  with~\eqref{eq:Hinfmins} and~\eqref{eq:Hinfsplit} we conclude
  \begin{equation} \label{eq:Hinffb}
    \sfrac{1}{n} \Hmin^{\eps}(\rho_{X^n B^n}|\rhot_{B^n}) 
  \geq
    H(\denss_{X B}) - H(\denss_B) 
    - h(\slfrac{r}{n})
    - \slfrac{r}{n} \Hmax(\rho_X) 
    - \sfrac{n-r}{n} \cdot \delta' \ .
  \end{equation}
  Moreover, from~\eqref{eq:epspbound},
  \[
    \sqrt{n-r} \cdot \delta'
  \leq 
    \bigl(2 \Hmax(\rho_X) + 3 \bigr) 
      \sqrt{\log (2/\eps) + n h(\slfrac{r}{n}) + \log 6 + 1} \ ,
  \]
  and hence, using the fact that $c \leq \sqrt{c}$, for any $c \leq
  1$,
  \[
    \sfrac{n-r}{n} \cdot \delta'
  \leq
    \sqrt{\sfrac{n-r}{n}} \cdot \delta' 
  \leq
    \bigl(2 \Hmax(\rho_X) + 3 \bigr)
      \sqrt{\sfrac{\log(\slfrac{2}{\eps})+4}{n} + h(\slfrac{r}{n})} \ .
  \]
  Finally, because $\frac{2 r}{n} \leq h(\slfrac{r}{n})$ and
  $h(\slfrac{r}{n}) \leq \sqrt{h(\slfrac{r}{n})}$, we find
  \begin{multline*}
    h(\slfrac{r}{n}) + \sfrac{r}{n} \Hmax(\rho_X) + \sfrac{n-r}{n} \cdot \delta' \\
  \leq
    \bigl(\sfrac{5}{2}\Hmax(\rho_X) + 4\bigr)
      \sqrt{\sfrac{2 \log(\slfrac{2}{\eps})+4}{n} + h(\slfrac{r}{n})} \ .    
  \end{multline*}
  Inserting this into~\eqref{eq:Hinffb} concludes the proof.
\end{proof}

\index{smooth~min-entropy|)}

\section{Statistics of symmetric states} \label{sec:symstat}

\index{frequency~distribution|(}

Let $z_1, \ldots, z_n$ be the outcomes of $n$ independent measurements
of a state $\ket{\theta} \in \cH$ with respect to a POVM $\cM =
\{M_z\}_{z \in \cZ}$. The law of large numbers tells us that, for
large $n$, the statistics $\freq{\bz}$ of the $n$-tuple $\bz = (z_1,
\ldots, z_n)$ is close to the probability distribution $P_Z$ defined
by $P_Z(z) := \tr(M_z \proj{\theta})$, for $z \in \cZ$.
Theorem~\ref{thm:symstat} below states that the same is true if the
$n$-tuple $\bz$ is the outcome of a product measurement $\cM^{\otimes
  n}$ applied to a state $\ket{\Psi}$ of the symmetric subspace of
$\cH^{\otimes n}$ along $\ket{\theta}^{\otimes n-r}$, for some small
$r \ll n$.

For the proof of this result, we need the following technical lemma.

\begin{lemma} \label{lem:sumprob}
  Let $\ket{\psi} = \sum_{x \in \cX} \ket{\psi^x}$ and let $\rho \in
  \NN(\cH)$.  Then
  \[
    \bra{\psi} \rho \ket{\psi}
  \leq
    |\cX| \sum_{x \in \cX} \bra{\psi^x} \rho \ket{\psi^x} \ .
  \]
\end{lemma}

\begin{proof}
  Let $\rho = \sum_{y \in \cY} p_y \proj{y}$ be a spectral
  decomposition of $\rho$. For any $y \in \cY$,
  \[
    |\spr{y}{\psi}|^2
  =
    \bigl|\sum_{x \in \cX} \spr{y}{\psi^x}\bigr|^2
  \leq
    \bigl(\sum_{x \in \cX} |\spr{y}{\psi^x}|\bigr)^2
  \leq
    |\cX| \sum_{x \in \cX} |\spr{y}{\psi^x}|^2 \ ,
  \]
  where we have used the Cauchy-Schwartz inequality in the last step.
  Consequently,
  \[
  \begin{split}
    \bra{\psi} \rho \ket{\psi}
  & =
    \sum_{y \in \cY} p_y |\spr{y}{\psi}|^2 \\
  & \leq 
    |\cX| \sum_{y \in \cY} \sum_{x \in \cX} 
      p_y |\spr{y}{\psi^x}|^2 \\
  & =
    |\cX| \sum_{x \in \cX} 
      \sum_{y \in \cY} p_y \spr{\psi^x}{y} \spr{y}{\psi^x} \\
  & =
    |\cX| \sum_{x \in \cX} \bra{\psi^x} \rho \ket{\psi^x} \ .\qedhere
  \end{split}
  \] 
\end{proof}

\begin{theorem} \label{thm:symstat}
  Let $0 \leq r \leq \frac{1}{2} n$, let $\ket{\theta} \in \cH$ and
  $\ket{\Psi} \in \SymR{\cH}{n}{\theta}{n-r}$ be normalized, let $\cM
  = \{M_z\}_{z \in \cZ}$ be a POVM on $\cH$, and let $P_Z$ be the
  probability distribution of the outcomes of the measurement $\cM$
  applied to $\proj{\theta}$. Then
  \[
    \Pr_{\bz}
      \Bigl[
        \| \freq{\bz} - P_Z \|_1 
      >
        2 \sqrt{
          \sfrac{\log(\slfrac{1}{\eps})}{n} + h(\slfrac{r}{n}) 
        + \sfrac{|\cZ|}{n} \log(\sfrac{n}{2}+1)
        }
      \Bigr] 
  \leq 
    \eps \ ,
  \]
  where the probability is taken over the outcomes $\bz=(z_1, \ldots,
  z_n)$ of the product measurement $\cM^{\otimes n}$ applied to
  $\proj{\Psi}$.
\end{theorem}

\begin{proof}
  According to Lemma~\ref{lem:symspacebin}, the vector $\ket{\Psi}$
  can be written as a superposition of orthonormal vectors
  $\ket{\Psi^s} \in \Vrep{\cH}{n}{\theta}{n-r}$, that is,
  \begin{equation} \label{eq:Psisumstat}
    \ket{\Psi}
  =
    \sum_{s \in \cS}
      \gamma_{s} \ket{\Psi^s} \ ,
  \end{equation}
  where $\cS$ is a set of size $|\cS| \leq 2^{n h(\sfrac{r}{n})}$ and
  where $\gamma_s$ are coefficients such that $\sum_{s \in \cS}
  |\gamma_s|^2 = 1$.
  
  Let now $s \in \cS$ be fixed. Because $\ket{\Psi^s} \in
  \Vrep{\cH}{n}{\theta}{n-r}$, there exists a permutation $\pi$ which
  maps $\ket{\Psi^s}$ to a vector which, on the first $n-r$
  subsystems, has the form $\ket{\theta}^{\otimes n-r}$. We can thus
  assume without loss of generality that $\ket{\Psi^s} =
  \ket{\theta}^{\otimes n-r} \otimes \ket{\Psih}$, for some
  $\ket{\Psih} \in \cH^{\otimes r}$.
  
  Let $\bz = (z_1, \ldots, z_n)$ be the outcome of the measurement
  $\cM^{\otimes n}$ applied to $\proj{\Psi^{s}}$ and define
  $\bzp:=(z_1, \ldots, z_{n-r})$ and $\bzpp:=(z_{n-r+1}, \ldots,
  z_n)$. Clearly, $\bzp$ is distributed according to the product
  distribution $P_Z^{n-r}$.  Hence, with high probability, $\bzp$ is a
  typical sequence\index{typical~sequence}, that is, by
  Corollary~\ref{cor:typsec},
  \begin{equation} \label{eq:typsecbin}
    \Pr_{\bzp}\Bigl[
      \| \freq{\bzp} - P_Z \|_1
    > 
      \sqrt{2 (\ln 2) \bigl(\delta + \sfrac{|\cZ| \log(n-r+1)}{n-r}\bigr)}
    \Bigr] 
  \leq
    2^{-(n-r) \delta} \ ,
  \end{equation}
  for any $\delta \geq 0$.  Moreover, because $\freq{\bz} =
  \frac{n-r}{n} \freq{\bzp} + \frac{r}{n} \freq{\bzpp}$, we can apply
  the triangle inequality which gives
  \[
    \bigl\| \freq{\bz} - P_Z \bigr\|_1
  \leq
    \sfrac{n-r}{n} \bigl\| \freq{\bzp} - P_Z \bigr\|_1
    + \sfrac{r}{n} \bigl\| \freq{\bzpp} - P_Z \bigr\|_1 
  \leq
    \bigl\| \freq{\bzp} - P_Z \bigr\|_1 + \sfrac{r}{n} \ .
  \]
  Using this inequality and the assumption $r \leq \frac{1}{2} n$,
  \eqref{eq:typsecbin} implies that
  \begin{equation} \label{eq:Psizmeasbound}
    \Pr_{\bz\leftarrow \ket{\Psi^s}}
      [\bz \in \cW_{\delta}]
  \leq 
    2^{- \frac{n \delta}{2}} \ ,
  \end{equation}
  where we write $\bz \leftarrow \ket{\Psi^s}$ to indicate that $\bz$
  is distributed according to the outcomes of the measurement applied
  to $\ket{\Psi^s}$ and where $\cW_{\delta}$ is the subset of $\cZ^n$
  defined by
  \[
    \cW_{\delta}
  :=
    \bigl\{
    \bz \in \cZ^n : \, 
      \| \freq{\bz} - P_Z \|_1
    > 
      \sqrt{2 (\ln 2)
        \bigl(\delta + \sfrac{2 |\cZ|}{n} \log(\sfrac{n}{2}+1)\bigr) } 
    + \sfrac{r}{n} 
    \bigr\} \ .
  \]
  
  Let $M_{\bz} := M_{z_1} \otimes \cdots \otimes M_{z_n}$, for $\bz =
  (z_1, \ldots, z_n) \in \cZ^n$, be the linear operators defined by
  the POVM $\cM^{\otimes n}$.  Then, using Lemma~\ref{lem:sumprob},
  \eqref{eq:Psisumstat}, and~\eqref{eq:Psizmeasbound} we get
  \[
  \begin{split}
    \Pr_{\bz \leftarrow \ket{\Psi}}[\bz \in \cW_{\delta}]
  & =
    \sum_{\bz \in \cW_{\delta}} \bra{\Psi} M_{\bz} \ket{\Psi} \\
  & \leq
    \sum_{\bz \in \cW_{\delta}} 
      |\cS| \sum_{s \in \cS} |\gamma_{s}|^2
        \bra{\Psi^{s}} M_{\bz} \ket{\Psi^{s}} \\
  & =
    |\cS| \sum_{s \in \cS} |\gamma_{s}|^2
      \Pr_{\bz \leftarrow \ket{\Psi^s}}
        [\bz \in \cW_{\delta}] \\
  & \leq
    2^{n h(\slfrac{r}{n})} 2^{-\frac{n \delta}{2}} \\
  & =
    2^{-n (\frac{\delta}{2} - h(\slfrac{r}{n}))} \ .
  \end{split} 
  \]
  Hence, with $\delta := \frac{2 \log(\slfrac{1}{\eps})}{n} + 2 h(\slfrac{r}{n})$,
  \[
    \Pr_{\bz}
    \Bigl[
      \|\freq{\bz}- P_Z \|_1
    > 
      \sqrt{4 (\ln 2)
        \bigl(
          \sfrac{\log(\slfrac{1}{\eps})}{n} + h(\slfrac{r}{n}) 
            + \sfrac{|\cZ|}{n} \log(\sfrac{n}{2}+1)
        \bigr) } 
    + \sfrac{r}{n} 
    \Bigr]
  \leq
    \eps \ .
  \]
  The assertion then follows from the fact that $\sqrt{c} +
  \frac{r}{n} \leq \sqrt{c + \frac{2 r}{n}}$, for any $c \geq 0$ with
  $c + \frac{2 r}{n} \leq 1$, and from $\frac{2 r}{n} \leq
  h(\slfrac{r}{n})$.
\end{proof}

\index{frequency~distribution|)}

\index{symmetric~state|)}
\index{permutation-invariant~state|)}

\chapter{Privacy Amplification} \label{ch:PA}

\index{privacy~amplification|(}

A fundamental problem in cryptography is to distill a secret
key\index{secret~key} from only partially secret data, on which an
adversary might have information encoded into the state of a quantum
system. In this chapter, we propose a general solution to this
problem, which is called \emph{privacy amplification}: We show that
the key computed as the output of a hash
function\index{two-universal~hashing} (chosen at random from a
two-universal\footnote{See Section~\ref{sec:twohash} for a
  definition.}  family of functions) is secure under the sole
condition that its length is smaller than the adversary's uncertainty
on the input, measured in terms of (smooth) min-entropy.

We start with the derivation of various technical results
(Sections~\ref{sec:normbound}--\ref{sec:twohash}). These are used for
the proof of the main statement, which is first formulated in terms of
min-entropy (Section~\ref{sec:pans}) and then generalized to
\emph{smooth} min-entropy (Section~\ref{sec:pasm}).

\section{Bounding the norm of hermitian operators} \label{sec:normbound}

In this section, we derive an upper bound on the trace norm for
hermitian operators (Lemma~\ref{lem:Schurnew}). The bound only
involves matrix multiplications, which makes it easy to evaluate.

\begin{lemma} \label{lem:Hermprod}
  Let $S$ and $T$ be hermitian operators on $\cH$. Then
  \[
    \tr(S T) \leq \sqrt{\tr(S^2) \tr(T^2)} \ .
  \]
\end{lemma}

\begin{proof}
  Let $S = \sum_{y \in \cY} \beta_y \proj{y}$ and $T = \sum_{z \in
    \cZ} \gamma_z \proj{z}$ be spectral decompositions of $S$ and $T$,
  respectively.  With the definition $a_{y,z} := |\spr{y}{z}|^2$, we
  have
  \[
    \tr(S T)
  =
    \sum_{y,z} \beta_y \gamma_z \tr\bigl(\proj{y} \cdot \proj{z}\bigr)  
  =
    \sum_{y,z} \beta_y \gamma_z a_{y,z} \ .
  \]
  On the other hand, $\tr(S^2) = \sum_{y} \beta_y^2 $ and $\tr(T^2) =
  \sum_{z} \gamma_z^2$. It thus suffices to show that
  \begin{equation} \label{eq:CSgen}
      \sum_{y,z} \beta_y \gamma_z a_{y,z}
    \leq
      \sqrt{
        \bigl(\sum_{y} \beta_y^2 \bigr) 
        \bigl(\sum_{z} \gamma_z^2 \bigr) 
      } \ .     
  \end{equation}
  
  It is easy to verify that $(a_{y,z})_{y \in \cY,z \in \cZ}$ is a
  bistochastic matrix. Hence, according to Birkhoff's theorem (cf.\ 
  Theorem~\ref{thm:Birkhoff}) there exist nonnegative coefficients
  $\mu_{\pi}$ parameterized by the bijections $\pi$ from $\cZ$ to
  $\cY$ such that $\sum_{\pi} \mu_{\pi} = 1$ and, for any $y \in \cY$,
  $z \in \cZ$, $a_{y,z} = \sum_{\pi} \mu_{\pi} \delta_{y, \pi(z)}$.
  We thus have
  \begin{equation} \label{eq:Birkd}
    \sum_{y,z} \beta_y \gamma_z a_{y,z}
  =
    \sum_{\pi} \mu_{\pi} 
      \sum_{y,z} \beta_y \gamma_z \delta_{y,\pi(z)} \ .
  \end{equation}
  
  Furthermore, by the Cauchy-Schwartz inequality, for any fixed
  bijection~$\pi$,
  \[
    \sum_{z} \beta_{\pi(z)} \gamma_z     
  \leq
    \sqrt{
      \bigl( \sum_{z} \beta_{\pi(z)}^2 \bigr) 
      \bigl(\sum_{z} \gamma_z^2 \bigr) 
    } \ .
  \]
  This can be rewritten as
  \[
    \sum_{y,z}
      \beta_y \gamma_z \delta_{y,\pi(z)}
  \leq
    \sqrt{
      \bigl(\sum_{y} \beta_y^2 \bigr) 
      \bigl(\sum_{z} \gamma_z^2 \bigr) 
    } \ .
  \]
  Inserting this into~\eqref{eq:Birkd} implies~\eqref{eq:CSgen} and
  thus concludes the proof.
\end{proof}

\begin{lemma} \label{lem:trABbound}
  Let $S$ be a hermitian operator on $\cH$ and let $\sigma$ be a
  nonnegative operator on $\cH$. Then
  \[
    \tr |\sqrt{\sigma} S \sqrt{\sigma} |
  \leq
    \sqrt{\tr(S^2) \tr(\sigma^2)} \ .
  \]
\end{lemma}

\begin{proof}
  Let $\{\ket{v}\}_{v \in \cV}$ be an eigenbasis of $\sqrt{\sigma} S
  \sqrt{\sigma}$ and let $S = \sum_{x \in \cX} \alpha_x \proj{x}$ be a
  spectral decomposition of $S$.  Then
  \[
  \begin{split}
    \tr |\sqrt{\sigma} S \sqrt{\sigma} |
  & =
    \sum_{v} \bigl| \bra{v} \sqrt{\sigma} S \sqrt{\sigma} \ket{v} \bigr| \\
  & =
    \sum_{v} 
      \bigl| 
        \sum_{x} \alpha_x \bra{v} \sqrt{\sigma} \proj{x} \sqrt{\sigma} \ket{v} 
      \bigr| \\
  & \leq
    \sum_{v} \sum_{x} 
      |\alpha_x| \bra{v} \sqrt{\sigma} \proj{x} \sqrt{\sigma} \ket{v} \\
  & =
    \sum_{v}
      \bra{v} \sqrt{\sigma} |S| \sqrt{\sigma} \ket{v} \\
  & =
    \tr(\sqrt{\sigma} |S| \sqrt{\sigma}) \ .
  \end{split}  
  \]
  Furthermore, by Lemma~\ref{lem:Hermprod}, 
  \[
    \tr(\sqrt{\sigma} |S| \sqrt{\sigma})
  =
    \tr(|S| \sigma)
  \leq
    \sqrt{\tr(|S|^2) \tr(\sigma^2)}
  =
    \sqrt{\tr(S^2) \tr(\sigma^2)} \ ,
  \]
  which concludes the proof.
\end{proof}

\begin{lemma} \label{lem:Schurnew}
  Let $S$ be a hermitian operator on $\cH$ and let $\sigma$ be a
  nonnegative operator on $\cH$. Then
  \[
    \|S\|_1 \leq \sqrt{\tr(\sigma) \tr(S \sigma^{-1/2} S \sigma^{-1/2})} \ .
  \]
\end{lemma}

\begin{proof}
  The assertion follows directly from Lemma~\ref{lem:trABbound} with
  $\bar{\sigma} := \sqrt{\sigma}$ and $\bar{S} := \bar{\sigma}^{-1/2}
  S \bar{\sigma}^{-1/2}$, that is, $\sigma = \bar{\sigma}^2$ and $S =
  \sqrt{\bar{\sigma}} \bar{S} \sqrt{\bar{\sigma}}$.
\end{proof}

\section{Distance from uniform} \label{sec:twodist}

According to the discussion on universal
security~\index{universal~security} in Section~\ref{sec:univdef}, the
security of a key is defined with respect to its \distance{} from a
perfect key which is uniformly distributed and independent of the
adversary's state (see~\eqref{eq:secdef}). This motivates the
following definition.

\begin{definition} \label{def:nonunif}
  Let $\rho_{A B} \in \NN(\cH_A \otimes \cH_B)$. Then the
  \emph{\distance{} from uniform of $\rho_{A B}$ given
    $B$}\index{\distance{}~from~uniform} is 
  \[
    d(\rho_{A B}|B)
  :=
    \bigl\|\rho_{A B} - \rho_U \otimes \rho_B \bigr\|_1 \ ,
  \]
  where $\rho_U := \frac{1}{\dim(\cH_A)} \id_{A}$ is the fully mixed
  state on $\cH_A$.
\end{definition}

For an operator $\rho_{X Z}$ defined by a classical probability
distribution $P_{X Z}$, $d(\rho_{X Z}|Z)$ is the expectation (over $z$
chosen according to $P_Z$) of the \distance{} between the conditional
distribution $P_{X|Z=z}$ and the uniform distribution.  This property
is generalized by the following lemma.

\begin{lemma} \label{lem:distexp}
  Let $\rho_{A B Z}$ be classical with respect to an orthonormal basis
  $\{\ket{z}\}_{z \in \cZ}$ of $\cH_Z$ and let $\rho_{A B}^z$, for $z
  \in \cZ$, be the corresponding (non-normalized) conditional
  operators. Then
  \[
    d(\rho_{A B Z}|B Z)
  =
    \sum_{z \in \cZ} d(\rho_{A B}^z|B) \ .
  \]
\end{lemma}

\begin{proof}
  Let $\rho_U$ be the fully mixed state on $\cH_A$. Then, by
  Lemma~\ref{lem:distclass},
  \[
  \begin{split}
    d(\rho_{A B Z}|B Z)
  & =
    \|\rho_{A B Z} - \rho_U \otimes \rho_{B Z}\|_1 \\
  & =
    \sum_{z \in \cZ} 
      \| \rho_{A B}^z - \rho_U \otimes \rho_B^z \|_1 \\
  & =
    \sum_{z \in \cX}d(\rho_{A B}^z|B) \ .
  \end{split}
  \]
\end{proof}

To derive our result on the security of privacy amplification, it is
convenient to consider an alternative measure for the distance from
uniform. Let $\rho_{A B} \in \NN(\cH_{A B})$ and $\sigma_B \in
\NN(\cH_B)$. The \emph{(conditional) $L_2$-distance from uniform of
  $\rho_{A B}$ relative to
  $\sigma_B$}\index{$L_2$-distance~from~uniform} is defined by
\[
  d_2(\rho_{A B}|\sigma_B)
:=
  \tr\Bigl(
    \bigl((\rho_{A B} - \rho_U \otimes \rho_B)
    (\id_A \otimes \sigma_B^{-1/2}) \bigr)^2
  \Bigr) \ ,
\]
where $\rho_U$ is the fully mixed state on $\cH_A$. Note that
$d_2(\rho_{A B}|\sigma_B)$ can equivalently be written as
\begin{equation} \label{eq:Dexpr}
  d_2(\rho_{A B}|\sigma_B)
= 
 \tr\Bigl(\bigl( 
    (\id_A \otimes \sigma_B^{-1/4}) 
    (\rho_{A B} - \rho_U \otimes \rho_B) 
    (\id_A \otimes \sigma_B^{-1/4})\bigr)^2\Bigr) \ ,
\end{equation}
which proves that $d_2(\rho_{A B}|\sigma_B)$ cannot be negative. 

The $L_2$-distance from uniform can be used to bound the \distance{}
from uniform.

\begin{lemma} \label{lem:disttwodist}
  Let $\rho_{A B} \in \NN(\cH_A \otimes \cH_B)$. Then, for any
  $\sigma_B \in \NN(\cH_B)$, 
  \[
    d(\rho_{A B}|B) 
  \leq 
    \sqrt{\dim(\cH_A) \tr(\sigma_B) d_2(\rho_{A B}|\sigma_B)} \ .
  \]
\end{lemma}

\begin{proof}
  The assertion follows directly from Lemma~\ref{lem:Schurnew} with
  $S:= \rho_{A B} - \rho_U \otimes \rho_B$ and $\sigma:=\id_A \otimes
  \sigma_B$, where $\rho_U$ is the fully mixed state on $\cH_A$.
\end{proof}

The following lemma provides an expression for the $L_2$-distance from
uniform for the case where the first subsystem is classical.

\begin{lemma} \label{lem:sqdistexpl}
  Let $\rho_{X B} \in \NN(\cH_X \otimes \cH_B)$ be classical with
  respect to an orthonormal basis $\{\ket{x}\}_{x \in \cX}$ of
  $\cH_X$, let $\rho_B^x$, for $x \in \cX$, be the corresponding
  (non-normalized) conditional operators, and let $\sigma \in
  \NN(\cH_B)$. Then
  \[
    d_2(\rho_{X B}|\sigma_B)
  =
    \sum_{x} \tr\bigl((\sigma_B^{-1/4} \rho_B^x \sigma_B^{-1/4})^2\bigr) 
      - \frac{1}{|\cX|} 
        \tr\bigl((\sigma_B^{-1/4} \rho_B \sigma_B^{-1/4})^2\bigr) \ .
  \]  
\end{lemma}

\begin{proof}
  Let $\rho_U$ be the fully mixed state on $\cH_X$.  Because $\rho_{X
    B}$ is classical on $\cH_X$, we have
  \[
    \rho_{X B} - \rho_U \otimes \rho_B
  =    
    \sum_{x} \proj{x} \otimes (\rho_B^x - \frac{1}{|\cX|} \rho_B) \ ,
  \]
  and thus
  \begin{multline*}
    (\id_X \otimes \sigma_B^{-1/4})
    (\rho_{X B} - \rho_U \otimes \rho_B) 
    (\id_X \otimes \sigma_B^{-1/4}) \\
  =    
    \sum_{x} \proj{x} 
      \otimes \Bigl(\sigma_B^{-1/4} \rho_B^x \sigma_B^{-1/4} 
               - \frac{1}{|\cX|} \sigma_B^{-1/4} \rho_B \sigma_B^{-1/4}\Bigr) 
    \ .
  \end{multline*}
  Hence, since $\{\ket{x}\}_{x \in \cX}$ is an orthonormal basis,
  \begin{multline*}
    \tr\Bigl(\bigl(
      (\id_X \otimes \sigma_B^{-1/4})
      (\rho_{X B} - \rho_U \otimes \rho_B)
      (\id_X \otimes \sigma_B^{-1/4})
      \bigr)^2\Bigr) \\
  =
    \sum_{x}
      \tr\Bigl(
        \bigl(\sigma_B^{-1/4} \rho_B^x \sigma_B^{-1/4} 
               - \frac{1}{|\cX|} \sigma_B^{-1/4} \rho_B \sigma_B^{-1/4}\bigr)^2
      \Bigr) \\
  =
    \sum_{x} \tr\bigl((\sigma_B^{-1/4} \rho_B^x \sigma_B^{-1/4})^2\bigr)
    - \frac{1}{|\cX|} 
        \tr\bigl((\sigma_B^{-1/4} \rho_B \sigma_B^{-1/4})^2 \bigr) \ ,
  \end{multline*}
  where the second equality holds because $\sum_{x} \rho_B^x =
  \rho_B$. The assertion then follows from~\eqref{eq:Dexpr}.
\end{proof}

\section{Collision entropy} \label{sec:coll}

\index{collision~entropy|(}

Definition~\ref{def:colentr} below can be seen as a generalization of
the well-known classical (conditional) collision entropy to quantum
states.

\begin{definition} \label{def:colentr}
  Let $\rho_{A B} \in \NN(\cH_A \otimes \cH_B)$ and $\sigma_B \in
  \NN(\cH_B)$. Then the \emph{collision entropy of $\rho_{A B}$
    relative to $\sigma_B$} is
\[
  H_2(\rho_{A B}|\sigma_B) 
:=
  - \log \frac{1}{\tr(\rho_{A B})}
      \tr\Bigl(\bigl(\rho_{A B} (\id_A \otimes \sigma_B^{-1/2})\bigr)^2 \Bigr) \ .
\]
\end{definition}

\begin{remark} \label{rem:Hmincol}
It follows immediately from Lemma~\ref{lem:maxcond} that
\[
  \Hmin(\rho_{A B}|\sigma_B) \leq H_2(\rho_{A B}|\sigma_B) \ . 
\]
\end{remark}

\begin{remark} \label{rem:Htworewr}
  If $\rho_{X B} \in \NN(\cH_X \otimes \cH_B)$ is classical with
  respect to an orthonormal basis $\{\ket{x}\}_{x \in \cX}$ of $\cH_X$
  such that the (non-normalized) conditional operators $\rho^x_B$ on
  $\cH_B$, for $x \in \cX$, are orthogonal then
  \[
    2^{-H_{2}(\rho_{X B}|\sigma_B)}
  =
    \frac{1}{\tr(\rho_{X B})} \sum_{x} \tr\bigl((
      \sigma_B^{-1/4} \rho_B^x \sigma_B^{-1/4}
    )^2\bigr)\ .
  \]
\end{remark}

\index{collision~entropy|)}

\section{Two-universal hashing} \label{sec:twohash}

\index{two-universal~hashing|(}

\begin{definition} \label{def:gentu}
  Let $\cF$ be a family of functions from $\cX$ to $\cZ$ and let $P_F$
  be a probability distribution on $\cF$. The pair $(\cF, P_F)$ is
  called \emph{two-universal} if $\Pr_{f}[f(x) = f(x')] \leq
  \frac{1}{|\cZ|}$, for any distinct $x, x' \in \cX$ and $f$ chosen at
  random from $\cF$ according to the distribution $P_F$.
\end{definition}

In accordance with the standard literature on two-universal hashing,
we will, for simplicity, assume that $P_F$ is the uniform distribution
on $\cF$.  In particular, the family $\cF$ is said to be
\emph{two-universal}\index{two-universal~hashing} if $(\cF, P_F)$, for
$P_F$ uniform, is two-universal. It is, however, easy to see that all
statements proven below also hold with respect to the general
definition where $P_F$ is arbitrary.

We will use the following lemma on the existence of two-universal
function families.

\begin{lemma} \label{lem:hashexists}
  Let $0 \leq \ell \leq n$. Then there exists a two-universal family
  of hash functions from $\{0,1\}^n$ to $\{0,1\}^\ell$.
\end{lemma}  

\begin{proof}
  For the proof of this statement we refer to~\cite{CarWeg79}
  or~\cite{WegCar81}, where explicit constructions of hash function
  families are given.
\end{proof}

Consider an operator $\rho_{X B}$ which is classical with respect to
an orthonormal basis $\{\ket{x}\}_{x \in \cX}$ of $\cH_X$ and assume
that $f$ is a function from $\cX$ to $\cZ$. The density operator
describing the classical function output together with the quantum
system $\cH_B$ is then given by
\begin{equation} \label{eq:funcdef}
  \rho_{f(X) B} 
:= 
  \sum_{z \in \cZ} 
    \proj{z} \otimes \rho_B^z \qquad 
  \text{for } \rho_B^z :=\sum_{x \in f^{-1}(z)}\rho_B^x \ ,
\end{equation}
where $\{\ket{z}\}_{z \in \cZ}$ is an orthonormal basis of $\cH_Z$.

Assume now that the function $f$ is randomly chosen from a family of
functions $\cF$ according to a probability distribution $P_F$.  The
function output $f(x)$, the state of the quantum system, and the
choice of the function $f$ is then described by the operator
\begin{equation} \label{eq:hashfuncdef}
  \rho_{F(X) B F} 
:= 
  \sum_{f \in \cF} P_F(f) \rho_{f(X) B} \otimes \proj{f}
\end{equation}
on $\cH_Z \otimes \cH_B \otimes \cH_F$, where $\cH_F$ is a Hilbert
space with orthonormal basis $\{\ket{f}\}_{f \in \cF}$.

The following lemma provides an upper bound on the expected
$L_2$-distance from uniform of a key computed by two-universal
hashing.

\begin{lemma} \label{lem:twodisthashbound}
  Let $\rho_{X B} \in \NN(\cH_X \otimes \cH_B)$ be classical on
  $\cH_X$, let $\sigma_B \in \NN(\cH_B)$, and let $\cF$ be a
  two-universal family of hash functions from $\cX$ to $\cZ$. Then
  \[
    \ExpE_{f}\Bigl[
      d_2(\rho_{f(X) B}|\sigma_B)
    \Bigr]
  \leq
    \tr(\rho_{X B}) 2^{-H_{2}(\rho_{X B}|\sigma_B)} ,
  \]
  for $\rho_{f(X) B} \in \NN(\cH_Z \otimes \cH_B)$ defined
  by~\eqref{eq:funcdef} and $f$ chosen uniformly from~$\cF$.
\end{lemma}

\begin{proof}
  Since $\rho_{f(X) B}$ is classical on $\cH_Z$, we have, according
  to Lemma~\ref{lem:sqdistexpl}, 
  \begin{equation} \label{eq:sqdistexpl}
    d_2(\rho_{f(X) B}|\sigma_B)
  =
    \sum_{z} \tr\bigl((\sigma_B^{-1/4} \rho_B^z \sigma_B^{-1/4})^2\bigr) 
      - \frac{1}{|\cZ|} 
        \tr\bigl((\sigma_B^{-1/4} \rho_B \sigma_B^{-1/4})^2\bigr) \ ,
  \end{equation}
  where $\rho_B^z$, for $z \in \cZ$, are the conditional operators
  defined by~\eqref{eq:funcdef}.  The first term on the right hand
  side of~\eqref{eq:sqdistexpl} can be rewritten as
  \begin{multline*}
    \sum_{z} \tr\bigl((\sigma_B^{-1/4} \rho_B^z \sigma_B^{-1/4})^2\bigr) \\
  =
    \sum_{z} \sum_{\substack{x \in f^{-1}(z) \\ x' \in f^{-1}(z)}}
      \tr\bigl(
        (\sigma_B^{-1/4} \rho_B^x \sigma_B^{-1/4})
        (\sigma_B^{-1/4} \rho_B^{x'} \sigma_B^{-1/4}) 
      \bigr)  \\
  =
    \sum_{x, x'} \delta_{f(x), f(x')}
      \tr\bigl(
        (\sigma_B^{-1/4} \rho_B^x \sigma_B^{-1/4})
        (\sigma_B^{-1/4} \rho_B^{x'} \sigma_B^{-1/4}) 
      \bigr) \ .
  \end{multline*}
  Similarly, for the second term of~\eqref{eq:sqdistexpl} we find
  \[
    \frac{1}{|\cZ|} \tr\bigl((\sigma_B^{-1/4} \rho_B \sigma_B^{-1/4})^2\bigr)
  =
    \sum_{x, x'} \frac{1}{|\cZ|}
      \tr\bigl(
        (\sigma_B^{-1/4} \rho_B^x \sigma_B^{-1/4})
        (\sigma_B^{-1/4} \rho_B^{x'} \sigma_B^{-1/4})
      \bigr) \ .
  \]
  Hence,
  \begin{multline} \label{eq:expDone}
    \ExpE_{f}\Bigl[
      d_2(\rho_{f(X) B} | \sigma_B)
    \Bigr] \\
  =
    \sum_{x, x'} 
      \ExpE_{f}\bigl[\delta_{f(x), f(x')} - \frac{1}{|\cZ|}\bigr]
      \cdot \tr\bigl(
        (\sigma_B^{-1/4} \rho_B^x \sigma_B^{-1/4})
        (\sigma_B^{-1/4} \rho_B^{x'} \sigma_B^{-1/4})
      \bigr) \ .
  \end{multline}
  Because $f$ is chosen at random from a two-universal family of hash
  functions from $\cX$ to $\cZ$, we have, for any $x \neq x'$,
  \[
    \ExpE_{f}\bigl[\delta_{f(x), f(x')} - \frac{1}{|\cZ|}\bigr] 
  =
    \Pr_{f}[f(x) = f(x')] - \frac{1}{|\cZ|}
  \leq 
    0 \ ,
  \]
  Since the trace $\tr(\sigma \sigma')$ of two nonnegative operators
  $\sigma, \sigma' \in \NN(\cH)$ cannot be negative (cf.\ 
  Lemma~\ref{lem:trprod}) the trace on the right hand side
  of~\eqref{eq:expDone} cannot be negative, for any $x, x' \in \cX$.
  Consequently, when omitting all terms with $x \neq x'$, the sum can
  only get larger, that is,
  \[
    \ExpE_{f}\Bigl[
      d_2(\rho_{f(X) B} | \sigma_B)
    \Bigr] 
  \leq
    \sum_{x} 
      \tr\bigl(
        (\sigma_B^{-1/4} \rho_B^x \sigma_B^{-1/4})^2
      \bigr)  \ .
  \] 
  The assertion then follows from Remark~\ref{rem:Htworewr}.
\end{proof}

\index{two-universal~hashing|)}

\section{Security of privacy amplification} \label{sec:pans}

We are now ready to state our main result on privacy amplification in
the context of quantum adversaries.  Let $X$ be a string and assume
that an adversary controls a quantum system $\cH_B$ whose state is
correlated with~$X$. Theorem~\ref{thm:pa} provides a bound on the
security of a key $f(X)$ computed from $X$ by two-universal hashing.
The bound only depends on the uncertainty of the adversary on $X$,
measured in terms of collision entropy\index{collision~entropy},
min-entropy\index{min-entropy} (cf.\ Corollary~\ref{cor:pa}), or
smooth min-entropy\index{smooth~min-entropy}
(Corollary~\ref{cor:pasmooth}), where the latter is (nearly) optimal
(see Section~\ref{sec:pasm}).

\begin{theorem} \label{thm:pa}
  Let $\rho_{X B} \in \NN(\cH_X \otimes \cH_B)$ be classical with
  respect to an orthonormal basis $\{\ket{x}\}_{x \in \cX}$ of
  $\cH_X$, let $\sigma_B \in \NN(\cH_B)$, and let $\cF$ be a
  two-universal family of hash function from $\cX$ to $\{0,1\}^\ell$.
  Then
  \[
    d(\rho_{F(X) B F}|B F)
  \leq
    \sqrt{\tr(\rho_{X B}) \cdot \tr(\sigma_B)}
      \cdot 2^{-\frac{1}{2} (H_{2}(\rho_{X B}|\sigma_B) - \ell)} \ ,
  \]
  for $\rho_{F(X) B F} \in \NN(\cH_Z \otimes \cH_B \otimes \cH_F)$
  defined by~\eqref{eq:hashfuncdef}.
\end{theorem}

\begin{proof}
  We use Lemma~\ref{lem:distexp} to write the \distance{} from uniform
  as an expectation value,
  \[
    d(\rho_{F(X) B F} | B F)  
  =
    \sum_{f \in \cF} P_F(f) \cdot d(\rho_{f(X) B}| B)
  =
    \ExpE_{f} \bigl[d(\rho_{f(X) B} | B) \bigr]  \ .
  \]
  With Lemma~\ref{lem:disttwodist}, the term in the expectation can be
  bounded in terms of the $L_2$-distance from uniform, that is, for
  any $\sigma_B \in \NN(\cH_B)$,
  \[
  \begin{split}
    d(\rho_{F(X) B F} | B F)
  & \leq
    \sqrt{2^\ell \, \tr(\sigma_B)} 
    \ExpE_{f} \bigl[\sqrt{d_2(\rho_{f(X) B} |\sigma_B)}\bigr] \\
  & \leq
    \sqrt{2^\ell \, \tr(\sigma_B)} 
    \sqrt{\ExpE_{f} \bigl[d_2(\rho_{f(X) B} | \sigma_B)\bigr]}  \ ,
  \end{split}
  \]
  where we have used Jensen's inequality. Finally, we apply
  Lemma~\ref{lem:twodisthashbound} to bound the $L_2$-distance from
  uniform in terms of the collision entropy, which gives
  \[
    d(\rho_{F(X) B F}| B F)
  \leq
    \sqrt{2^\ell \, \tr(\sigma_B)} 
      \sqrt{\tr(\rho_{X B}) 2^{-H_{2}(\rho_{X B}|\sigma_B)}} \ . \qedhere
  \]
\end{proof}

\begin{corollary} \label{cor:pa}
  Let $\rho_{X B} \in \NN(\cH_X \otimes \cH_B)$ be classical with
  respect to an orthonormal basis $\{\ket{x}\}_{x \in \cX}$ of $\cH_X$
  and let $\cF$ be a two-universal family of hash functions from $\cX$
  to $\{0,1\}^\ell$. Then
  \[
    d(\rho_{F(X) B F}|B F)
  \leq
    \sqrt{\tr(\rho_{X B})} \cdot 2^{-\frac{1}{2} (\Hmin(\rho_{X B}|B) - \ell)} 
  \ ,
  \]
  for $\rho_{F(X) B F} \in \NN(\cH_Z \otimes \cH_B \otimes \cH_F)$
  defined by~\eqref{eq:hashfuncdef}.
\end{corollary}

\begin{proof}
  The assertion follows directly from Theorem~\ref{thm:pa} and
  Remark~\ref{rem:Hmincol}.
\end{proof}

\section{Characterization using smooth min-entropy} \label{sec:pasm}

The characterization of privacy amplification in terms of the
collision entropy or min-entropy is not optimal.\footnote{This also
  holds for the classical result, as observed in~\cite{BBCM95}. In
  fact, depending on the probability distribution $P_X$ of the initial
  string $X$, it might be possible to extract a key whose length
  exceeds the collision entropy of $P_X$.}  Because of
Remark~\ref{rem:Hmincol}, the same problem arises if we replace the
collision entropy by the min-entropy (as in Corollary~\ref{cor:pa}).
However, as we shall see, the statement of Theorem~\ref{thm:pa} still
holds if the uncertainty is measured in terms of \emph{smooth}
min-entropy\index{smooth~min-entropy}.  That is, the key generated
from $X$ by two-universal hashing is secure if its length is slightly
smaller than roughly $\Hmin^\eps(\rho_{X B}|B)$, where $\rho_{X B}$ is
the joint state of the initial string $X$ and the adversary's
knowledge.  This is essentially optimal, i.e., $\Hmin^\eps(\rho_{X
  B}|B)$ is also an upper bound on the maximum number of key bits that
can be generated from $X$.\footnote{To see this, let $F$ be an
  arbitrary hash function.  It follows from Lemma~\ref{lem:classnn}
  that the smooth min-entropy cannot increase when applying a function
  on $X$, i.e., $\Hmin^\eps(\rho_{X B}|B) \geq \Hmin^\eps(\rho_{F(X) B
    F}|B F)$.  Moreover, it is easy to verify that the smooth
  min-entropy of a secret key given the adversary's information is
  roughly equal to its length.  Hence, if $F(X)$ is a secret key of
  length $\ell$, we have $\Hmin^\eps(\rho_{F(X) B F}|B F) \geq \ell$.
  Combining this with the above gives $\Hmin^\eps(\rho_{X B}|B) \geq
  \ell$.}

\begin{corollary} \label{cor:pasmooth}
  Let $\rho_{X B} \in \NN(\cH_X \otimes \cH_B)$ be a density operator
  which is classical with respect to an orthonormal basis
  $\{\ket{x}\}_{x \in \cX}$ of $\cH_X$, let $\cF$ be a two-universal
  family of hash functions from $\cX$ to $\{0,1\}^{\ell}$, and let
  $\eps \geq 0$.  Then
  \[
    d(\rho_{F(X) B F} | B F)
  \leq
    2 \eps + 2^{-\frac{1}{2} (\Hmin^\eps(\rho_{X B}|B) - \ell)} \ ,
  \]
  for $\rho_{F(X) B F} \in \NN(\cH_Z \otimes \cH_B \otimes \cH_F)$
  defined by~\eqref{eq:hashfuncdef}.
\end{corollary}

\begin{proof}
  Consider an arbitrary operator $\rhob_{X B} \in \cB^\eps(\rho_{X
    B})$ and let $\rhob_{F(X) B F} \in \NN(\cH_Z \otimes \cH_B \otimes
  \cH_F)$ be the corresponding operator defined
  by~\eqref{eq:hashfuncdef}. Because the \distance{} cannot increase
  when applying a trace-preserving quantum operation (cf.\ 
  Lemma~\ref{lem:distdecr}), we have $\rhob_{F(X) B F} \in
  \cB^\eps(\rho_{F(X) B F})$.  Hence, by the triangle inequality,
  \begin{multline*}
    d(\rho_{F(Z) B F} | B F)
  = 
    \bigl\|\rho_{F(X) B F} - \rho_U \otimes \rho_{B F}\bigr\|_1 \\
  \leq
    \bigl\|\rho_{F(X) B F} - \rhob_{F(X) B F}\bigr\|_1
  +
    \bigl\|\rhob_{F(X) B F} - \rho_U \otimes \rhob_{B F} \bigr\|_1
  +
    \bigl\|\rhob_{B F} - \rho_{B F}\bigr\|_1 \\
  \leq
    2 \eps 
    +  \bigl\|\rhob_{F(X) B F} - \rho_U \otimes \rhob_{B F} \bigr\|_1
  =
    2 \eps + d(\rhob_{F(Z) B F} | B F)
    \ ,
  \end{multline*}
  where $\rho_U$ is the fully mixed state on $\cH_Z$.
  Corollary~\ref{cor:pa}, applied to $\rhob_{X B}$, gives
  \[
    d(\rho_{F(X) B F} |B F)
  \leq
    2 \eps 
    + \sqrt{\tr(\rhob_{X B})} \cdot 2^{-\frac{1}{2} 
      (\Hmin(\rhob_{X B}|B) - \ell)} \ .
  \]
  Because this holds for any $\rhob_{X B} \in \cB^\eps(\rhob_{X B})$,
  the assertion follows by the definition of smooth min-entropy.
\end{proof}

\index{privacy~amplification|)}

\chapter{Security of QKD} \label{ch:QKD}

In this chapter, we use the techniques developed in
Chapters~\ref{ch:smooth}--\ref{ch:PA} to prove the security of
QKD.\footnote{As discussed in Chapter~\ref{ch:intro}, we actually
  consider quantum key \emph{distillation}, which is somewhat more
  general than quantum key \emph{distribution} (QKD).}  (The reader is
referred to Section~\ref{sec:secsum} for a high-level description of
the material presented in the following, including a sketch of the
security proof.) Typically, a QKD protocol is built from several
subprotocols, e.g., for parameter
estimation\index{parameter~estimation}, information
reconciliation\index{information~reconciliation}, or privacy
amplification\index{privacy~amplification}. We first describe and
analyze these subprotocols (Sections~\ref{sec:PE}--\ref{sec:PP}) and
then put the parts together to get a general security criterion for
quantum key \emph{distillation}\index{quantum~key~distillation}
(Section~\ref{sec:eQKD}), which directly implies the security of
quantum key \emph{distribution}\index{quantum~key~distribution} (QKD)
(Section~\ref{sec:qkeydist}).

\section{Preliminaries}

\subsection{Two-party protocols} \label{sec:protocols}

\index{two-party~protocol|(}

A \emph{protocol}\index{protocol} $\cP$ between two parties, Alice and
Bob, is specified by a sequence of operations, called \emph{(protocol)
  steps}\index{protocol~step}, to be performed by each of the parties.
In the first protocol step, Alice and Bob might take (classical or
quantum) \emph{inputs} $A$ and $B$, respectively (e.g., some
correlated data).  In each of the following steps, Alice and Bob
either perform local computations or exchange messages (using a
classical or a quantum communication channel).  Finally, in the last
protocol step, Alice and Bob generate \emph{outputs} $A'$ and $B'$,
respectively (e.g., a pair of secret keys).


We will mostly (except for Section~\ref{sec:qkeydist}) be concerned
with the analysis of protocols $\cP$ that only use communication over
a classical and authentic channel\index{authentic~channel}.  In this
case, Alice and Bob's outputs as well as the transcript of the
communication do not depend on the attack of a potential adversary.
Let $\rho_{A B}$ and $\rho_{A' B' C}$ be the density operators
describing Alice and Bob's inputs $A$ and $B$ as well as their outputs
$A'$ and $B'$ together with the communication transcript $C$,
respectively.  The mapping that brings $\rho_{A B}$ to $\rho_{A' B'
  C}$, in the following denoted by $\cE^{\cP}_{A' B' C \leftarrow A
  B}$, is then uniquely defined by the protocol $\cP$.  Moreover,
because it must be physically realizable, $\cE^{\cP}_{A' B' C
  \leftarrow A B}$ is a CPM (see Section~\ref{sec:qdefs}).

To analyze the security of a protocol $\cP$, we need to include Eve's
information in our description. Let $\rho_{A B E}$ be the state of
Alice and Bob's inputs as well as Eve's initial information.
Similarly, let $\rho_{A' B' E'}$ be the state of Alice and Bob's
outputs together with Eve's information after the protocol execution.
As Eve might get a transcript $C$ of the messages sent over the
classical channel, the CPM that maps $\rho_{A B E}$ to $\rho_{A' B'
  E'}$ is given by
\[
  \cE^{\cP}_{A' B' E' \leftarrow A B E}
:=
  \cE^{\cP}_{A' B' C \leftarrow A B} \otimes \id_E \ ,
\]
where $\cH_{E'} := \cH_C \otimes \cH_E$.

\index{two-party~protocol|)}

\subsection{Robustness of protocols}

\index{robustness|(}

Depending on its input, a protocol might be unable to produce the
desired output. For example, if a key distillation protocol starts
with uncorrelated randomness, it cannot generate a pair of secret
keys.  In this case, the best we can hope for is that the protocol
recognizes this situation and aborts\footnote{Technically, the
  protocol might output a certain predefined symbol which indicates
  that it is unable to accomplish the task.}  (instead of generating
an insecure result).

Clearly, one is interested in designing protocols that are successful
on certain inputs. This requirement is captured by the notion of
\emph{robustness}\index{robustness}.

\begin{definition}
  Let $\cP$ be a two-party protocol and let $\rho_{A B} \in \NN(\cH_A
  \otimes \cH_B)$. We say that $\cP$ is \emph{$\eps$-robust on
    $\rho_{A B}$}\index{$\eps$-robustness} if, for inputs defined by
  $\rho_{A B}$, the probability that the protocol aborts is at most
  $\eps$.
\end{definition}


Mathematically, we represent the state that describes the situation
after an abortion of the protocol as a zero operator. The CPM $\cE_{A'
  B' C \leftarrow A B}^{\cP}$ (as defined in
Section~\ref{sec:protocols}) is then a projection onto the space that
represents the outputs of successful protocol executions (i.e., where
it did not abort). The probability that the protocol is successful
when starting with an initial state $\rho_{A B}$ is thus equal to the
trace $\tr(\rho_{A' B' E})$ of the operator $\rho_{A' B' E} = \cE_{A'
  B' C \leftarrow A B}^{\cP}(\rho_{A B})$.  In particular, if $\cP$ is
$\eps$-robust\index{$\eps$-robustness} on a density operator $\rho_{A
  B}$ then $\tr(\rho_{A' B' E}) \geq 1-\eps$.

\index{robustness|)}

\subsection{Security definition for key distillation} \label{sec:KAsec}

\index{security~definition|(}

\newcommand*{\KA}{\mathsf{KD}}

A \emph{(quantum) key distillation
  protocol}\index{quantum~key~distillation} $\KA$ is a two-party
protocol with classical communication where Alice and Bob take inputs
from $\cH_A$ and $\cH_B$, respectively, and either output classical
keys $s_A, s_B \in \cS$, where $\cS$ is called the \emph{key
  space}\index{key~space} of $\KA$, or abort the protocol.

\begin{definition}  
  Let $\KA$ be a key distillation
  protocol\index{quantum~key~distillation} and let $\rho_{A B E} \in
  \NN(\cH_A \otimes \cH_B \otimes \cH_E)$. We say that $\KA$ is
  \emph{$\eps$-secure on $\rho_{A B E}$} if $\rho_{S_A S_B E'} :=
  \cE^{\KA}_{S_A S_B E' \leftarrow A B E}(\rho_{A B
    E})$\index{$\eps$-security} satisfies
  \[
    \sfrac{1}{2} \bigl\| \rho_{S_A S_B E'} - \rho_{U U} \otimes \rho_{E'} \bigr\|_1
  \leq
    \eps \ ,
  \]
  where $\rho_{U U} := \sum_{s \in \cS} \frac{1}{|\cS|} \proj{s}
  \otimes \proj{s}$, for some family $\{\ket{s}\}_{s \in \cS}$ of
  orthonormal vectors representing the values of the key space $\cS$.
  
  Moreover, we say that $\KA$ is \emph{$\eps$-fully
    secure}\index{$\eps$-full~security} if it is $\eps$-secure on all
  density operators $\rho_{A B E} \in \NN(\cH_A \otimes \cH_B \otimes
  \cH_E)$.
\end{definition}

According to the discussion on universal
security\index{universal~security} in
Section~\ref{sec:univdef},\footnote{\label{fn:ip}If a key $S$ is
  $\eps$-secure, one could define a perfectly secure (independent and
  uniformly distributed) key $U$ such that $\Pr[s \neq u] \leq \eps$
  (see also Proposition~\ref{pro:vardistevent}).}  this definition has
a very intuitive interpretation: If the protocol is $\eps$-fully
secure then, for any arbitrary input, the probability of the event
that Alice and Bob do not abort \emph{and} the adversary gets
information on the key pair\footnote{According to
  Footnote~\ref{fn:ip}, one could say that the adversary \emph{gets
    information on a key $S$} whenever the value of $S$ is not equal
  to the value of a perfect key $U$.}  is at most
$\eps$.\footnote{Note that the adversary's information on the key,
  \emph{conditioned} on the event that Alice and Bob generate a key,
  is not necessarily small.  In fact, if, for a certain input, the
  probability that Alice and Bob generate a key is very small (e.g.,
  smaller than $\eps$) then|conditioned on this rare event|the key
  might be insecure (see also the discussion in~\cite{BBBMR05}).}  In
other words, except with probability $\eps$, Alice and Bob either
abort or generate a pair of keys which are identical to a perfect
key\index{perfect~key}.

\begin{remark} \label{rem:CKAsec}
  The above security definition for key distillation protocols $\KA$
  can be subdivided into two parts:
  \begin{itemize}
  \item \emph{$\eps'$-correctness:}\index{$\eps$-correctness} $\Pr[s_A
    \neq s_B] \leq \eps'$,\footnote{$\Pr[s_A \neq s_B]$ is the
      probability of the event that Alice and Bob do not abort
      \emph{and} the generated keys $s_A$ and $s_B$ are different.}
    for $s_A$ and $s_B$ chosen according to the distribution defined
    by $\rho_{S_A S_B}$.
  \item \emph{$\eps''$-secrecy of Alice's key:}\index{$\eps$-secrecy}
    $\sfrac{1}{2} d(\rho_{S_A C E}|C E) \leq \eps''$.\footnote{See
      Definition~\ref{def:nonunif}.}
  \end{itemize}
  In particular, if $\KA$ is $\eps'$-correct and $\eps''$-secret on
  $\rho_{X Y E}$ then it is $(\eps'+\eps'')$-secure on $\rho_{X Y E}$.
\end{remark}

\index{security~definition|)}

\section{Parameter estimation} \index{parameter~estimation} \label{sec:PE}

\index{parameter~estimation|(}

\newcommand*{\PE}{\mathsf{PE}}

The purpose of a parameter estimation is to decide whether the input
given to the protocol can be used for a certain task, e.g. to distill
a secret key.  Technically, a \emph{parameter estimation
  protocol}\index{parameter~estimation} $\PE$ is simply a two-party
protocol where Alice and Bob take inputs from $\cH_A$ and $\cH_B$,
respectively, and either output ``accept'' or abort the protocol.

\begin{definition}  
  Let $\PE$ be a parameter estimation protocol and let $\rho_{A B} \in
  \NN(\cH_A \otimes \cH_B)$. We say that $\PE$ \emph{$\eps$-securely
    filters}\index{$\eps$-secure~filtering} $\rho_{A B}$ if, on input
  $\rho_{A B}$, the protocol aborts except with probability $\eps$.
\end{definition}

\newcommand{\figtop}{\vspace{0.4ex}}

\begin{protocolfloat}
\figtop
Parameters: \\[1ex]
\begin{tabular}{ll}
  $\cM$: & bipartite POVM $\{M_w\}_{w \in \cW}$ on $\cH_A \otimes \cH_B$ \\
  $\cQ$: & set of frequency distributions on $\cW$
\end{tabular}
\tabprotsep
\begin{protocol}{Alice}{Bob} 
  \protno{input space: $\cH_A^{\otimes n}$}{input space:
  $\cH_B^{\otimes n}$} \\
  \protleftright{$\cH_A^{\otimes n}$}{meas.\ $\cM^{\otimes n}$}{$\cH_A^{\otimes n}$ \\ $\quad \rightarrow \bw = (w_1, \ldots, w_n)$}
  \protno{}{if $\freq{\bw} \notin \cQ$ \\ $\quad$ then abort \\
  $\quad$ else output ``acc.''}
\end{protocol}
\caption{
  Parameter estimation protocol $\PE_{\cM, \cQ}$.} \label{pr:PE}
\end{protocolfloat}

A typical and generic example for parameter estimation is the protocol
$\PE_{\cM, \cQ}$ depicted in Fig.~\ref{pr:PE}. Alice and Bob take
inputs from an $n$-fold product space. Then they measure each of the
$n$ subspaces according to a POVM $\cM = \{M_{w}\}_{w \in
  \cW}$.\footnote{$\cM$ might be an arbitrary measurement that can be
  performed by two distant parties connected by a classical channel.}
Finally, they output ``accept'' if the frequency distribution
$\freq{\bw}$ of the measurement outcomes $\bw = (w_1, \ldots, w_n)$ is
contained in a certain set $\cQ$.

For the analysis of this protocol, it is convenient to consider the
set $\Gamma_{\cM, \cQ}^{\leq \distPE}$ of density operators $\sigma_{A
  B}$ for which the measurement $\cM$ leads to a distribution which
has distance \emph{at most} $\distPE$ to the set $\cQ$.  Formally,
\begin{equation} \label{eq:Gammadef}
  \Gamma_{\cM, \cQ}^{\leq \distPE}
:=
  \bigl\{ 
    \sigma_{A B} : \, \min_{Q \in \cQ} \| P_W^{\sigma_{A B}} - Q \|_1 \leq \distPE
  \bigr\} \ ,
\end{equation}
where $P_W^{\sigma_{A B}}$ denotes the probability distribution of the
outcomes when measuring $\sigma_{A B}$ according to $\cM$, i.e.,
$P_W(w) = \tr(M_w \sigma_{A B})$, for any $w \in \cW$.

Assume that the protocol $\PE_{\cM, \cQ}$ takes as input a product
state $\rho_{A^n B^n} = \sigma_{A B}^{\otimes n}$. Then, by the law of
large numbers, the measurement statistics $\lambda_{\bw}$ must be
close to $\cM(\sigma_{A B})$. In particular, if the protocol accepts
with non-negligible probability (i.e., $\lambda_{\bw}$ is contained in
$\cQ$) then $\sigma_{A B}$ is likely to be contained in $\Gamma_{\cM,
  \cQ}^{\leq \distPE}$, for some small $\distPE > 0$. In other words,
the protocol aborts with high probability if $\sigma_{A B}$ is not an
element of the set $\Gamma_{\cM, \cQ}^{\leq \distPE}$.  The following
lemma generalizes this statement to permutation-invariant inputs.

\begin{lemma} \label{lem:PEsec}
  Let $\cM := \{M_w\}_{w \in \cW}$ be a POVM on $\cH_A \otimes \cH_B$,
  let $\cQ$ be a set of frequency distributions on $\cW$, let $0 \leq
  r \leq \frac{1}{2}n$, and let $\eps \geq 0$. Moreover, let
  $\ket{\theta} \in \cH_{A B E} := \cH_A \otimes \cH_B \otimes \cH_E$
  and let $\rho_{A^n B^n E^n}$ be a density operator on $\SymR{\cH_{A
      B E}}{n}{\theta}{n-r}$. If $\tr_E(\proj{\theta})$ is not
  contained in the set $\Gamma_{\cM, \cQ}^{\leq \distPE}$ defined
  by~\eqref{eq:Gammadef}, for
  \[
    \distPE 
  := 2 \sqrt{ \sfrac{\log(\slfrac{1}{\eps})}{n} +
      h(\slfrac{r}{n}) + \sfrac{|\cW|}{n} \log(\sfrac{n}{2}+1)} \ ,
  \]
  then the protocol $\PE_{\cM, \cQ}$ defined by Fig.~\ref{pr:PE}
  $\eps$-securely filters $\rho_{A^n B^n}$. 
\end{lemma}

\begin{proof}
  The assertion follows directly from Theorem~\ref{thm:symstat}.
\end{proof}


Similarly to~\eqref{eq:Gammadef}, we can define a set
$\overline{\Gamma}_{\cM, \cQ}^{\geq\distPE}$ containing all density
operators $\sigma_{A B}$ for which the measurement $\cM$ leads to a
distribution which has distance \emph{at least} $\distPE$ to the
\emph{complement} of $\cQ$.  Formally,
\begin{equation} \label{eq:Gammabardef}
  \overline{\Gamma}_{\cM, \cQ}^{\geq \distPE}
:=
  \bigl\{ 
    \sigma_{A B} : \, \min_{Q \notin \cQ} \| P_W^{\sigma_{A B}} - Q \|_1 \geq \distPE
  \bigr\} \ .
\end{equation} 

Analogously to the above argument, one can show that the protocol
$\PE_{\cM, \cQ}$ defined by Fig.~\ref{pr:PE} is $\eps$-robust on
product operators $\sigma_{A B}^{\otimes n}$, for any $\sigma_{A B}
\in \overline{\Gamma}_{\cM, \cQ}^{\geq \distPE}$.

\index{parameter~estimation|)}

\section{Information reconciliation} \label{sec:IR}

\index{information~reconciliation|(}

Assume that Alice and Bob hold weakly correlated classical values $x$
and $y$, respectively. The purpose of an information reconciliation
protocol is to transform $x$ and $y$ into a pair of fully correlated
strings, while leaking only a minimum amount of information (on the
final strings) to an eavesdropper (see, e.g., \cite{BraSal94}).

\newcommand*{\dec}{\mathrm{dec}}

\newcommand*{\IR}{\mathsf{IR}}
\newcommand*{\IRit}{\mathsf{IR}}
\newcommand*{\IRcomp}{\mathsf{IR}_{\mathrm{comp}}}
\newcommand*{\abort}{\mathrm{abort}}

\subsection{Definition}

We focus on information reconciliation schemes where Alice keeps her
input value $x$ and where Bob outputs a guess $\hat{x}$ for $x$.
Hence, technically, an \emph{information reconciliation
  protocol}\index{information~reconciliation} $\IR$ is a two-party
protocol where Alice and Bob take classical inputs $x \in \cX$ and $y
\in \cY$, respectively, and where Bob outputs a classical value
$\hat{x} \in \cX$ or aborts.

\begin{definition}  
  Let $P_{X Y} \in \NN(\cX \times \cY)$ and let $\eps \geq 0$.  We say
  that an information reconciliation protocol $\IR$ is
  \emph{$\eps$-secure on $P_{X Y}$}\index{$\eps$-security} if, for
  inputs $x$ and $y$ chosen according to $P_{X Y}$, the probability
  that Bob's output $\hat{x}$ differs from Alice's input $x$ is at
  most $\eps$, i.e., $\Pr[\hat{x} \neq x] \leq \eps$.\footnote{We
    denote by $\Pr[\hat{x} \neq x]$ the probability of the event that
    the protocol does not abort \emph{and} $\hat{x}$ is different from
    $x$.}
  
  Moreover, we say that $\IR$ is \emph{$\eps$-fully
    secure}\index{$\eps$-full~security} if it is $\eps$-secure on all
  probability distributions $P_{X Y} \in \NN(\cX \times \cY)$.
\end{definition}

The communication transcript of an information reconciliation scheme
$\IR$ generally contains useful information on Alice and Bob's values.
If the communication channel is insecure, this information might be
leaked to Eve.  Clearly, in the context of key agreement, one is
interested in information reconciliation schemes for which this
leakage is minimal.

\begin{definition}
  Let $\IR$ be an information reconciliation protocol where Alice and
  Bob take inputs from $\cX$ and $\cY$, respectively. Let $\cC$ be the
  set of all possible communication transcripts $c$ and let
  $P_{C|X=x,Y=y}$ be the distribution of the transcripts $c \in \cC$
  conditioned on inputs $(x,y) \in \cX \times \cY$.  Then the
  \emph{leakage of $\IR$}\index{leakage} is
  \[
    \leak_{\IR}
  := 
    \log |\cC| - \inf_{x, y} \Hmin(P_{C|X=x,Y=y}) \ ,
  \]
  where the infimum ranges over all $(x,y) \in \cX \times \cY$. 
\end{definition}

Note that the leakage is independent of the actual distribution $P_{X
  Y}$ of Alice and Bob's values.

\subsection{Information reconciliation with minimum leakage}

A typical information reconciliation protocol is the protocol
$\IRit_{\hat{\cX}, \cF}$ defined by Fig.~\ref{pr:IRit}. It is a
so-called \emph{one-way}\index{one-way~protocol} protocol where only
Alice sends messages to Bob. We show that the leakage\index{leakage}
of this protocol, for appropriately chosen parameters, is roughly
bounded by the max-entropy of $X$ given~$Y$ (Lemma~\ref{lem:errcorr}).
This statement can be extended to smooth max-entropy
(Lemma~\ref{lem:errcorrsmooth}), which turns out to be optimal, i.e.,
the minimum leakage of an information reconciliation protocol for
$P_{X Y}$ is exactly characterized by $\Hmax^\eps(X|Y)$.  In
particular, for the special case where the input is chosen according
to a product distribution, we get an asymptotic expression in terms of
Shannon entropy (Corollary~\ref{cor:errcorr}), which corresponds to
the Shannon coding theorem.

\begin{protocolfloat}
\figtop
Parameters:\\[1ex]
\begin{tabular}{ll}
  $\hat{\cX}$: & family of sets
  $\hat{\cX}_y \subseteq \cX$ parameterized by $y \in \cY$. \\
  $\cF$: & family of hash functions from  $\cX$ to $\cZ$.
\end{tabular}

\vspace{2ex}

\begin{protocol}{Alice}{Bob}
  \protno{input: $x \in \cX$}{input: $y \in \cY$} \\
\protright{$f \in_R
    \cF$ \\ $z := f(x)$}{$f, z$}{$\hat{\cD}:=$ \\ $\quad \{\hat{x} \in \hat{\cX}_y : \,
    f(\hat{x}) = z\}$}
  \protno{}{if $\hat{\cD} \neq \emptyset$\\
    $\quad$ then $\hat{x} \in_R \hat{\cD}$ \\$\quad$ else
    $\abort$} \protno{}{output $\hat{x}$}
\end{protocol}
\caption{
  Information reconciliation protocol $\IRit_{\hat{\cX},
    \cF}$.}\label{pr:IRit}
\end{protocolfloat}

\begin{lemma} \label{lem:errcorr}
  Let $P_{X Y} \in \NN(\cX \times \cY)$ and let $\eps > 0$.  Then the
  information reconciliation protocol $\IRit_{\hat{\cX}, \cF}$ defined
  by Fig.~\ref{pr:IRit}, for an appropriate choice of the parameters
  $\hat{\cX}$ and $\cF$, is $0$-robust on $P_{X Y}$, $\eps$-fully
  secure, and has leakage\index{leakage}
  \[
    \leak_{\IRit_{\hat{\cX}, \cF}} \leq \Hmax(P_{X Y}|Y) 
      + \log(\slfrac{2}{\eps}) \ .
  \]
\end{lemma}

\begin{proof}
  Let $k := \lceil \Hmax(P_{X Y}|Y) + \log(\slfrac{1}{\eps})\rceil$
  and let $\cF$ be a two-universal family of hash functions from $\cX$
  to $\cZ := \{0,1\}^{k}$ (which exists according to
  Lemma~\ref{lem:hashexists}).  Furthermore, let $\hat{\cX} =
  \{\hat{\cX}_y\}_{y \in \cY}$ be the family of sets defined by
  $\hat{\cX}_y := \supp(P^y_X)$, where $\supp(P^y_X)$ denotes the
  support of the function $P^y_X: \, x \mapsto P_{X Y}(x,y)$.
  
  For any pair of inputs $x$ and $y$ and for any communication $(f,z)
  = (f, f(x))$ computed by Alice, Bob can only output a wrong value if
  the set $\hat{\cX}_y = \supp(P^y_X)$ contains an element $\hat{x}
  \neq x$ such that $f(\hat{x}) = z$.  Because $f$ is chosen uniformly
  at random from the family of two-universal hash functions $\cF$, we
  have $\Pr_f\bigl[f(\hat{x}) = f(x) \bigr] \leq \frac{1}{|\cZ|} =
  2^{-k}$, for any $\hat{x} \neq x$.  Hence, by the union bound, for
  any fixed $(x, y) \in \cX \times \cY$,
  \[
  \begin{split}
    \Pr[\hat{x} \neq x ]
  & \leq
    \Pr_{f}\bigl[
      \exists \hat{x} \in \supp(P^y_X) : \, \hat{x} \neq x \, \wedge \,  f(\hat{x}) = f(x)
    \bigr] \\
  & \leq
    \bigl|\supp(P^y_X) \bigr| \cdot 2^{-k} \ .
  \end{split}
  \]
  Because, by Remark~\ref{rem:classminmaxentr}, $\max_{y'}
  |\supp(P^{y'}_X) | = 2^{\Hmax(P_{X Y}|Y)}$, we conclude
  \[
    \Pr[\hat{x} \neq x]
  \leq
    2^{\Hmax(P_{X Y}|Y) - \lceil \Hmax(P_{X Y}|Y) + \log(\slfrac{1}{\eps})  \rceil}
  \leq
    \eps \ ,
  \]
  that is, $\IRit_{\hat{\cX}, \cF}$ is $\eps$-secure on any
  probability distribution.
  
  Moreover, if $(x,y)$ is chosen according to the distribution $P_{X
    Y}$, then, clearly, $x$ is always contained in $\hat{\cX}_y =
  \supp(P^y_X)$, that is, Bob never aborts. This proves that the
  protocol is $0$-robust.
  
  Since $f$ is chosen uniformly at random and independently of $x$
  from the family of hash-functions $\cF$, all nonzero probabilities
  of the distribution $P_{C|X=x}$ are equal to $\frac{1}{|\cF|}$.
  Hence, using the fact that $\cC = \cF \times \cZ$,
  \[
  \begin{split}
    \leak_{\IRit_{\hat{\cX}, \cF}}
  & =
    \log |\cC| - \inf_{x \in \cX} \Hmin(P_{C|X=x}) \\
  & = 
    \log |\cF \times \cZ| -  \log |\cF|
  \leq
    \log |\cZ|
  =
    k \ .
  \end{split}
  \]
  The claimed bound on the leakage then follows by the definition of
  $k$.
\end{proof}

\begin{lemma} \label{lem:errcorrsmooth}
  Let $P_{X Y} \in \NN(\cX \times \cY)$ and let $\eps,\eps' \geq 0$.
  Then the information reconciliation protocol $\IRit_{\hat{\cX},
    \cF}$ defined by Fig.~\ref{pr:IRit}, for an appropriate choice of
  the parameters $\hat{\cX}$ and $\cF$, is $\eps'$-robust on $P_{X
    Y}$, $\eps$-fully secure, and has leakage
  \[
    \leak_{\IRit_{\hat{\cX}, \cF}}
  \leq 
    \Hmax^{\eps'}(P_{X Y}|Y) + \log(\slfrac{2}{\eps}) \ .
  \]
\end{lemma}

\begin{proof}
  For any $\nu > 0$ there exists $\Pb_{X Y} \in \NN(\cH_X \otimes
  \cH_Y)$ such that
  \begin{equation} \label{eq:encprobdiff}
    \bigl\| P_{X Y} - \Pb_{X Y} \bigr\|_1
  \leq
    \eps'
  \end{equation}
  and
  \begin{equation} \label{eq:encentrbnd}  
    \Hmax(\Pb_{X Y}|Y) \leq \Hmax^{\eps'}(P_{X Y}|Y) + \nu \ .
  \end{equation}
  According to Lemma~\ref{lem:errcorr}, there exists $\hat{\cX}$ and
  $\cF$ such that $\IRit_{\hat{\cX}, \cF}$ is $\eps$-fully secure,
  $0$-robust on $\Pb_{X Y}$, and has leakage
  \[
    \leak_{\IRit_{\hat{\cX}, \cF}}
  \leq 
    \Hmax(\Pb_{X Y}|Y) + \log(\slfrac{2}{\eps}) \ .
  \]
    
  The stated bound on the leakage follows immediately from this
  inequality and~\eqref{eq:encentrbnd}.  Moreover, the bound on the
  robustness is a direct consequence of the
  bound~\eqref{eq:encprobdiff} and the fact that the protocol is
  $0$-robust on $\Pb_{X Y}$.
\end{proof}

\begin{corollary} \label{cor:errcorr}
  Let $P_{X Y} \in \NN(\cX \times \cY)$ be a probability distribution,
  let $n \geq 0$, and let $\eps \geq 0$.  Then there exists an
  information reconciliation protocol $\IR$ which is $\eps$-fully
  secure, $\eps$-robust on the product distribution $P_{X^n Y^n} :=
  (P_{X Y})^n$, and has leakage
  \[
    \sfrac{1}{n} \leak_{\IR}
  \leq 
     H(X|Y) + \sqrt{\sfrac{3 \log(\slfrac{2}{\eps})}{n}} \log (|\cX| + 3)  \ .
  \]
\end{corollary}

\begin{proof}
  Using Lemma~\ref{lem:errcorrsmooth} (with $\eps = \eps'$) and
  Theorem~\ref{thm:classprodentr}, we find
  \[  
    \sfrac{1}{n} \leak_{\IR}
  \leq 
     H(X|Y) + \sfrac{\log(\slfrac{2}{\eps})}{n} 
    + \sqrt{\sfrac{2 \log(\slfrac{1}{\eps})}{n}} \log (|\cX| + 3) \ .
  \]
  Let $a:=\frac{\log(\slfrac{2}{\eps})}{n}$ and $b:=\log (|\cX| + 3)$.
  The last two terms on the right hand side of this inequality are
  then upper bounded by $a + \sqrt{2 a} b \leq (\frac{a}{2}+\sqrt{2
    a})b$, which holds because $b \geq 2$.  We can assume without loss
  of generality that $3 a \leq 1$ (otherwise, the statement is
  trivial).  Then $\frac{a}{2} + \sqrt{2 a} \leq \sqrt{3 a}$. The last
  two terms in the above inequality are thus bounded by $\sqrt{ 3 a}
  b$, which concludes the proof.
\end{proof}

For practical applications, we are interested in protocols where Alice
and Bob's computations can be done
\emph{efficiently}\index{computational~efficiency} (e.g., in time that
only depends polynomially on the length of their inputs).  This is,
however, not necessarily the case for the information reconciliation
protocol $\IRit_{\hat{\cX}, \cF}$ described above.  While Alice's
task, i.e., the evaluation of the hash function, can be done in
polynomial time,\footnote{Recall that Alice only has to evaluate a
  function which is randomly chosen from a two-universal family of
  functions.  For most known constructions of such families (see,
  e.g., \cite{CarWeg79,WegCar81}), this can be done efficiently.} no
efficient algorithm is known for the decoding operation of Bob.
Nevertheless, based on a specific encoding scheme, one can show that
there exist information reconciliation protocols which only require
polynomial-time computations and for which the statement of
Corollary~\ref{cor:errcorr} (asymptotically) still holds (see
Appendix~\ref{app:effIR}).

\index{information~reconciliation|)}

\section{Classical post-processing} \label{sec:PP}

\index{classical~post-processing|(}

Classical post-processing is used to transform an only partially
secure\footnote{That is, $x$ and $y$ are only weakly correlated and
  partially secret strings.} pair of raw keys\index{raw~key} $x$ and
$y$ held by Alice and Bob, respectively, into a fully secure key pair.
A classical post-processing protocol is thus actually a key
distillation protocol that starts with classical randomness.


In this section, we analyze the security of the generic
post-processing protocol depicted in Fig.~\ref{pr:PP}. It consists of
an information reconciliation\index{information~reconciliation}
subprotocol (see Section~\ref{sec:IR}) followed by privacy
amplification\index{privacy~amplification} (see Chapter~\ref{ch:PA}).

\newcommand*{\CKA}{\mathsf{PP}}

\begin{protocolfloat} 
\figtop
Parameters: \\[1ex]
\begin{tabular}{ll}
  $\IR$: & information reconciliation protocol. \\
  $\cF$: & family of hash functions from $\cX$ to $\{0, 1\}^{\ell}$.   
\end{tabular}
\tabprotsep
\begin{protocol}{Alice}{Bob}
  \protno{input: $x \in \cX$}{input: $y \in \cY$} \\
  \protleftright{$x$}{$\IR$}{$y \rightarrow \hat{x}$} \\
  \protright{$f \in_R \cF$}{$f$}{}
  \protno{output $s_A :=  f(x)$}{output $s_B:= f(\hat{x})$}
\end{protocol}
\caption{%
  Classical post-processing protocol $\CKA_{\IR,\cF}$.} \label{pr:PP}
\end{protocolfloat}

\begin{lemma} \label{lem:ppsec}
  Let $\IR$ be an information reconciliation protocol and let $\cF$ be
  a two-universal family of hash functions from $\cX$ to
  $\{0,1\}^{\ell}$. Additionally, let $\rho_{X Y E} \in \NN(\cH_X
  \otimes \cH_Y \otimes \cH_E)$ be a density operator which is
  classical on $\cH_X \otimes \cH_Y$ and let $\eps', \eps'' \geq 0$.
  If $\IR$ is $\eps'$-secure on the distribution defined by $\rho_{X
    Y}$ and if
  \[
    \ell
  \leq 
    \Hmin^{\eps}(\rho_{X E} | E)
    - \leak_{\IR}
    - 2 \log(\slfrac{1}{\eps}) \ ,
  \]
  for $\eps := \frac{2}{3} \eps''$, then the key distillation protocol
  $\CKA_{\IR, \cF}$ defined by Fig.~\ref{pr:PP} is
  $(\eps'+\eps'')$-secure on $\rho_{X Y E}$.
\end{lemma}

\begin{proof}  
  For simplicity, we assume in the following that the protocol $\IR$
  is one-way. It is straightforward to generalize this argument to
  arbitrary protocols.

  Note first that the keys $s_A$ and $s_B$ generated by Alice and Bob
  can only differ if $\hat{x} \neq x$. Hence, because the information
  reconciliation protocol $\IR$ is $\eps'$-secure on the distribution
  defined by $\rho_{X Y}$, the classical post-processing protocol
  $\CKA_{\IR, \cF}$ is $\eps'$-correct on $\rho_{X Y E}$. According to
  Remark~\ref{rem:CKAsec}, it thus remains to show that Alice's key is
  $\eps''$-secret.
  
  For this, we use the result on the security of privacy
  amplification\index{privacy~amplification} by two-universal
  hashing\index{two-universal~hashing} presented in
  Chapter~\ref{ch:PA}.  Because $f$ is chosen from a two-universal
  family of hash functions, Corollary~\ref{cor:pasmooth} implies that
  the key computed by Alice is $\eps''$-secret if
  \begin{equation} \label{eq:HXlastbound}
    \Hmin^{\eps}(\rho_{X C' E}|C' E)  \geq 2 \log(\slfrac{1}{\eps}) + \ell \ ,
  \end{equation}
  where $\rho_{X \hat{X} C' E} := (\cE^\IR \otimes \id_E) (\rho_{X Y
    E})$ is the operator describing the situation after the execution
  of the information reconciliation protocol $\IR$ (where $C'$ is the
  transcript of $\IR$). It thus suffices to verify that the bound on
  the entropy~\eqref{eq:HXlastbound} holds.
  
  Using the chain rule (cf.~\eqref{eq:fchain} of
  Theorem~\ref{thm:Hmincalc}), the left hand side
  of~\eqref{eq:HXlastbound} can be bounded by
  \[
    \Hmin^{\eps}(\rho_{X C' E} | C' E)
  \geq 
    \Hmin^{\eps}(\rho_{X C' E} | E) - \Hmax(\rho_{C'}) \ .
  \]
  Moreover, because the communication $c'$ is computed only from $x$,
  the conditional operators $\rho_{C' E}^x$ have product form and thus
  (cf.~\eqref{eq:fdataproc} of Theorem~\ref{thm:Hmincalc})
  \begin{equation} \label{eq:HminCEbound}
    \Hmin^{\eps}(\rho_{X C' E} | C' E)
  \geq
    \Hmin^{\eps}(\rho_{X E} | E) + \Hmin(\rho_{C' X}|\rho_X)
    -  \Hmax(\rho_{C'}) \ .
  \end{equation}
  Using the fact that $\Hmax(\rho_{C'}) = \log \rank(\rho_{C'})$ and
  Lemma~\ref{lem:Hminclasscondr}, the last two terms in the above
  expression can be bounded by
  \[
    \Hmax(\rho_{C'}) - \Hmin(\rho_{C' X}|\rho_X)
  \leq
    \log \rank(\rho_{C'}) - \inf_{x \in \cX} \Hmin(\rhob_{C'}^x) \ ,
  \]
  where, for any $x \in \cX$, $\rhob_{C'}^x$ is the normalized
  conditional operator defined by $\rho_{C' X}$. Hence, by the
  definition of leakage,\index{leakage}
  \[
    \Hmax(\rho_{C'}) - \Hmin(\rho_{C' X}|\rho_X)
   \leq
     \leak_{\IR} \ .
  \]
  Combining this with~\eqref{eq:HminCEbound}, we find
  \[
    \Hmin^{\eps}(\rho_{X C' E} | C' E)
  \geq
    \Hmin^{\eps}(\rho_{X E} | E) - \leak_{\IR} \ ,
  \]
  which, by the assumption on the length of the final key $\ell$,
  implies~\eqref{eq:HXlastbound} and thus concludes the proof.
 \end{proof}

\index{classical~post-processing|)}

\section{Quantum key distillation} \label{sec:eQKD}

\index{quantum~key~distillation|(}

\newcommand*{\eQKD}{\mathsf{QKD}}

\newcommand*{\Bl}{\mathsf{Bl}}
\newcommand*{\Meas}{\mathsf{Meas}}

We are now ready to describe and analyze a general quantum key
distillation protocol\index{quantum~key~distillation}, which uses
parameter estimation and classical post-processing as discussed above.
(For a high-level description of the content of this section, we refer
to Section~\ref{sec:secsum}.)

\subsection{Description of the protocol}

\begin{protocolfloat} 
\figtop
Parameters: \\[1ex]
\begin{tabular}{ll}
  $\PE$: & parameter estimation protocol on
  $\cH_A^{\otimes m} \otimes \cH_B^{\otimes m}$.
  \\
  $\Bl$: & subprotocol on $\cH_A^{\otimes b} \otimes
  \cH_B^{\otimes b}$ with classical output in  $\cX \times \cY$.  \\
  $\CKA$: & classical post-processing protocol on
  $\cX^n \times \cY^n$. \\
  $N$: & Number of input systems ($N \geq b n + m$)
\end{tabular}
\tabprotsep
\begin{protocol}{Alice}{Bob}
  \protno{input space: $\cH_A^{\otimes N}$}{input space:
  $\cH_B^{\otimes N}$}
   \\[1.4ex]
   \protright{$\pi \in_R \cS_{N}$}{$\pi$}{}
   \protno{permute subsyst.}{permute subsyst.}\\[1.4ex]
  \protleftright{$\cH_A^{\otimes
  m}$}{$\PE$}{$\cH_B^{\otimes m}$ \\ $\quad \rightarrow \mathrm{acc./abort}$}\\ 
  \protleftright{$(\cH_A^{\otimes
      b})^{\otimes n}$ \\ $\quad \rightarrow (x_1, \ldots, x_{n})$}{$\Bl^{\otimes n}$}{$(\cH_B^{\otimes b})^{\otimes n}$ \\ $\quad \rightarrow (y_1, \ldots, y_{n}$)} \\ 
  \protleftright{$(x_1, \ldots, x_n) \rightarrow s_A$}{$\CKA$}{$(y_1,
    \ldots, y_n) \rightarrow s_B$} \\ \protno{output
    $s_A$}{output $s_B$}
\end{protocol}
\caption{Quantum key distillation protocol $\eQKD_{\PE, \Bl,
    \CKA}$.} \label{pr:eQKD}
\end{protocolfloat}

\newcommand*{\distvN}{\delta'}
\newcommand*{\epss}{\bar{\eps}}

\begin{tabularfloat}
\begin{tabular}{ll}
\\[-2.4ex]
$\PE := \PE_{\cM, \cQ}$ & prot.\ on $\cH_A^{\otimes m} \otimes \cH_B^{\otimes m}$  defined by Fig.~\ref{pr:PE} \\[0.5ex]
$\quad \cM = \{M_w\}_{w \in \cW}$ &  POVM on $\cH_A \otimes \cH_B$ \\
$\quad \cQ$ & set of freq.\ dist.\ on $\cW$ \\[0.5ex]
\hline
\\[-2ex]
$\Bl$ & prot.\ on $\cH_A^{\otimes b} \otimes \cH_B^{\otimes b}$ with cl.\ output in $\cX \times \cY$ \\[0.5ex]
\hline
\\[-2ex]
$\CKA := \CKA_{\IR, \cF}$ & prot.\ on $\cX^n \times \cY^n$ defined by Fig.~\ref{pr:PP}  \\[0.5ex]
$\quad \IR$ & inf.\ rec.\ prot.\ on $\cX^n \times \cY^n$ \\
$\quad \cF$ & two-univ.\ fam.\ of hash func.\ from $\cX^n$ to $\{0,1\}^\ell$ \\[0.5ex]
\end{tabular}
\caption{Subprotocols used for $\eQKD_{\PE, \Bl,
    \CKA}$ (cf. Fig.~\ref{pr:eQKD}).} \label{tab:par}
\end{tabularfloat}

Consider the quantum key distillation protocol $\eQKD_{\PE, \Bl,
  \CKA}$ depicted in Fig.~\ref{pr:eQKD}.  Alice and Bob take inputs
from product spaces $\cH_A^{\otimes N}$ and $\cH_B^{\otimes N}$,
respectively.  Then, they subsequently run the following subprotocols
(see also Table~\ref{tab:par}):
\begin{itemize}
\item \emph{Random permutation of the
    subsystems:}\index{permutation-invariant~protocol} Alice and Bob
  reorder their subsystems according to a commonly chosen random
  permutation $\pi$.
\item \emph{Parameter estimation}\index{parameter~estimation} ($\PE$):
  Alice and Bob sacrifice $m$ subsystems to perform some statistical
  checks. We assume that they do this using a protocol of the form
  $\PE_{\cM, \cQ}$ (see Fig.~\ref{pr:PE}), which is characterized by a
  POVM $\cM = \{M_w\}_{w \in \cW}$ on $\cH_A \otimes \cH_B$ and a set
  $\cQ$ of valid frequency distributions on $\cW$.
\item \emph{Block-wise measurement and
    processing}\index{measurement}\index{block-wise~processing}
  ($\Bl^{\otimes n}$): In order to obtain classical data, Alice and
  Bob apply a measurement to the remaining $b \cdot n$ subsystems,
  possibly followed by some further processing (e.g., advantage
  distillation\index{advantage~distillation}). We assume here that
  Alice and Bob group their $b \cdot n$ subsystems in $n$ blocks of
  size $b$ and then process each of these blocks independently,
  according to some subprotocol, denoted $\Bl$.  Each application of
  $\Bl$ to a block $\cH_A^{\otimes b} \otimes \cH_B^{\otimes b}$
  results in a pair of classical outputs $x_i$ and $y_i$.
\item \emph{Classical
    post-processing}\index{classical~post-processing} ($\CKA$): Alice
  and Bob transform their classical strings $(x_1, \ldots, x_n)$ and
  $(y_1, \ldots, y_n)$ into a pair of secret keys.  For this, they
  invoke a post-processing subprotocol of the form $\CKA_{\IR, \cF}$
  (see Fig.~\ref{pr:PP}), for some (arbitrary) information
  reconciliation scheme $\IR$ and a two-universal
  family\index{two-universal~hashing} of hash functions~$\cF$ for
  privacy amplification\index{privacy~amplification}.
\end{itemize}

\subsection{Robustness}

\index{robustness|(}

The usefulness of a key distillation protocol depends on the set of
inputs for which it is robust, i.e., from which it can successfully
distill secret keys. Obviously, the described protocol $\eQKD_{\PE,
  \Bl, \CKA}$ is robust on all inputs for which none of its
subprotocols $\PE$, $\Bl$, or $\CKA$ aborts. Note that the
post-processing $\CKA = \CKA_{\IR, \cF}$ only aborts if the underlying
information reconciliation scheme $\IR$ aborts.

Typically, the subprotocols $\Bl$ and $\IR$ are chosen in such a way
that they are robust on any of the input states accepted by $\PE$. In
this case, the key distillation protocol $\eQKD_{\PE, \Bl, \CKA}$ is
successful whenever it starts with an input for which $\PE$ is robust.
According to the discussion in Section~\ref{sec:PE}, the protocol $\PE
= \PE_{\cM, \cQ}$ is robust on product states $\sigma_{A B}^{\otimes
  m}$ if $\sigma_{A B}$ is contained in the set
$\overline{\Gamma}^{\geq\distPE}_{\cM, \cQ}$ defined
by~\eqref{eq:Gammabardef}.  Consequently, $\eQKD_{\PE, \Bl, \CKA}$ is
robust on all inputs of the form $\sigma_{A B}^{\otimes N}$, for
$\sigma_{A B} \in \overline{\Gamma}^{\geq\distPE}_{\cM, \cQ}$.

\index{robustness|)}

\subsection{Security}

The following is a generic criterion for the security of QKD.

\index{security~criterion} \index{security~proof|(}

\begin{theorem} \label{thm:main}
  Let $\eQKD_{\PE, \Bl, \CKA}$ be the quantum key distillation
  protocol defined by Fig.~\ref{pr:eQKD} and Table~\ref{tab:par}, let
  $\eps, \eps' \geq 0$, let $\delta$, $\distPE$ be defined by
  Table~\ref{tab:sec}, and let $\Gamma_{\cM, \cQ}^{\leq \distPE}$ be
  defined by~\eqref{eq:Gammadef}.  Then $\eQKD_{\PE, \Bl, \CKA}$ is
  $(\eps + \eps')$-fully secure if the underlying information
  reconciliation protocol $\IR$ is $\eps'$-fully secure and if
  \[
    \ell 
  \leq
      n \min_{\sigma_{A B} \in \Gamma_{\cM, \cQ}^{\leq \distPE}}            
            H(X|\Eb)
    - \leak_{\IR}
    - n \delta \ ,
  \]
  where the entropy in the minimum is evaluated on 
  \[
    \sigma_{X Y \Eb} 
  =
    \cE^{\Bl}_{X Y \Eb \leftarrow A^b B^b E^b}
      (\sigma_{A B E}^{\otimes b}) \ ,
  \]
  for a purification $\sigma_{A B E}$ of $\sigma_{A B}$.
\end{theorem}

\begin{tabularfloat}
\begin{tabular}{ll}
\\[-2.4ex]
  $N$ & $b n + m + k$ \\
  $r$ & $\sfrac{N}{k} \bigl(2 \log(\slfrac{9}{\eps}) + \dim(\cH_A \otimes
  \cH_B)^2 \ln k \bigr)$ \\
 $\distvN$ & $(\frac{5}{2} \log |\cX| + 4) \sqrt{h(\slfrac{r}{n}) + \sfrac{2}{n}
  \log(\slfrac{18}{\eps})}$ \\
  $\distPE$ & $2 \sqrt{ h(\slfrac{r}{m}) +  \frac{1}{m} \bigl(\log(\slfrac{9}{2 \eps}) + |\cW| \log(\sfrac{m}{2} + 1) \bigr)}$ \\ 
  $\delta$ & $\distvN + \sfrac{2(m+k)}{n} \log \dim(\cH_A \otimes \cH_B) +\sfrac{2}{n} \log(\slfrac{3}{2\eps})$ 
\end{tabular}
\caption{Security parameters for $\eQKD_{\PE, \Bl,
    \CKA}$ (cf. Fig.~\ref{pr:eQKD}).} \label{tab:sec}
\end{tabularfloat} 

\begin{proof}  
  Let $\rho_{A^N B^N}$ be any state held by Alice and Bob after they
  have applied the random permutation $\pi$ (averaged over all
  possible choices of $\pi$). Because, obviously, $\rho_{A^N B^N}$ is
  permutation-invariant, Lemma~\ref{lem:sympurification} implies that
  there exists a purification\index{symmetric~purification} $\rho_{A^N
    B^N E^N}$ of $\rho_{A^N B^N}$ on the symmetric subspace of $(\cH_A
  \otimes \cH_B \otimes \cH_E)^{\otimes N}$. We show that the
  remaining part of the protocol is secure on $\rho_{A^N B^N E^N}$.
  This is sufficient because any density operator $\rho_{A^N B^N
    \tilde{E}}$ which has the property that taking the partial trace
  over $\cH_{\tilde{E}}$ gives $\rho_{A^N B^N}$ can be obtained from
  the pure state $\rho_{A^N B^N E^N}$ by a trace-preserving CPM which
  only acts on Eve's space.
  
  Let $\rho_{A^{b n + m} B^{b n + m} E^{b n + m}}$ be the operator
  obtained by taking the partial trace (over $k$ subsystems $\cH_A
  \otimes \cH_B \otimes \cH_E$) of $\rho_{A^N B^N E^N}$. It describes
  the joint state on the $m$ subsystems used for parameter estimation
  and the $b \cdot n$ subsystems which are given as input to
  $\Bl^{\otimes n}$.  According to the de Finetti
  representation\index{de~Finetti~representation} theorem
  (Theorem~\ref{thm:symmix}) this density operator is approximated by
  a convex combination of density operators, where each of them is on
  the symmetric subspace along vectors $\ket{\theta} \in \cH_A \otimes
  \cH_B \otimes \cH_E$.  More precisely, with $\epss :=
  \sfrac{2\eps}{9}$,
  \begin{equation} \label{eq:orgdist}
    \Bigl\|
      \rho_{A^{b n + m} B^{b n + m} E^{b n + m}}
      - \int_{\cS_1} 
          \rho^{\ket{\theta}}_{A^{b n + m} B^{b n + m} E^{b n + m}}
        \nu(\ket{\theta})
    \Bigr\|_1
  \leq 
    \epss \ ,
  \end{equation}
  where the integral runs over the set $\cS_1 := \cS_1(\cH_A \otimes
  \cH_B \otimes \cH_E)$ of normalized vectors in $\cH_A \otimes \cH_B
  \otimes \cH_E$ and where, for any $\ket{\theta} \in \cH_A \otimes
  \cH_B \otimes \cH_E$,
  \begin{equation} \label{eq:orgsym}
    \rho^{\ket{\theta}}_{A^{b n + m} B^{b n + m} E^{b n + m}}
  \in
    \NN\bigl(
      \SymR{(\cH_A \otimes \cH_B \otimes \cH_E)}{b n + m}{\theta}{b n
  +  m - r}
    \bigr) \ .
  \end{equation}
  
  We first analyze the situation after the parameter estimation is
  completed. Let $\cE^{\PE}_{A^m B^m}$ be the CPM which maps all
  density operators on $(\cH_A \otimes \cH_B)^{\otimes m}$ either to
  the scalar $0$ or $1$, depending on whether the parameter estimation
  protocol $\PE_{\cM, \cQ}$ accepts or aborts. Moreover, define
  \begin{align*}
    \rho^{\PE}_{A^{b n} B^{b n} E^N}
  & :=
    (\id_{A^{b n} B^{b n}} \otimes \cE^{\PE}_{A^m B^m} \otimes \id_{E^N}) 
      (\rho_{A^{b n + m} B^{b n + m} E^N}) 
  \\
    \rho^{\ket{\theta}, \PE}_{A^{b n} B^{b n} E^{b n}}
  & :=
    (\id_{A^{b n} B^{b n}} \otimes \cE^{\PE}_{A^m B^m} \otimes \id_{E^{ b n}})
       (\rho^{\ket{\theta}}_{A^{b n + m} B^{b n + m} E^{b n }}) \ .
  \end{align*}   
  Because of~\eqref{eq:orgsym}, we have
  \begin{equation} \label{eq:PEsymop}
    \rho^{\ket{\theta},\PE}_{A^{b n} B^{b n} E^{b n}}
  \in
    \NN\bigl(
      \SymR{(\cH_A \otimes \cH_B \otimes \cH_E)}{b n}{\theta}{b n-r}
    \bigr) \ ,
  \end{equation}
  for any $\ket{\theta} \in \cH_A \otimes \cH_B \otimes \cH_E$.
  Moreover, from~\eqref{eq:orgdist} and the fact that the \distance{}
  cannot increase when applying a quantum operation
  (Lemma~\ref{lem:distdecr}) we have
  \begin{equation} \label{eq:symbound}
    \Bigl\|
      \rho^{\PE}_{A^{b n} B^{b n} E^{b n}}
      - \int_{\cS_1} 
          \rho^{\ket{\theta}, \PE}_{A^{b n} B^{b n} E^{b n}}
        \nu(\ket{\theta})
    \Bigr\|_1
  \leq 
    \epss \ .
  \end{equation}   
  According to Lemma~\ref{lem:PEsec}, the parameter
  estimation\index{parameter~estimation} $\PE_{\cM, \cQ}$
  $\epss$-securely filters all states $\rho^{\ket{\theta}, \PE}_{A^{b
      n} B^{b n} E^{b n}}$ for which $\ket{\theta}$ is not contained
  in the set
  \[
    \cV^{\distPE} 
  := 
    \bigl\{ 
      \ket{\theta} \in \cS_1 : \, 
        \tr_E(\proj{\theta}) \in  \Gamma_{\cM, \cQ}^{\leq \distPE}
    \bigr\} \ .
  \]
  We can thus restrict the integral in~\eqref{eq:symbound} to the set
  $\cV^{\distPE}$, thereby only losing terms with total weight at most
  $\epss$, i.e.,
  \begin{equation} \label{eq:rhoPEdist}
    \Bigl\|
      \rho^{\PE}_{A^{b n} B^{b n}E^{b n}}
    -
      \int_{\cV^{\distPE}} \rho^{\ket{\theta}, \PE}_{A^{b n} B^{b n} E^{b n}}
        \nu(\ket{\theta})
    \Bigr\|_1
  \leq
    2 \epss \ .
  \end{equation}
  
  To describe the situation after the measurement and blockwise
  processing $\Bl^{\otimes n}$, we define
  \begin{align*}
    \rho_{X^n Y^n \Eb^n E^{m + k}}
  & :=
    \bigl((\cE_{X Y \Eb \leftarrow A^b B^b E^b}^{\Bl})^{\otimes n} \otimes \id_{E^{m + k}}\bigr)
      (\rho^{\PE}_{A^{b n} B^{b n } E^{b n + m + k}})
\\
    \rho^{\ket{\theta}}_{X^n Y^n \Eb^{n}}
  & :=
    (\cE_{X Y \Eb \leftarrow A^b B^b E^b}^{\Bl})^{\otimes n}
      (\rho^{\ket{\theta}, \PE}_{A^{b n} B^{b n} E^{b n}}) \ .
  \end{align*}
  Using once again the fact that the \distance{} cannot decrease under
  quantum operations (Lemma~\ref{lem:distdecr}), we conclude
  from~\eqref{eq:rhoPEdist} that
  \begin{equation} \label{eq:rhoXYdist}
    \Bigl\|
      \rho_{X^n Y^n \Eb^n}
    -
      \int_{\cV^{\distPE}} \rho^{\ket{\theta}}_{X^n Y^n \Eb^{n}}
        \nu(\ket{\theta})
    \Bigr\|_1
  \leq
    2 \epss \ .   
  \end{equation}
  
  According to~\eqref{eq:PEsymop}, the density operator
  $\rho^{\ket{\theta}, \PE}_{A^{b n} B^{b n} E^{b n}}$ lies in the
  symmetric subspace of the $(b \cdot n)$-fold product space $(\cH_A
  \otimes \cH_B \otimes \cH_E)^{\otimes b n}$ along $\ket{\theta}^{b n
    -r}$, i.e., it has product form except on $r$
  subsystems.\index{product~state} Equivalently, we can view
  $\rho^{\ket{\theta}, \PE}_{A^{b n} B^{b n} E^{b n}}$ as a density
  operator on the $n$-fold product of subsystems $\cH_{A^b B^b E^b} :=
  \cH_A^{\otimes b} \otimes \cH_B^{\otimes b} \otimes \cH_E^{\otimes
    b}$. It then has product form an all but (at most) $r$ of these
  subsystems. That is, $\rho^{\ket{\theta}, \PE}_{A^{b n} B^{b n} E^{b
      n}}$ is contained in the symmetric subspace of $\cH_{A^b B^b
    E^b}^{\otimes n}$ along $\ket{\theta^b}^{\otimes n-r}$, where
  $\ket{\theta^b}:=\ket{\theta}^{\otimes b} \in \cH_{A^b B^b E^b}$.
  This allows us to apply Theorem~\ref{thm:Renyisym} in order to bound
  the entropy of the symmetric states $\rho^{\ket{\theta}}_{X^n Y^n
    \Eb^{n}}$. With the definition
  \[
    \sigma_{X Y \Eb}^{\ket{\theta}} 
  := 
    \cE_{X Y \Eb \leftarrow A^b B^b E^b}^{\Bl}(\sigma_{A B E}^{\otimes
  b}) \ ,
  \]
  where $\sigma_{A B E} := \proj{\theta}$, we
  obtain\index{smooth~min-entropy}
  \[
    \Hmin^{\epss}(\rho^{\ket{\theta}}_{X^n \Eb^n }|\Eb^n )
  \geq
    n\bigl( H(\sigma^{\ket{\theta}}_{X \Eb}) -
  H(\sigma^{\ket{\theta}}_{\Eb}) - \distvN \bigr) \ .
  \]
  Consequently, using~\eqref{eq:rhoXYdist} together with the
  inequalities~\eqref{eq:fsubadd} and~\eqref{eq:fclasscond} of
  Theorem~\ref{thm:Hmincalc},
  \[
    \Hmin^{3 \epss}(\rho_{X^n \Eb^n}|\Eb^n) 
  \geq
    n  \min_{\ket{\theta} \in \cV^{\distPE}} \bigl( H(\sigma_{X
  \Eb}^{\ket{\theta}}) - H(\sigma_{\Eb}^{\ket{\theta}}) - \distvN \bigr)
    \ .
  \]
  Moreover, by the chain rule for smooth min-entropy
  (cf.~\eqref{eq:fchain} of Theorem~\ref{thm:Hmincalc})
  \begin{multline*}
    \Hmin^{3 \epss}(\rho_{X^n \Eb^n E^{m + k}}|\Eb^n E^{m + k}) \\
  \geq 
    n  \min_{\ket{\theta} \in \cV^{\distPE}} \bigl( H(\sigma_{X
  \Eb}^{\ket{\theta}}) - H(\sigma_{\Eb}^{\ket{\theta}})  - \distvN \bigr)
    - 2 \Hmax(\rho_{E^{m+k}}) \ .
  \end{multline*}
  
  Finally, we use Lemma~\ref{lem:ppsec} which provides a criterion on
  the maximum length $\ell$ such that the secret key computed by the
  post-processing subprotocol $\CKA$ is $(\eps+\eps')$-secure,
  \[
    \ell
  \leq
    n  
      \min_{\ket{\theta} \in \cV^{\distPE}} 
        \bigl( H(\sigma_{X \Eb}^{\ket{\theta}}) - H(\sigma_{\Eb}^{\ket{\theta}}) - \distvN
    \bigr)
    - 2 \Hmax(\rho_{E^{m+k}}) 
    - \leak_{\IR} - 2 \log(\slfrac{3}{2\eps}) \ .
   \]
   The assertion then follows from 
   \[
     \Hmax(\rho_{E^{m+k}}) \leq (m+k) \log \dim(\cH_A \otimes \cH_B) \ ,
   \]
   the fact that $\ket{\theta} \in \cV^{\distPE}$ if and only if the
   trace $\sigma_{A B}$ of $\sigma_{A B E} := \proj{\theta}$ is
   contained in the set $\Gamma_{\cM, \cQ}^{\leq \distPE}$, and the
   definition of $\delta$ (cf.\ Table~\ref{tab:sec}).
\end{proof}

Note that the protocol $\eQKD_{\PE, \Bl, \CKA}$ takes as input $N$
subsystems and generates a key of a certain fixed length $\ell$.  In
order to make asymptotic statements, we need to consider a family
$\{\eQKD_{\PE, \Bl, \CKA}^{N}\}_{N \in \bbN}$ of such protocols,
where, for any $N \in \bbN$, the corresponding protocol takes $N$
input systems and generates a key of length $\ell(N)$. The \emph{rate}
of the protocol family is then defined
by\index{secret~key~rate@secret-key~rate}
\[
  \rate := \lim_{N \to \infty} \frac{\ell(N)}{N} \ .
\]

\begin{corollary} \label{cor:main}
  Let $\delta, \distPE > 0$, a protocol $\Bl$ acting on blocks of
  length $b$, a POVM $\cM = \{M_w\}_{w \in \cW}$, and a set $\cQ$ of
  probability distributions on $\cW$ be fixed, and let $\Gamma_{\cM,
    \cQ}^{\leq \distPE}$ be the set defined by~\eqref{eq:Gammadef}.
  Then there exist $\gamma>0$ and parameters $n = n(N), m = m(N), \ell
  = \ell(N)$ such that the class of protocols $\eQKD_{\PE, \Bl,
    \CKA}^N$ (parameterized by $N \in \bbN$) defined by
  Fig.~\ref{pr:eQKD} and Table~\ref{tab:par} has rate
  \[
    \rate
  =
    \frac{1}{b} 
      \min_{\sigma_{A B} \in \Gamma_{\cM, \cQ}^{\leq \distPE}} 
            H(X|\Eb) - H(X|Y) - \delta \ ,
  \]
  where the entropies in the minimum are evaluated on
  \[
    \sigma_{X Y \Eb} 
  =
    \cE^{\Bl}_{X Y \Eb \leftarrow A^b B^b E^b}
      (\sigma_{A B E}^{\otimes b}) \ ,
  \]
  for a purification $\sigma_{A B E}$ of $\sigma_{A B}$. Moreover, for
  any $N \geq 0$, the protocol $\eQKD_{\PE, \Bl, \CKA}^N$ is
  $e^{-\gamma N}$-fully secure.
\end{corollary}

\begin{proof}
  The statement follows directly from Theorem~\ref{thm:main} combined
  with Corollary~\ref{cor:errcorr}.
\end{proof}

\index{quantum~key~distillation|)}

\index{security~proof|)}

\section{Quantum key distribution} \label{sec:qkeydist}

\index{quantum~key~distribution|(}

As described in Section~\ref{sec:QKD}, one can think of a quantum key
\emph{distribution}\index{quantum~key~distribution} (QKD) protocol as
a two-step process where Alice and Bob first use the quantum channel
to distribute entanglement and then apply a quantum key
\emph{distillation} scheme to generate the final key pair.  To prove
security of a QKD protocol, it thus suffices to verify that the
underlying key distillation protocol is secure on any input.  Hence,
the security results for key distillation protocols derived in the
previous section (Theorem~\ref{thm:main} and Corollary~\ref{cor:main})
directly apply to QKD protocols.

We can, however, further improve these results by taking into account
that the way Alice and Bob use the quantum channel in the first step
imposes some additional restrictions on the possible inputs to the
distillation protocol. For example, if Alice locally prepares
entangled states and then sends parts of them to Bob (note that this
is actually the case for most QKD protocols, viewed as
entanglement-based schemes), it is impossible for the adversary to
tamper with the part belonging to Alice. Formally, this means that the
partial state on Alice's subsystem is independent of Eve's attack.

Using this observation, we can restrict the set $\Gamma_{\cM,
  \cQ}^{\leq \distPE}$ of states $\sigma_{A B}$ (as defined
by~\eqref{eq:Gammadef}) over which the minimum is taken in the
criterion of Theorem~\ref{thm:main} and Corollary~\ref{cor:main}.  In
fact, it follows directly from Remark~\ref{rem:addcond} that it
suffices to consider states $\sigma_{A B}$ such that $\sigma_A =
\tr_B(\sigma_{A B})$ is fixed.

\index{quantum~key~distribution|)}

\chapter{Examples} \label{ch:examples}

To illustrate the general results of the previous chapter, we analyze
certain concrete QKD protocols. We first specialize the formula for
the rate\index{secret~key~rate@secret-key~rate} (cf.\ 
Corollary~\ref{cor:main}) to protocols based on \emph{two-level}
quantum systems (Section~\ref{sec:twolevel}).  Then, as an example, we
analyze different variants of the six-state protocol and compute
explicit values for their rates (Section~\ref{sec:sixstate}).

\section{Protocols based on two-level systems} \label{sec:twolevel}

A large class of QKD protocols, including the well-known BB84
protocol\index{BB84~protocol} or the six-state
protocol\index{six-state~protocol}, are based on an encoding of binary
classical values using the state of a \emph{two-level} quantum system,
such as the the spin of a photon.  For the corresponding key
distillation protocol (see Fig.~\ref{pr:eQKD}), this means that Alice
and Bob take inputs from (products of) two-dimensional Hilbert spaces
on which they apply binary measurements. In the following, we analyze
different variants of such protocols.

\subsection{One-way protocols} \label{sec:oneway}

\index{secret~key~rate@secret-key~rate|(} \index{one-way~protocol|(}

We start with a basic key distillation protocol which only uses
information reconciliation and privacy amplification (as described in
Section~\ref{sec:PP}) to transform the raw key pair into a pair of
secret keys. More precisely, after the measurement of their
subsystems, Alice and Bob immediately invoke an information
reconciliation protocol (e.g., the protocol~$\IRit_{\hat{\cX}, \cF}$
depicted in Fig.~\ref{pr:IRit}) such that Bob can compute a guess of
Alice's values; the final key is then obtained by two-universal
hashing.  Because this post-processing only requires communication
from Alice to Bob, such protocols are also called \emph{one-way key
  distillation protocols}\index{one-way~protocol}.\footnote{Note,
  however, that bidirectional communication is always needed for the
  parameter estimation step.}

Clearly, the one-way key distillation protocol described above is a
special case of the general protocol $\eQKD_{\PE, \Bl, \CKA}$ depicted
in Fig~\ref{pr:eQKD}, where $\Bl := \Meas$ is the subprotocol
describing the measurement operation of Alice and Bob. Additionally,
assume that the parameter estimation subprotocol $\PE$ is the protocol
$\PE_{\cM, \cQ}$ depicted in Fig.~\ref{pr:PE}, where $\cM$ is a POVM
and $\cQ$ is the set of statistics for which the protocol does not
abort. We can then use Corollary~\ref{cor:main} to compute the rate of
the protocol, that is,\index{secret~key~rate@secret-key~rate}
\begin{equation} \label{eq:ratesimple}
  \rate
= 
  \min_{\sigma_{A B} \in \Gamma} 
    H(X|E) - H(X|Y) \ .
\end{equation}
Here, the minimum ranges over the set
\begin{equation} \label{eq:Gammasdef}
  \Gamma 
:= 
  \bigl\{ \sigma_{A B} : \, P_W^{\sigma_{A B}} \in \cQ \bigr\} 
\end{equation}
of all density operators $\sigma_{A B}$ on the $2 \times
2$-dimensional Hilbert space $\cH_A \otimes \cH_B$ such that the
measurement with respect to $\cM$ gives a probability distribution
$P_W^{\sigma_{A B}}$ which is contained in the set $\cQ$.  Moreover,
the von Neumann (or Shannon) entropies $H(X|E)$ and $H(X|Y)$ are
evaluated for the operators
\[
  \sigma_{X Y E} := (\cE^{\Meas}_{X Y \leftarrow A B} \otimes
  \id_E)(\sigma_{A B E}) \ ,
\]
where $\sigma_{A B E}$ is a purification of $\sigma_{A B}$.

Let $\{\ket{0}_A, \ket{1}_A\}$ and $\{\ket{0}_B, \ket{1}_B\}$ be the
bases that Alice and Bob use for the measurement
$\Meas$.\footnote{$\Meas$ describes the measurement that generates the
  data used for the computation of the final key. It might be
  different from the measurement $\cM$ which is used for parameter
  estimation. } Lemma~\ref{lem:Hdiffnonoise} below provides an
explicit lower bound on the entropy difference on the right hand side
of~\eqref{eq:ratesimple} as a function of $\sigma_{A B E}$. The bound
only depends on the diagonal values of $\sigma_{A B}$ with respect to
the \emph{Bell basis}\index{Bell~state}, which is defined by the
vectors
\begin{align*}
  \ket{\Phi_0} 
& := 
  \sfrac{1}{\sqrt{2}} \ket{0,0} + \sfrac{1}{\sqrt{2}} \ket{1,1} 
\\
  \ket{\Phi_1} 
& := 
  \sfrac{1}{\sqrt{2}} \ket{0,0} - \sfrac{1}{\sqrt{2}} \ket{1,1}
\\
  \ket{\Phi_2} 
& := 
  \sfrac{1}{\sqrt{2}} \ket{0,1} + \sfrac{1}{\sqrt{2}} \ket{1,0}
\\
  \ket{\Phi_3} 
& := 
  \sfrac{1}{\sqrt{2}} \ket{0,1} - \sfrac{1}{\sqrt{2}} \ket{1,0} \ ,
\end{align*}
where $\ket{x,y} := \ket{x}_A \otimes \ket{y}_B$.

\begin{lemma} \label{lem:Hdiffnonoise}
  Let both $\cH_A$ and $\cH_B$ be two-dimensional Hilbert spaces, let
  $\sigma_{A B E} \in \NN(\cH_A \otimes \cH_B \otimes \cH_E)$ be a
  density operator, and let $\sigma_{X Y E}$ be obtained from
  $\sigma_{A B E}$ by applying orthonormal measurements on $\cH_A$ and
  $\cH_B$.  Then
  \begin{multline*}
    H(X|E) - H(X|Y) \\
  \geq
    1- (\lambda_0 + \lambda_1)h\bigl(\sfrac{\lambda_0}{\lambda_0 +
  \lambda_1}\bigr) - (\lambda_2 + \lambda_3) h\bigl(\sfrac{\lambda_2}{\lambda_2
  + \lambda_3}\bigr)
    - h(\lambda_0 + \lambda_1) \ ,
  \end{multline*}
  where $\lambda_i := \bra{\Phi_i} \sigma_{A B} \ket{\Phi_i}$ are the
  diagonal values of $\sigma_{A B}$ with respect to the Bell basis
  (defined relative to the measurement basis).
\end{lemma}

\begin{proof}
  Let $\cD$ be the CPM defined by
  \[
    \cD(\sigma_{A B})
  :=
    \sfrac{1}{4} \hspace{-1em}
      \sum_{\tau \in \{\id, \sigma_x, \sigma_y, \sigma_z \}} 
      \hspace{-1em}  \tau^{\otimes 2} \, \sigma_{A B} \, \tau^{\otimes 2} \ ,
  \]
  where $\sigma_x, \sigma_y, \sigma_z$ are the Pauli operators 
  \begin{equation}
  \begin{matrix}
   \sigma_x
  :=
    \left( \begin{matrix} 0 & 1 \\ 1 & 0 \end{matrix} \right) \,
  & \,
   \sigma_y
  :=
    \left( \begin{matrix} 0 & -i \\ i & 0 \end{matrix} \right) \,
  & \,
   \sigma_z
  :=
    \left( \begin{matrix} 1 & 0 \\ 0 & -1 \end{matrix} \right)  \ ,
  \end{matrix}
  \end{equation}
  and let $\sigmat_{A B E}$ be a purification of $\sigmat_{A B} :=
  \cD(\sigma_{A B})$.  Moreover, let $\sigmat_{A B E}$ be an arbitrary
  purification of $\sigmat_{A B}$ with auxiliary system $\cH_E$ and
  define
  \[
    \sigmat_{X Y E} 
  := 
    (\cE^{\Meas}_{X Y \leftarrow A B} \otimes \id_E)(\sigmat_{A B E}) \ .
  \]
  
  A straightforward calculation shows that the operator $\sigmat_{A
  B}$ has the form
  \[
    \sigmat_{A B} = \sum_{i=0}^3 \lambda_i \proj{\Phi_i} \ ,
  \]
  i.e., it is diagonal with respect to the Bell basis.  Moreover,
  because $\cD$ commutes with the measurement operation on $\cH_A
  \otimes \cH_B$, it is easy to verify that the entropy $H(X|Y)$
  evaluated for $\sigma_{X Y}$ is upper bounded by the corresponding
  entropy for $\sigmat_{X Y}$.  Similarly, because $\sigmat_{A B E}$
  is a purification of $\sigmat_{A B}$, the entropy $H(X|E)$ evaluated
  for $\sigma_{X E}$ is lower bounded by the entropy of $\sigmat_{X
    E}$.  It thus suffices to show that the inequality of the lemma
  holds for the operator $\sigmat_{X Y E}$, which is obtained from the
  diagonal operator $\sigmat_{A B}$.
    
  Let $\ket{e_i}_{i}$ be an orthonormal basis of a $4$-dimensional
  Hilbert space $\cH_E$. Then the operator $\sigmat_{A B E} :=
  \proj{\Psi} \in \NN(\cH_A \otimes \cH_B \otimes \cH_E)$ defined by
  \[
    \ket{\Psi}
  :=
    \sum_{i} \sqrt{\lambda_i} \ket{\Phi_i}_{A B} \otimes \ket{e_i}_E
  \]
  is a purification of $\sigmat_{A B}$. With the definition
  \begin{align*}
    \ket{f_{0,0}} 
  & := 
    \sqrt{\sfrac{\lambda_0}{2}} \ket{e_0} 
    + \sqrt{\sfrac{\lambda_1}{2}} \ket{e_1}
  \\
    \ket{f_{1,1}} 
  & := 
    \sqrt{\sfrac{\lambda_0}{2}} \ket{e_0} 
    - \sqrt{\sfrac{\lambda_1}{2}} \ket{e_1}
  \\
    \ket{f_{0,1}} 
  & := 
    \sqrt{\sfrac{\lambda_2}{2}} \ket{e_2} 
    + \sqrt{\sfrac{\lambda_3}{2}} \ket{e_3}
  \\
    \ket{f_{1,0}} 
  & :=  
    \sqrt{\sfrac{\lambda_2}{2}} \ket{e_2} 
    - \sqrt{\sfrac{\lambda_3}{2}} \ket{e_3} \ ,
  \end{align*}
  the state $\ket{\Psi}$ can be rewritten as
  \[
    \ket{\Psi}
  =
    \sum_{x, y} \ket{x,y} \otimes \ket{f_{x,y}} \ .
  \]
  Because the operator $\sigmat_{X Y E}$ is obtained from $\sigmat_{A
    B E}$ by orthonormal measurements on $\cH_A$ and $\cH_B$, we
  conclude
  \[
    \sigmat_{X Y E}
  =
    \sum_{x,y} \proj{x} \otimes \proj{y} \otimes \sigmat_{E}^{x,y}
  \]
  where $\sigmat_{E}^{x,y} := \proj{f_{x,y}}$.   
  
  Using this representation of the operator $\sigmat_{X Y E}$, it is
  is easy to see that
  \begin{align*}
    H(\sigmat_{X E}) 
  & = 
    1 + h(\lambda_0 + \lambda_1)
  \\
    H(\sigmat_{E}) 
  & = 
    h(\lambda_0 + \lambda_1)
    + (\lambda_0 + \lambda_1)
      h\bigl(\sfrac{\lambda_0}{\lambda_0 + \lambda_1} \bigr)
    + (\lambda_2 + \lambda_3)
      h\bigl(\sfrac{\lambda_2}{\lambda_2 + \lambda_3} \bigr)     
  \\
    H(X|Y)
  & =
    h(\lambda_0 + \lambda_1) \ ,
  \end{align*}
  from which the assertion follows.  
\end{proof}

\newcommand*{\diag}{\mathrm{diag}}

Using Lemma~\ref{lem:Hdiffnonoise}, we conclude that the above
described one-way protocol can generate secret-key bits at rate
\begin{multline} \label{eq:rateow}
  \rate 
\geq
  \min_{(\lambda_0, \ldots, \lambda_3) \in \diag(\Gamma)}
  1- (\lambda_0 + \lambda_1)h\bigl(\sfrac{\lambda_0}{\lambda_0 +
  \lambda_1}\bigr) \\ - (\lambda_2 + \lambda_3) h\bigl(\sfrac{\lambda_2}{\lambda_2
  + \lambda_3}\bigr)
    - h(\lambda_0 + \lambda_1) \ ,
\end{multline}
where $\diag(\Gamma)$ denotes the $4$-tuples of diagonal entries
(relative to the Bell basis) of the operators $\sigma_{A B} \in
\Gamma$, for $\Gamma$ defined by~\eqref{eq:Gammasdef}.

\subsection{One-way protocols with noisy preprocessing} \label{sec:preproc}

\index{noisy~preprocessing|(}

The efficiency of the basic QKD protocol described in
Section~\ref{sec:oneway} can be increased in different ways. We
consider an extension of the protocol where, before starting with
information reconciliation, Alice applies some local
\emph{preprocessing}\index{preprocessing} operation to her raw key. A
very simple|but surprisingly useful|variant of preprocessing is to add
noise, i.e., Alice flips each of her bits independently with some
probability~$q$. In the following, we call this~\emph{noisy
  preprocessing}\index{noisy~preprocessing}.

To compute the rate of the one-way protocol enhanced with this type of
preprocessing, we need a generalization of
Lemma~\ref{lem:Hdiffnonoise}.

\begin{lemma} \label{lem:twonoise}
  Let both $\cH_A$ and $\cH_B$ be two-dimensional Hilbert spaces, let
  $\sigma_{A B E} \in \NN(\cH_A \otimes \cH_B \otimes \cH_E)$ be a
  density operator, and let $\sigma_{X Y E}$ be obtained from
  $\sigma_{A B E}$ by applying orthonormal measurements on $\cH_A$ and
  $\cH_B$ where, additionally, the outcome of the measurement on
  $\cH_A$ is flipped with probability $q \in [0,1]$.  Then
  \begin{multline*}
    H(X|E) - H(X|Y) \\
  \geq  
  1 
  - (\lambda_0 + \lambda_1) \bigl( 
      h(\alpha)
      - \hb(\alpha, q)       
    \bigr)
  - (\lambda_2 + \lambda_3) \bigl(
      h(\beta)
      - \hb(\beta, q) 
    \bigr) \\
  - h\bigl((\lambda_0 + \lambda_1) q + (\lambda_2 + \lambda_3) (1-q)
  \bigr) \ ,
  \end{multline*}
  where $\lambda_i := \bra{\Phi_i} \sigma_{A B} \ket{\Phi_i}$, $\alpha
  := \sfrac{\lambda_0}{\lambda_0 + \lambda_1}$, $\beta :=
  \sfrac{\lambda_2}{\lambda_2 + \lambda_3}$, and
 \[
   \hb(p, q)
 :=
   h\bigl(\sfrac{1}{2} \pm \sfrac{1}{2} 
     \sqrt{ 1- 16 p (1-p) q (1-q)}
   \bigr) \ .
 \]
\end{lemma}

\begin{proof}
  The statement follows by a straightforward extension of the proof of
  Lemma~\ref{lem:Hdiffnonoise}.
\end{proof}

Similarly to formula~\eqref{eq:rateow}, the rate of the one-way
protocol with noisy preprocessing|where Alice additionally flips her
bits with probability $q$|is given by the expression provided by
Lemma~\ref{lem:twonoise}, minimized over all $4$-tuples $(\lambda_0,
\ldots, \lambda_3) \in \diag(\Gamma)$.  It turns out that this rate is
generally larger than the rate of the corresponding one-way protocol
without preprocessing (see Section~\ref{sec:sixstate} below).

\index{noisy~preprocessing|)}
\index{one-way~protocol|)}

\newcommand{\deltab}{\eps}

\newcommand*{\AD}{\mathsf{AD}}
\newcommand*{\advacc}{\mathrm{acc}}

\subsection{Protocols with advantage distillation} \label{sec:AD}

\index{advantage~distillation|(}

To further increase the efficiency of the key distillation protocol
described above, one might insert an additional \emph{advantage
  distillation}\index{advantage~distillation} step after the
measurement $\Meas$, i.e., before the classical one-way
post-processing.\footnote{The concept of advantage distillation has
  first been introduced in a purely classical context~\cite{Maurer93},
  where a secret key is generated from some predistributed correlated
  data.}  Its purpose is to identify subsets of highly correlated bit
pairs such as to separate these from only weakly correlated
information.

\begin{protocolfloat} 
\figtop
Parameters: \\[1ex]
\begin{tabular}{ll}
  $b$: & block length
\end{tabular}
\tabprotsep
\begin{protocol}{Alice}{Bob} 
  \protno{input: $(x_1, \ldots, x_b)$}{input: $(y_1, \ldots, y_b)$} \\
  \protno{$r \in_R \{0,1\}$}{}
  \protright{$(c_1, \ldots, c_b) :=$ \\ $\quad (x_1 \oplus r, \ldots, x_b \oplus r)$}{$(c_1, \ldots, c_b)$}{if $ (y_1 \oplus c_1, \ldots, y_b \oplus
  c_b)$ \\ $\qquad \in \{\mathbf{0}, \mathbf{1}\}$ \\ $\quad$ then $\advacc := \mathrm{true}$}
 \protleft{}{$\advacc$}{} 
 \protno{if $\advacc$ \\ $\quad$ then output $x_1$ \\ $\quad$ else
  output $\Delta$}{if $\advacc$ \\ $\quad$ then output $y_1$ \\
  $\quad$ else output $\Delta$}
\end{protocol}
\caption{
  Advantage distillation protocol $\AD_b$.} \label{pr:AD}
\end{protocolfloat}

\newcommand*{\psucc}{p_{\mathrm{succ}}}

A typical advantage distillation protocol is depicted in
Fig.~\ref{pr:AD}: Alice and Bob split their bitstrings into blocks
$(x_1, \ldots, x_b)$ and $(y_1, \ldots, y_b)$ of size $b$.  Then,
depending on a randomly chosen binary value $r$, Alice announces to
Bob either $(x_1, \ldots, x_b)$ or $(x_1 \oplus 1, \ldots, x_b \oplus
1)$ (where $\oplus$ denotes the bitwise xor). Bob compares this
information with his block $(y_1, \ldots, y_b)$ and accepts if it
either differs in none or in all positions, i.e., if the difference
equals either $\mathbf{0} := (0, \ldots, 0)$ or $\mathbf{1} := (1,
\ldots, 1)$. In this case, Alice and Bob both keep the first bit of
their initial string.  Otherwise, they output some dummy symbol
$\Delta$.\footnote{As suggested in~\cite{Maurer93}, the efficiency of
  this advantage distillation protocol is further increased if Alice
  and Bob, instead of acting on large blocks at once, iteratively
  repeat the described protocol step on very small blocks (consisting
  of only $2$ or $3$ bits).}  Obviously, if the error probability per
bit (i.e., the error rate of the channel) is $e$\index{error~rate}
then the probability $\psucc$ that advantage distillation on a block
of length $b$ is successful (i.e., Alice and Bob keep their bit) is
$\psucc = e^b + (1-e)^b$.

Let us now consider the general protocol $\eQKD_{\PE, \Bl, \CKA}$
where the subprotocol $\Bl$ consists of $b$ binary measurements
$\Meas$ of Alice and Bob followed by the advantage distillation
protocol $\AD_b$ described in Fig.~\ref{pr:AD}, i.e., 
\begin{equation} \label{eq:ADop}
  \cE^{\Bl}_{X Y \Eb \leftarrow A^b B^b E^b} 
= 
  \cE^{\AD}_{X Y \Eb \leftarrow X^b Y^b E^b} 
  \circ (\cE^{\Meas}_{X Y \leftarrow A B} \otimes \id_{E} )^{\otimes b}
\end{equation}
It is easy to see that the subprotocol $\AD_b$ commutes with the
measurement $\Meas$, that is, \eqref{eq:ADop} can be rewritten as
\[
  \cE^{\Bl}_{X Y \Eb \leftarrow A^b B^b E^b} 
= 
  ((\cE^{\Meas}_{X Y \leftarrow A B })^{\otimes b} \otimes \id_{\Eb})
  \circ \cE^{\AD}_{A B \Eb \leftarrow A^b B^b E^b} \ .
\]
Moreover, a straightforward computation\footnote{For this computation,
  it is convenient to use the mapping $\cD$ defined above, which
  allows to restrict the argument to the special case where $\sigma_{A
    B}$ is Bell diagonal.} shows that, if $\sigma_{A B}$ has diagonal
entries $\lambda_0, \ldots, \lambda_3$ with respect to the Bell basis
then, with probability
\[
  \psucc := (\lambda_0 + \lambda_1)^b + (\lambda_2 + \lambda_3)^b \ ,
\]
the advantage distillation $\AD_b$ is successful and the operation
$\cE^{\AD}_{A B \leftarrow A^b B^b}$ induced by $\AD_b$ (conditioned
on the event that it is successful) maps $\sigma_{AB}^{\otimes b}$ to
an operator $\sigmat_{A B} $ with diagonal entries
\begin{align*}
  \lambdat_0 
& = 
  \frac{(\lambda_0+\lambda_1)^b +(\lambda_0-\lambda_1)^b}{2 \psucc}
\\
  \lambdat_1 
& = 
  \frac{(\lambda_0+\lambda_1)^b-(\lambda_0-\lambda_1)^b}
    {2 \psucc}
\\
  \lambdat_2 
& = 
  \frac{(\lambda_2+\lambda_3)^b+(\lambda_2-\lambda_3)^b}
    {2 \psucc}
\\
  \lambdat_3 
& = 
  \frac{(\lambda_2+\lambda_3)^b-(\lambda_2-\lambda_3)^b}
    {2 \psucc} 
  \ .
\end{align*}

Inserting these coefficients into the expressions provided by
Lemma~\ref{lem:Hdiffnonoise} gives a bound on the entropy difference
which can be inserted into the formula for the
rate~\eqref{eq:ratesimple}.\footnote{Note that, conditioned on the
  event that $AD_b$ is not successful (i.e., Alice and Bob's outputs
  are $\Delta$), the entropy difference is zero.}  We conclude that
the key distillation protocol enhanced with advantage distillation on
blocks of length $b$ can generate key bits at rate
\begin{multline} \label{eq:rateAD}
  \rate
\geq
  \frac{1}{b} \min_{(\lambda_0, \ldots, \lambda_3) \in \diag(\Gamma)}
  \psucc 
  \cdot \bigl( 1- (\lambdat_0 + \lambdat_1)h\bigl(\sfrac{\lambdat_0}{\lambdat_0 +
  \lambdat_1}\bigr) \\ - (\lambdat_2 + \lambdat_3) h\bigl(\sfrac{\lambdat_2}{\lambdat_2
  + \lambdat_3}\bigr)
    - h(\lambdat_0 + \lambdat_1) \bigr) \ ,
\end{multline}
where $\Gamma$ is the set defined by~\eqref{eq:Gammasdef}.  Note that,
in the special case where the block size $b$ equals $1$, the advantage
distillation is trivial, that is, $\lambdat_i = \lambda_i$,
and~\eqref{eq:rateAD} reduces to~\eqref{eq:rateow}.

Similarly to the discussion in Section~\ref{sec:preproc}, one might
enhance the protocol with noisy
preprocessing\index{noisy~preprocessing} on Alice's side, i.e., Alice
flips her bits with some probability $q$ after the advantage
distillation step. The rate is then given by a formula similar
to~\eqref{eq:rateAD}, where the expression in the minimum is replaced
by the bound on the entropy difference provided by
Lemma~\ref{lem:twonoise}, evaluated for the coefficients~$\lambdat_i$.

Note that, as the block size $b$ increases, the coefficients
$\lambdat_2$ and $\lambdat_3$ approach zero, while $\lambdat_0$ and
$\lambdat_1$ both tend to $\frac{1}{2}$. To get an approximation, it is
thus sufficient to evaluate the expression of Lemma~\ref{lem:twonoise}
up to small orders in $\lambda_2$ and $\lambda_3$.

\begin{lemma} \label{lem:ratezerocrit}
  Let $\lambda_0, \cdots \lambda_3$ and $\sigma_{X Y \Eb}$ be defined
  as in Lemma~\ref{lem:twonoise}, where $\lambda_0 = (1-\delta)
  \sfrac{1+ \deltab}{2}$, $\lambda_1 = (1-\delta)
  \sfrac{1-\deltab}{2}$, $\lambda_2 = \lambda_3 = \sfrac{\delta}{2}$
  for some $\delta, \deltab \geq 0$. Then
  \begin{multline*}
    H(X|\Eb) - H(X|Y) \\
  \geq
    \sfrac{4}{\ln 8}  (1-\delta) \bigl(\deltab^2 - 6 \delta\bigr) 
    \bigl(\sfrac{1}{2} -q \bigr)^2 
    + O\bigl(\delta^3 + \deltab^3 + (\sfrac{1}{2} -q)^3 \bigr)\ .
  \end{multline*}
  In particular, this quantity is positive if $ \deltab^2 \geq 6
  \delta$.
\end{lemma}

\begin{proof}
  The assertion follows immediately from a series expansion of the
  bound provided by Lemma~\ref{lem:twonoise} about $\eps=0$ and
  $\delta = 0$.
\end{proof}

Lemma~\ref{lem:ratezerocrit} can be used to compute a bound on the
rate of the protocol described above (advantage distillation followed
by noisy preprocessing).  Under the assumption that the coefficients
$\lambdat_0, \ldots, \lambdat_3$ are of the form
\begin{align*}
  \lambdat_0 & = (1-\delta) \sfrac{1+ \deltab}{2} \\
  \lambdat_1 & = (1-\delta) \sfrac{1-\deltab}{2} \\
  \lambdat_2 = \lambdat_3 & = \sfrac{\delta}{2} \ ,
\end{align*}
for some small $\delta, \deltab \geq 0$, we get, analogously
to~\eqref{eq:rateAD},
\begin{multline} \label{eq:rateADnoise}
  \rate
\geq
  \frac{1}{b} \min_{(\lambda_0, \ldots, \lambda_3) \in
    \diag(\Gamma)}  \psucc \cdot \Bigl( \sfrac{4}{\ln 8}
  (1-\delta) \bigl(\deltab^2 - 6 \delta\bigr) 
    \bigl(\sfrac{1}{2} -q \bigr)^2 \\
    + O\bigl(\delta^3 + \deltab^3 + (\sfrac{1}{2} -q)^3 \bigr) \Bigr) \ .
\end{multline}

\index{advantage~distillation|)}

\index{secret~key~rate@secret-key~rate|)}

\section{The six-state protocol} \label{sec:sixstate}

\index{six-state~protocol|(}

To illustrate the results of Section~\ref{sec:twolevel}, we apply them
to different variants of the six-state QKD protocol, for which we
explicitly compute the rate and the maximum tolerated channel noise.
The six-state protocol is one of the most efficient QKD schemes based
on two-level systems, that is, the rate at which secret key bits can
be generated per channel use is relatively close to the theoretical
maximum.  On the other hand, it is not very suitable for practical
implementations, as it requires devices for preparing and measuring
two-level quantum systems with respect to six different states.

\subsection{Description} \label{sec:sixdesc}


Instead of describing the actual six-state QKD protocol, we specify
the underlying key distillation scheme: Alice and Bob take as input
entangled two-level systems and measure each of them using at random
one of three mutually unbiased bases\index{mutually~unbiased~bases},
which results in a pair of raw keys\index{raw~key}.\footnote{Because
  each of the three bases consists of two orthonormal vectors, the
  information is encoded into six different states, which explains the
  name of the protocol.} Usually, these are the
\emph{rectilinear}\index{rectilinear~basis} or \emph{$z$-basis}
$\{\ket{0}_z, \ket{1}_z\}$, the \emph{diagonal}\index{diagonal~basis}
or \emph{$x$-basis} $\{\ket{0}_x, \ket{1}_x\}$, and the
\emph{circular}\index{circular~basis} or \emph{$y$-basis}
$\{\ket{0}_y, \ket{1}_y\})$, which are related by
\begin{equation*}
\begin{matrix}
  \ket{0}_x & = & \frac{1}{\sqrt{2}} (\ket{0}_z + \ket{1}_z) & & 
  \ket{0}_y & = & \frac{1}{\sqrt{2}} (\ket{0}_z + i \ket{1}_z) \\[1ex]
  \ket{1}_x & = & \frac{1}{\sqrt{2}} (\ket{0}_z - \ket{1}_z) & & 
  \ket{1}_y & = & \frac{1}{\sqrt{2}} (\ket{0}_z - i \ket{1}_z) & .
\end{matrix}
\end{equation*}
Next, in a \emph{sifting}\index{sifting} step, Alice and Bob compare
their choices of bases and discard all outcomes for which these do not
agree. Note that, if Alice and Bob choose one of the bases with
probability almost one, they only have to discard a small fraction of
their raw keys (see discussion in~Section~\ref{sec:QKD}).

In the parameter estimation step, Alice and Bob compare the bit values
of their raw keys for a small fraction of randomly chosen positions.
They abort if the \emph{error rate}\index{error~rate} $e$|i.e., the
fraction of positions for which their bits differ|is larger than some
threshold.  For the following analysis, we assume that Alice and Bob
additionally check whether the error $e$ is equally distributed among
the different choices of the measurement bases and symmetric under
bitflips.

Finally, Alice and Bob use the remaining part of their raw key to
generate a pair of secret keys. For this, they might invoke different
variants of advantage distillation and one-way post-processing
subprotocols, as described in Section~\ref{sec:twolevel}.

\subsection{Analysis}

\index{secret~key~rate@secret-key~rate|(}

To compute the rate of the six-state protocol (for different variants
of the post-processing) we use the formulas derived in
Section~\ref{sec:twolevel}. The set $\Gamma$, as defined
by~\eqref{eq:Gammasdef}, depends on the error rate $e$.  For any
fixed~$e$, we get six conditions on the operators $\sigma_{A B}$
contained in $\Gamma$, namely
\begin{equation}
  (\bra{b}_u \otimes \bra{b'}_u) \sigma_{A B} (\ket{b}_u \otimes \ket{b'}_u)
=
  \frac{e}{2} \ ,
\end{equation}
for any $u \in \{x, y, z\}$ and $b, b' \in \{0,1\}$ with $b \neq b'$.
It is easy to verify that the only density operator that satisfies
these equalities is Bell-diagonal and has eigenvalues $\lambda_0 =
1-\frac{3 e}{2}$, $\lambda_1 = \lambda_2 = \lambda_3 = \frac{e}{2}$.
$\Gamma$ is thus the set of all density operators of the form (with
respect to the Bell basis)
\[
  \sigma_{A B}
= 
  \left( \begin{matrix} 1-\sfrac{3 e}{2} & 0 & 0 & 0 \\ 0 &
  \sfrac{e}{2} & 0 & 0 \\ 0 & 0 & \sfrac{e}{2} & 0 \\ 0 & 0 & 0 &
  \sfrac{e}{2} \end{matrix} \right) \ ,
\]
for any $e \geq 0$ below some threshold.

\subsubsection{One-way six-state protocol}

\index{maximum~tolerated~channel~noise|(}

In a basic version of the six-state QKD protocol, Alice and Bob apply
post-processing (i.e., information reconciliation followed by privacy
amplification) directly to their measured data, as described in
Section~\ref{sec:oneway}. The rate of this protocol can be computed
using~\eqref{eq:rateow} where, according to the above discussion,
$\lambda_0 = 1-\frac{3 e}{2}$ and $\lambda_1 = \lambda_2 = \lambda_3 =
\frac{e}{2}$. Plot~\ref{pl:SixSnoNoise} shows the result of a
numerical evaluation of this formula. In particular, the maximum
tolerated channel noise for which the key rate is nonzero is $12.6
\%$.

Next, we consider the one-way six-state protocol enhanced with
additional noisy preprocessing\index{noisy~preprocessing} as described
in Section~\ref{sec:preproc}. That is, before the information
reconciliation step, Alice applies random bitflips with probability
$q$ to her measurement outcomes. The rate of this protocol can be
computed with Lemma~\ref{lem:twonoise}.  A little bit surprisingly, it
turns out that noisy preprocessing increases its performance (see
Plot~\ref{pl:SixSoptNoise}).  As shown in Plot~\ref{pl:SixSoptNoiseq},
the optimal value of the bit-flip probability $q$ depends on the error
rate of the channel $e$.  The protocol can tolerate errors up to $14.1
\%$ and thus beats the basic version (without noisy preprocessing)
described above.  Note that this result also improves on the
previously best known lower bound for the maximum error tolerance of
the six-state protocol with one-way processing, which was~$12.7
\%$~\cite{Lo00}. (Similarly, the same preprocessing can be applied to
the BB84 protocol\index{BB84~protocol}, in which case we get an error
tolerance of~$12.4 \%$, compared to the best known value of $11.0
\%$~\cite{ShoPre00}.)

\begin{picturefloat}
  \includegraphics[scale=0.8]{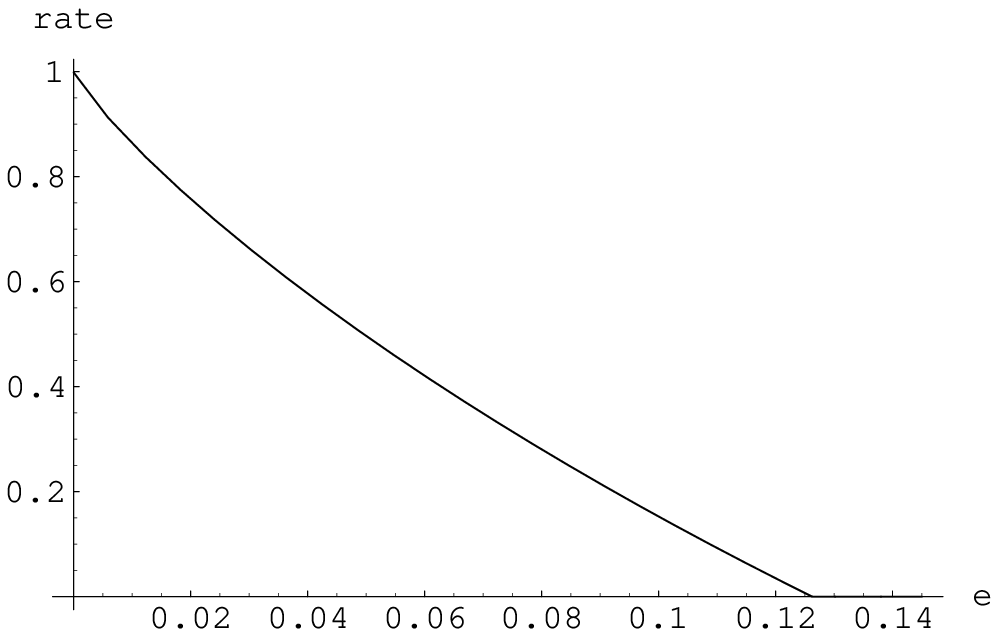}
  \caption{Rate of the basic one-way six-state protocol (without noisy preprocessing) as a function of the error rate $e$.}
  \label{pl:SixSnoNoise}
\end{picturefloat}

\begin{picturefloat}
  \includegraphics[scale=0.8]{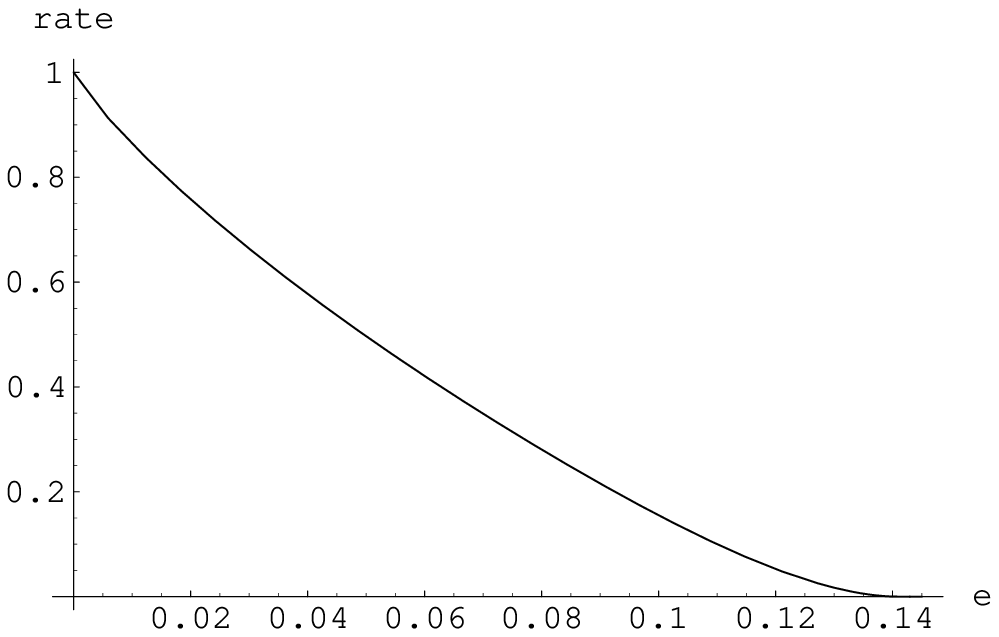}
  \caption{Rate of the one-way six-state protocol with noisy preprocessing (where Alice flips her bits with probability $q$ as depicted in Plot~\ref{pl:SixSoptNoiseq}).}
  \label{pl:SixSoptNoise}
\end{picturefloat}

\begin{picturefloat}
  \includegraphics[scale=0.8]{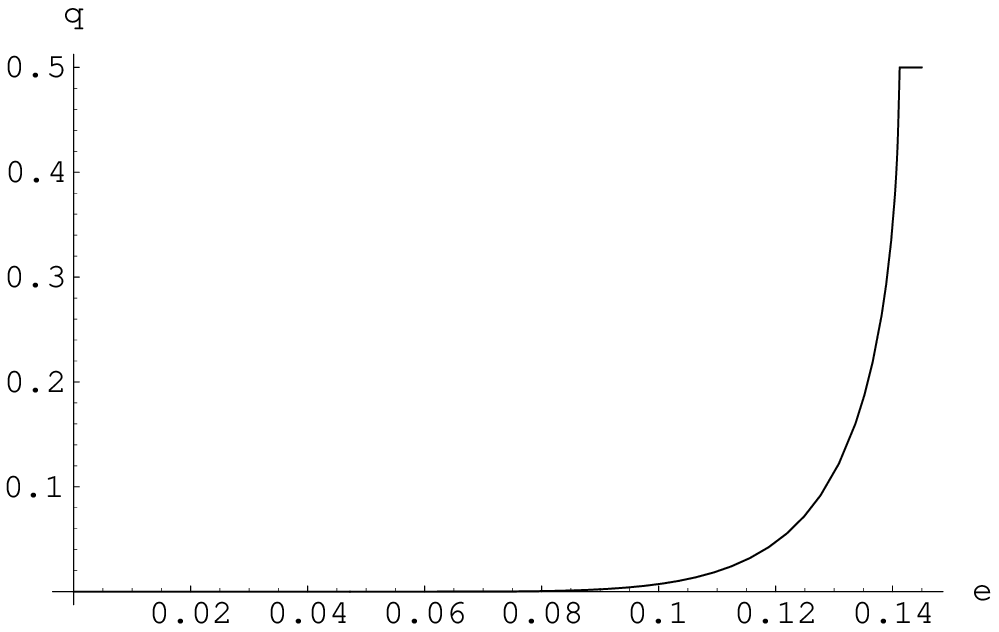}
  \caption{Optimal value of the bit-flip probability $q$ for the noisy preprocessing used in the one-way six-state protocol.}
  \label{pl:SixSoptNoiseq}
\end{picturefloat}

\index{advantage~distillation|(}

\subsubsection{Six-state protocol with advantage distillation} \label{sec:ad}

The performance of the six-state protocol is increased if Alice and
Bob additionally use advantage distillation as described in
Section~\ref{sec:AD}. For example, Alice and Bob might invoke the
protocol $\AD_b$ depicted in Fig.~\ref{pr:AD} to process their
measurement outcomes before the information reconciliation and privacy
amplification step. The rate of the protocol is then given
by~\eqref{eq:rateAD}. Because $\lambda_0 = 1-\frac{3 e}{2}$ and
$\lambda_1 = \lambda_2 = \lambda_3 = \frac{e}{2}$, the coefficients
$\lambdat_i$ occurring in this formula are
\begin{align*}
  \lambdat_0 & = \frac{(1-e)^b + (1-2 e)^b}{2 \psucc} \\
  \lambdat_1 & = \frac{(1-e)^b - (1-2 e)^b}{2 \psucc} \\
  \lambdat_2 & = \frac{e^b}{2 \psucc} \\
  \lambdat_3 & = \frac{e^b}{2 \psucc} \ ,
\end{align*}
where $\psucc = (1-e)^b + e^b$. Plot~\ref{pl:sixstateAD} shows the result
of this computation for a block size of $b = 4$.

\begin{picturefloat}
  \includegraphics[scale=0.8]{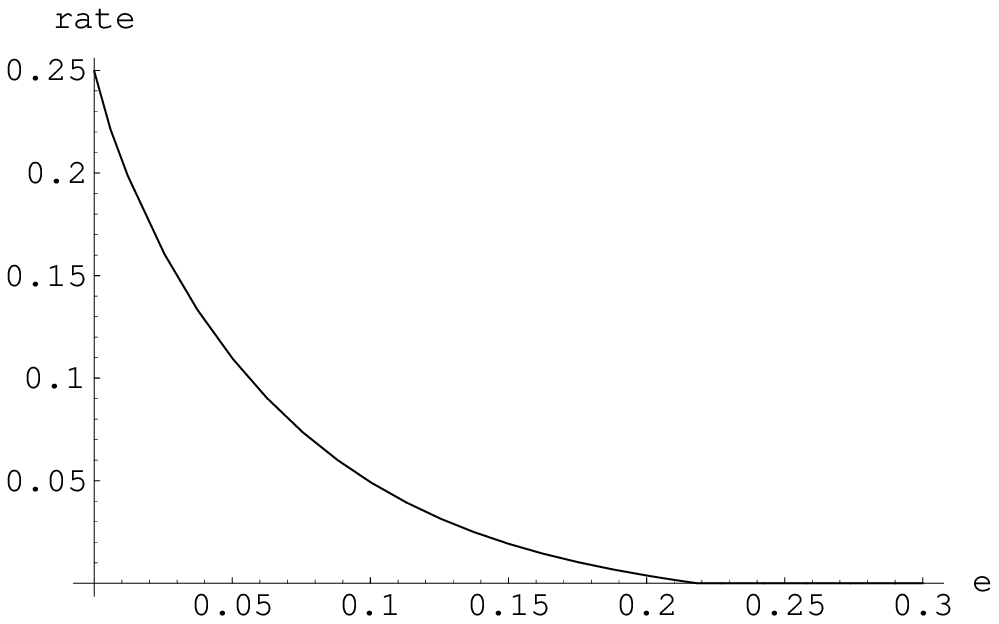}
  \caption{Rate of the six-state protocol with advantage distillation
    on blocks of length $4$.}
  \label{pl:sixstateAD}
\end{picturefloat}

Finally, we have a look at an extended protocol which combines
advantage distillation\index{advantage~distillation} and noisy
preprocessing\index{noisy~preprocessing}. That is, after the advantage
distillation $\AD_b$, Alice flips her bits with probability $q$ (see
Plot~\ref{pl:sixSoptrate}).  For large block sizes $b$, the rate of
the protocol is given by~\eqref{eq:rateADnoise}, for
\begin{align*}
  \delta & = \frac{e^b}{(1-e)^b + e^b} \\
  \deltab & = \Bigl(\frac{1-2 e}{1-e}\Bigr)^b \ .
\end{align*}
In particular, for $b$ approaching infinity, the secret-key rate is
positive if (see Lemma~\ref{lem:ratezerocrit})
\[
  \Bigl(\frac{1-2 e}{1-e}\Bigr)^{2 b}
\geq
  6 \frac{e^b}{(1-e)^b + e^b} \ .
\]
Some simple analysis shows that this inequality is satisfied (for
large $b$) if $e \leq \frac{1}{2} - \frac{\sqrt{5}}{10} \approx
0.276$. We conclude that the protocol can tolerate errors up to $27.6
\%$.

Note that this value coincides with the corresponding error tolerance
of another variant of the six-state protocol due to Chau~\cite{Chau02}
and is actually optimal for this class of protocols (cf.\ 
\cite{ABBBMMT04}). However, compared to Chau's protocol, the above
described variant of the six-state protocol is
simpler\footnote{Instead of adding noise, Chau's protocol uses xor
  operations between different bits of the raw key.} and has a higher
key rate.

\begin{picturefloat}
  \includegraphics[scale=0.8]{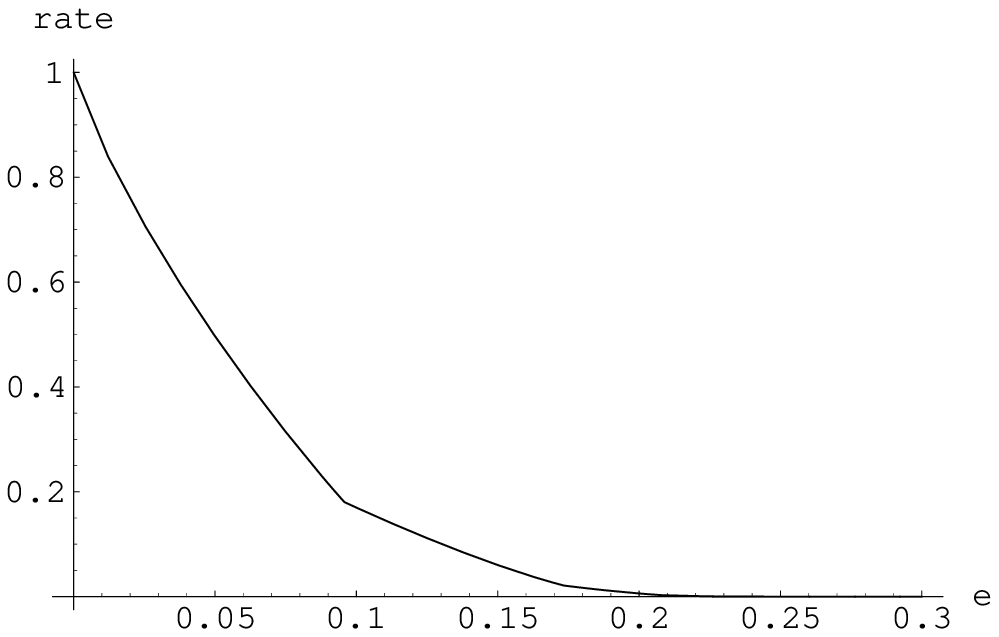}
  \caption{Rate of the six-state protocol with advantage distillation (on blocks of optimal length) followed by (optimal) random bit-flips on Alice's side.}
  \label{pl:sixSoptrate}
\end{picturefloat}

\index{six-state~protocol|)} \index{maximum~tolerated~channel~noise|)}
\index{secret~key~rate@secret-key~rate|)}

\index{advantage~distillation|)}

\appendix


\chapter{Distance measures}

\section{Fidelity}

\index{fidelity|(}

The fidelity between two (not necessarily normalized) states $\rho,
\rho' \in \NN(\cH)$ is defined by
\[
F(\rho, \rho') := \tr\sqrt{\rho^{1/2} \rho' \rho^{1/2}} \ .
\]
In particular, if $\rho = \proj{\psi}$ and $\rho' = \proj{\psi'}$ are
pure states,
\[
  F(\rho, \rho')
=
  |\spr{\psi}{\psi'}| \ .
\]

\begin{remark} \label{rem:fidscal}
  For any $\alpha, \beta \in \bbR^+$,
  \[
    F(\alpha \rho, \beta \rho') 
  =
    \sqrt{\alpha \beta} F(\rho, \rho') \ .
  \]
\end{remark}

\subsubsection{Fidelity of purifications}

Uhlmann's theorem states that the fidelity between two operators is
equal to the maximum fidelity of their
purifications.\index{Uhlmann's~Theorem}

\begin{theorem}[Uhlmann] \label{thm:Uhlmann}
  Let $\rho, \rho' \in \NN(\cH)$ and let $\proj{\psi}$ be a
  purification of $\rho$. Then
  \[
    F(\rho,\rho') 
  =
    \max_{\proj{\psi'}} F(\proj{\psi}, \proj{\psi'})
  \]
  where the maximum is taken over all purifications $\proj{\psi'}$ of
  $\rho'$.
\end{theorem}

\begin{proof}
  The assertion follows directly from the corresponding statement for
  normalized density operators (see, e.g., Theorem~9.4
  in~\cite{NieChu00}) and Remark~\ref{rem:fidscal}.
\end{proof}

\begin{remark} \label{lem:Uhlmann}
  Because the fidelity $F(\proj{\psi}, \proj{\psi'})$ does not depend
  on the phase of the vectors, the vector $\ket{\psi'}$ which
  maximizes the expression of Theorem~\ref{thm:Uhlmann} can always be
  chosen such that $\spr{\psi}{\psi'}$ is real and nonnegative.
\end{remark}

\subsubsection{Fidelity and quantum operations}

The fidelity between two density operators is equal to the minimum
fidelity between the distributions of the outcomes resulting from a
measurement.

\begin{lemma} \label{lem:fidmeas}
  Let $\rho, \rho' \in \NN(\cH)$. Then
  \[
    F(\rho, \rho')
  =
    \min_{\{M_z\}_z} F(P_Z, P'_Z)
  \]
  where the minimum ranges over all POVMs $\{M_z\}_{z \in \cZ}$ on
  $\cH$ and where $P_Z, P'_Z \in \NN(\cZ)$ are defined by $P_Z(z) =
  \tr(\rho M_z)$ and $P'_Z(z) = \tr(\rho' M_z)$, respectively.
\end{lemma}

\begin{proof}
  The statement follows directly from the corresponding statement for
  normalized density operators (cf.\ formula~(9.74)
  in~\cite{NieChu00}) and Remark~\ref{rem:fidscal}.
\end{proof}

The fidelity between two operators cannot decrease when applying the
same quantum operation to both of them.

\begin{lemma} \label{lem:fidincr}
  Let $\rho, \rho' \in \NN(\cH)$ and let $\cE$ be a trace-preserving
  CPM on~$\cH$. Then
  \[
    F(\cE(\rho), \cE(\rho')) \geq F(\rho, \rho') \ .
  \]
\end{lemma}

\begin{proof}
  See Theorem 9.6 of~\cite{NieChu00} and Remark~\ref{rem:fidscal}.
\end{proof}

\index{fidelity|)}

\section{ \distance}

\index{\distance{}|(}

\subsubsection{\distance{} and quantum operations}

The \distance{} between two density operators cannot increase when
applying the same (trace-preserving) quantum operation to both of
them.

\begin{lemma} \label{lem:distdecr}
  Let $\rho, \rho' \in \NN(\cH)$ and let $\cE$ be a CPM such that
  $\tr(\cE(\sigma)) \leq \tr(\sigma)$ for any $\sigma \in \NN(\cH)$.
  Then
  \[
    \|\cE(\rho) - \cE(\rho') \|_1 \leq \|\rho-\rho'\|_1 \ .
  \]
\end{lemma}

\begin{proof}
  
  It suffices to show that $\|\cE(T)\|_1 \leq \|T\|_1$, for any
  hermitian operator $T$. The assertion then follows with $T:= \rho -
  \rho'$ because $\cE$ is linear.
  
  For any hermitian operator $S$, let
  $\|S\|_\infty:=\sup_{\ket{\phi} \in\cH: \, \|\ket{\phi}\| \leq
    1} \|S \ket{\phi}\|$ be the \emph{$L_\infty$-operator norm}.
  Note that the $L_\infty$-operator norm can equivalently be written
  as
  \[
    \|S\|_\infty
  =
    \sup_{\sigma\in\NN(\cH) : \, \tr(\sigma) \leq 1}\tr(S \sigma) \ .
  \]  
  Moreover, it is easy to see that for any hermitian operator $T$
  \begin{equation}\label{eq:supdefinitiontrace}
    \|T\|_1=\sup_{S:\, \|S\|_\infty \leq 1} |\tr(S T)| \ .
  \end{equation}
 
  Let $\{E_k\}_{k}$ be the family of linear operators from $\cH$ to
  $\cH'$ defined by the CPM $\cE$, i.e., $\cE(\sigma) = E_k \sigma
  E^\dagger_k$, for any $\sigma \in \NN(\cH)$.  Moreover, let
  $\cE^\dagger$ be the CPM defined by
  $\cE^\dagger(S'):=\sum_{k}E_k^\dagger S' E_k$, for any hermitian
  operator $S'$ on $\cH'$. We then have the identity
  \[
    \tr(\cE^\dagger(S')\sigma)
  =
    \tr(S' \cE(\sigma)) \ .
  \]
  Hence 
  \begin{equation} \label{eq:distinfbound}
  \begin{split}
    \|\cE^\dagger(S') \|_\infty
  & =
    \sup_{\sigma\in\NN(\cH): \, \tr(\sigma) \leq 1}
      \tr(\cE^\dagger(S')\sigma) \\
  & =
    \sup_{\sigma\in\NN(\cH): \, \tr(\sigma) \leq 1} \tr(S'\cE(\sigma)) \\
  & \leq 
    \|S'\|_\infty \ ,
  \end{split}
  \end{equation}
  where the inequality holds because $\cE(\sigma)\in\NN(\cH')$ and
  $\tr(\cE(\sigma)) \leq \tr(\sigma) = 1$, for any
  $\sigma\in\NN(\cH)$.  Using~\eqref{eq:supdefinitiontrace}, this
  implies that
  \[
  \begin{split}
    \|\cE(T)\|_1
  & =
    \sup_{S':\, \|S'\|_\infty \leq 1}| \tr(\cE(T) S')| \\
  & =
    \sup_{S':\, \|S'\|\infty \leq 1} |\tr(T \cE^\dagger(S'))| \\
  & \leq 
    \sup_{S:\,\|S\|_\infty \leq 1}\tr(T S) \\
  & =
    \| T \|_1 \ ,
  \end{split}
  \]
  where the inequality follows from~\eqref{eq:distinfbound}.
  
\end{proof}



\subsubsection{\distance{} of mixtures}

\begin{lemma} \label{lem:distclass}
  Let $\rho_{A Z}$ and $\rhob_{A Z}$ be classical with respect to an
  orthonormal basis $\{\ket{z}\}_{z \in \cZ}$ of $\cH_Z$ and let
  $\{\rho_{A}^z\}_{z \in \cZ}$ and $\{\rhob_{A}^z\}_{z \in \cZ}$ be
  the corresponding conditional operators. Then
  \[
    \bigl\| \rho_{A Z} - \rhob_{A Z} \bigr\|_1
  =
    \sum_{z \in \cZ} \bigl\| \rho_{A}^z - \rhob_{A}^z \bigr\|_1 \ .
  \]
\end{lemma} 

\begin{proof}
  For any $z \in \cZ$, let $\{\ket{\phi_x^z}\}_{x \in \cX}$ be an
  eigenbasis of $\rho_A^z - \rhob_A^z$. Then, the family
  $\{\ket{\phi_x^z} \otimes \ket{z}\}_{(x,z) \in \cX \times \cZ}$ is
  an eigenbasis of $\rho_{A Z} - \rhob_{A Z}$. Hence,
  \[
  \begin{split}
    \bigl\| \rho_{A Z} - \rhob_{A Z} \bigr\|_1
  & =
    \sum_{z' \in \cZ} \sum_{x \in \cX} \bigl|
      (\bra{\phi_x^{z'}} \otimes \bra{z'})
      \bigl(
      \sum_{z \in \cZ} (\rho_A^{z} - \rhob_A^{z}) \otimes \proj{z}
      \bigr)
      (\ket{\phi_x^{z'}} \otimes \ket{z'}) 
    \bigr| \\
  & =
    \sum_{z \in \cZ} \sum_{x \in \cX}
      \bigl| \bra{\phi_x^z} \rho_A^z - \rhob_A^z \ket{\phi_x^z} \bigr| \\
  & =
    \sum_{z \in \cZ} \bigl\|\rho_A^z - \rhob_A^z \bigr\|_1 \ . \qedhere
  \end{split} 
  \]
\end{proof}

\subsubsection{\distance{} of pure operators in terms of vector distance}

The scalar product of a Hilbert space $\cH$ induces a canonical norm,
defined by $\|\ket{\phi}\| := \sqrt{\spr{\phi}{\phi}}$, for any
$\ket{\phi} \in \cH$. In particular, the norm of the difference
between two vectors $\ket{\psi}$ and $\ket{\psi'}$, $\|\ket{\psi} -
\ket{\psi'}\|$, is a metric on $\cH$.

The following lemma relates the \distance{} between two pure states
$\proj{\psi}$ and $\proj{\psi'}$ to the vector distance $\| \ket{\psi}
- \ket{\psi'}\|$.

\begin{lemma} \label{lem:puredist}
  Let $\ket{\psi}, \ket{\psi'} \in \cH$ such that $\spr{\psi}{\psi'}$
  is real.  Then
  \[
    \bigl\|\proj{\psi} - \proj{\psi'} \bigr\|_1
  = 
    \bigl\|\ket{\psi}-\ket{\psi'}\bigr\| \cdot \bigl\|\ket{\psi} + \ket{\psi'}\bigr\| \ .
  \]
\end{lemma}

\begin{proof}
  Define $\ket{\alpha} := \ket{\psi} + \ket{\psi'}$, $\ket{\beta} :=
  \ket{\psi} - \ket{\psi'}$ and let $a := \|\ket{\alpha}\|$, $b:=
  \|\ket{\beta}\|$. We then have
  \[
    \bigl\| \proj{\psi} - \proj{\psi'} \bigr\|_1 
  =
    \tr \bigl| \proj{\psi} - \proj{\psi'} \bigr|
  =
    \frac{1}{2}
      \tr
        \bigl| \ket{\alpha}\bra{\beta} + \ket{\beta}\bra{\alpha} \bigr| \ .
  \]
  Moreover, because $\spr{\psi}{\psi'}$ is real, the scalar product
  $\spr{\alpha}{\beta} = \spr{\psi}{\psi} - \spr{\psi'}{\psi'}$ is
  real as well.  Using this, it is easy to verify that $b \ket{\alpha}
  + a \ket{\beta}$ and $b \ket{\alpha} - a \ket{\beta}$ are
  eigenvectors of $\ket{\alpha}\bra{\beta} + \ket{\beta}\bra{\alpha}$
  with eigenvalues $\spr{\alpha}{\beta} + a b$ and
  $\spr{\alpha}{\beta} - a b$, respectively.  Hence,
  \[
    \tr
      \bigl| \ket{\alpha}\bra{\beta} + \ket{\beta}\bra{\alpha} \bigr|
  =
    \bigl|\spr{\alpha}{\beta} + a b\bigr| + \bigl|\spr{\alpha}{\beta} - a b\bigr|
  =
    2 a b \ ,
  \]
  where the last equality holds because the Cauchy-Schwartz inequality
  implies $|\spr{\alpha}{\beta}| \leq a b$.
\end{proof}

\subsubsection{Upper bound on \distance{} in terms of fidelity}

\begin{lemma} \label{lem:distfidbound}
  Let $\rho, \rho \in \NN(\cH)$. Then
  \[
    \|\rho - \rho' \|_1
  \leq
    \sqrt{
      \bigl(
        \tr(\rho) + \tr(\rho')
      \bigr)^2
      - 4 F(\rho, \rho')^2} \ .
  \]
\end{lemma}

\begin{proof}
  It follows from Uhlmann's theorem (see Theorem~\ref{thm:Uhlmann} and
  remark thereafter) that there exist purifications $\proj{\psi}$ and
  $\proj{\psi'}$ of $\rho$ and $\rho'$, respectively, such that
  $\spr{\psi}{\psi'}$ is nonnegative and $F(\rho, \rho') =
  F(\proj{\psi}, \proj{\psi'})$.  Using Lemma~\ref{lem:puredist}, a
  simple calculation leads to
  \[
    \bigl\| \proj{\psi} - \proj{\psi'}\bigr\|
  =
    \sqrt{
      (\spr{\psi}{\psi} + \spr{\psi'}{\psi'})^2
      - 4 \spr{\psi}{\psi'}^2} \ .
  \]
  Since $\spr{\psi}{\psi} = \tr(\proj{\psi}) = \tr(\rho)$,
  $\spr{\psi'}{\psi'} = \tr(\rho')$, and $F(\proj{\psi}, \proj{\psi'})
  = \spr{\psi}{\psi'}$, this identity can be rewritten as
  \[
    \bigl\| \proj{\psi} - \proj{\psi'}\bigr\|
  = 
    \sqrt{ 
      \bigl( \tr(\rho) + \tr(\rho') \bigr)^2 
       - 4 F(\rho, \rho')^2
    } \ .
  \]
  The assertion then follows from the fact that the \distance{} can
  only decrease when taking the partial trace (cf.\ 
  Lemma~\ref{lem:distdecr}).
\end{proof}

\subsubsection{Upper bound on \distance{} in terms of vector distance}

The following lemma is a generalization of one direction of
Lemma~\ref{lem:puredist} to mixed states.

\begin{lemma} \label{lem:distpurebound}
  Let $\rho, \rho' \in \NN(\cH)$ and let $\proj{\psi}$ and
  $\proj{\psi'}$ be purifications of $\rho$ and $\rho'$, respectively.
  Then
  \[
    \|\rho-\rho'\|_1
  \leq 
    \bigl(\sqrt{\tr(\rho)} + \sqrt{\tr(\rho')}\bigr) 
      \cdot \bigl\|\ket{\psi} - \ket{\psi'}\bigr\| \ .
  \]
\end{lemma}

\begin{proof}
  Let $\nu \in [0,2\pi]$ such that $e^{i \nu}\spr{\psi}{\psi'}$ is
  nonnegative and define $\ket{\tilde{\psi}'} := e^{i \nu}
  \ket{\psi'}$.  Then, from Lemma~\ref{lem:puredist},
  \begin{equation} \label{eq:distpureboundcalc}
  \begin{split}
    \bigl\|\proj{\psi} - \proj{\tilde{\psi}'}\bigr\|_1
  & =
      \bigl\|\ket{\psi}-\ket{\tilde{\psi}'}\bigr\| \cdot 
      \bigl\|\ket{\psi} + \ket{\tilde{\psi}'}\bigr\| \\
  & \leq
      \bigl\|\ket{\psi}-\ket{\tilde{\psi}'}\bigr\| \cdot
      \bigl(\big\|\ket{\psi}\bigr\| + \big\|\ket{\psi'}\bigr\|\bigr)
  \end{split}
  \end{equation}
  where the inequality follows from the triangle inequality for the
  norm $\|\cdot \|$ and $\bigl\|\ket{\tilde{\psi}'}\bigr\| =
  \bigl\|\ket{\psi'}\bigr\|$. Moreover, since
  $\spr{\psi}{\tilde{\psi}'}$ is nonnegative, it cannot be smaller
  than the real value of the scalar product $\spr{\psi}{\psi'}$, that
  is, $\Re(\spr{\psi}{\tilde{\psi}'}) = |\spr{\psi}{\tilde{\psi}'}| =
  |\spr{\psi}{\psi'}| \geq \Re(\spr{\psi}{\psi'})$, and thus
  \[
  \begin{split}
    \bigl\|\ket{\psi}-\ket{\tilde{\psi}'}\bigr\| 
  & =
    \sqrt{\spr{\psi}{\psi} + \spr{\tilde{\psi}'}{\tilde{\psi}'}
      -2 \Re(\spr{\psi}{\tilde{\psi}'})} \\
  & \leq
    \sqrt{\spr{\psi}{\psi} + \spr{\psi'}{\psi'}
      -2 \Re(\spr{\psi}{\psi'})} \\
  & =
    \bigl\|\ket{\psi}-\ket{\psi'}\bigr\| \ .
  \end{split}
  \]  
  Combining this with~\eqref{eq:distpureboundcalc} gives
  \[
  \begin{split}
    \bigl\|\proj{\psi} - \proj{\psi'}\bigr\|
  & =
    \bigl\|\proj{\psi} - \proj{\tilde{\psi}'} \bigr\| \\
  & \leq
      \bigl(\big\|\ket{\psi}\bigr\| + \big\|\ket{\psi'}\bigr\|\bigr)
      \cdot \bigl\|\ket{\psi}-\ket{\psi'}\bigr\| \ .
  \end{split}
  \]
  The assertion follows from the fact that the \distance{} cannot
  increase when taking the partial trace (cf.\ 
  Lemma~\ref{lem:distdecr}).
\end{proof}

\subsubsection{Lower bound on \distance{} in terms of fidelity}

The following statement is the converse of
Lemma~\ref{lem:distfidbound}.

\begin{lemma} \label{lem:distfidrel} 
  Let $\rho, \rho' \in \NN(\cH)$. Then
  \[
    \tr(\rho) + \tr(\rho') - 2 F(\rho, \rho') 
  \leq 
    \| \rho - \rho' \|_1 \ .
  \]
\end{lemma}

The proof is a direct generalization of an argument given
in~\cite{NieChu00} (see formula~(9.109) of~\cite{NieChu00}).

\begin{proof}  
  According to Lemma~\ref{lem:fidmeas}, there exists a POVM $\cM =
  \{M_z\}_{z \in \cZ}$ such that
  \[
    F(\rho, \rho') = F(P_Z, P'_Z) \ ,
  \]
  for $P_Z$ and $P'_Z$ defined by $P_Z(z) = \tr(\rho M_z)$ and
  $P'_Z(z) = \tr(\rho' M_z)$.  Using the abbreviation $p_z := P_Z(z)$
  and $p'_z := P_{Z'}$, we observe that
  \begin{equation} \label{eq:classsumeq}
  \begin{split}
    \sum_{z \in \cZ} \bigl( \sqrt{p_z} - \sqrt{p'_z} \bigr)^2 
  & = 
    \sum_{z \in \cZ} \bigl( p_z + p'_z 
    - 2 \sqrt{ p_z p'_z} \bigr) \\
  & =
    \tr(\rho) + \tr(\rho') - 2 F(\rho, \rho') \ .
  \end{split}
  \end{equation}
  Moreover, because $\bigl| \sqrt{p_z} - \sqrt{p'_z} \bigr| \leq
  \sqrt{p_z} + \sqrt{p'_z}$,
  \[
  \begin{split}
    \sum_{z \in \cZ} \bigl( \sqrt{p_z} - \sqrt{p'_z} \bigr)^2
  & \leq
    \sum_{z \in \cZ}
        \bigl| \sqrt{p_z} - \sqrt{p'_z} \bigr| 
      \cdot \bigl( \sqrt{p_z} + \sqrt{p'_z} \bigr) \\
  & =
    \sum_{z \in \cZ} \bigl| p_z - p'_z \bigr| \\
  & \leq
    \|\rho - \rho'\|_1 \ ,
  \end{split}
  \]
  where the last inequality follows from the fact that the trace
  distance cannot increase when applying a POVM (cf.\ 
  Lemma~\ref{lem:distdecr}). The assertion then follows by combining
  this with~\eqref{eq:classsumeq}.
\end{proof}

\subsubsection{Lower bound on \distance{} in terms of vector distance}

The following statement can be seen as the converse of
Lemma~\ref{lem:distpurebound}.

\begin{lemma} \label{lem:puredistbound}
  Let $\rho, \rho' \in \NN(\cH)$ and let $\proj{\psi}$ be a
  purification of $\rho$. Then there exists a purification
  $\proj{\psi'}$ of $\rho'$ such that
  \[
    \bigl\| \ket{\psi} - \ket{\psi'} \bigr\|
  \leq
    \sqrt{\| \rho - \rho'\|_1} \ .
  \]  
\end{lemma}

\begin{proof}
  Uhlmann's theorem (see Theorem~\ref{thm:Uhlmann} and remark
  thereafter) implies that there exists a purification $\proj{\psi'}$
  of $\rho'$ such that $F(\rho, \rho') = \spr{\psi}{\psi'}$.  Hence,
  \[
  \begin{split}
    \bigl\| \ket{\psi} - \ket{\psi'} \bigr\|
  & =
    \sqrt{ 
      \spr{\psi}{\psi} + \spr{\psi'}{\psi'}
      - \spr{\psi}{\psi'} - \spr{\psi'}{\psi} } \\
  & =
    \sqrt{
      \tr(\rho) + \tr(\rho') - 2 F(\rho, \rho')} \ .
  \end{split}
  \]
  The assertion then follows from Lemma~\ref{lem:distfidrel}.
\end{proof}

\subsubsection{\Distance{} and trace}

A slightly different variant of the following statement is known as
the \emph{Gentle Measurement
  Lemma}\index{gentle~measurement}~\cite{Winter99}.

\begin{lemma} \label{lem:disttracebound}
  Let $\rho, \rhob \in \NN(\cH)$ such that $\rhob =P \rho P$ for some
  projector $P$ on $\cH$. Then,
  \[
    \| \rho - \rhob \|_1 
  \leq 
    2 \sqrt{\tr(\rho)\bigl(\tr(\rho)-\tr(\rhob)\bigr)} \ .
  \]
\end{lemma}

\begin{proof}
  We first show that the assertion holds if $\rho$ is normalized
  (i.e., $\tr(\rho) = 1$) and pure, that is, $\rho = \proj{\phi}$ for
  some normalized vector $\ket{\phi}$.  Since $P$ is a projector, the
  vector $\ket{\phi}$ can be written as a weighted sum of two
  orthonormal vectors $\ket{a}$ and $\ket{b}$, $\ket{\phi} = \alpha
  \ket{a} + \beta \ket{b}$, for $\alpha, \beta \geq 0$, such that $P
  \ket{a} = \ket{a}$ and $P \ket{b} = 0$.  In particular, $\rhob =
  \alpha^2 \proj{a}$.  A straightforward calculation then shows that
  \[
  \begin{split}
    \| \rho - \rhob \|_1
  & =
    \bigl|
      (\alpha \ket{a} + \beta \ket{b})( \alpha \bra{a} + \beta \bra{b})
      - \alpha^2 \proj{a} \bigr\|_1 \\
  & \leq
    2 \beta
  =
    2 \sqrt{1-\tr(\rhob)} 
  \end{split}
  \]
  which concludes the proof for normalized pure states $\rho$.
  
  To show that the assertion holds for general operators $\rho \in
  \NN(\cH)$, let $\rho = \sum_{x \in \cX} p_x \proj{x}$ be a spectral
  decomposition of $\rho$.  In particular, $\sum_{x \in \cX} p_x =
  \tr(\rho)$. Define $\rho_x := \proj{x}$ and $\rhob_x := P \rho_x P$.
  By linearity, we have
  \[
    \rhob = P \rho P = \sum_{x \in \cX} p_x \rhob_x \ .
  \]
  Hence, using the triangle inequality and the fact that the assertion
  holds for the normalized pure states $\rho_x$, we find
  \[
    \| \rho - \rhob \|_1
  \leq
    \sum_{x \in \cX} p_x \| \rho_x - \rhob_x \|_1
  \leq
    2 \sum_{x \in \cX} p_x \sqrt{1-\tr(\rhob_x)} \ .
  \]
  Moreover, with Jensen's inequality we find
  \[
  \begin{split}
    \sum_{x \in \cX} p_x \sqrt{1-\tr(\rhob_x)} 
  & =
    \tr(\rho) \sum_{x \in \cX} \frac{p_x}{\tr(\rho)} \sqrt{1-\tr(\rhob_x)} \\
  & \leq
    \tr(\rho) 
    \sqrt{ \sum_{x \in \cX} \frac{p_x}{\tr(\rho)} 
      \bigl(1 - \tr(\rhob_x)\bigr)} \\
  & =
    \sqrt{ \tr(\rho) \bigl(\tr(\rho) - \tr(\rhob)\bigr) } \ ,
  \end{split}
  \]
  which concludes the proof.
\end{proof}

\index{\distance{}|)}

\chapter{Various Technical Results}

\section{Combinatorics}

For proofs of the following statements, we refer to the standard
literature on combinatorics.

\begin{lemma} \label{lem:cQsize}
  The set $\freqset{\cX}{n}$ of types with denominator $n$ on a set
  $\cX$ has cardinality\index{type}
  \[
    |\freqset{\cX}{n}| = \binom{n+|\cX|-1}{n} \ .
  \]
\end{lemma}

\begin{lemma} \label{lem:Lambdasize}
  Let $Q \in \freqset{\cX}{n}$ be a type with denominator $n$ on a set
  $\cX$. Then the type class $\ntuplefreq{Q}{n}$ has cardinality\index{type~class}
  \[
    | \ntuplefreq{Q}{n} |= \frac{n!}{\prod_{x \in \cX} (n Q(x))!} \ .
  \]
\end{lemma}

\begin{lemma} \label{lem:binsize}
  A set of cardinality $n$ has at most $2^{n h(\slfrac{r}{n})}$
  subsets of cardinality $r$.
\end{lemma}

\begin{proof}
  A set of cardinality $n$ has exactly $\binom{n}{r}$ subsets of
  cardinality $r$. The assertion thus follows from the
  inequality\footnote{See, e.g., \cite{CovTho91}, Formula~(12.40).} $
  \binom{n}{r} \leq 2^{n h(\slfrac{r}{n})}$.
\end{proof}

\section{Birkhoff's Theorem}

\index{Birkhoff's~Theorem|(}

\begin{definition}
  A matrix $(a_{x,y})_{x \in \cX, y \in \cY}$ is
  \emph{bistochastic}\index{bistochastic~matrix} if $a_{x,y} \geq 0$,
  for any $x \in \cX$, $y \in \cY$, and $\sum_{y \in \cY} a_{x,y} =
  \sum_{x \in \cX} a_{x,y} = 1$.
\end{definition}

It is easy to see that a matrix $(a_{x,y})_{x \in \cX, y \in \cY}$ can
only be bistochastic if $|\cX| = |\cY|$. The following theorem due to
Birkhoff~\cite{Birkho46} states that any bistochastic matrix can be
written as a mixture of permutation matrices. (See, e.g.,
\cite{HorJoh85} for a proof.)

\begin{theorem}[Birkhoff's theorem] \label{thm:Birkhoff}
  Let $(a_{x,y})_{x \in \cX, y \in \cY}$ be a bistochastic matrix.
  Then there exist nonnegative coefficients $\mu_\pi$, parameterized
  by the bijections $\pi$ from $\cY$ to $\cX$, such that $\sum_{\pi}
  \mu_{\pi} = 1$ and, for any $x \in \cX$, $y \in
  \cY$,\footnote{$\delta_{x, \pi(y)}$ denotes the Kronecker symbol
    which equals one if $x = \pi(y)$ and zero otherwise.} 
  \[
    a_{x,y} = \sum_\pi \mu_\pi \delta_{x, \pi(y)} \ .
  \]
\end{theorem}

It follows immediately from Birkhoff's theorem that any sum of the
form
\[
  S = \sum_{x, y} a_{x,y} S_{x,y}
\]
can be rewritten as
\[
  S 
= 
  \sum_{x, y} \sum_{\pi} \mu_{\pi} \delta_{x, \pi(y)} S_{x,y}
=
  \sum_{\pi} \mu_\pi \sum_{y} S_{\pi(y), y} \ .
\]

\index{Birkhoff's~Theorem|)}

\section{Typical sequences}

\index{typical~sequence|(}

Let $\bx$ be an $n$-tuple chosen according to an $n$-fold product
distribution $(P_X)^n$. Then, with probability almost one, $\bx$ is a
\emph{typical sequence}\index{typical~sequence}, i.e., its frequency
distribution $\freq{\bx}$ is close to the distribution $P_X$.

\begin{theorem} \label{thm:typsec}
  Let $P_X$ be a probability distribution on $\cX$ and let $\bx$ be
  chosen according to the $n$-fold product distribution $(P_X)^n$.
  Then, for any $\delta \geq 0$,
  \[
    \Pr_{\bx}\bigl[D(\freq{\bx}\|P_X) > \delta\bigr] 
  \leq 
    2^{-n(\delta-|\cX|\frac{\log(n+1)}{n})} \ .
  \]
\end{theorem}

\begin{proof}
  See Theorem 12.2.1 of~\cite{CovTho91}.
\end{proof}

Theorem~\ref{thm:typsec} quantifies the distance between $\freq{\bx}$
and $P_X$ with respect to the relative entropy. To obtain a statement
in terms of the \distance{}, we need the following lemma.

\begin{lemma} \label{lem:distrelentr}
  Let $P$ and $Q$ be probability distributions. Then
  \[
    \| P - Q\|_1
  \leq 
    \sqrt{2 (\ln 2) D(P\|Q)} \ .
  \]
\end{lemma}

\begin{proof}
  See Lemma~12.6.1 of~\cite{CovTho91}.
\end{proof}

\begin{corollary} \label{cor:typsec}
  Let $P_X$ be a probability distribution on $\cX$ and let $\bx$ be
  chosen according to the $n$-fold product distribution $(P_X)^n$.
  Then, for any $\delta \geq 0$,
  \[
    \Pr_{\bx}\bigl[\|\freq{\bx} - P_X \|_1  > \delta \bigr] 
  \leq 
    2^{-n(\frac{\delta^2}{2 \ln 2}-|\cX|\frac{\log(n+1)}{n})} \ .
  \]
\end{corollary}

\begin{proof}
  The assertion follows directly from Theorem~\ref{thm:typsec}
  combined with Lemma~\ref{lem:distrelentr}.
\end{proof}

\index{typical~sequence|)}

\section{Product spaces}

\begin{lemma} \label{lem:tensprodimage}
  Let $\rho_{A B} \in \NN(\cH_A \otimes \cH_B)$. Then
  \[
    \im(\rho_{A B}) \subseteq \im(\rho_{A}) \otimes \im(\rho_{B}) \ .
  \]
\end{lemma}

\begin{proof}
  Assume first that $\rho_{A B}$ is pure, i.e., $\rho_{A B} =
  \proj{\Psi}$.  Let $\ket{\Psi} = \sum_{z \in \cZ} \alpha_z
  \ket{\phi^z} \otimes \ket{\psi^z}$ be a Schmidt decomposition of
  $\ket{\Psi}$, i.e., $\{\ket{\phi^z}\}_{z \in \cZ}$ and
  $\{\ket{\psi^z}\}_{z \in \cZ}$ are families of orthonormal vectors
  in $\cH_A$ and $\cH_B$, respectively. Then
  \[
    \im(\rho_{A B})
  =
    \{ \ket{\Psi} \}
  \subseteq
    \spanv\{\ket{\phi^z}\}_{z \in \cZ} \otimes
      \spanv\{\ket{\psi^z}\}_{z \in \cZ} \ .
  \]
  Because $\spanv\{\ket{\phi^z}\}_{z \in \cZ} = \im(\rho_{A})$ and
  $\spanv\{\ket{\psi^z}\}_{z \in \cZ} = \im(\rho_B)$ the assertion
  follows.
  
  To show that the statement also holds for mixed states, let $\rho_{A
    B} = \sum_{x \in \cX} \rho_{A B}^x$ be a decomposition of $\rho_{A
    B}$ into pure states $\rho_{A B}^x$, for $x \in \cX$. Then,
  because the lemma holds for the states $\rho_{A B}^x$,
  \[
  \begin{split}
    \im(\rho_{A B})
  & =
    \spanv \bigcup_{x \in \cX} \im(\rho_{A B}^x) \\
  & \subseteq
    \spanv \bigcup_{x \in \cX} \im(\rho_{A}^x) \otimes \im(\rho_{B}^x) \\
  & \subseteq
   \Bigl( \spanv \bigcup_{x \in \cX} \im(\rho_{A}^x) \Bigr) 
   \otimes \Bigl( \spanv \bigcup_{x \in \cX} \im(\rho_{B}^x) \Bigr) \\
  & =
    \im(\rho_{A}) \otimes \im(\rho_{B}) \ .
  \end{split}
  \]
\end{proof}

\begin{lemma} \label{lem:tensimage}
  Let $\rho_{A B}, \rhob_{A B} \in \NN(\cH_A \otimes \cH_B)$ such that
  $\im(\rhob_{A B}) \subseteq \im(\rho_{A B})$. Then $\im(\rhob_A)
  \subseteq \im(\rho_{A})$.
\end{lemma}

\begin{proof}
  Assume first that $\rhob_{A B}$ is pure, i.e., $\rhob_{A B} =
  \proj{\Psi}$. Let $\ket{\Psi} = \sum_{z \in \cZ} \alpha_z
  \ket{\phi^z} \otimes \ket{\psi^z}$ be a Schmidt decomposition of
  $\ket{\Psi}$, i.e., $\{\ket{\phi^z}\}_{z \in \cZ}$ and
  $\{\ket{\psi^z}\}_{z \in \cZ}$ are families of orthonormal vectors
  in $\cH_A$ and $\cH_B$, respectively. Then $\im(\rhob_{A B}) =
  \{\ket{\Psi}\}$.  Moreover, by Lemma~\ref{lem:tensprodimage},
  \[
    \im(\rhob_{A B})
  \subseteq
    \im(\rho_{A B})
  \subseteq
    \im(\rho_{A}) \otimes \im(\rho_{B}) \ ,
  \]
  i.e., $\ket{\Psi} \in \im(\rho_A) \otimes \im(\rho_B)$. This implies
  $\ket{\phi^z} \in \im(\rho_A)$, for any $z \in \cZ$, and thus
  $\spanv\{\ket{\phi^z}\}_{z \in \cZ} \subseteq \im(\rho_A)$.  The
  assertion then follows because $\spanv\{\ket{\phi^z}\}_{z \in \cZ} =
  \im(\rhob_{A})$.
  
  To show that the statement holds for mixed states, let $\rhob_{A B}
  = \sum_{x \in \cX} \rhob_{A B}^x$ be a decomposition of $\rhob_{A
    B}$ into pure states $\rhob_{A B}^x$, for $x \in \cX$. We then
  have $\im(\rhob_{A B}^x) \subseteq \im(\rho_{A B})$, for any $x \in
  \cX$, and thus, because the lemma holds for pure states,
  $\im(\rhob_A^x) \subseteq \im(\rho_A)$.  Consequently,
  \[
    \im(\rhob_{A})
  =
    \spanv \bigcup_{x \in \cX} \im(\rhob_{A}^x)
  \subseteq
   \im(\rho_{A}) \ .
  \]
\end{proof}

\section{Nonnegative operators}

\index{nonnegative~operator|(}

\begin{lemma} \label{lem:nnprod}
  Let $\rho \in \NN(\cH)$ and let $S$ be a hermitian operator on
  $\cH$. Then $S \rho S$ is nonnegative.
\end{lemma}

\begin{proof}
  Let $\rho = \sum_{x \in \cX} p_x \proj{x}$ be a spectral
  decomposition of $\rho$. Then, for any vector $\ket{\theta} \in
  \cH$,
  \[
    \bra{\theta} S \rho S \ket{\theta} 
  = 
    \sum_{x \in \cX} p_x \bra{\theta} S \proj{x} S \ket{\theta} 
  = 
    \sum_{x \in \cX} p_x |\bra{\theta} S \ket{x}|^2 \geq 0 \ .
  \]
  The assertion then follows because $S \rho S$ is hermitian.
\end{proof}

\begin{lemma} \label{lem:trprod}
  Let $\rho, \sigma \in \NN(\cH)$. Then $\tr(\rho \sigma) \geq 0$.
\end{lemma}

\begin{proof}
  The assertion is an immediate consequence of the fact that $\tr(\rho
  \sigma) = \tr(\sigma^{1/2} \rho \sigma^{1/2})$ and
  Lemma~\ref{lem:nnprod}.
\end{proof}

\begin{lemma} \label{lem:maxcond}
  Let $\rho, \sigma \in \NN(\cH)$ such that $\sigma$ is invertible.
  Then the operator $\lambda \cdot \sigma - \rho$ is nonnegative if
  and only if
  \[
    \lambda_{\max}(\sigma^{-1/2} \rho \sigma^{-1/2}) \leq \lambda \ .
  \]
\end{lemma}

\begin{proof}
  With $D := \lambda \cdot \id - \sigma^{-1/2} \rho \sigma^{-1/2}$, we
  have $\lambda \cdot \sigma - \rho = \sigma^{1/2} D \sigma^{1/2}$.
  Because of Lemma~\ref{lem:nnprod}, this operator is nonnegative if
  and only if $D$ is nonnegative, which is equivalent to say that all
  eigenvalues of $\sigma^{-1/2} \rho \sigma^{-1/2}$ are upper bounded
  by $\lambda$.
\end{proof}

\begin{lemma} \label{lem:prodpos}
  Let $\rho, \sigma \in \NN(\cH)$ such that $\lambda \cdot \sigma -
  \rho$ is nonnegative and $\sigma$ is invertible. Then
  \[
    \lambda_{\max}(\rho^{1/2} \sigma^{-1} \rho^{1/2})
  \leq 
    \lambda \ .
  \]
\end{lemma}

\begin{proof}
  Assume without loss of generality that $\rho$ is invertible
  (otherwise, the statement follows by continuity). Because the
  operator $\lambda \cdot \sigma - \rho$ is nonnegative, the same
  holds for $\rho^{-1/2} (\lambda \cdot \sigma-\rho) \rho^{-1/2} =
  \lambda \cdot \rho^{-1/2} \sigma \rho^{-1/2} - \id$ (cf.\ 
  Lemma~\ref{lem:nnprod}).  Hence, all eigenvalues of $\rho^{-1/2}
  \sigma \rho^{-1/2}$ are at least $\lambda^{-1}$.  Consequently, the
  eigenvalues of the inverse $\rho^{1/2} \sigma^{-1} \rho^{1/2}$
  cannot be larger than~$\lambda$.
\end{proof}

\index{nonnegative~operator|)}

\section{Properties of the function $r_t$} \label{sec:rt}

The class of functions $r_t: z \mapsto z^t - t \ln z -1$, for $t \in
\bbR$, is used in Section~\ref{sec:smoothprod} for the proof of a
Chernoff style bound. In the following, we list some of its
properties.


\begin{lemma} \label{lem:rtincr}
  For any $t \in \bbR$, the function $r_t$ is monotonically increasing
  on the interval $[1, \infty)$.
\end{lemma}

\begin{proof}
  The first derivative of $r_t$ is given by
  \[
    \sfrac{d}{d z} r_t(z) 
  = 
    t z^{t-1} - \sfrac{t}{z} = \sfrac{t}{z} (z^t - 1) \ .
  \]
  The assertion follows because the term on the right hand side is
  nonnegative for any $z \in [1, \infty)$.
\end{proof}

\begin{lemma} \label{lem:rtzz}
  For any $t \in \bbR$ and $z \in (0, \infty)$,
  \[
    r_t(z) \leq r_{|t|}(z+ \sfrac{1}{z}) \ .
  \]
\end{lemma}

\begin{proof}
  Observe first that $r_t(z) = r_{-t}(\frac{1}{z})$.  It thus suffices
  to show that the statement holds for $t \geq 0$. If $z \geq 1$, the
  assertion follows directly from Lemma~\ref{lem:rtincr}. For the case
  where $t \geq 0$ and $z < 1$, let $v := -t \ln z$. Then
  $r_t(\frac{1}{z}) = e^v - v - 1$ and $r_t(z) = e^{-v} + v - 1$.
  Because $v \geq 0$, we have $e^v - e^{-v} \geq 2 v$, which implies
  $r_t(z) \leq r_t(\frac{1}{z})$.  The assertion then follows again
  from Lemma~\ref{lem:rtincr}.
\end{proof}

\begin{lemma} \label{lem:rtconc}
  For any $t \in [-\frac{1}{2}, \frac{1}{2}]$, the function $r_t$ is
  concave on the interval $[4, \infty]$.
\end{lemma}

\begin{proof}
  We show that $\frac{d^2}{dz^2} r_t(z) \leq 0$ for any $z \geq 4$.
  Because $\frac{d^2}{dz^2} r_t(z) = t (t-1) z^{t-2} + \frac{t}{z^2}$,
  this is equivalent to $t(1-t) z^t \geq t$.  It thus suffices to
  verify that
  \[
    z \geq \left(\frac{1}{1-t}\right)^{\frac{1}{t}} \ ,
  \]
  for any $z \geq 4$. Using some simple analysis, it is easy to see
  that the term on the right hand side is monotonically increasing in
  $t$ on the interval $[-\frac{1}{2}, \frac{1}{2}]$ and thus takes its
  maximum at $t=\frac{1}{2}$, in which case it equals $4$.
\end{proof}

\begin{lemma} \label{lem:rtbound}
  For any $z \in [1, \infty)$ and $t \in [-\frac{1}{\log z},
  \frac{1}{\log z }]$,
   \[
     r_t(z) \leq \bigl(1-\ln 2 \bigr) (\log z)^2 t^2 \ .
   \]
\end{lemma}

\begin{proof}
  Let $v := t \ln z$. Then
  \begin{equation} \label{eq:rtsecond}
    \frac{r_t(z)}{t^2}
  =
    \frac{e^{t \ln z} - t \ln z -1}{t^2}
  =
    \frac{e^v - v - 1}{v^2} (\ln z)^2 \ .
  \end{equation}
  We first show that the term on the right hand side
  of~\eqref{eq:rtsecond} is monotonically increasing in $v$, that is,
  \[
    \frac{d}{d v} \frac{e^v - v - 1}{v^2}
  =
    \frac{e^v - 1}{v^2} - 2 \frac{e^v - v -1}{v^3}
  \geq 
    0 \ .
  \]
  A simple calculation shows that this inequality can be rewritten as
  \[
    1 \geq \frac{2}{v}\frac{e^{v/2}-e^{-v/2}}{e^{v/2}+e^{-v/2}} \ ,
  \]
  which holds because, for any $v \in \bbR$, 
  \[
    \Bigl| \frac{e^{v/2}-e^{-v/2}}{e^{v/2}+e^{-v/2}} \Bigr| 
  = 
    \bigl| \tanh \frac{v}{2} \bigr| 
  \leq 
    \frac{|v|}{2} \ . 
  \]
  
  Hence, in order to find an upper bound on~\eqref{eq:rtsecond}, it is
  sufficient to evaluate the right hand side of~\eqref{eq:rtsecond}
  for the maximum value of $v$. By assumption, we have $v \leq \ln 2$,
  i.e.,
  \[
    \frac{e^v - v - 1}{v^2} (\ln z)^2 
  \leq
    (1-\ln 2) (\log z)^2 \ ,
  \]
  which concludes the proof.
\end{proof}

\chapter[Efficient Information Reconciliation]{Computationally Efficient Information Reconciliation} \label{app:effIR}

\index{information~reconciliation|(} \index{computational~efficiency}

In Section~\ref{sec:IR}, we have proposed a general one-way
information reconciliation scheme which is optimal with respect to its
information leakage\index{leakage}.  The scheme, however, requires the
receiver of the error-correcting information to perform some decoding
operation for which no efficient algorithm is known. In the following,
we propose an alternative information reconciliation scheme based on
error-correcting codes where all computations can be done efficiently.

\section{Preliminaries}

To describe and analyze the protocol, we need some terminology and
basic results from the theory of channel coding. Let $\mathfrak{C}$ be
a discrete memoryless channel which takes inputs from a set $\cU$ and
gives outputs from a set $\cV$.\footnote{A \emph{discrete memoryless
    channel}\index{discrete~memoryless~channel} $\mathfrak{C}$ from
  $\cU$ to $\cV$ is defined by the conditional probability
  distributions $P_{V|U=u}$ on $\cV$, for any $u \in \cU$.}  An
\emph{encoding scheme for $\mathfrak{C}$}\index{encoding~scheme} is a
family of pairs $(\cC_n, \dec_n)$ parameterized by $n \in \bbN$ where
$\cC_n$ is a \emph{code on $\cU$ of length $n$}\index{code}, i.e., a
set of $n$-tuples $\bu \in \cU^n$, called
\emph{codewords}\index{codeword}, and $\dec_n$ is a \emph{decoding
  function}\index{decoding~function}, i.e., a mapping from $\cV^n$ to
$\cC_n$.  The \emph{rate}\index{rate~of~a~code} of the code $\cC_n$ is
defined by $\rate(\cC_n) := \frac{1}{n} \log |\cC_n|$.  Moreover, the
\emph{maximum error probability}\index{maximum~error~probability} of
$(\cC_n, \dec_n)$ is defined by
\[
  \eps_{\max}(C_n, \dec_n)
:=
  \max_{\bu \in \cC_n} \Pr_{\bv}[\bu \neq \dec(\bv)] \ ,
\] 
where, for any $\bu = (u_1, \ldots, u_n) \in \cC_n$, the probability
is over all outputs $\bv = (v_1, \ldots, v_n)$ of $n$ parallel
invocations of $\mathfrak{C}$ on input $\bu$.

We will use the following fundamental theorem for channel coding (cf.,
e.g., \cite{CovTho91}, Section~8.7).

\begin{proposition} \label{pr:channelcode}
  Let $\mathfrak{C}$ be a discrete memoryless channel from $\cU$ to
  $\cV$ and let $\delta > 0$. Then there exists an encoding scheme
  $\{(\cC_n, \dec_n)\}_{n \in \bbN}$ for $\mathfrak{C}$ such that the
  following holds:
  \begin{itemize}
  \item $\rate(\cC_n) \geq \max_{P_U} H(U) - H(U|V) - \delta$, for any
    $n \in \bbN$. (The entropies in the maximum are computed for the
    distribution $P_{U V}$ of an input/output pair $(u,v)$ of
    $\mathfrak{C}$, where $u$ is chosen according to $P_U$.)
  \item $\lim_{n \to \infty} \eps_{\max}(\cC_n, \dec_n) = 0$.
  \end{itemize}
\end{proposition}

\section{Information reconciliation based on codes}

Let us now consider an information reconciliation protocol based on
channel coding. For this, we assume that Alice's and Bob's inputs are
strings $\bx = (x_1, \ldots, x_n)$ and $\by = (y_1, \ldots, y_n)$,
respectively. Our protocol shall be secure if the inputs $\bx, \by$
are distributed according to a product distribution $P_{X^n Y^n} =
(P_{X Y})^n$.

Let $\mathfrak{C}$ be the channel which maps any $u \in \cX$ to $v :=
(x \oplus u, y)$, where the pair $(x,y)$ is chosen according to the
probability distribution $P_{X Y}$ and where $\oplus$ is a group
operation on $\cX$. For any $n \in \bbN$, let $\IR_{\cC_n, \dec_n}$ be
the information reconciliation protocol specified by
Fig.~\ref{pr:IRcomp}, where $\cC_n$ is the code and $\dec_n$ the
decoding function defined by Proposition~\ref{pr:channelcode}.

It is easy to see that $\bxh = \bx$ holds whenever $\dec_n$ decodes to
the correct value $\buh = \bu$. Hence, the information reconciliation
protocol $\IR_{\cC_n, \dec_n}$ is $\eps_n$-secure, for $\eps_n :=
\eps_{\max}(\cC_n, \dec_n)$. Because, by
Proposition~\ref{pr:channelcode}, the maximum error probability
$\eps_{\max}(\cC_n, \dec_n)$ of $(\cC_n, \dec_n)$ goes to zero, for
$n$ approaching infinity, the protocol $\IR_{\cC_n, \dec_n}$ is
asymptotically secure.

\begin{protocolfloat}
\figtop
Parameters: \\[1ex]
\begin{tabular}{ll}
  $\cC_n$: & set of codewords from $\cX^n$ \\
  $\dec_n$: & decoding function from $\cX^n \times \cY^n$ to $\cC_n$
  \\
  $\oplus$: & group operation on $\cX$ (with inverse $\ominus$). 
\end{tabular}
\tabprotsep
\begin{protocol}{Alice}{Bob}
  \protno{input: $\bx \in \cX^n$}{input: $\by \in \cY^n$}
  \protright{$\bu \in_R \cC_n$ \\ $\bc := \bx \oplus \bu$}{$\bc$}{$\buh :=
  \dec_n(\bc, \by)$}
  \protno{}{if decoding not succ.\\ $\quad$ then abort}
  \protno{}{output $\bxh := \bc \ominus \buh$}
\end{protocol}
\caption{
  Information reconciliation protocol $\IR_{\cC_n, \dec_n}$.}
\label{pr:IRcomp}
\end{protocolfloat}

Moreover, by Proposition~\ref{pr:channelcode}, 
\[
  \rate(\cC_n) 
\geq
  \max_{P_U}
  H(U) - H(U|X \oplus U, Y) - \delta \ .
\]
Using the fact that the input $u$ is chosen independently of the
randomness of the channel $(x,y)$, a simple information-theoretic
computation shows that the entropy difference in the maximum can be
rewritten as $H({X \oplus U}|Y) - H(X|Y)$.  Hence, because
$\max_{P_{U}} H(X \oplus U|Y) = \Hmax(P_U) = \log |\cX|$, we find
\begin{equation} \label{eq:compleak}
  \sfrac{1}{n} \log |\cC_n|
=
  \rate(\cC_n) 
\geq 
  \log |\cX| - H(X|Y) - \delta \ .
\end{equation}
The communication $\bc$ of the protocol is contained in the set
$\cX^n$. Furthermore, because $\bu$ is chosen uniformly at random from
$\cC_n$, the distribution $P_{\bC|X^n=\bx}$ of the communication
$\bc$, conditioned on any input $\bx \in \cX^n$, is uniform over a set
of size $|\cC_n|$. The leakage of $\IR_{\cC_n, \dec_n}$ is thus given
by\index{leakage}
\[
  \leak_{\IR_{\cC_n, \dec_n}}
= 
  \log |\cX^n| - \min_{\bx} \Hmin(P_{\bC|X^n=\bx})
=
  n \log |\cX| - \log |\cC_n| \ .
\]
Combining this with~\eqref{eq:compleak} we conclude
\[
  \sfrac{1}{n} \leak_{\IR_{\cC_n,\dec_n}} \leq H(X|Y) + \delta \ .
\]

Because Proposition~\ref{pr:channelcode} also holds for
efficient\footnote{An encoding scheme $\{(\cC_n, \dec_n)\}_{n \in
    \bbN}$ is said to be
  \emph{efficient}\index{computational~efficiency} if there exist
  polynomial-time algorithms (in $n$) for sampling a codeword from the
  set $\cC_n$ and for evaluating the decoding function $\dec_n$.}
encoding schemes (see, e.g., \cite{Dumer98}),
Corollary~\ref{cor:errcorr} is asymptotically still true if we
restrict to computationally efficient protocols (see
also~\cite{HolRen05}).  More precisely, this result can be formulated
as follows.

\begin{proposition}
  Let $P_{X Y} \in \NN(\cX \times \cY)$ be a probability distribution
  and let $\delta > 0$.  Then there exists a family of computationally
  efficient information reconciliation protocols $\IR_{\cC_n,\dec_n}$
  (parameterized by $n \in \bbN$) which are $\eps_n$-fully secure,
  $\eps_n$-robust on the product distribution $(P_{X Y})^n$, and have
  leakage $\frac{1}{n} \leak_{\IR_{\cC_n,\dec_n}} \leq H(X|Y) +
  \delta$, for any $n \in \bbN$, where $\lim_{n \to \infty} \eps_n =
  0$.
\end{proposition}

\index{information~reconciliation|)}










\chapter{Notation} 

\newcommand{\tabstart}[1]{\noindent \begin{tabular}{p{2.95cm}p{8.7cm}}
    \multicolumn{2}{l}{{\bf #1}} \\ \hline \\[-2.5ex] } 

\newcommand{\tabstop}{\\ \hline \end{tabular}}

\newcommand{\tabinter}{\vspace{2ex}}

\tabstart{General}
  $\log$ & binary logarithm  \\
  $\ln$ & natural logarithm \\
  $\delta_{x,y}$ & Kronecker symbol: $\delta_{x,y} \in \{0,1\}$, $\delta_{x,y}=1$ iff $x=y$ \\
  $\overline{c}$ & complex conjugate of $c$ \\
  $\Re(c)$ & real value of $c$ \\
  $\NN(\cX)$ & set of nonnegative functions on the set $\cX$ \\  
  $\cS_n$ & set of permutations on the set $\{1, \ldots, n\}$ \\
  $\ExpE_{x}[f(x)]$ & expectation of $f(x)$ over random choices of $x$ \\
  $\supp(f)$ & support of the function $f$ \\
  $[a,b]$ & set of real numbers $r$ such that $a \leq r \leq b$ \\
  $[a,b)$ & set of real numbers $r$ such that $a \leq r < b$
\tabstop

\tabinter

\tabstart{Frequency distributions and types}
$\freq{\bx}$ & frequency distribution of the $n$-tuple $\bx$ \\
$\freqset{\cX}{n}$ & set of types with denominator $n$ on the set $\cX$ \\
$\ntuplefreq{Q}{n}$ & type class of the type $Q$ with denominator $n$
\tabstop

\tabinter

\tabstart{Vectors}  
  $\spanv \cV$ & space spanned by the set of vectors $\cV$ \\
  $\spr{\phi}{\psi}$ & scalar product of the vectors $\ket{\phi}$ and $\ket{\psi}$ \\
  $\|\ket{\phi}\|$ & norm of the vector $\ket{\phi}$ \\
  $\proj{\phi}$ & projector onto the vector $\ket{\phi}$ \\ 
  $\cS_1(\cH)$ & set of normalized vectors on $\cH$
\tabstop

\tabinter

\tabstart{Operators}
$\NN(\cH)$ & set of nonnegative operators on $\cH$  \\
$\id$ & identity \\
$\tr(S)$ & trace of the hermitian operator $S$ \\
$\im(S)$ & support of the hermitian operator $S$ \\
$\rank(S)$ & rank of the hermitian operator $S$ \\
$\lambda_{\max}(S)$ & maximum eigenvalue of the hermitian operator $S$ \\
$\| S \|_1$ & trace norm of the hermitian operator $S$ 

\tabstop

\tabinter

\tabstart{Distance measures for operators}
$\|\rho - \rho'\|_1$ & \distance{} between $\rho$ and $\rho'$ \\
$F(\rho, \rho')$ & fidelity between $\rho$ and
$\rho'$. \\
$d(\rho_{A B}|B)$ & \distance{} from uniform of $\rho_{A B}$ given $B$ \\
$d_2(\rho_{A B}|\sigma_B)$ & $L_2$-distance from uniform of $\rho_{A B}$
relative to $\sigma_B$ 
\tabstop



\tabinter

\tabstart{Entropies}
  $H(P_X)$ & Shannon entropy of the probability distribution $P_X$ \\
  $h(p)$ & binary Shannon entropy with bias $p$ \\
  $H(\rho_A)$ & von Neumann entropy of the density operator $\rho_A$ \\
  $H(A|B)$ &  conditional entropy $H(\rho_{A B}) - H(\rho_B)$ \\
  $D(\rho\|\sigma)$ & relative entropy of $\rho$ to $\sigma$ \\
  $\Hmin(\rho_{A B}|\sigma_B)$ & min-entropy of $\rho_{A B}$ relative to $\sigma_B$\\
  $\Hmax(\rho_{A B}|\sigma_B)$ & max-entropy of $\rho_{A B}$ relative to $\sigma_B$\\
  $\Hmin^\eps(\rho_{A B}|\sigma_B)$ & $\eps$-smooth min-entropy of $\rho_{A B}$ relative to $\sigma_B$\\
  $\Hmax^\eps(\rho_{A B}|\sigma_B)$ & $\eps$-smooth max-entropy of $\rho_{A B}$ relative to $\sigma_B$\\
  $\Hmin^\eps(\rho_{A B}|B)$ & $\eps$-smooth min-entropy of $\rho_{A B}$ given $\cH_B$\\
  $\Hmax^\eps(\rho_{A B}|B)$ & $\eps$-smooth max-entropy of $\rho_{A
  B}$ given $\cH_B$\\
  $\Hmin^\eps(A|B)$ & abbreviation for $\Hmin^\eps(\rho_{A B}|B)$ \\
  $\Hmax^\eps(A|B)$ & abbreviation for $\Hmax^\eps(\rho_{A B}|B)$ \\
  $H_2(\rho_{A B}|\sigma_B)$ & collision entropy of $\rho_{A B}$ relative to $\sigma_B$
\tabstop

\tabinter

\tabstart{Symmetric spaces}
  $\Sym{\cH}{n}$ & Symmetric subspace of
  $\cH^{\otimes n}$ \\
  $\SymR{\cH}{n}{\theta}{m}$ & Symmetric subspace of $\cH^{\otimes
  n}$ along $\ket{\theta}^{\otimes m}$
\tabstop

\bibliography{qkd}

\newcommand{\etalchar}[1]{$^{#1}$}
\begin{thebibliography}{BOHL{\etalchar{+}}05}

\bibitem[ABB{\etalchar{+}}04]{ABBBMMT04}
A.~Ac\'in, J.~Bae, E.~Bagan, M.~Baig, Ll. Masanes, and R.~{Mu\~{n}oz}-Tapia.
\newblock Secrecy content of two-qubit states.
\newblock \url{http://arxiv.org/abs/quant-ph/0411092}, November 2004.

\bibitem[Bar05]{Bariska05}
A.~Bariska.
\newblock On a weakness of common security definitions for {QKD} protocols.
\newblock Master's thesis, ETH Z\"urich, March 2005.
\newblock See also \url{http://arxiv.org/abs/quant-ph/0512021}.

\bibitem[BB84]{BenBra84}
C.~H. Bennett and G.~Brassard.
\newblock Quantum cryptography: Public-key distribution and coin tossing.
\newblock In {\em Proceedings of IEEE International Conference on Computers,
  Systems and Signal Processing}, pages 175--179, 1984.

\bibitem[BBB{\etalchar{+}}02]{BBBGM02}
E.~Biham, M.~Boyer, G.~Brassard, J.~van~de Graaf, and T.~Mor.
\newblock Security of quantum key distribution against all collective attacks.
\newblock {\em Algorithmica}, 34:372--388, 2002.

\bibitem[BBB{\etalchar{+}}05]{BBBMR05}
E.~Biham, M.~Boyer, P.~O. Boykin, T.~Mor, and V.~Roychowdhury.
\newblock A proof of the security of quantum key distribution.
\newblock {\em Journal of Cryptology}, 2005.
\newblock to appear.

\bibitem[BBCM95]{BBCM95}
C.~H. Bennett, G.~Brassard, C.~Cr{\'e}peau, and U.~Maurer.
\newblock Generalized privacy amplification.
\newblock {\em IEEE Transaction on Information Theory}, 41(6):1915--1923, 1995.

\bibitem[BBM92]{BeBrMe92}
C.~H. Bennett, G.~Brassard, and N.~D. Mermin.
\newblock Quantum cryptography without {Bell's} theorem.
\newblock {\em Phys.\ Rev.\ Lett.}, 68:557--559, 1992.

\bibitem[BBP{\etalchar{+}}96]{BBPSSW96}
C.~H. Bennett, G.~Brassard, S.~Popescu, B.~Schumacherand~J. Smolin, and
  W.~Wootters.
\newblock Purification of noisy entanglement and faithful teleportation via
  noisy channels.
\newblock {\em Phys.\ Rev.\ Lett.}, 76:722--726, 1996.

\bibitem[BBR88]{BeBrRo88}
C.~H. Bennett, G.~Brassard, and J.-M. Robert.
\newblock Privacy amplification by public discussion.
\newblock {\em SIAM Journal on Computing}, 17(2):210--229, 1988.

\bibitem[Ben92]{Bennet92}
C.~H. Bennett.
\newblock Quantum cryptography using any two nonorthogonal states.
\newblock {\em Phys.\ Rev.\ Lett.}, 68(21):3121--3124, 1992.

\bibitem[Bir46]{Birkho46}
G.~Birkhoff.
\newblock Three observations on linear algebra.
\newblock {\em Univ.\ Nac.\ Tucum\'an. Rev. Ser. A}, 5:147--151, 1946.

\bibitem[BM97a]{BihMor97b}
E.~Biham and T.~Mor.
\newblock Bounds on information and the security of quantum cryptography,.
\newblock {\em Phys.\ Rev.\ Lett.}, 79:4034--4037, 1997.

\bibitem[BM97b]{BihMor97a}
E.~Biham and T.~Mor.
\newblock Security of quantum cryptography against collective attacks.
\newblock {\em Phys.\ Rev.\ Lett.}, 78(11):2256--2259, 1997.

\bibitem[BMS96]{BeMoSm96}
C.~H. Bennett, T.~Mor, and J.~A. Smolin.
\newblock The parity bit in quantum cryptography.
\newblock {\em Phys.\ Rev.\ A}, 54:2675--2684, 1996.

\bibitem[BOHL{\etalchar{+}}05]{BHLMO05}
M.~Ben-Or, M.~Horodecki, D.~W. Leung, D.~Mayers, and J.~Oppenheim.
\newblock The universal composable security of quantum key distribution.
\newblock In {\em Second Theory of Cryptography Conference {TCC}}, volume 3378
  of {\em Lecture Notes in Computer Science}, pages 386--406. Springer, 2005.
\newblock Also available at \url{http://arxiv.org/abs/quant-ph/0409078}.

\bibitem[BOM04]{BenMay04}
M.~Ben-Or and D.~Mayers.
\newblock General security definition and composability for quantum and
  classical protocols.
\newblock \url{http://arxiv.org/abs/quant-ph/0409062}, September 2004.

\bibitem[BPG99]{BecGis99}
H.~Bechmann-Pasquinucci and N.~Gisin.
\newblock Incoherent and coherent eavesdropping in the six-state protocol of
  quantum cryptography.
\newblock {\em Phys.\ Rev.\ A,}, 59:4238, 1999.

\bibitem[Bru98]{Bruss98}
D.~Bruss.
\newblock Optimal eavesdropping in quantum cryptography with six states.
\newblock {\em Phys.\ Rev.\ Lett.}, 81:3018, 1998.

\bibitem[BS94]{BraSal94}
G.~Brassard and L.~Salvail.
\newblock Secret-key reconciliation by public discussion.
\newblock In {\em Advances of Cryptology --- \mbox{EUROCRYPT '93}}, Lecture
  Notes in Computer Science, pages 410--423. Springer, 1994.

\bibitem[Can01]{Canetti01}
R.~Canetti.
\newblock Universally composable security: A new paradigm for cryptographic
  protocols.
\newblock In {\em Proc.\ 42nd IEEE Symposium on Foundations of Computer Science
  (FOCS)}, pages 136--145, 2001.

\bibitem[CFS02]{CaFuSh02}
C.~M. Caves, C.~A. Fuchs, and R.~Schack.
\newblock Unknown quantum states: The quantum de {Finetti} representation.
\newblock {\em Journal of Mathematical Physics}, page 4537, 2002.

\bibitem[Cha02]{Chau02}
H.~F.\ Chau.
\newblock Practical scheme to share a secret key through a quantum channel with
  a $27.6\%$ bit error rate.
\newblock {\em Phys.\ Rev.\ A}, 66:060302, 2002.

\bibitem[CRE04]{ChReEk04}
M.~Christandl, R.~Renner, and A.~Ekert.
\newblock A generic security proof for quantum key distribution.
\newblock \url{http://arxiv.org/abs/quant-ph/0402131}, February 2004.

\bibitem[CT91]{CovTho91}
T.~M. Cover and J.~A. Thomas.
\newblock {\em Elements of Information Theory}.
\newblock Wiley Series in Telecommunications. Wiley, New York, 1991.

\bibitem[CW79]{CarWeg79}
J.~L. Carter and M.~N. Wegman.
\newblock Universal classes of hash functions.
\newblock {\em Journal of Computer and System Sciences}, 18:143--154, 1979.

\bibitem[DFSS05]{DFSS05}
I.~Damgaard, S.~Fehr, L.~Salvail, and C.~Schaffner.
\newblock Cryptography in the bounded quantum-storage model.
\newblock In {\em 46th Annual Symposium on Foundations of Computer Science
  (FOCS)}, pages 449--458, 2005.

\bibitem[DHL{\etalchar{+}}04]{DHLST04}
D.~P. DiVincenzo, M.~Horodecki, D.~W. Leung, J.~A. Smolin, and B.~M. Terhal.
\newblock Locking classical correlation in quantum states.
\newblock {\em Phys.\ Rev.\ Lett.}, 92:067902, 2004.

\bibitem[DM04]{DziMau04}
S.~Dziembowski and U.~Maurer.
\newblock Optimal randomizer efficiency in the bounded-storage model.
\newblock {\em Journal of Cryptology}, 17(1):5--26, 2004.
\newblock Conference version appeared in Proc. of \mbox{STOC '02}.

\bibitem[Dum98]{Dumer98}
Ilya~I. Dumer.
\newblock Concatenated codes and their multilevel generalizations.
\newblock In Vera~Saine Pless and W.~C. Huffman, editors, {\em The Handbook of
  Coding Theory}, volume~2, chapter~23, pages 1191--1988. North-Holland,
  Elsevier, 1998.

\bibitem[DW05]{DevWin05}
I.~Devetak and A.~Winter.
\newblock Distillation of secret key and entanglement from quantum states.
\newblock {\em Proc.\ R.\ Soc.\ Lond. A}, 461:207--235, 2005.

\bibitem[Eke91]{Ekert91}
A.~K. Ekert.
\newblock Quantum cryptography based on {Bell's} theorem.
\newblock {\em Phys.\ Rev.\ Lett.}, 67:661, 1991.

\bibitem[GL03]{GotLo03}
D.~Gottesman and H.-K. Lo.
\newblock Proof of security of quantum key distribution with two-way classical
  communications.
\newblock {\em IEEE Transactions on Information Theory}, 49(2):457--475, 2003.

\bibitem[GN93]{GemNao93}
P.~Gemmell and N.~Naor.
\newblock Codes for interactive authentication.
\newblock In {\em Advances in Cryptology --- \mbox{CRYPTO '93}}, volume 773 of
  {\em Lecture Notes in Computer Science}, pages 355--367. Springer, 1993.

\bibitem[Gro05]{Grossha05}
F.~Grosshans.
\newblock Collective attacks and unconditional security in continuous variable
  quantum key distribution.
\newblock {\em Phys.\ Rev.\ Lett}, 94:020504, 2005.

\bibitem[HJ85]{HorJoh85}
R.~A. Horn and C.~R. Johnson.
\newblock {\em Matrix analysis}.
\newblock Cambridge University Press, 1985.

\bibitem[HLSW04]{HLSW04}
P.~Hayden, D.~Leung, P.~W. Shor, and A.~Winter.
\newblock Randomizing quantum states: Constructions and applications.
\newblock {\em Communications in Mathematical Physics}, 250(2):371--391, 2004.

\bibitem[HM76]{HudMoo76}
R.~L. Hudson and G.~R. Moody.
\newblock Locally normal symmetric states and an analogue of de {Finetti's}
  theorem.
\newblock {\em Z.\ Wahrschein. verw. Geb.}, 33:343--351, 1976.

\bibitem[HR05]{HolRen05}
T.~Holenstein and R.~Renner.
\newblock One-way secret-key agreement and applications to circuit polarization
  and immunization of public-key encryption.
\newblock In {\em Advances in Cryptology --- \mbox{CRYPTO '05}}, Lecture Notes
  in Computer Science, pages 478--493. Springer, 2005.

\bibitem[ILL89]{ImLeLu89}
R.~Impagliazzo, L.~A. Levin, and M.~Luby.
\newblock Pseudo-random generation from one-way functions (extended abstract).
\newblock In {\em Proceedings of the Twenty-First Annual ACM Symposium on
  Theory of Computing}, pages 12--24, 1989.

\bibitem[ILM01]{InLuMa01}
H.~Inamori, N.~L\"utkenhaus, and D.~Mayers.
\newblock Unconditional security for practial quantum key distribution.
\newblock \url{http://arxiv.org/abs/quant-ph/0107017}, July 2001.

\bibitem[KGR05]{KrGiRe05}
B.~Kraus, N.~Gisin, and R.~Renner.
\newblock Lower and upper bounds on the secret key rate for {QKD} protocols
  using one-way classical communication.
\newblock {\em Phys.\ Rev.\ Lett.}, 95:080501, 2005.

\bibitem[KMR05]{KoMaRe05}
R.~K\"onig, U.~Maurer, and R.~Renner.
\newblock On the power of quantum memory.
\newblock {\em IEEE Transactions on Information Theory}, 51(7):2391--2401,
  2005.
\newblock A preliminary version appeared 2003 on the eprint archive
  \url{http://arxiv.org/abs/quant-ph/0305154}.

\bibitem[KR05]{KoeRen05}
R.~K\"onig and R.~Renner.
\newblock A de {Finetti} representation for finite symmetric quantum states.
\newblock {\em Journal of Mathematical Physics}, 46:122108, 2005.

\bibitem[LC99]{LoCha99}
H.-K. Lo and H.~F. Chau.
\newblock Unconditional security of quantum key distribution over arbitrarily
  long distances.
\newblock {\em Science}, 283:2050--2056, 1999.

\bibitem[LCA05]{LoChAr05}
H.-K. Lo, H.~F. Chau, and M.~Ardehali.
\newblock Efficient quantum key distribution scheme and a proof of its
  unconditional security.
\newblock {\em Journal of Cryptology}, 18(2):133--165, 2005.

\bibitem[Lo00]{Lo00}
H.-K. Lo.
\newblock Proof of unconditional security of six-state quantum key distribution
  scheme.
\newblock {\em Quantum Information and Computation}, 1(2):81, 2000.

\bibitem[Lo05]{Lo05}
H.-K. Lo.
\newblock Getting something out of nothing.
\newblock \url{http://arxiv.org/abs/quant-ph/0503004}, March 2005.

\bibitem[Mau93]{Maurer93}
U.~M. Maurer.
\newblock Secret key agreement by public discussion from common information.
\newblock {\em IEEE Transactions on Information Theory}, 39(3):733--742, 1993.

\bibitem[May96]{Mayers96}
D.~Mayers.
\newblock Quantum key distribution and string oblivious transfer in noisy
  channels.
\newblock In {\em Advances in Cryptology --- \mbox{CRYPTO '96}}, volume 1109 of
  {\em Lecture Notes in Computer Science}, pages 343--357. Springer, 1996.

\bibitem[May01]{Mayers01}
D.~Mayers.
\newblock Unconditional security in quantum cryptography.
\newblock {\em Journal of the ACM}, 48(3):351--406, 2001.

\bibitem[MC93]{deFine93}
P.~Monari and D.~Cocchi, editors.
\newblock {\em Introduction to {Bruno} de {Finetti}'s ``Probabili\'a e
  Induzione''}.
\newblock Cooperativa Libraria Universitaria Editrice, Bologna, 1993.

\bibitem[MRH04]{MaReHo04}
U.~Maurer, R.~Renner, and C.~Holenstein.
\newblock Indifferentiability, impossibility results on reductions, and
  applications to the random oracle methodology.
\newblock In {\em First Theory of Cryptography Conference {TCC}}, volume 2951
  of {\em Lecture Notes in Computer Science}, pages 21--39. Springer, 2004.

\bibitem[NA05]{NavAci05}
M.~Navascu\'es and A.~Ac\'in.
\newblock Security bounds for continuous variables quantum key distribution.
\newblock {\em Phys.\ Rev.\ Lett.}, 94:020505, 2005.

\bibitem[NC00]{NieChu00}
M.~A. Nielsen and I.~L. Chuang.
\newblock {\em Quantum computation and quantum information}.
\newblock Cambridge University Press, 2000.

\bibitem[PW00]{PfiWai00a}
B.~Pfitzmann and M.~Waidner.
\newblock Composition and integrity preservation of secure reactive systems.
\newblock In {\em 7th ACM Conference on Computer and Communications Security},
  pages 245--254. ACM Press, 2000.

\bibitem[RGK05]{ReGiKr05}
R.~Renner, N.~Gisin, and B.~Kraus.
\newblock Information-theoretic security proof for quantum key distribution
  protocols.
\newblock {\em Phys.\ Rev.\ A}, 72:012332, 2005.

\bibitem[RK05]{RenKoe05}
R.~Renner and R.~K\"onig.
\newblock Universally composable privacy amplification against quantum
  adversaries.
\newblock In {\em Second Theory of Cryptography Conference {TCC}}, volume 3378
  of {\em Lecture Notes in Computer Science}, pages 407--425. Springer, 2005.
\newblock Also available at \url{http://arxiv.org/abs/quant-ph/0403133}.

\bibitem[RW03]{RenWol03c}
R.~Renner and S.~Wolf.
\newblock Unconditional authenticity and privacy from an arbitrarily weak
  secret.
\newblock In {\em Advances in Cryptology --- \mbox{CRYPTO '03}}, Lecture Notes
  in Computer Science, pages 78--95. Springer, 2003.

\bibitem[RW04]{RenWol04}
R.~Renner and S.~Wolf.
\newblock The exact price for unconditionally secure asymmetric cryptography.
\newblock In {\em Advances in Cryptology --- \mbox{EUROCRYPT '04}}, Lecture
  Notes in Computer Science, pages 109--125. Springer, 2004.

\bibitem[SARG04]{SARG04}
V.~Scarani, A.~Acin, G.~Ribordy, and N.~Gisin.
\newblock Quantum cryptography protocols robust against photon number splitting
  attacks for weak laser pulse implementations.
\newblock {\em Phys.\ Rev.\ Lett.}, 92:057901, 2004.

\bibitem[Sha49]{Shanno49}
Claude~E. Shannon.
\newblock Communication theory of secrecy systems.
\newblock {\em Bell Systems Technical Journal}, 28:656--715, 1949.

\bibitem[SP00]{ShoPre00}
P.~Shor and J.~Preskill.
\newblock Simple proof of security of the {BB84} quantum key distribution
  protocol.
\newblock {\em Phys.\ Rev.\ Lett.}, 85:441, 2000.

\bibitem[Sti91]{Stinso91}
D.~R. Stinson.
\newblock Universal hashing and authentication codes.
\newblock In {\em Advances in Cryptology --- \mbox{CRYPTO '91}}, volume 576 of
  {\em Lecture Notes in Computer Science}, pages 74--85, 1991.

\bibitem[TKI03]{TaKoIm03}
K.~Tamaki, M.~Koashi, and N.~Imoto.
\newblock Unconditionally secure key distribution based on two nonorthogonal
  states.
\newblock {\em Phys.\ Rev.\ Lett.}, 90:167904, 2003.

\bibitem[Unr04]{Unruh04}
D.~Unruh.
\newblock Simulatable security for quantum protocols.
\newblock \url{http://arxiv.org/abs/quant-ph/0409125}, September 2004.

\bibitem[Ver26]{Vernam26}
G.~S. Vernam.
\newblock Cipher printing telegraph systems for secret wire and radio
  telegraphic communications.
\newblock {\em J. Am. Inst. Elec. Eng.}, 55:109--115, 1926.

\bibitem[WC81]{WegCar81}
M.~N. Wegman and J.~L. Carter.
\newblock New hash functions and their use in authentication and set equality.
\newblock {\em Journal of Computer and System Sciences}, 22:265--279, 1981.

\bibitem[WG00]{WalGoo00}
N.~R. Wallach and R.~Goodman.
\newblock {\em Representations and Invariants of the Classical Groups}.
\newblock Cambridge University Press, 2000.

\bibitem[Wie83]{Wiesner83}
S.~Wiesner.
\newblock Conjugate coding.
\newblock {\em Sigact News}, 15(1):78--88, 1983.

\bibitem[Win99]{Winter99}
A.~Winter.
\newblock Coding theorem and strong converse for quantum channels.
\newblock {\em IEEE Transactions on Information Theory}, 45(7), 1999.

\end{thebibliography}

\bibliographystyle{alpha}

\sloppy

\printindex











\end{document}